\documentclass[b5paper,10pt]{book}

\usepackage{phdthesis}
\usepackage[
paperwidth=17cm
,paperheight=24cm
,inner=2.6cm
,outer=2.3cm
,top=2.3cm
,bottom=2.3cm
]{geometry}

\usepackage{bibentry}
\nobibliography*

\usepackage{graphicx,times}
\usepackage[hyphens]{url}
\usepackage{tabularx,amsmath}
\usepackage{multirow}
\usepackage{booktabs}
\usepackage{amssymb}
\usepackage[square,comma,numbers,sort&compress,sectionbib]{natbib}
\usepackage{wrapfig}
\usepackage{enumerate}
\usepackage{supertabular}
\usepackage{multicol}
\usepackage{setspace}
\usepackage{import}
\usepackage{booktabs} 
\usepackage{url}
\usepackage{times}
\usepackage{microtype}
\usepackage{amssymb}
\usepackage{multirow}
\usepackage{color}
\usepackage[usenames,dvipsnames,table]{xcolor}
\usepackage[normalem]{ulem}
\usepackage[english]{babel}
\usepackage{statex}
\usepackage[noend]{algorithmic}
\usepackage{algorithm}
\usepackage{paralist}
\usepackage{flushend}
\usepackage{enumerate}
\usepackage[inline]{enumitem}
\usepackage{setspace}
\usepackage{textcomp}
\usepackage{listings}
\usepackage{keyval}
\usepackage{rotating}
\usepackage{acronym}
\usepackage[backref=page]{hyperref}
\hypersetup{%
pdfborder = {0 0 0},
pdftitle = {Optimizing Ranking System Online as Bandits},
pdfsubject = {PhD Thesis},
pdfkeywords = {Machine learning, information retrieval, online learning to rank, bandits, implicit feedback},
pdfauthor = {Chang Li}
}
\usepackage{bookmark}

\usepackage[final]{pdfpages}

\usepackage{dsfont,amsthm}
\usepackage[all]{xy}
\usepackage{epstopdf}
\usepackage{numprint}
\usepackage{epstopdf}
\usepackage[capitalize]{cleveref}

\crefname{assumption}{Assumption}{Assumptions}

\creflabelformat{myrq}{#2\textbf{\textup{#1}}#3}
\crefname{myrq}{\textbf{RQ}}{\textbf{RQs}}

\usepackage{CJKutf8}

\usepackage{cancel}

\usepackage{amsmath,amsthm,amssymb}

\usepackage{color}
\definecolor{Blue}{rgb}{0.9,0.3,0.3}
\usepackage{cancel}

\newcommand{\squishlist}{
   \begin{list}{$\bullet$}
    { \setlength{\itemsep}{0pt}      \setlength{\parsep}{3pt}
      \setlength{\topsep}{3pt}       \setlength{\partopsep}{0pt}
      \setlength{\leftmargin}{1.5em} \setlength{\labelwidth}{1em}
      \setlength{\labelsep}{0.5em} } }

\newcommand{\squishlisttwo}{
   \begin{list}{$\bullet$}
    { \setlength{\itemsep}{0pt}    \setlength{\parsep}{0pt}
      \setlength{\topsep}{0pt}     \setlength{\partopsep}{0pt}
      \setlength{\leftmargin}{2em} \setlength{\labelwidth}{1.5em}
      \setlength{\labelsep}{0.5em} } }

\newcommand{\squishend}{
    \end{list}  }

\newtheorem{thm}{Theorem}[chapter]
\newtheorem{theorem}{Theorem}[chapter]
\newtheorem{lemma}{Lemma}[chapter]

\newtheorem{assumption}{Assumption}[chapter]

\newcommand{\real}{\mathbb{R}}
\newcommand{\myvec}[1]{\mathbf{#1}}
\newcommand{\myvecsym}[1]{\boldsymbol{#1}}

\newcommand{\valpha}{\myvecsym{\alpha}}
\newcommand{\vbeta}{\myvecsym{\beta}}

\newcommand{\vomega}{\myvecsym{\omega}}

\newcommand{\vphi}{\myvecsym{\phi}}

\newcommand{\vtheta}{\myvecsym{\theta}}

\newcommand{\vc}{\myvec{c}}

\newcommand{\vr}{\myvec{r}}

\newcommand{\vu}{\myvec{u}}

\newcommand{\vw}{\myvec{w}}

\newcommand{\vx}{\myvec{x}}

\newcommand{\vy}{\myvec{y}}

\newcommand{\vz}{\myvec{z}}

\newcommand{\vA}{\myvec{A}}
\newcommand{\vB}{\myvec{B}}

\newcommand{\vF}{\myvec{F}}
\newcommand{\vG}{\myvec{G}}
\newcommand{\vH}{\myvec{H}}
\newcommand{\vI}{\myvec{I}}

\newcommand{\vM}{\myvec{M}}

\newcommand{\vN}{\myvec{N}}
\newcommand{\vO}{\myvec{O}}

\newcommand{\vR}{\myvec{R}}

\newcommand{\vT}{\myvec{T}}
\newcommand{\vU}{\myvec{U}}

\newcommand{\vX}{\myvec{X}}

\newcommand{\expect}[1]{\mathbb{E}\left[ {#1} \right]}

\newcommand{\be}{\begin{equation}}
\newcommand{\ee}{\end{equation}}
\newcommand{\bea}{\begin{eqnarray}}
\newcommand{\eea}{\end{eqnarray}}
\newcommand{\beaa}{\begin{eqnarray*}}
\newcommand{\eeaa}{\end{eqnarray*}}

\DeclareMathAlphabet{\mathpzc}{OT1}{pzc}{m}{n}

\newcommand{\qqquad}{\qquad\qquad}

\renewcommand*{\backref}[1]{} 
\renewcommand*{\backrefalt}[4]{%
\ifcase #1
\or (Cited on page~#2.)  %
\else 
(Cited on pages~#2.)  
\fi
}

\let\svthefootnote\thefootnote
\newcommand\blankfootnote[1]{%
  \let\thefootnote\relax\footnotetext{#1}%
  \let\thefootnote\svthefootnote%
}
\let\svfootnote\footnote
\renewcommand\footnote[2][?]{%
  \if\relax#1\relax%
    \blankfootnote{#2}%
  \else%
    \if?#1\svfootnote{#2}\else\svfootnote[#1]{#2}\fi%
  \fi
}

\newcommand{\myparagraph}[1]{\medskip\noindent\textbf{#1.}}

\newcommand{\Comment}[1]{\COMMENT{#1}}

\newcommand{\Input}{\REQUIRE}
\newcommand{\Output}{\ENSURE}

\newcounter{myrqcounter}

\newcommand{\norm}[1]{\left\lVert#1\right\rVert_2}

\DeclareMathOperator*{\argmin}{arg\,min}
\DeclareMathOperator*{\argmax}{arg\,max}
\newcommand{\cDsi}{\mathcal{D}^*_e}
\newcommand{\cDci}{\bar{\mathcal{D}}_e}
\newcommand{\cDst}{\mathcal{D}^*_t}
\newcommand{\cDct}{\bar{\mathcal{D}}_t}
\newcommand{\cD}{\mathcal{D}}
\newcommand{\cE}{\mathcal{E}}
\newcommand{\ccE}{\overline{\cE}}
\newcommand{\cF}{\mathcal{F}}
\newcommand{\cP}{\mathcal{P}}
\newcommand{\cR}{\mathcal{R}}
\newcommand{\cV}{\mathcal{V}}

\mathchardef\mhyphen="2D
\newcommand{\barvalphat}{\bar{\valpha}_t}
\newcommand{\cT}{\mathcal{T}}
\newcommand{\set}[1]{\left\{#1\right\}}
\newcommand{\abs}[1]{\left|#1\right|}

\newcommand{\rnd}[1]{\bm{#1}}
\newcommand{\floors}[1]{\left\lfloor#1\right\rfloor}
\newcommand{\Et}[1]{\mathbb{E}_t \left[#1\right]}
\newcommand{\Etp}[1]{\mathbb{E}_{t - 1} \left[#1\right]}
\newcommand{\condE}[2]{\mathbb{E} \left[#1 \,\middle|\, #2\right]}
\newcommand{\condEsub}[3]{\mathbb{E}_{#3} \! \left[#1 \,\middle|\, #2\right]}

\newcommand{\ucb}{{\sf UCB1}}

\newcommand{\lsbgreedy}{\sf LSBGreedy}

\newcommand{\recurrank}{\sf RecurRank}
\newcommand{\ccucb}{\sf C^2UCB}

\newcommand{\mergedts}{\sf MergeDTS}

\newcommand{\cascadehybrid}{\sf CascadeHybrid}
\newcommand{\cascadelinucb}{\sf CascadeLinUCB}
\newcommand{\cascadelsb}{\sf CascadeLSB}
\newcommand{\cascadelinucbfull}{\sf CascadeLinUCBFull}
\newcommand{\cascadelsbfull}{\sf CascadeLSBFull}

\newcommand{\cascadeducb}{{\sf CascadeDUCB}}
\newcommand{\cascadeswucb}{{\sf CascadeSWUCB}}
\newcommand{\batchrank}{{\sf BatchRank}}
\newcommand{\bubblerank}{{\sf BubbleRank}}
\newcommand{\cascadeklucb}{{\sf CascadeKL\mhyphen UCB}}
\newcommand{\cascadeucb}{{\sf CascadeUCB1}}
\newcommand{\rankedexpthree}{\sf RankedEXP3}
\newcommand{\ducb}{{\sf DUCB}}
\newcommand{\swucb}{{\sf SWUCB}}
\newcommand{\expthree}{{\sf EXP3}}

\newcommand{\toprank}{{\tt TopRank}}
\newcommand{\baseline}{{\tt Baseline}}

\acrodef{IR}{Information Retrieval}
\acrodef{OLTR}{Online Learning to Rank}
\acrodef{LTR}{Learning to Rank}

\acrodef{FL}{federated learning}
\acrodef{GDPR}{General Data Protection Regulation}

\acrodef{CM}{Cascade Model}
\acrodef{DCM}{Dependent Click Model}

\acrodef{MAB}{Multi-Armed Bandits}
\acrodef{UCB}{Upper Confidence Bound}
\acrodef{LCB}{Lower Confidence Bound}
\acrodef{TS}{Thompson Sampling}
\acrodef{MDB}{Multi-Dueling Bandits}

\acrodef{CB}{Cascading Bandits}

\acrodef{PI}{Probabilistic Interleaving}
\acrodef{SOSM}{Sample-Only Scored Multileave}

\acrodef{MergeDTS}{Merge Double Thompson Sampling}
\acrodef{MergeRUCB}{Merge Relative Upper Confidence Bound}
\acrodef{DTS}{Double Thompson Sampling}
\acrodef{SBM}{Singleton Bandits Machines}
\acrodef{RUCB}{Relative Upper Confidence Bound}
\acrodef{RMED1}{Relative Minimum Empirical Divergence}
\acrodef{DMED}{Deterministic Minimum Empirical Divergence}
\acrodef{REX3}{Relative Exponential-weight algorithm for Exploration and Exploitation}
\acrodef{EXP3}{Exponential-weight algorithm for Exploration and Exploitation}

\acrodef{CHB}{Cascade Hybrid Bandits}

\graphicspath{{figures/recsys20/}{figures/tois/}{figures/ijcai19/}{figures/uai19/}}

\begin{document}
\frontmatter

{\pagestyle{empty}
\newcommand{\printtitle}{%
{\Huge\bf Optimizing Ranking Systems Online as Bandits  \\[0.8cm]
}}

\begin{titlepage}
\par\vskip 2cm
\begin{center}
\printtitle
\vfill
{\LARGE\bf Chang Li}
\vskip 2cm
\end{center}
\end{titlepage}

\mbox{}\newpage
\setcounter{page}{1}

\clearpage
\par\vskip 2cm
\begin{center}
\printtitle
\par\vspace {4cm}
{\large \sc Academisch Proefschrift}
\par\vspace {1cm}
{\large ter verkrijging van de graad van doctor\\
aan de Universiteit van Amsterdam\\
op gezag van de Rector Magnificus\\
prof.\ dr.\ ir.\ K.I.J. Maex\\
ten overstaan van een door het College voor Promoties ingestelde \\
commissie, in het openbaar te verdedigen in de Agnietenkapel\\
op donderdag 4 maart 2021, te 13.00 uur \\ } 
\par\vspace {1cm} {\large door}
\par \vspace {1cm}
{\Large Chang Li}
\par\vspace {1cm}
{\large geboren te Hebei} 
\end{center}

\clearpage
\noindent%
\textbf{Promotiecommissie} \\\\
\begin{tabular}{@{}l l l}
Promotor: 
& prof.\ dr.\ M.\ de Rijke & Universiteit van Amsterdam \\  
\\
Copromotor: 
&dr. M. Zoghi & Google \\
\\
Overige leden: 
& {prof. dr. T. Gevers} & {Universiteit van Amsterdam} \\
& {prof. dr. ir. D. Hiemstra} & {Radboud Universiteit} \\
& {dr. H.C. van Hoof} & {Universiteit van Amsterdam} \\
& {prof. dr. E. Kanoulas} & {Universiteit van Amsterdam} \\
& {dr. M. Lalmas} & {Spotify} \\
\end{tabular}

\bigskip\noindent%
Faculteit der Natuurwetenschappen, Wiskunde en Informatica\\

\vfill

\noindent
The research was supported by the Netherlands Organisation for Scientific Research~(NWO) under project number
612.\-001.\-551.\\
\bigskip

\noindent
Copyright \copyright~2021 Chang Li, Amsterdam, The Netherlands\\
Cover by Feng Li\\
Printed by Off Page, Amsterdam\\
\\
ISBN:  978-94-93197-46-6\\

\clearpage
}

{\pagestyle{empty}

{
\begin{center}

\noindent
\textbf{Acknowledgements} \\ \vskip .5cm
\end{center}
}

\noindent%
With the aim to experience a different type of life and continue working on machine learning, I submitted my application for a PhD position at the University of Amsterdam. 
After four years and 9 months, I notice that it has been one of the best decisions I have made ever, not only because I received the offer from my supervisor Maarten de Rijke, but also because doing a Ph.D. is way more interesting, meaningful and, of course, challenging than what I expected. 

I used to think that doing a Ph.D. is nothing but producing some papers :P.
After I joined the group, I realized that there are more things in Ph.D. life than papers.
Thank you to all ILPSers for teaching me this. 
It has been a really interesting journey with all of you around. 
Special thanks to Maurits Bleeker for helping me translate the summary into a \emph{samenvatting}. 

Maarten has been always trying to tell me that a Ph.D. thesis is more meaningful that a stack of papers, in several ways. 
Unfortunately, I did not fully get his points and understand how important it is until the beginning of my third year. 
One day, I was asked by one of my co-authors, Branislav Kventon, what will your thesis topic be? 
All of a sudden, Maarten's words appeared, and I noticed that I should be thinking in a more structured way.
Thank you Maarten, for the guidance and help during my study! 

A frustrating part of my Ph.D. has been that my papers got rejected eleven times in a row. 
Thank you to all my co-authors for helping me get through this. 
However, in my mind, a bigger challenge beyond this is how to make my research meaningful. 
Fortunately, I had Maarten as my supervisor, who supported me and helped to connect me to other researchers in the community. 
This is how I got to know Masrour Zoghi (who became my co-promoter) and Brano, two awesome collaborators. 
Thank you both for helping me with my research. 
Last but not least, I want to thank Ilya Markov for helping me think scientifically in the early stages of my research. 

I was fortunate enough to enjoy two internships. 
Thank you, Haoyun and Xinran at Bloomberg. 
That summer in New York was fantastic! 
Thank you, Hua at Apple.
The internship in Cupertino was also amazingly special.
I also want to thank Jeremy and Derek from the HR team at Apple for helping me relocate to the US during the COVID-19 pandemic.  

Feng, I owe you a big thank you. 
Because of you, I chose Europe.
Without you, it would have been hard for me to reach this point. 
I really feel lucky and full of joy to have you as my girlfriend. 
And, of course, thank you for creating the cover of my thesis. 

\begin{CJK*}{UTF8}{gbsn}
最后，感谢来自父母的支持。读博这几年跟你们在一起的时间也就一个多月，感谢你们的理解，让我可以安心的在国外。
\end{CJK*}

Although a bit sad, it is the time to conclude my student career. 
Thank you to all my classmates and teachers. 

\hspace{2pt}

\noindent 
Chang Li \\ Amsterdam\\ December 2020

 \vskip .3cm

\clearpage
}
\tableofcontents

\mainmatter

\chapter{Introduction}
\label{chapter:introduction}

Ranking is widely utilized in daily interactive systems, where a service provider responds to a user's request by using a sorted list of items. 
A typical scenario is web search: a user issues a query to a search engine, and then the search engine returns a list of items. 
In this thesis, we focus on optimizing these interactive ranking systems.  
We simplify the ranking system as a ranking function, also called ranker, which, given a context, (e.g., query), scores candidate items, and then sorts them in decreasing order according to the scores. 
The optimization objective is to find or learn an optimal ranker that displays the most relevant candidate items at the top. 
The problem is challenging and one of the fundamental challenges is the notion of \emph{relevance}~\citep{saracevic75relevance}. 

About a decade ago, this problem was widely studied as an offline supervised learning problem~\citep{liu09learning}. 
Service providers, such as Microsoft and Yahoo, would generate big collections of queries and documents, and ask human workers to judge the relevance of each query-document pair~\citep{qin10letor,chapelle10yahoo,qin13introducing}. 
The collected datasets would then be used to optimize rankers, and the optimal ranker was the one that had the highest offline ranking metric, e.g., NDCG~\citep{jaervelin02cumulated}.  
This approach is still largely the industrial standard~\citep{lucchese19learning}. 
However, one disadvantage is that annotated relevance is not consistent with user preferences~\citep{joachims02optimizing,yue09interactively,hofmann11balancing}. 
Thus, even an optimized ranker can be suboptimal in production~\citep{zoghi16click}.  

As an alternative, optimizing rankers with interactive feedback has drawn great attention in the past decade~\citep{hofmann13balancing,zoghi17dueling}. 
In these approaches,  the interactive user feedback (e.g., clicks) is used as a label, and the optimal ranker is the one that attracts the highest number of clicks.
Compared to human annotated feedback, there are three advantages in using interactive feedback~\cite{joachims02optimizing,hofmann11probabilistic}: 
\begin{enumerate*}[label=(\arabic*)]
\item Interactive feedback is abundant in interactive systems and easily obtained.
\item Interactive feedback is better aligned with the user's information needs.
\item Interactive feedback is more timely and can capture shifts in user preferences. 
\end{enumerate*}
However, interactive feedback is noisy and highly affected by the presentation order in the system, i.e., position bias. 
To use this type of relatively low-quality feedback for ranker optimization, new learning algorithms need to be designed. 

In this thesis, we focus on learning from this feedback. 
Even though the quality of interactive feedback is low, its advantages have the potential to assist in designing new and powerful ranking systems. 
For instance, non-stationarity is known to exist in user preference~\citep{jagerman19when}. 
In traditional supervised learning, the annotated data may not be enough to capture shifts in user preferences. 
But, with interactive feedback, which is more timely than human annotated data, we can design a ranker that changes its ranking policy according to  the change in user preferences, capturing the non-stationarity. 
In this thesis, we study the problems of online ranker optimization, and translate our ideas into new algorithms that work for novel \ac{OLTR} problems.  
More specifically, we formulate the problem of learning from noisy and biased feedback in ranking systems as different bandit problems~\citep{thompson33likelihood,lattimore20bandit}, and propose new bandit algorithms to solve the proposed problems. 
Thus, the title of this thesis is \emph{Optimizing Ranking Systems Online as Bandits}.


\section{Research Outline and Questions}
\label{introduction:rqs}

Through out the thesis, we want to answer the following question: how should we use implicit feedback to optimize ranking systems online? 
As implicit feedback is typically noisy, we face the challenge of \emph{exploration versus exploitation}. 
Bandit algorithms are suitable for addressing this problem. 
In this thesis, we consider four different research directions in ranking system optimization and formulate each of them as a research question.  
The first one is about ranker evaluation and the other three are related to \ac{OLTR}. 
\begin{description}
\item[\textbf{RQ1}]  How to conduct effective large-scale online ranker evaluation? 
\end{description}
Online ranker evaluation is one of the key challenges in information retrieval. 
While the user preferences for ranker over another can be inferred by using interleaving methods~\citep{zoghi15mergerucb}, the problem of how to effectively choose the ranker pair that generates the interleaved list without degrading the user experience too much is still challenging. 
On the one hand, if two rankers have not been compared enough, the inferred preference can be noisy and inaccurate. 
On the other hand, if two rankers are compared too many times, the interleaving process inevitably hurts user experience.  
This dilemma is known as the \emph{exploration versus exploitation} dilemma. 
It is captured by the $K$-armed dueling bandit problem~\citep{zoghi17dueling}, which is a variant of the $K$-armed bandit problem,  where feedback comes in the form of pairwise preferences. 
Today's deployed search systems evaluate  large numbers of rankers concurrently, and scaling effectively in the presence of numerous rankers is a critical aspect of nline ranker evaluation. 

We provide an answer to \textbf{RQ1} in \cref{chapter:tois} by introducing the MergeDTS algorithm. 
MergeDTS uses a divide-and-conquer strategy to localize the comparisons carried out by the algorithm to small batches of rankers, and then employs \ac{TS} to reduce the comparisons between suboptimal rankers inside these small batches.  
We conduct extensive experiments on three large-scale datasets to assess the performance of MergeDTS. 
The experimental results demonstrate that MergeDTS outperforms the state-of-the-art baselines. 
\begin{description}
	\item[\textbf{RQ2}] How to achieve safe online learning to re-rank? 
\end{description}
Learning to rank has been studied in online and offline settings~\citep{liu09learning,grotov16online}. In the offline setting, rankers are typically learned from relevance labels created by judges. 
This approach has become standard in industrial applications of ranking, such as web search and recommender systems~\citep{liu17cascade}. 
However, this approach lacks exploration and thus is limited by the information contained in the offline training data. 
In the online setting, an algorithm can experiment with ranked lists and learn from feedback on them in a sequential fashion. 
Bandit algorithms are well-suited for this setting but they tend to learn user preferences from scratch (i.e., the ``cold-start'' problem), which results in a high initial cost of exploration. 
This poses an additional challenge of \emph{safe} exploration in ranked lists. 

\textbf{RQ2} is set up to address the safe exploration problem in \ac{OLTR}. 
We address this question by introducing the $\bubblerank$ algorithm in \cref{chapter:uai19}. 
$\bubblerank$ can be viewed as a combination of offline and online \ac{LTR} algorithms. 
It uses the offline trained ranker, e.g., the production ranker, to obtain the initial ranked list, and then conducts safe online pairwise exploration to improve this list. 
The safety comes from the fact that $\bubblerank$ explores the ranked lists by randomly exchanging items with their neighbors. 
Thus, during exploration, an item never moves too far from its original position. 
We analyze the performance as well as the safety of $\bubblerank$ theoretically, and then conduct experiments on a large-scale click dataset to {evaluate} $\bubblerank$ empirically.
\begin{description}
	\item[\textbf{RQ3}] How to conduct online learning to rank when users change their preferences constantly? 
\end{description}
Non-stationarity appears in many online applications such as web search and recommender systems~\citep{jagerman19when}. 
User preferences are effected by different factors and may change constantly. 
An example is that user preferences are effected by abrupt incidents. 
Let us take the query ``corona'' as an example. 
Before January of 2020, the query was mainly related to a beer brand. 
But after the onset of the COVID-19 pandemic, the meaning of this word has shifted to a deadly virus. 
If the learning algorithm is not able to capture this change in user preferences, it would continue responding with beer-related items, which would hurt the user experience.

We provide one solution to \textbf{RQ3} in \cref{chapter:ijcai19}. 
Particularly, we consider \emph{abruptly changing environments} where user preferences remain constant in certain time periods, named \emph{epochs}, but change occurs abruptly at unknown moments called \emph{breakpoints}. 
We introduce {cascading non-stationary bandits}, an online variant of the \ac{CM}~\citep{craswell08experimental} with the goal of identifying the $K$ most attractive items in a non-stationary environment, and propose two algorithms: $\cascadeducb$ and $\cascadeswucb$ to solve this bandit problem. 
The performance of the proposed algorithms are analyzed, and two gap-dependent upper bounds on their $n$-step regret are derived, respectively.  
We also establish a lower bound on the regret of the non-stationary \ac{OLTR} problem, and show that both algorithms match the lower bound up to a logarithmic factor.
At the end of the chapter, we evaluate the performance of the proposed algorithms on a real-world web search click dataset.
\begin{description}
	\item[\textbf{RQ4}] How to learn a ranker online considering both relevance and diversity? 
\end{description}
Relevance ranking and result diversification are two important aspects in modern recommender systems.  
Relevance ranking aims at putting relevant items at the top of result lists, while result diversification focuses on generating ranked lists covering a broad range of topics. 
The former is achieved by displaying a list whose items are sorted in decreasing order of relevance.
However, this sorting process often makes the result list less diverse since items with the highest relevance generally come from the same topic. 
On the other hand, submodular functions are used for result diversification~\citep{yue11linear,hiranandani19cascading}, but this alone is not enough to guarantee  items at the top of list to be relevant. 

In \cref{chapter:recsys20}, we provide a solution to \textbf{RQ4} by formulating the \ac{OLTR} for relevance and diversity as the \acf{CHB}. 
The $\cascadehybrid$ algorithm is then proposed to solve \ac{CHB}. 
We first provide a gap free bound on the $n$-step regret of $\cascadehybrid$, and then evaluate it on two real-world recommendation tasks: movie recommendation and music recommendation.


\section{Main Contributions}
\label{section:introduction:contributions}

We list the main contributions of the thesis in this section. 
The contributions of this thesis come in three groups: algorithmic contributions, theoretical contributions and empirical contributions.

\paragraph{Algorithmic contributions}
We proposed five novel bandit algorithms to solve four online optimization problems in ranking systems:
\begin{itemize}
	\item A \ac{TS}-based dueling bandit algorithm, MergeDTS, which solves the  large-scale online ranker evaluation problem,  \cref{chapter:tois}. 
	\item A safe online learning to re-rank algorithm, $\bubblerank$, which is inspired by bubble sort, \cref{chapter:uai19}.  
	To the best of our knowledge, $\bubblerank$ is the first safe online learning to re-rank algorithm. 
	\item The $\cascadeducb$ and $\cascadeswucb$ algorithms for solving the non-station\-ary \ac{OLTR} problem,  \cref{chapter:ijcai19}.
	To the best of our knowledge, the proposed $\cascadeducb$ and $\cascadeswucb$ are the first bandit algorithms that solve the non-stationary \ac{OLTR} problem. 
	\item A hybrid algorithm, $\cascadehybrid$, for solving the \ac{OLTR} problem,  where both relevance ranking and result diversification are critical to users,  \cref{chapter:recsys20}. 
	To the best of our knowledge, $\cascadehybrid$  is the first bandit algorithm that considers both relevance and diversity in the ranking problem. 
\end{itemize}

\paragraph{Theoretical contributions}
We theoretically analyze the performance of all proposed algorithms. 
Our theoretical analyses do not aim to show superior performance of the algorithms, but to provide worst-case guarantees on their performance when  deployed online. We provide:
\begin{itemize}
	\item A theoretical guarantee on the performance of MergeDTS,  \cref{chapter:tois}. 
	\item A theoretical analysis of $\bubblerank$ and the safety  of  $\bubblerank$,  \cref{chapter:uai19}.
	\item The first theoretical analysis for the non-stationary \ac{OLTR} problem, and an analysis of  the proposed $\cascadeducb$ and $\cascadeswucb$ algorithms,  \cref{chapter:ijcai19}. 
	\item A theoretical analysis of $\cascadehybrid$ for solving the \ac{OLTR} for both relevance and diversity, \cref{chapter:recsys20}. 
\end{itemize}

\paragraph{Evaluation contributions}
To complement our theoretical findings, we conduct extensive experiments to evaluate the proposed algorithms for the considered tasks:
\begin{itemize}
	\item An empirical evaluation of MergeDTS, and the comparison against the state of the art in large-scale online ranker evaluation, \cref{chapter:tois}. 
	\item An empirical comparison of $\bubblerank$ against baselines and a sanity check on the proven theoretical results, \cref{chapter:uai19}.
	\item An empirical evaluation of $\cascadeducb$ and $\cascadeswucb$ for the non-stationary \ac{OLTR} task, \cref{chapter:ijcai19}. 
	\item An empirical comparison of $\cascadehybrid$ against baselines in \ac{OLTR} for relevance and diversity, \cref{chapter:recsys20}.
\end{itemize}


\section{Thesis Overview}
\label{section:introduction:overview}

In this section, we provide a brief overview of the remaining chapters. 
In \cref{chapter:tois,chapter:uai19,chapter:ijcai19,chapter:recsys20}, we study four online ranking optimization tasks, formulate each of them as a bandit problem, and provide  corresponding algorithms for solving each problem.
The proposed algorithms are theoretically analyzed and empirically compared against the state of the art.

More precisely, we organize them in the following order: 

\begin{itemize}
	\item \cref{chapter:tois} treats  large-scale online ranker evaluation as a dueling bandits problem, and proposes the MergeDTS algorithm to solve it with theoretical guarantees on the performance. 
	Experimental evaluation reveals that MergeDTS outperforms the state of the art baselines, e.g., MergeRUCB~\citep{zoghi15mergerucb}, DTS~\citep{wu16double} and Self-Sparring~\citep{sui17multi}. 
	
	\item \cref{chapter:uai19} studies the safe  \ac{OLTR} problem, and presents the $\bubblerank$ algorithm.  We analyze the performance of $\bubblerank$, and  theoretically show that it is a safe algorithm. 
	Then, we empirically compare $\bubblerank$ with $\cascadeklucb$~\citep{kveton15cascading}, $\batchrank$~\citep{zoghi17online} and $\toprank$~\citep{lattimore18toprank}, and show that the performance of 
	$\bubblerank$ is comparable to those algorithms while only $\bubblerank$ satisfies the theoretical safety constraint. 

	\item \cref{chapter:ijcai19} studies the non-stationary \ac{OLTR} problem, where the user may change their preference over time. 
	We formulate this problem as  cascade non-stationary bandits and propose $\cascadeducb$ and $\cascadeswucb$ to solve the problem. 
	Both algorithms have theoretical guarantees on their performance, and the experimental evaluation validates the theoretical findings. 
	\item \cref{chapter:recsys20} presents the $\cascadehybrid$ algorithm, which is designed to solve the \ac{OLTR} problem for relevance and diversity. 
	The proposed $\cascadehybrid$ algorithm is theoretically sound and empirically outperforms $\cascadelinucb$~\citep{zong16cascading} and $\cascadelsb$~\citep{hiranandani19cascading}. 
\end{itemize}
 
\noindent%
In \cref{chapter:conclusions}, we summarize all main findings of this thesis and provide several directions for future work. 

We have tried to keep the research chapters as close as possible to their published version (see below for the publication details).
This implies that some research chapters have some overlap, especially concerning the related work and background material that they present.
But we believe that this disadvantage is outweighed by the advantage that the research chapters are self-contained and by the fact that sticking close to the published versions of the research chapters prevents us from creating alternative versions of published work.

\cref{chapter:tois} discusses online ranker evaluation. 
\cref{chapter:uai19,chapter:ijcai19,chapter:recsys20} consider the online learning to rank problem. 
It is recomended for the chapters to be read in order: \cref{chapter:tois}, \ref{chapter:uai19}, \ref{chapter:ijcai19} and \ref{chapter:recsys20}.


\section{Origins}
\label{section:introduction:origins}

The material in this thesis comes from the following publications: 

\begin{itemize}
\item \cref{chapter:tois} is based on \bibentry{li20mergedts}~\citep{li20mergedts}.
	
	CL designed the algorithm, worked on the theory, and conducted the experiments. MZ helped with algorithm design and theory. All authors contributed to the writing. 
	
\item \cref{chapter:uai19} is based on \bibentry{li19bubblerank}~\citep{li19bubblerank}.
	
	CL designed the algorithm, and conducted the experiments. BK worked on the theory. BK and MZ helped with the algorithm design. CL, TL, CS and MZ helped with the theory. All authors contributed to the writing. 

\item \cref{chapter:ijcai19} is based on  \bibentry{li19cascading}~\citep{li19cascading}.
	
	CL designed the algorithm, worked on the theory, and conducted the experiments. CL and MdR contributed to the writing. 
	
\item \cref{chapter:recsys20} is based on \bibentry{li20cascading}~\citep{li20cascading}.
	
	CL designed the algorithm, worked on the theory and conducted the experiments. CL,  HF and MdR contributed to the writing. 
\end{itemize}
	
\noindent%
There are other publications that indirectly contributed to this thesis:

\begin{itemize}
\item \bibentry{li21federated}~\citep{li21federated}.

\item \bibentry{li18incremental}~\citep{li18incremental}. 

\item \bibentry{jiang19probabilistic}~\citep{jiang19probabilistic}.

\item \bibentry{li18online}~\citep{li18online}.
\end{itemize}


\chapter{Effective Large-Scale Online Ranker Evaluation}
\label{chapter:tois}
\footnote[]{This chapter was published as~\citep{li20mergedts}.}

\acresetall

This chapter is set up to address \textbf{RQ1}: 
\begin{description}
	\item[\textbf{RQ1}]  How to conduct effective large-scale online ranker evaluation? 
\end{description}
Particularly, we focus on solving the large-scale online ranker evaluation problem under the so-called Condorcet assumption, where there exists an optimal ranker that is preferred to all other rankers. 
We propose the MergeDTS algorithm, which first utilizes a divide-and-conquer strategy that localizes the comparisons carried out by the algorithm to small batches of rankers, and then employs \ac{TS} to reduce the comparisons between suboptimal rankers inside these small batches.  
The effectiveness (regret) and efficiency (time complexity) of MergeDTS are extensively evaluated using examples from the domain of online evaluation for web search. 
Our main finding is that for large-scale Condorcet ranker evaluation problems, MergeDTS outperforms the state-of-the-art dueling bandit algorithms.

\section{Introduction}
\label{tois:sec:introduction}

Online ranker evaluation concerns the task of determining the ranker with the best performance out of a finite set of rankers.	
It is an important challenge for information retrieval systems~\citep{ibrahim16comparing,moffat17incorporating}.
In the absence of an oracle judge who can tell the preferences between all rankers, the best ranker is usually inferred from user feedback on the result lists produced by the rankers~\citep{hofmann16online}. 
Since user feedback is known to be noisy~\citep{joachims03evaluating,joachims07evaluating,nelissen18swipe,goldberg18further}, how to infer ranker quality and when to stop evaluating a ranker are two important challenges in online ranker evaluation.

The former challenge, i.e., how to infer the quality of a ranker, is normally addressed by \emph{interleaving} methods~\citep{chapelle12large,hofmann11probabilistic,hofmann13fidelity,chuklin15comparative}. 
Specifically, an interleaving method interleaves the result lists generated by two rankers for a given query and presents the interleaved list to the user.  
Then it infers the preferred ranker based on the user's click feedback. 
As click feedback is noisy, the interleaved comparison of two rankers has to be repeated many times so as to arrive at a reliable outcome of the comparison.

Although interleaving methods address the first challenge of online ranker evaluation (how to infer the quality of a ranker), they give rise to another challenge, i.e., which rankers to compare and when to stop the comparisons.  
Without enough comparisons, we may mistakingly infer the wrong ranker preferences. But with too many comparisons we may degrade the user experience since we continue showing results from sub-optimal rankers. 
Based on previous work~\citep{,zoghi14relativea,zoghi15mergerucb,brost16multi}, the challenge of choosing and comparing rankers can be formalized as a $K$-armed dueling bandit problem~\citep{yue12k}, which is an important variant of the \ac{MAB} problem, where feedback is given in the form of pairwise preferences. 
In the $K$-armed dueling bandit problem, a ranker is defined as an arm and the best ranker is the arm that has the highest expectation to win the interleaving game against other candidates. 

A number of dueling bandit algorithms have been proposed; cf.~\citep{zoghi17dueling,busafekete14survey} for an overview. 
However, the study of these algorithms has mostly been limited to small-scale dueling bandit problems, with the state-of-the-art being \ac{DTS}~\citep{wu16double}. By ``small-scale'' we mean that the number of arms being compared is small. 
But, in real-world online ranker evaluation problems, experiments involving hundreds or even thousands of rankers are commonplace~\citep{kohavi13online}.
Despite this fact, to the best of our knowledge, the only work that address this particular scalability issue is \ac{MergeRUCB}~\citep{zoghi15mergerucb}. 
As we demonstrate in this chapter, the performance of \ac{MergeRUCB} can be improved upon substantially.

In this chapter, we propose and evaluate a novel algorithm, named \ac{MergeDTS}. 
The main idea of \ac{MergeDTS} is to combine the benefits of \ac{MergeRUCB}, which is the state-of-the-art algorithm for large-scale dueling bandit problems, and the benefits of \ac{DTS}, which is the state-of-the-art algorithm for small-scale problems, and attain improvements in terms of effectiveness (as measured in terms of regret) and efficiency (as measured in terms of time complexity). 
More specifically, what we borrow from \ac{MergeRUCB} is the divide and conquer idea used to group rankers into small batches to avoid global comparisons. 
On the other hand, from \ac{DTS} we import the idea of using  \acf{TS} \citep{thompson33likelihood}, rather than using uniform randomness as in \ac{MergeRUCB}, to choose the arms to be played. 

We analyze the performance of \ac{MergeDTS}, and demonstrate that the soundness of  \ac{MergeDTS} can be guaranteed if the time step $T$ is known and the exploration parameter $\alpha > 0.5$ (\cref{tois:th:bound}).
Finally, we conduct extensive experiments to evaluate the performance of \ac{MergeDTS} in the scenario of online ranker evaluation on three widely used real-world datasets: 
Microsoft, Yahoo!~Learning to Rank, and ISTELLA~\citep{qin13introducing,chapelle10yahoo,lucchese16post}. 
We show that with tuned parameters \ac{MergeDTS} outperforms \ac{MergeRUCB} and \ac{DTS} in large-scale online ranker evaluation under the Condorcet assumption, i.e., where there is a ranker preferred to all other rankers.\footnote{
	Our theoretical analysis is rather conservative and the regret bound only holds for the parameter values within a certain range. This is because our bound is proven using Chernoff-Hoeffding bound~\citep{hoeffding63probability} together with the union bound~\citep{casella02statistical}, both of which, in our case, introduce gaps between theory and practice. 
	In our experiments, we show that the parameter values outside of the theoretical regime can boost up the performance of \ac{MergeDTS} as well as that of the baselines.
	Thus, our experimental results of \ac{MergeDTS} are not restricted to the parameter values within the theoretical regime.
}
Moreover, we demonstrate the potential of using \ac{MergeDTS} beyond the Condorcet assumption, i.e., where there might be multiple best rankers.

In summary, the main contributions of this chapter are as follows:
\begin{enumerate}[label=(\arabic*)]
	\item 
		We propose a novel $K$-armed dueling bandits algorithm for large-scale online ranker evaluation, called \ac{MergeDTS}. 
		We use the idea of divide and conquer together with Thompson sampling to reduce the number of comparisons of arms. 
	\item 
		We analyze the performance of \ac{MergeDTS} and theoretically demonstrate that the soundness of \ac{MergeDTS} can be guaranteed in the case of known time horizon and parameter values in the theoretical regime. 
	\item 
		We evaluate \ac{MergeDTS} experimentally on the Microsoft, Yahoo! Learning to Rank and ISTELLA datasets, and show that, with the tuned parameters,
		\ac{MergeDTS} outperforms baselines in most of the large-scale online ranker evaluation configurations. 
\end{enumerate}

\noindent
The rest of the chapter is organized as follows. 
In \cref{tois:sec:problem setting}, we detail the definition of the dueling bandit problem. 
We discuss prior work in \cref{tois:sec:related work}. 
MergeDTS is proposed in \cref{tois:sec:mergedts}. 
Our experimental setup is detailed in \cref{tois:sec:experimental setup} and the results are presented in \cref{tois:sec:experimental results}.
We conclude the chapter in \cref{tois:sec:conclusion}. 

\section{Problem Setting}
\label{tois:sec:problem setting}

In this section, we first describe in more precise terms the $K$-armed dueling bandit problem, which is a variation of the \acf{MAB} problem. 
The latter can be described as follows: 
given $K$ choices, called ``arms'' and denoted by $a_1,\ldots,a_K$, we are required to choose one arm at each step;
choosing arm $a_i$ generates a reward which is drawn i.i.d.\ from a random variable with mean, denoted by $\mu_i$, and our goal is to maximize the expected total reward accumulated by our choices of arms over time.
This objective is more commonly formulated in terms of the \emph{cumulative regret} of the \ac{MAB} algorithm, where regret at step $t$ is the difference between the reward of the chosen arm, e.g., $a_j$, and the reward of the best arm, e.g., $a_k$, in hindsight, and the average regret of arm $a_j$ is defined to be 
$\mu_k-\mu_j$:
cumulative regret is defined to be the sum of the instantaneous regret over time~\citep{auer02finite,li19bubblerank}. 

The dueling bandit problem differs from the above setting in that at each step we can  choose up to two arms, $a_i$ and $a_j$ ($a_i$ and $a_j$ can be the same); the feedback is either $a_i$ or $a_j$, as the winner of the comparison between the two arms (rather than an absolute reward), where $a_i$ is chosen as the winner with \emph{preference probability} $p_{ij}$ and $a_j$ with probability $p_{ji}=1-p_{ij}$. 
These probabilities form the entries of a $K\times K$ \emph{preference matrix} $\mathbf{P}$, which defines the dueling bandit problem but is not revealed to the dueling bandit algorithm.

In a similar fashion to the \ac{MAB} setting, we evaluate a dueling bandit algorithm based on its \textit{cumulative regret}, which is the total regret incurred by choosing suboptimal arms comparing to the best arm over time~\citep{busafekete14survey,zoghi17dueling,sui18advancements}.
However, the definition of regret is less clear-cut in the dueling bandit setting, due to the fact that our dueling bandit problem might not contain a clear winner that is preferred to all other arms, i.e., an arm $a_C$, called the \emph{Condorcet winner}, such that $p_{Cj} > 0.5$ for all $j \neq C$. 
There are numerous proposals in the literature for alternative notions of winners in the absence of a Condorcet winner, e.g., Borda winner \citep{urvoy13generic,jamieson15sparse}, Copeland winner \citep{zoghi15copeland,komiyama16copeland}, von Neumann winner \citep{zoghi15copeland}, with each definition having its own disadvantages as well as practical settings where its use is appropriate.

MergeDTS, like most of the other dueling bandits algorithms~\citep{busafekete14survey, zoghi14relativea,zoghi14relative,zoghi15mergerucb,urvoy13generic}, relies on the existence of a  Condorcet winner, 
in which case the Condorcet winner is the clear choice for the best arm, since it is preferred to all other arms, and with respect to which regret can be defined.
We pose, as an interesting direction for future work, the task of extending the method proposed in this chapter to each of the other notions of winner listed above.

In order to simplify the notation in the rest of the chapter, we re-label the arms such that $a_1$ is the Condorcet winner, although this is not revealed to the algorithm.
We define the \emph{regret} incurred by comparing $a_i$ and $a_j$ at time $t$ to be  
\begin{equation}
r_t = (\Delta_{1i} + \Delta_{1j})/2 ,
\end{equation} 
where $\Delta_{1k} := p_{1k} - 0.5$ for each $k$. 
Moreover, the \emph{cumulative regret} after $T$ steps is defined to be 
\begin{equation}
\mathcal{R}(T) = \sum_{t=1}^{T} r_t, 
\end{equation}
where $r_t$ is the regret incurred by our choice of arms at time $t$.

Let us translate the online ranker evaluation problem into  the dueling bandit problem.
The input, a finite set of arms, consists of a set of rankers, e.g., based on different ranking models or based on the same model but with different parameters~\citep{kohavi13online}. 
The Condorcet winner is the ranker that is preferred, by the majority of users, over suboptimal rankers.  
More specifically, a result list from the Condorcet winner is expected to receive the highest number of clicks from users when compared to a list from a suboptimal ranker.  
The preference matrix $\mathbf{P}$ records the users'  relative preferences for all rankers. 
Regret measures the user frustration incurred by showing the interleaved list from suboptimal rankers instead of the Condorcet winner.
In the rest of the chapter,  we use the  term  \emph{ranker} to indicate the term \emph{arm} in $K$-armed dueling bandit problems since we focus on the online ranker evaluation task. 

\section{Related Work}
\label{tois:sec:related work}

There are two main existing approaches for solving  dueling bandit problems: 
\begin{enumerate*}[label=(\arabic*)]
\item reducing the problem to a \ac{MAB} problem, e.g., Sparring~\citep{ailon14reducing}, Self-Sparring~\citep{sui17multi} and \ac{REX3}~\citep{gajane15relative};
\item generalizing existing MAB algorithms to the dueling bandit setting, e.g., \ac{RUCB}~\citep{zoghi14relative}, \ac{RMED1}~\citep{komiyama15regret} and \ac{DTS}~\citep{wu16double}.
\end{enumerate*}
The advantage of the latter group of algorithms is that they come equipped with theoretical guarantees, proven for a broad class of problems.
The first group, however, have guarantees that either only hold for a restricted class of problems, where the dueling bandit problem is obtained by comparing the arms of an underlying MAB problem (a.k.a. utility-based dueling bandits), e.g., Self-Sparring,  \ac{REX3} and Sparring T-INF~\citep{zimmert19optimal}, or have substantially suboptimal instance-dependent regret bounds as in the case of Sparring EXP3, which has a regret bound of the form $O(\sqrt{KT})$, as opposed to $O(K\log T)$.
Indeed, as our experimental results below demonstrate, Sparring-type algorithms can perform poorly when the dueling bandit problem does not arise from a \ac{MAB} problem. 

Below, we describe some of these algorithms to provide context for our work. 
Sparring~\citep{ailon14reducing} uses two \ac{MAB} algorithms, e.g., \ac{UCB}, to choose rankers.  
At each step, Sparring asks each \ac{MAB} algorithm to output a ranker to be compared. 
The two rankers are then compared and the \ac{MAB} algorithm that proposed the winning ranker gets a reward of $1$ and the other a reward of $0$.

Self-Sparring~\citep{sui17multi} improves upon Sparring by employing a single \ac{MAB} algorithm, but at each step samples twice to choose rankers.
More precisely, \citet{sui17multi} use \acf{TS} as the MAB algorithm.  
Self-Sparring assumes that the problem it solves arises from an MAB; it can perform poorly when there exists a cycle relation in rankers, i.e., if there are rankers $a_i$, $a_j $ and $a_k$ with $p_{ij} > 0.5$, $p_{jk}>0.5$ and $p_{ki} > 0.5$. 
As Self-Sparring does not estimate confidence intervals of the comparison results, it does not eliminate rankers.

Another extension of Sparring is \ac{REX3}~\citep{gajane15relative}, which is designed for the adversarial setting.
\ac{REX3} is inspired by the \ac{EXP3}~\citep{auer03nonstochastic}, an algorithm for adversarial bandits, and has a regret bound of the form $O( \sqrt{K \ln{(K)}T)}$. 
Note that the regret bound grows as the square-root of time-steps, but sublinearly in the number of rankers, which shows the potential for improvement in the case of large-scale problems. 

\acf{RUCB}~\citep{zoghi14relative} extends \ac{UCB} to dueling bandits using a matrix of optimistic estimates of the relative preference probabilities. 
At~each step, \ac{RUCB} chooses the first ranker to be one that beats all other rankers based on the idea of \emph{optimism in the face of uncertainty}. 
Then it chooses the second ranker to be the ranker that beats the first ranker with the same idea of optimism in the face of uncertainty, which translates to pessimism for the first ranker. 
The cumulative regret of \ac{RUCB} after $T$ steps is upper bounded by an expression of the form $O(K^2 + K \log T)$. 

\ac{RMED1}~\citep{komiyama15regret} extends an asymptotically optimal \ac{MAB} algorithm, called \ac{DMED} \citep{honda11asymptotically}, by first proving an asymptotic lower bound on the cumulative regret of all dueling bandit algorithms, which has the order of $\Omega(K \log T)$, and pulling each pair of rankers the minimum number of times prescribed by the lower bound.
\ac{RMED1} outperforms \ac{RUCB} and Sparring. 

\acf{DTS}~\citep{wu16double} improves upon \ac{RUCB} by using \ac{TS} to break ties when choosing the first ranker. 
Specifically, it uses one \ac{TS} to choose the first ranker from a set of candidates that are pre-chosen by UCB. 
Then it uses another \ac{TS} to choose the second ranker that performs the best compared to the first one. 
The cumulative regret of \ac{DTS} is upper bounded by $O(K \log T +  K^2 \log \log T)$.
{Note that the bound of \ac{DTS} is higher than that of \ac{RUCB}. We hypothesize that this is because the bound of  \ac{DTS} is rather loose.} 
\ac{DTS} outperforms other dueling bandits algorithms empirically and is the state-of-the-art in the case of small-scale dueling bandit problems~\citep{wu16double,sui17multi}. 
As discussed in \cref{tois:sec:experimental results}, for computational reasons \ac{DTS} is not suitable for large-scale problems. 

The work that is the closest to ours is by~\citet{zoghi15mergerucb}.
They propose \ac{MergeRUCB}, which is the state-of-the-art for large-scale dueling bandit problems.
\ac{MergeRUCB} partitions rankers into small batches and compares rankers within each batch. 
A ranker is eliminated from a batch once we realize that even according to the most optimistic estimate of the preference probabilities it loses to another ranker in the batch. 
Once enough rankers have been eliminated, \ac{MergeRUCB} repartitions the remaining rankers and continues as before. 
Importantly, \ac{MergeRUCB} does not require global pairwise comparisons between all pairs of rankers, and so it reduces the computational complexity and increases the time efficiency, as shown in \cref{tois:sec:computational}.
The cumulative regret of \ac{MergeRUCB} can be upper bounded by $O(K \log T)$~\citep{zoghi15mergerucb}, i.e., with no quadratic dependence on the number of rankers.  
This upper bound has the same order as the lower bound proposed by \citet{komiyama15regret} in terms of $K \log{T}$, but it is not optimal in the sense that it has large constant coefficients. 
As we demonstrate in our experiments,  
 \ac{MergeRUCB} can be improved by making use of \ac{TS} to reduce the amount of randomness in the choice of rankers. 
More precisely, the cumulative regret of \ac{MergeRUCB} is almost twice as large as that of \ac{MergeDTS} in the large-scale setup shown in~\cref{tois:sec:experimental results}. 

A recent extension of dueling bandits is called \emph{multi-dueling bandits}~\citep{brost16multi,sui17multi,saha18battle}, where more than two rankers can be compared at each step. 
\ac{MDB} is the first proposed algorithm in this setting, which is specifically designed for online ranker evaluation. 
It maintains two UCB estimators for each pair of rankers, a looser confidence bound and a tighter one. 
At each step, if there is more than one ranker that is valid for the tighter UCB estimators, \ac{MDB} compares all the rankers that are valid for the looser UCB estimators.   
\ac{MDB} is outperformed by Self-Sparring, the state-of-the-art \emph{multi-dueling bandit} algorithm,  significantly~\citep{sui17multi}. 
In this chapter, we do not focus on the \emph{multi-dueling bandit} setup.
The reasons are two-fold.
First, to the best of our knowledge, there are no theoretical results in the multi-dueling setting that allow for the presence of cyclical preference relationships among the rankers. 
Second, \citet{saha18battle} state that ``(perhaps surprisingly) [\ldots] the flexibility of playing size-$k$ subsets does not really help to gather information faster than the corresponding dueling case ($k=2$),  at least for the current subset-wise feedback choice model.''
This statement demonstrates that there is no clear advantage to using multi-dueling comparisons over pairwise dueling comparisons at this moments.

\section{Algorithm}
\label{tois:sec:mergedts}
In this section, we propose the \ac{MergeDTS} algorithm, and explain the intuition behind it.
Then, we provide theoretical guarantees bounding the regret of \ac{MergeDTS}. 

\subsection{$\mergedts$}

The MergeDTS algorithm combines the benefits of both the elimination-based divide and conquer strategy of MergeRUCB and the sampling strategy of DTS, producing an effective scalable dueling bandit algorithm.

The pseudo-code for \ac{MergeDTS} is provided in \cref{tois:alg:mergedts,tois:alg:phase1,tois:alg:phase2}, 
with the notation summarized in \cref{tois:tb:notation} for the reader's convenience. 
The input parameters are the exploration parameter $\alpha$, the size of a batch $M$ and the  failure probability $\epsilon \in (0, 1)$. 
The algorithm records the outcomes of the past comparisons in matrix $\mathbf{W}$, whose element $w_{ij}$ is the number of times ranker $a_i$ has beaten ranker $a_j$ so far.
\ac{MergeDTS} stops when only one ranker remains and then returns that ranker, which it claims to be the Condorcet winner.\footnote{In the online ranker evaluation application, we can stop \ac{MergeDTS} once it finds the best ranker. However, in our experiments, we keep \ac{MergeDTS} running by comparing the remaining ranker with itself. If the remaining ranker is the Condorcet winner, there will be no regret. }

\begin{table}
	\caption{Notation used in this chapter.}
	\label{tois:tb:notation}
	\begin{tabular}{p{0.15\columnwidth} p{0.75\columnwidth}}
		\toprule	
		\bf Notation & \bf Description \\
		\midrule
		$K$ & Number of rankers \\
		$a_i$ & The $i$-th ranker\\
		$p_{ij}$ & Probability of $a_i$ beating $a_j$ \\
		$M$ & Size of a batch \\ 
		$\alpha$ & Exploration parameter, $\alpha > 0.5$ \\
		$\epsilon$ & Probability of failure \\
		$\mathbf{W}$ & The comparison matrix \\
		$w_{ij}$ & Number of times $a_i$ has beaten $a_j$ \\
		$s$ & Stage of the algorithm \\
		$\mathcal{B}_s$ & Set of batches at the $s$-th stage \\
		$b_s$ & Number of batches in $\mathcal{B}_s$ \\
		$\theta_{ij}$  & Sampled probability of $a_i$ beating $a_j$ \\
		$a_c$ & Ranker chosen in Phase~I of MergeDTS \\
		$\phi_{i}$ & Sampled probability of $a_i$ beating $a_c$ \\
		$a_d$ & Ranker chosen in Phase~II of MergeDTS \\
		$u_{ij}$ & Upper confidence bound (UCB): $\frac{w_{ij}}{w_{ij} + w_{ji}} + \sqrt{\frac{\alpha \log{(t+C(\epsilon))}}{w_{ij}+w_{ji}}}$ \\
		$\Delta_{ij}$ & $|p_{ij} - 0.5|$ \\
		$\Delta_{\min}$ & $\min_{\Delta_{ij} > 0} \Delta_{ij}$ \\
		$\Delta_{B, min}$ & $ \min_{a_i, a_j \in B \text{ and }i\neq j}\Delta_{ij}$\\
		$C(\epsilon)$ & $ \left(\frac{(4\alpha -1)K^2}{(2\alpha-1)\epsilon}\right) ^{\frac{1}{2\alpha-1}} $  \\
		\bottomrule
	\end{tabular}
\end{table}

\begin{algorithm}[!t]
	\begin{algorithmic}[1]
		\setcounter{ALC@unique}{0}
		\REQUIRE $K$ rankers $a_1, a_2, \ldots, a_K$; partition size $M$; exploration parameter $\alpha>0.5$; running time steps $T$;  probability of failure $\epsilon = 1/T$.
		\ENSURE The Condorcet winner. 
		\STATE $\mathbf{W} \leftarrow \mathbf{0}_{K,K}$ \hfill \Comment{\emph{The comparison matrix}}
		\STATE  \label{tois:alg:C} $C(\epsilon) = \left(\frac{(4\alpha-1)K^2}{(2\alpha-1)\epsilon}\right)^{\frac{1}{2\alpha-1}}$
		\STATE $s=1$ \hfill \Comment{\emph{The stage of the algorithm}}
		\STATE $\mathcal{B}_s = \big\{\underbrace{[a_1, \ldots, a_M]}_{B_1}, \ldots, \underbrace{[a_{(b_1-1)M+1}, \ldots, a_K]}_{B_{b_1}}\big\}$ \label{tois:alg:group_batches}
		\hfill \Comment{\emph{Disjoint batches of rankers, with $b_1 = \lceil\frac{K}{M}\rceil $}}\nonumber
		
		\FOR{$t = 1,2,\ldots T $} 
			\STATE $m = t \mod b_s$ \hfill \Comment{\emph{Index of the batches}}
			\IF{$b_s = 1$ and $|B_m|=1$  \Comment{One ranker left}}\label{tois:alg:oneranker} 
			\STATE\label{tois:alg:bestranker} Return the remaining ranker $a\in B_m$. 
			\ENDIF			
			\STATE $\mathbf{U} = \frac{\mathbf{W}}{\mathbf{W} + \mathbf{W}^{T}}+ \sqrt{\left(\frac{\alpha \log(t + C(\epsilon))}{\mathbf{W} + \mathbf{W}^{T}}\right)}$ \label{tois:alg:ubound}
			\hfill \Comment{\emph{UCB estimators: operations are element-wise and $\frac{x}{0} := 1$}}\nonumber
			
			\STATE Remove $a_i$ from $B_m$ if $u_{ij} < 0.5$ for any $a_j \in B_m$.
			\label{tois:alg:remove}
			\IF{$b_s > 1$ and $|B_m|=1$} 
				\STATE Merge $B_m$ with the next batch and decrement $b_s$.%
				\label{tois:alg:merge}
			\ENDIF
			
			\hspace{-2mm}\Comment{\textbf{Phase~I}: Choose the first candidate $a_c$}\nonumber
			\STATE $a_c = $ SampleTournament($\mathbf{W}$, $B_m$) \hfill \Comment {\emph{See \cref{tois:alg:phase1}}}\label{tois:alg:first_tournament}

			\hspace{-2mm}\Comment{\textbf{Phase~II}: Choose the second candidate $a_d$}
			\STATE $a_d$ = RelativeTournament($\mathbf{W}$, $B_m$, $a_c$) 
			\hfill \Comment{\emph{See \cref{tois:alg:phase2}}}
			
			\hspace{-2mm}\Comment{\textbf{Phase~III}: Compare candidates and update batches}
			\STATE Compare pair $(a_c, a_d)$ and increment $w_{cd}$ if $a_c$ wins otherwise increment $w_{dc} $.
			\label{tois:alg:update}

			\hspace{-2mm}\Comment{\textbf{Phase~IV}: Update batch set}
			\IF{$\sum_m|B_m| \leq \frac{K}{2^s}$} \label{tois:alg:update_batch}
			\STATE Pair the larger size batches with the smaller ones, making sure the size of every batch is in $[0.5M, 1.5M]$.
			\STATE $s = s+1$
			\STATE Update $\mathcal{B}_s$, $b_s = |\mathcal{B}_s|$.
			\ENDIF\label{tois:alg:update_batch_end} 
		\ENDFOR
	\end{algorithmic}
	\caption{MergeDTS (Merge Double Thompson Sampling)}
	\label{tois:alg:mergedts}
\end{algorithm}

\ac{MergeDTS} begins by grouping rankers into small batches (line~\ref{tois:alg:group_batches}). 
At each time-step, \ac{MergeDTS} checks whether there is more than one ranker remaining (line~\ref{tois:alg:oneranker}). 
If so, \ac{MergeDTS} returns that ranker, the potential Condorcet winner. 
If not, \ac{MergeDTS} considers one batch $B_m$ and, using optimistic estimates of the preference probabilities~(line~\ref{tois:alg:ubound}), it purges any ranker that loses to another ranker even with an optimistic boost in favor of the former (line~\ref{tois:alg:remove}).

If, as a result of the above purge, $B_m$ becomes a single-element batch, it is merged with the next batch $B_{m+1}$ (line~\ref{tois:alg:merge}).
Here, $m+1$ is interpreted as modulo $b_s$, where $b_s$ is the number of batches in the current stage.
This is done to avoid comparing a suboptimal ranker against itself, since if there is more than one batch, the best  ranker in any given batch is unlikely to be the Condorcet winner of the whole dueling bandit problem.
As we will see again below, MergeDTS takes great care to avoid comparing suboptimal rankers against themselves because it results in added regret, but yields no extra information, since we know that each ranker is tied with itself.

After the above elimination step, the algorithm proceeds in four phases: 
choosing the first ranker (Phase~I), choosing the second ranker based on the first ranker (Phase II), comparing the two rankers and updating the statistics (Phase III), and repartitioning the rankers at the end of each stage (Phase~IV).
Of the four phases, Phase~I and Phase~II are  the major reasons that lead to a boost in effectiveness of \ac{MergeDTS} when compared to \ac{MergeRUCB}. 
We will elaborate both phases in the remainder of this section. 

In Phase~I, the method \emph{SampleTournament} (\cref{tois:alg:phase1}) chooses the first candidate ranker: \ac{MergeDTS} samples preference probabilities $\theta_{ij}$ from the posterior distributions to estimate the true preference probabilities $p_{ij}$ for all pairs of rankers in the batch $B_m$ (lines~\ref{tois:alg:first_sample}--\ref{tois:alg:first_sample_end}, the first \ac{TS}).
Based on these sampled probabilities, \ac{MergeDTS} chooses the first candidate $a_c$ so that it beats most of the other rankers according to the sampled preferences (line~\ref{tois:alg:first_arm}). 

\begin{algorithm}[!h]
	\begin{algorithmic}[1]
		\setcounter{ALC@unique}{0}
		\Input  The comparison matrix $\mathbf{W}$ and the current batch $B_m$.
		\Output The first candidate  $a_c$.
		\FOR{$a_i, a_j \in B_m$ and $i < j$}	\label{tois:alg:first_sample}
		\STATE Sample $\theta_{ij} \sim Beta(w_{ij}+1, w_{ji}+1)$ 
		\STATE $\theta_{ji} = 1 - \theta_{ij}$	
		\ENDFOR\label{tois:alg:first_sample_end}
		
		\STATE $\kappa_i = \frac{1}{|B_m|-1} \sum_{a_j \in B_m, j\neq i} \mathds{1}(\theta_{ij}>0.5)$
		\STATE $a_c = \underset{a_i \in B_m}{\argmax} ~ \kappa_i$; \emph{breaking ties randomly} \hfill \Comment{\emph{First candidate}} \label{tois:alg:first_arm}
		
	\end{algorithmic}
	\caption{SampleTournament }
	\label{tois:alg:phase1}
\end{algorithm}

In Phase~II, the method \emph{RelativeTournament} (\cref{tois:alg:phase2}) chooses the second candidate ranker: \ac{MergeDTS} samples another set of preference probabilities $\phi_j$ from the posteriors of $p_{jc}$ for all rankers $a_j$ in $B_m \setminus \{a_c\}$ (lines~\ref{tois:alg:second_sample}--\ref{tois:alg:second_sample_end}, the second \ac{TS}).
Moreover, we set $\phi_c$ to be~$1$ (line~\ref{tois:alg:ac_nonfinal}). 
This is done to avoid self-comparisons between suboptimal rankers for the reasons that were described above.

\begin{algorithm}[!h]
	\begin{algorithmic}[1]
		\setcounter{ALC@unique}{0}
		\Input  The comparison matrix $\mathbf{W}$, the current batch $B_m$ and the first candidate $a_c$.
		\Output The second candidate  $a_d$.
			\FOR{$a_j \in B_m$ and  $j \neq c$}\label{tois:alg:second_sample}
				\STATE Sample $\phi_j \sim Beta(w_{jc} +1, w_{cj} +1)$
			\ENDFOR\label{tois:alg:second_sample_end}
			\STATE $\phi_c = 1$ \hfill \Comment{\emph{Avoid self-comparison}}
			\label{tois:alg:ac_nonfinal}
			\STATE $a_d = \underset{a_j \in B_m}{\argmin} ~ \phi_j$; \emph{breaking ties randomly}\hfill \Comment{\emph{Second candidate}}
			\label{tois:alg:second_arm_worst}
	\end{algorithmic}
	\caption{RelativeTournament}
	\label{tois:alg:phase2}
\end{algorithm}

\noindent%
Once the probabilities $\phi_j$ have been sampled, we choose the ranker $a_d$ that is going to be compared against $a_c$, using the following strategy. 
The worst ranker according to the sampled probabilities $\phi_j$ is chosen as the second candidate $a_d$ (line~\ref{tois:alg:second_arm_worst}).
The rationale for this discrepancy is that we would like to eliminate rankers as quickly as possible, so rather than using the upper confidence bounds to explore when choosing $a_d$, we use the lower confidence bounds to knock the weakest link out of the batch as quickly as possible.

In Phase~III (line~\ref{tois:alg:update}) of \cref{tois:alg:mergedts}, \ac{MergeDTS} plays $a_c$ and $a_d$ and updates the comparison matrix $\mathbf{W}$ based on the observed feedback. 

Finally, in Phase~IV (lines~\ref{tois:alg:update_batch}--\ref{tois:alg:update_batch_end}), if the number of remaining rankers in the current stage is half of the rankers of the previous stage (line~\ref{tois:alg:update_batch}), \ac{MergeDTS} enters the next stage, before which it repartitions the rankers. 
Following the design of \ac{MergeRUCB}, this is done by merging batches of rankers such that the smaller sized batches are combined with the larger sized batches; 
we enforce that the number of rankers in the new batches is kept in the range of $[0.5M, 1.5M]$.

\subsection{Theoretical guarantees}


In this section, we state and prove a high probability upper bound on the regret accumulated by MergeDTS after $T$ steps, under the assumption that the dueling bandit problem contains a Condorcet winner. 
Since the theoretical analysis of MergeDTS is based on that of \ac{MergeRUCB}, we start by listing two assumptions that we borrow from \ac{MergeRUCB} in~\cite[Section~7]{zoghi15mergerucb}.
\begin{assumption}
	\label{tois:assumption:1}
	There is no repetition in rankers. All ranker pairs $(a_i, a_j)$ with $i \neq j$ are distinguishable, i.e., $p_{ij} \neq 0.5$, unless both of them are ``uninformative'' rankers that provide random ranked lists and cannot beat any other rankers. 
\end{assumption}

\begin{assumption}
	\label{tois:assumption:2}
	The uninformative rankers are at most one third of the full set of rankers.  
\end{assumption}

\noindent%
These assumptions arise from the Yahoo! Learning to Rank Challenge dataset, where there are $181$ out of $700$ rankers that always provide random ranked lists. 
\cref{tois:assumption:1} ensures that each informative ranker is distinguishable from other rankers. 
\cref{tois:assumption:2} restricts the maximal  percentage of uninformative rankers and thus ensures that the probability of triggering the merge condition (line~\ref{tois:alg:update_batch} in \cref{tois:alg:mergedts}) is larger than $0$.\footnote{
In practice, \ac{MergeDTS} works without \cref{tois:assumption:2} because the Condorcet winner eliminates all other arms eventually with $O(K^2\log{T})$ comparisons. 
We keep \cref{tois:assumption:2}  to ensure that \ac{MergeDTS} also works in cases where we have the $O(K\log(T))$ guarantee. 
We refer readers to \citep{zoghi15mergerucb} for a detailed discussion.
}
Moreover, we emphasize that \cref{tois:assumption:1,tois:assumption:2} are milder than the assumptions made in Self-Sparring and DTS, where indistinguishability is simply not allowed.

\noindent
We now state our main theoretical result:

\begin{thm}
	\label{tois:th:bound}
	With the known time step $T$, applying \ac{MergeDTS} with $\alpha > 0.5$, $M \geq 4$ and $\epsilon = 1/T$ to a $K$-armed Condorcet dueling bandit problem under
	\cref{tois:assumption:1,tois:assumption:2}, with probability $1-\epsilon$ the cumulative regret $\mathcal{R}(T)$ after $T$ steps is bounded by: 
	\begin{equation}\label{tois:eq:bound}
	\mathcal{R}(T) < \frac{8\alpha M K \ln(T + C(\epsilon))}{\Delta_{\min}^2}, 
	\end{equation}
where
\begin{equation}\label{tois:eq:Delta_min}
\Delta_{\min} := \min_{\Delta_{ij} > 0} \Delta_{ij},
\end{equation}
is the minimal gap of two distinguishable rankers and $C(\epsilon) =  \left(\frac{(4\alpha -1)K^2}{(2\alpha-1)\epsilon}\right) ^{\frac{1}{2\alpha-1}} $.
\end{thm}

\noindent%
The upper bound on the $T$-step cumulative regret of \ac{MergeDTS} is $O(K\ln{(T)} /\Delta_{\min}^2)$. 
In other words, the cumulative regret grows linearly with the number of rankers, $K$.   
This is the most important advantage of \ac{MergeDTS}, which states the potential of applying it to the large-scale online evaluation. 
We emphasize that for most of the $K$-armed dueling bandit algorithms in the literature, the upper bounds contain a $K^2$ term, which renders them unsuitable for large-scale online ranker evaluation.
By the definition of $\Delta_{\min}$ in~\cref{tois:eq:Delta_min} we have $\Delta_{\min} > 0$, and so our bound is well-defined. 
However, the performance of \ac{MergeDTS} may degrade severely when $\Delta_{\min}$ is small. 
$\alpha$ is a common parameter in UCB-type algorithms, called the exploration parameter. 
$\alpha$ controls the trade-off between exploitation and exploration: larger $\alpha$ results in more exploration, whereas smaller $\alpha$ makes the algorithm more exploitative. 
Theoretically, $\alpha$ should be larger than $0.5$.
However, as shown in our experiments, using some values of $\alpha$ that are outside the theoretical regime can lead to a boost in the effectiveness of \ac{MergeDTS}. 

\cref{tois:th:bound} provides a finite-horizon high probability bound. 
From a practical point of view, this type of bound is of great utility. 
In practice, bandit algorithms are always deployed and evaluated within limited user iterations~\citep{yue09interactively,li10contextual}. 
Here, each time step is one user interaction. 
As the number of interactions is provided, we can choose a reasonable step $T$ to make sure the high probability bound holds. 
We can also get an expected regret bound of \ac{MergeDTS} at step $T$ by setting $\epsilon = 1/T$ and adding $1$ to the right-hand side of~\cref{tois:eq:bound}: 
this is because $\mathbb{E}[\mathcal{R}(T)]$ can be bounded by
\begin{equation}
 \frac{1}{T} \cdot T + \frac{T-1}{T} \cdot \frac{8\alpha M K \ln(T + C(\epsilon))}{\Delta_{\min}^2} \leq 1 + \frac{8\alpha M K \ln(T + C(\epsilon))}{\Delta_{\min}^2}. 
\end{equation}
We note that the above expected bound holds only at time-step $T$ and so the horizonless version of \ac{MergeDTS} does not possess an expected regret bound.

The proof of \cref{tois:th:bound} relies on the Lemma~3 in~\citep{zoghi15mergerucb}. We repeat it here for the reader's convenience. 
\begin{lemma}[Lemma~3 in \citep{zoghi15mergerucb}]
\label{tois:lm:mergerucb}
	Given any pair of distinguishable rankers $a_i, a_j \in B$ and $\epsilon \in [0, 1]$, with the probability of $1-\epsilon$, the maximum number of comparisons that could have been carried out between these two rankers in the first $T$ time-steps before a merger between B and another batch occurs, is bounded by 
	\begin{equation}
			\frac{4\alpha \ln (T + C(\epsilon))}{\Delta_{B, min}^2}, 
	\end{equation}
	where $\Delta_{B, min} = \min_{a_i, a_j \in B ~and ~i\neq j}\Delta_{ij}$ is the minimal gap of two distinguishable rankers in batch $B$.  
\end{lemma}

\begin{proof}[Proof of \cref{tois:th:bound}]
\cref{tois:lm:mergerucb} states that with probability $1-\epsilon$ the number of comparisons between a pair of distinguishable rankers $(i, j) \in B$ is bounded by 
\begin{equation}
\frac{4\alpha \ln (T + C(\epsilon))}{\Delta_{B, min}^2},
\end{equation}
regardless of the way the rankers are selected, as long as the same criterion as \ac{MergeRUCB} is used for eliminating rankers.
Since the elimination criterion for MergeDTS is the same as that of MergeRUCB, we can apply the same argument used to prove Theorem~1 in \citep{zoghi15mergerucb} to get a bound of 
\begin{equation}
\frac{8\alpha M K \ln(T+ C(\epsilon)) }{\Delta_{\min}^2}
\end{equation}
on the regret accumulated by MergeDTS. 
Here we use the fact that $\Delta_{B, min} \geq \Delta_{min}$ and thus $\frac{4\alpha \ln (T + C(\epsilon))}{\Delta_{B, min}^2} \leq \frac{4\alpha \ln (T + C(\epsilon))}{\Delta_{min}^2}$. 
\qedhere
\end{proof}

\subsection{Discussion}

The prefix ``merge'' in \ac{MergeDTS} signifies the fact that it uses a similar divide-and-conquer strategy as merge sort. 
It partitions the $K$-arm set into small batches of size $M$. 
The comparisons only happen between rankers in the same batch, which, in turn, avoids global  pairwise comparisons and gets rid of the $O(K^2)$ dependence in the cumulative regret, which is the main limitation for using dueling bandits for large datasets. 

In contrast to sorting, \ac{MergeDTS} needs a large number of comparisons before declaring a difference between rankers since the feedback is stochastic. 
The harder two rankers are to distinguish or in other words the closer $p_{ij}$ is to $0.5$, the more comparisons are required. 
Moreover, if a batch only contains the uninformative rankers, the comparisons between those rankers will not stop, which incurs infinite regret. 
\ac{MergeDTS} reduces the number of comparisons between hardly distinguishable rankers as follows:
\begin{enumerate}[label=(\arabic*)]
	\item \ac{MergeDTS} compares the best ranker in the batch to the worst to avoid comparisons between hardly distinguishable rankers; 
	\item When half of the rankers of the previous stage are eliminated, \ac{MergeDTS} pairs larger batches to smaller ones that contain at least one informative ranker and enters the next stage. 
\end{enumerate}
The second point is borrowed from the design of \ac{MergeRUCB}. 

\ac{MergeDTS} and  \ac{MergeRUCB} follow the same ``merge'' strategy.
The difference between these two algorithms is in their strategy of choosing rankers, i.e., \cref{tois:alg:phase1,tois:alg:phase2}. 
\ac{MergeDTS} employs a sampling strategy to choose the first ranker inside the batch and then uses another 
sampling strategy to choose the second ranker that is potentially beaten by the first one. 
As stated above, this design comes from the fact that \ac{MergeDTS} is  carefully designed to reduce the comparisons between barely distinguishable rankers. 
In contrast to \ac{MergeDTS}, \ac{MergeRUCB} randomly chooses the first ranker and chooses the second ranker to be the one that is the most likely to beat the first ranker, as discussed in \cref{tois:sec:related work}.
The uniformly random strategy inevitably increases the number of comparisons between those barely distinguishable rankers. 

In summary, the double sampling strategy used by \ac{MergeDTS} is the major factor that leads to the superior performance of \ac{MergeDTS} as demonstrated by our experiments. 

\section{Experimental Setup}
\label{tois:sec:experimental setup}

In this chapter, we investigate the application of dueling bandits to the large-scale online ranker evaluation setup. 
Our experiments are designed to answer \textbf{RQ1}, which we map into a set of six more refined research questions.

\subsection{Research questions}


\begin{description}
	\item[\textbf{RQ1.1}] In the large-scale online ranker evaluation task, does \ac{MergeDTS} outperform the state-of-the-art large-scale dueling bandits algorithms in terms of cumulative regret, i.e., effectiveness?
\end{description}
In the bandit literature~\citep{busafekete14survey,zoghi17dueling}, \emph{regret} is a measure of the rate of convergence to the Condorcet winner in hindsight. 
Mapping this to the online ranker evaluation setting, \textbf{RQ1.1} asks whether  \ac{MergeDTS} hurts the user experience less than baselines while it is being used for large-scale online ranker evaluation.

\begin{description}
		\item[\textbf{RQ1.2}] How do \ac{MergeDTS} and the baselines scale computationally?
\end{description}
What is the time complexity of \ac{MergeDTS}? 
Does \ac{MergeDTS} require less running time than the baselines? 

\begin{description}
	\item[\textbf{RQ1.3}] How do different levels of noise in the feedback signal affect cumulative regret of \ac{MergeDTS} and the baselines?
\end{description}
In particular, will we still observe the same results in \textbf{RQ1.1} after a (simulated) user changes its behavior? 
How sensitive are \ac{MergeDTS} and the baselines to noise? 

\begin{description}
	\item[\textbf{RQ1.4}] How do \ac{MergeDTS} and the baselines perform when the Condorcet dueling bandit problem contains cycles?
\end{description}
Previous work has found that cyclical preference relations between rankers are abundant in online ranker comparisons~\citep{zoghi14relativea,zoghi15mergerucb}. 
Can \ac{MergeDTS} and the baselines find the Condorcet winner when the experimental setup features a large number of cyclical relations between rankers?

\begin{description}
	\item[\textbf{RQ1.5}] How does \ac{MergeDTS} perform when the dueling bandit problem violates the Condorcet assumption?
\end{description}
We focus on the Condorcet dueling bandit task in this chapter. 
Can \ac{MergeDTS} be applied to dueling bandit tasks without the existence of a Condorcet winner? 
%

\begin{description}
\item[\textbf{RQ1.6}] What is the parameter sensitivity of \ac{MergeDTS}?
\end{description}
Can we improve the performance of \ac{MergeDTS} by tuning its parameters, such as the exploration parameter $\alpha$, the size of a batch $M$, and the probability of failure~$\epsilon$? 

\subsection{Datasets}
\label{tois:sec:dataset}
To answer our research questions, we use two types of dataset: three real-world datasets and a synthetic dataset.
First, to answer 
\textbf{RQ1.1} to \textbf{RQ1.3}, 
we run experiments on three large-scale datasets: the Microsoft Learning to Rank (MSLR) WEB30K dataset~\citep{qin13introducing}, the Yahoo!\ Learning to Rank Challenge Set 1 (Yahoo)~\citep{chapelle10yahoo} and the ISTELLA dataset~\citep{lucchese16post}.\footnote{
	We omit the Yahoo Set 2 dataset because it contains far fewer queries than the Yahoo Set 1 dataset.}
These datasets contain a large number of features based on unsupervised ranking functions, such as BM25, TF-IDF, etc.
In our experiments, we take the widely used setup in which each individual feature is regarded as a ranker~\citep{zoghi15mergerucb,hofmann11probabilistic}.  
This is different from a real-world setup, where a search system normally ranks documents using a well trained learning to rank algorithm that combines multiple features.
However, the difficulty of a dueling bandit problem comes from the relative quality of pairs of rankers and not from their absolute quality.  
In other words, evaluating rankers with similar and possibly low performance is as hard as evaluating state-of-the-art rankers, e.g., LambdaMART~\citep{burges10ranknet}. 
Therefore, we stick to the standard setup of~\citep{zoghi15mergerucb,hofmann11probabilistic}, treating each feature as a ranker
and each ranker as an arm in the $K$-armed dueling bandit problem. 
We leave experiments aimed at comparing different well trained learning to rank algorithms as future work. 
As a summary, the MSLR dataset contains $136$ rankers, the Yahoo dataset contains $700$ rankers and the ISTELLA dataset contains $220$ rankers.
Compared to the typical $K$-armed dueling bandit setups, where $K$ is generally substantially smaller than $100$~\citep{sui17multi,yue11beat,ailon14reducing,wu16double}, these are large numbers of rankers. 

Second, to answer 
\textbf{RQ1.4}, 
we use a synthetic dataset, generated by~\citet{zoghi15mergerucb}, which contains cycles (called the \emph{Cycle dataset} in the rest of the chapter). 
The Cycle dataset has $20$ rankers with one Condorcet winner, $a_1$, and $19$ suboptimal rankers, $a_2, \ldots, a_{20}$.  
The Condorcet winner beats the other $19$ suboptimal rankers. 
And those $19$ rankers have a cyclical preference relationship between them. 
More precisely, following~\citet{zoghi15mergerucb}, the estimated probability $p_{1j}$ of $a_1$ beating $a_j$  ($j=2, \ldots, 20$ is set to $p_{1j} = 0.51$, and the preference relationships between the suboptimal rankers are described as follows: visualize the $19$ rankers $a_2, \ldots, a_{20}$ sitting at a round table, then each ranker beats every ranker to its left with probability $1$ and loses to every ranker to its right with probability $1$.
In this way we obtain the Cycle dataset.

Note that this is a difficult setup for Self-Sparring, because Self-Sparring chooses rankers based on their Borda scores,
and the Borda scores ($\sum_{j=1}^{K}p_{ij}$ for each ranker $a_i$~\citep{urvoy13generic}) are close to each other in the Cycle dataset.
For example, the Borda score of the Condorcet winner is $10.19$, while the Borda score of a suboptimal ranker is $9.99$.
This makes it hard for Self-Sparring to distinguish between rankers.
In order to be able to conduct a fair comparison, we generate the \emph{Cycle2 dataset}, where each suboptimal ranker beats every ranker to its left with a probability of $0.51$ and the Condorcet winner beats all others with probability $0.6$. 
Now, in the Cycle2 dataset, the Borda score of the Condorcet winner is $11.90$, while the Borda score of a suboptimal ranker is $9.9$. 
Thus, it is an easier setup for Self-Sparring. 
 
Furthermore, to answer 
\textbf{RQ1.5}, 
we use the MSLR-non-Condorcet dataset from~\citep{wu16double}, which is a subset of the MSLR dataset that does not contain a Condorcet winner. 
This datasets has $32$ rankers with two Copeland winners (instead of one), each of which beats the other $30$ rankers. 
A \emph{Copeland winner} is a ranker that beats the largest number of other rankers; every dueling bandit dataset contains at least one Copeland winner~\citep{zoghi15copeland}.

Finally, we use the MSLR dataset with the navigational configuration (described in \cref{tois:sec:clicksimulation}) to assess the parameter sensitivity of \ac{MergeDTS}~\textbf{RQ1.6}.

\subsection{Evaluation methodology}
To evaluate dueling bandit algorithms, we follow the proxy approach from~\citep{zoghi15mergerucb}.
It first uses an interleaving algorithm to obtain a preference matrix, i.e., a matrix that for each pair of rankers contains the probability that one ranker beats the other.
More precisely, for each pair of rankers $a_i$ and $a_j$, $p_{ij}$ is the estimation that $a_i$ beats $a_j$ in the simulated interleaved comparisons. 
Then, this obtained preference matrix is used to evaluate dueling bandit algorithms: 
for two rankers $a_i$ and $a_j$ chosen by a dueling bandit algorithm, we compare them by drawing a sample from a Bernoulli distribution with mean $p_{ij}$, i.e., $1$ means that $a_i$ beats $a_j$ and vice versa. 
This is a standard approach to evaluating dueling bandit algorithms~\citep{yue11beat,wu16double,zoghi14relativea}. 
Moreover, the proxy approach has been shown to have the same quality as interleaving in terms of evaluating dueling bandit algorithms~\citep{zoghi15mergerucb}.

In this chapter, we adopt the procedure described by \citet{zoghi15mergerucb} and obtain preference matrices for MSLR, Yahoo and ISTELLA datasets. 
Specifically, we use Probabilistic Interleave~\citep{hofmann11probabilistic} to obtain preference matrices.\footnote{We use the implementation of Probabilistic Interleave in the LEROT software package~\citep{schuth13lerot}.}
The numbers of comparisons for every pair of rankers in MSLR, Yahoo and ISTELLA datasets are \numprint{400000}, \numprint{60000} and \numprint{400000}. 
The reason for the discrepancy is pragmatic: the Yahoo dataset has roughly $27$ times as many pairs of rankers to be compared.

\subsection{Click simulation}
\label{tois:sec:clicksimulation}
As the interleaved comparisons mentioned above are carried out using click feedback, we follow~\citet{hofmann13fidelity} and simulate clicks using three configurations of a click model~\citep{chuklin15click}: namely \emph{perfect}, \emph{navigational} and \emph{informational}.
The perfect configuration simulates a user who checks every document and clicks on a document with a probability proportional to the query-document relevance. 
This configuration is the easiest one for dueling bandit algorithms to find the best ranker, because it contains very little noise. 
The navigational configuration mimics a user who seeks specific information, i.e., who may be searching for the link of a website, and is likely to stop browsing results after finding a relevant document. 
The navigational configuration contains more noise than the perfect configuration and is harder for dueling bandit algorithms to find the best ranker. 
Finally, the informational configuration represents a user who wants to gather all available information for a query and may click on documents that are not relevant with high probability. 
In the informational configuration the feedback contains more noise than in the perfect and navigational configurations, which makes it the most difficult configuration for dueling bandit algorithms to determine the best ranker, which, in turn, may result in the highest cumulative regret among the three configurations.

To answer the research questions that concern large-scale dueling bandit problems, namely 
\textbf{RQ1.1}, \textbf{RQ1.2} and \textbf{RQ1.6},
we use the navigational configuration, which represents a reasonable middle ground between the perfect and informational configurations~\citep{hofmann11probabilistic}.
The corresponding experimental setups are called MSLR-Navigational, Yahoo-Navigational and ISTELLA-Navigational.
To answer 
\textbf{RQ1.3} 
regarding the effect of feedback with different levels of noise,
we use all three configurations on MSLR, Yahoo and ISTELLA datasets: namely MSLR-Perfect, MSLR-Navigational and MSLR-Informational; Yahoo-Perfect, Yahoo-Navigational and Yahoo-Informational; ISTELLA-Perfect, ISTELLA-Navigational and ISTELLA-Informational. 
Thus, we have nine large-scale setups in total.

\subsection{Baselines}
We compare \ac{MergeDTS} to five state-of-the-art dueling bandit algorithms: \break
\ac{MergeRUCB}~\citep{zoghi15mergerucb}, \ac{DTS}~\citep{wu16double}, \ac{RMED1}~\citep{komiyama15regret}, Self-Sparring~\citep{sui17multi}, and REX3~\citep{gajane15relative}. 
Among these algorithms, \ac{MergeRUCB} is designed for large-scale online ranker evaluation and is the state-of-the-art large-scale dueling bandit algorithm. 
\ac{DTS} is the state-of-the-art small-scale dueling bandit algorithm.
\ac{RMED1} is motivated by the lower bound of the Condorcet dueling bandit problem and matches the lower bound up to a factor of $O(K^2)$, which indicates that \ac{RMED1} has low regret in small-scale problems but may have large regret when the number of rankers is large. 
Self-Sparring is a more recently proposed dueling bandit algorithm that is the state-of-the-art algorithm in the multi-dueling setup, with which multiple rankers can be compared in each step.
REX3 is proposed for the adversarial dueling bandit problem but also performs well for the large-scale stochastic dueling bandit problem~\citep{gajane15relative}. 
We do not include \ac{RUCB}~\citep{zoghi14relative} and Sparring~\citep{ailon14reducing} in our experiments since they have been outperformed by more than one of our baselines~\citep{zoghi15mergerucb,sui17multi,gajane15relative}. 

\subsection{Parameters}
\label{tois:sec:parameters setup}

Recall that \cref{tois:th:bound} is based on Lemma~3 in~\citep{zoghi15mergerucb}.
The latter provides a high probability guarantee that the confidence intervals will not mislead the algorithm into eliminating the Condorcet winner by mistake.
However, this result is proven using the Chernoff-Hoeffding \citep{hoeffding63probability} bound together with an application of the union bound \citep{casella02statistical}, both of which introduce certain gaps between theory and practice.
That is, the analysis of regret mainly considers the worst-case scenario rather than the average-case scenario, which makes regret bounds much looser than they could have been.
We conjecture that the expression for $C(\epsilon)$, which derives its form from Lemma~3 in~\citep{zoghi15mergerucb}, is excessively conservative.
Put differently, \cref{tois:th:bound} specifies a sufficient condition for the proper functioning of \ac{MergeDTS}, not a necessary one. 
So, a natural question that arises is the following: to what extent can restrictions imposed by our theoretical results be violated without the algorithm failing in practice?
In short, what is the gap between theory and practice and what is the parameter sensitivity of \ac{MergeDTS}? 

To address these questions and answer 
\textbf{RQ1.6}, 
we conduct extensive parameter sensitivity analyses in the MSLR-Navigational setup with the following parameters: 
$\alpha \in \{0.8 ^{0}$, $0.8^{1}$, \dots, $0.8^{9}\}$,
$C \in \{4 \times 10^2$, $4\times 10^3, \ldots, 4 \times 10^6, \numprint{4726908}\}$, and $M \in \{2$, $4$,  $8$, $16\}$. 
$C$ is short for $C(\epsilon)$,	where $C(\epsilon)$ is the exploration bonus added to the confidence intervals. 
According to \cref{tois:tb:notation}, $C(\epsilon)$ is a function of $\alpha$ and $\epsilon$. 
However, to simplify our experimental setup, we consider $C$ as an individual parameter rather than a function parameterized by $\alpha$ and $\epsilon$, and study the impact of $C$ directly. 
The details about  the choice of the values are explained in the following paragraph.

When choosing candidate values for $\alpha$, we want them to cover the optimal theoretical value $\alpha > 1$, the lowest theoretically legal value $\alpha > 0.5$, and for smaller values of $\alpha$ we want to decrease the differences between two consecutive $\alpha$'s. 
This last condition is imposed because smaller values of $\alpha$ may mislead \ac{MergeDTS} to eliminate the Condorcet winner. 
So we shrink the search space for smaller values of $\alpha$.
The powers of $0.8$ from $0$ to $9$ seem to satisfy the above conditions, particularly $0.8^3 \approx 0.5$ and obviously $0.8^0 = 1$ with the difference between $0.8^n$ and $0.8^{n+1}$ becoming smaller with larger $n$.
The value $C = \numprint{4726908}$ is calculated from the definition of $C(\epsilon)$ with the default $\alpha=1.01$ and $M=4$ (see \cref{tois:tb:notation}), noting that the MSLR-Navigational setup contains $136$  rankers, i.e., $K=136$.
As discussed before, the design of $C(\epsilon)$ may be too conservative. 
So, we only choose candidate values smaller than $\numprint{4726908}$.
We use the log-scale of $C(\epsilon)$ because the upper bound is logarithmic with $C(\epsilon)$.

The sensitivity of parameters is analyzed by following the order of their importance to \cref{tois:th:bound}, i.e., $\alpha$ and $M$ have a linear relation to the cumulative regret and $C$ has a logarithmic relation to the cumulative regret.
We first evaluate the sensitivity of $\alpha$ with the default values of $M$ and $C$.
Then we use the best value of $\alpha$ to test a range of values of $M$ (with default $C$).
Finally, we analyze the impact of $C$ using the best values of $\alpha$ and $M$. 

We discover the practically optimal parameters for \ac{MergeDTS} to be $\alpha=0.8^6$, $M=16$ and $C=\numprint{4000000}$, in  \cref{tois:sec:parameter}. 
We repeat the procedure for \ac{MergeRUCB} and \ac{DTS}, and use their optimal parameter values in our experiments,
which are $\alpha = 0.8^6$, $M=8$, $C=\numprint{400000}$ for \ac{MergeRUCB} and $\alpha = 0.8^7$ for \ac{DTS}.
Then, we use these values to answer 
\textbf{RQ1.1} to \textbf{RQ1.5}.
Self-Sparring does not have any parameters, so further analysis and tuning are not needed here.

	The shrewd readers may notice that the parameters are somewhat overtuned in the MSLR-Navigational setup, and \ac{MergeDTS} with the tuned parameters does not enjoy the theoretical guarantees in \cref{tois:th:bound}. 
	However, because of the existence of the gap between theory and practice, we want to answer the question whether we can improve the performance of MergeDTS as well as that of the baselines by tuning the parameters outside of their theoretical limits. 
	We also want to emphasize that the parameters of MergeDTS and baselines are only tuned in the MSLR-Navigational setup, but MergeDTS is compared to baselines in nine setups. 
	If, with the tuned parameters, MergeDTS outperforms baselines in other eight setups, we can also show the potential of improving MergeDTS in an empirical way. 

\subsection{Metrics}
In our experiments, we assess the efficiency (time complexity) and effectiveness (cumulative regret) of \ac{MergeDTS} and baselines. 
The metric for efficiency is the running time in days. 
We compute the running time from the start of the first step to the end of the $T$-th step, where $T = 10^8$ in our experiments.  
A commercial search engine may serve more than $1$ billion search requests per day~\citep{gomes10google}, and each search request can be used to conduct one dueling bandit comparison. 
The total number of time steps, i.e. $T$, considered in our experiments is about $1\%$ of the one-week traffic of a commercial search engine.

We use cumulative regret in $T$ steps to measure the effectiveness of algorithms, which is computed as follows:
\begin{equation}
    \mathcal{R}(T) = \sum_{t = 1}^{T} r(t) =   \sum_{t = 1}^{T}  \frac{1}{2}(\Delta_{1,c_t} + \Delta_{1,d_t}),
\end{equation}
where $r(t)$ is the regret at step $t$, $c_t$ and $d_t$ are the indices of rankers chosen at step $t$, and without loss of generality, we assume $a_1$ to be the Condorcet winner. 
The regret $r(t)$ arises from the comparisons between the two suboptimal rankers at $t$ step. 
It is the average sub-optimality of comparing two rankers $a_i$ and $a_j$ with respect to the Condorcet winner $a_1$, i.e., $\frac{p_{1i} + p{1j}}{2} - 0.5$. 
In a real-world scenario, we have a fixed time period to conduct our online ranker evaluation, and thus, the number of steps $T$ can be estimated beforehand. 
In our online ranker evaluation task, the $T$-step cumulative regret is related to the drop in user satisfaction during the evaluation process, i.e. higher regret means larger degradation of user satisfaction, because the preference $p_{1i}$ can be interpreted as the probability of the Condorcet winner being preferred to ranker $i$.

Unless stated differently, the results in all experiments are averaged over $50$ and $100$ independent runs on large- and small-scale datasets respectively, where both numbers are either equal to or larger than the choices in previous studies~\citep{wu16double,sui17multi,zoghi15mergerucb}. 
In the effectiveness experiments, we also report the standard error of the average cumulative regret, which measures the differences between the average of samples and the expectation of the sample population. 

All experiments on the MSLR and Yahoo datasets are conducted on a server with Intel(R) Xeon(R) CPU E5-2650  $2.00$GHz ($32$ Cores) and $64$ Gigabyte.
All experiments on the ISTELLA dataset are conducted on servers with Intel(R) Xeon(R) Gold 5118 CPU @ $2.30$GHz ($48$ Cores) and $256$ Gigabyte. 
To be precise, an individual run of  each algorithm is conducted on a single core with 1 Gigabyte.

\section{Experimental Results}
\label{tois:sec:experimental results}

\newcommand{\plotheight}{0.98}
\newcommand{\wideplotheight}{0.68}

In this section, we analyze the results of our experiments.
In \cref{tois:sec:large-scale}, we compare the effectiveness (cumulative regret) of \ac{MergeDTS} and the baselines in three large-scale online evaluation setups. 
In \cref{tois:sec:computational}, we compare and analyze the efficiency (time complexity) of \ac{MergeDTS} and the baselines. 
In \cref{tois:sec:noise}, we study the impact of different levels of noise in the click feedback on the algorithms. 
In \cref{tois:sec:cycle} and \cref{tois:sec:noncondorcet}, we evaluate \ac{MergeDTS} and the baselines in two alternative setups: the cyclic case and the non-Condorcet case, respectively. 
In \cref{tois:sec:parameter}, we analyze the parameter sensitivity of \ac{MergeDTS}.

\subsection{Large-scale experiments}
\label{tois:sec:large-scale}

\begin{figure}[t]
	\centering
	\includegraphics{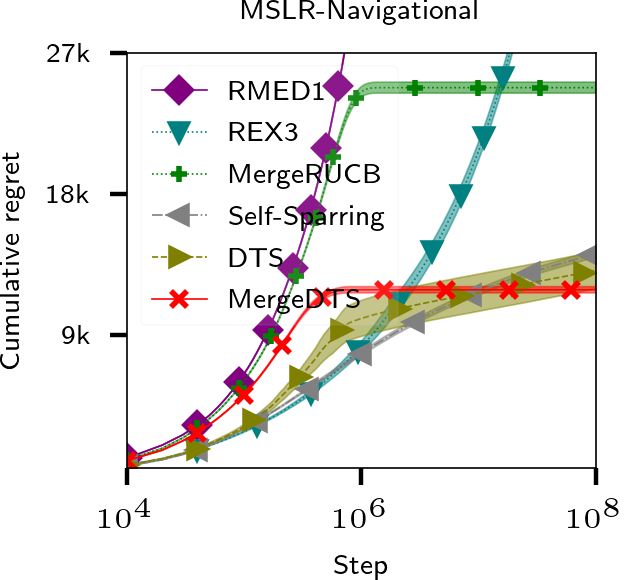}
	\includegraphics{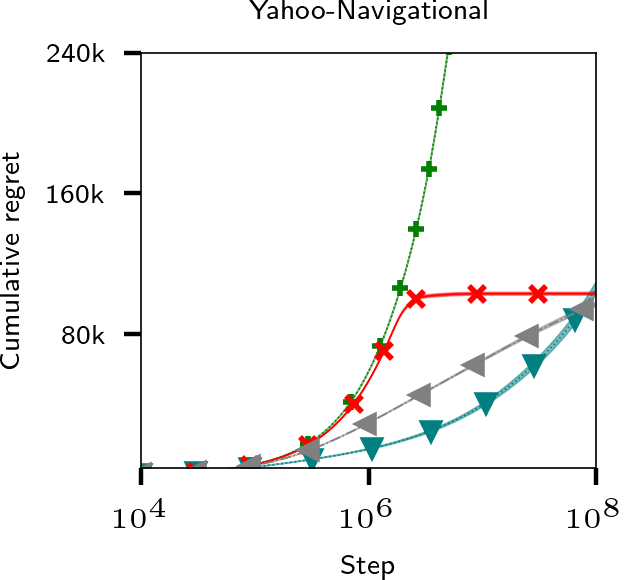}
	\includegraphics{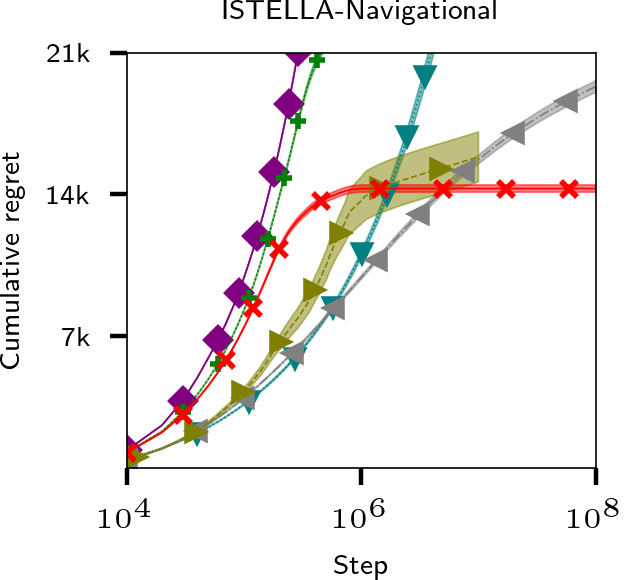}	
	\caption{Cumulative regret on large-scale online ranker evaluation: lower is better. 
	Note that the scales of the y-axes are different. 
	The shaded areas are $\pm$ standard error. 
	The results are averaged over $50$ independent runs.
	}  
	\label{tois:fig:large}	
\end{figure}

To answer 
\textbf{RQ1.1}, 
we compare \ac{MergeDTS} to the large-scale state-of-the-art baseline, \ac{MergeRUCB}, as well as the more recently proposed Self-Sparring, in three large-scale online evaluation setups, namely MSLR-Navigational, Yahoo-Navigational and ISTELLA-Navigational. 
The results are reported in \cref{tois:fig:large}, which depicts the cumulative regret of each algorithm, averaged over $50$ independent runs. 
As mentioned in \cref{tois:sec:experimental setup}, cumulative regret measures the rate of convergence to the Condorcet winner in hindsight and thus lower regret curves are better.
\ac{DTS} and \ac{RMED1} are not considered in the Yahoo-Navigational setup, and the cumulative regret of \ac{DTS} is reported with in $10^7$ steps on the ISTELLA dataset, because of the computational issues,
which are further discussed in \cref{tois:sec:computational}.
\cref{tois:fig:large} shows that MergeDTS  outperforms the large-scale state-of-the-art MergeRUCB with large gaps. 
Regarding the comparison with Self-Spar\-ring and DTS, we would like to point out the following facts:
\begin{inparaenum}[(1)]
\item Merge\-DTS outperforms \ac{DTS} and Self-Sparring in MSLR-Navigational and ISTELLA-Navigational setups; 
\item 
	In the Yahoo-Navigational setup, \ac{MergeDTS} has slightly higher cumulative regret than Self-Sparring, but \ac{MergeDTS} converges to the Condorcet winner after three million steps while the cumulative regret of Self-Sparring is still growing after $100$ million steps.
\end{inparaenum}
Translating these facts into real-world scenarios, 
\ac{MergeDTS} has higher regret compared to DTS and Self-Sparring in the early steps, but \ac{MergeDTS} eventually outperforms DTS and Self-Sparring with longer experiments.
As for \ac{REX3}, we see that it has a higher order of regret than other algorithms since \ac{REX3} is designed for the adversarial dueling bandits and the regret of \ac{REX3} is $O(\sqrt{T})$. 
In this chapter, we consider the stochastic dueling bandits, and the regret of the other algorithms is $\log{(T)}$. 
\ac{MergeDTS} outperforms the baselines in most setups, but we need to emphasize that the performance of \ac{MergeDTS} here cannot be guaranteed by \cref{tois:th:bound}.
This is because we use the parameter setup outside of the theoretical regime, as discussed in \cref{tois:sec:parameters setup}. 

\subsection{Computational scalability}
\label{tois:sec:computational}

\begin{table}[h]
	\caption{
	Average running time in days of each algorithm on large-scale problems for $10^8$ steps averaged over $50$ independent runs. 
	The running time of $\ac{DTS}$ in the ISTELLA-Navigational setup is estimated based on the running time with $10^7$ steps multiplied by $10$. 
 	The running time of \ac{DTS} and \ac{RMED1} in the Yahoo-Navigational setup is estimated by multiplying the average running time at $10^5$ steps by $10^3$.
 		The experiments on the ISTELLA dataset are conducted on different computer clusters from those on the MSLR and Yahoo datasets.
 		The speed of the former ones is about one time faster than the latter ones. 
 		Therefore the numbers may not be directly compared. 
	}
	\label{tois:tb:running_time}
	\centering
	\begin{tabular}{lrrr}
		\toprule
		& MSLR & ISTELLA & Yahoo  \\
		\midrule
		\# Rankers & $136$  &\phantom{10}$220$ & \phantom{10}$700$ \\
		\midrule
		\ac{MergeDTS} & $ 0.08$ &\phantom{10} $0.03$ & \phantom{10}$0.11$ \\
		\ac{MergeRUCB} & $0.08$ &\phantom{10} $0.03$& \phantom{10}$0.11$ \\
		Self-Sparring & $0.18$ &\phantom{10} $0.22$ & \phantom{10}$0.90$ \\
		\ac{DTS}& $5.23$ &\emph{9.88} & $\emph{100.27}$ \\
		\ac{RMED1}& $0.36$ &\phantom{10} $ 0.19$ & \phantom{0}$\emph{18.39}$ \\
		\ac{REX3}& $0.25$ &\phantom{10} $ 0.11$  & \phantom{10}$0.27$ \\				
		\bottomrule
	\end{tabular}
\end{table}

\noindent%
To address
\textbf{RQ1.2}, 
we report in \cref{tois:tb:running_time} the average running time (in days) of each algorithm in three large-scale dueling bandit setups, namely MSLR-Navigational, ISTELLA and Yahoo-Navigational.
As before, each algorithm is run for $10^8$ steps.
An individual run of  DTS  and \ac{RMED1} in the Yahoo-Navigational setup takes around $100.27$ and $18.39$ days, respectively, which is simply impractical for our experiments; therefore, the running time of DTS and \ac{RMED1} in this setup is estimated by multiplying the average running time at $10^5$ steps by $10^3$.
For a similar reason, we estimate the running time of DTS on the ISTELLA-Navigational setup by multiplying the average running time at $10^7$ by $10$. 

\cref{tois:tb:running_time} shows that \ac{MergeDTS} and \ac{MergeRUCB} have very low running times. 
This is due to the fact that they perform computations inside batches
and their computational complexity is $O(TM^2)$, where $T$ is the number of steps and $M$ is the size of batches. 
Moreover, \ac{MergeDTS} is considerably faster in the MSLR-Navigational setup, because there it finds the best ranker with fewer steps, as can be seen in \cref{tois:fig:large}.
After finding this best ranker, the size of batches $M$ becomes $1$ and, from that moment on, \ac{MergeDTS} does not perform any extra computations.
The running time of Self-Sparring is also low, but grows with the number of rankers, roughly linearly.
This is because at each step Self-Sparring draws a sample from the posterior distribution of each ranker and its running time is $\Omega(TK)$,
where $K$ is the number of rankers.
\ac{DTS} is orders of magnitude slower than other algorithms and its running time grows roughly quadratically, because \ac{DTS} requires a sample for each pair of rankers at each step and its running time is $\Omega(TK^2)$. 
 
Large-scale commercial search systems process over a billion queries a day \citep{gomes10google}, and run hundreds of different experiments \citep{kohavi13online} concurrently, in each of which two rankers are compared.  
The running time for \ac{DTS} and \ac{RMED1} that appears in \cref{tois:tb:running_time} is far beyond the realm of what might be considered reasonable to process on even $20\%$ of one day's traffic: note that one query in the search setting corresponds to one step for a dueling bandit algorithm, since each query could be used to compare two rankers by performing an interleaved comparison. 
Given the estimated running times listed in \cref{tois:tb:running_time}, we exclude \ac{DTS} and \ac{RMED1} from our experiments on the large-scale datasets for practical reasons.

\begin{figure}[h]
	\centering
	\includegraphics{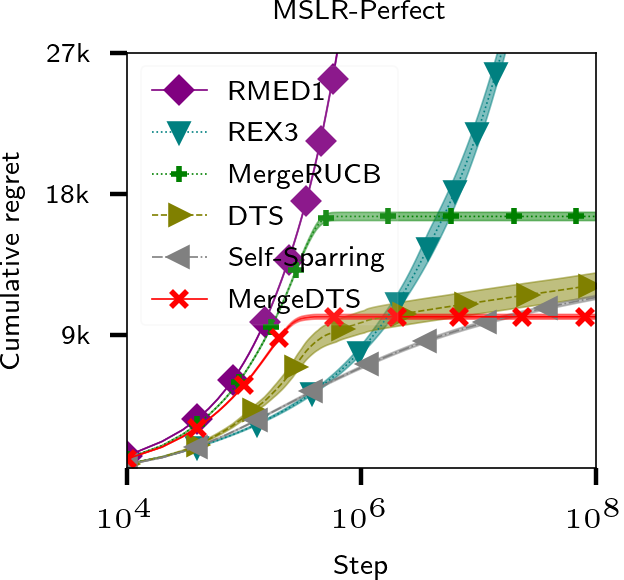}
	\includegraphics{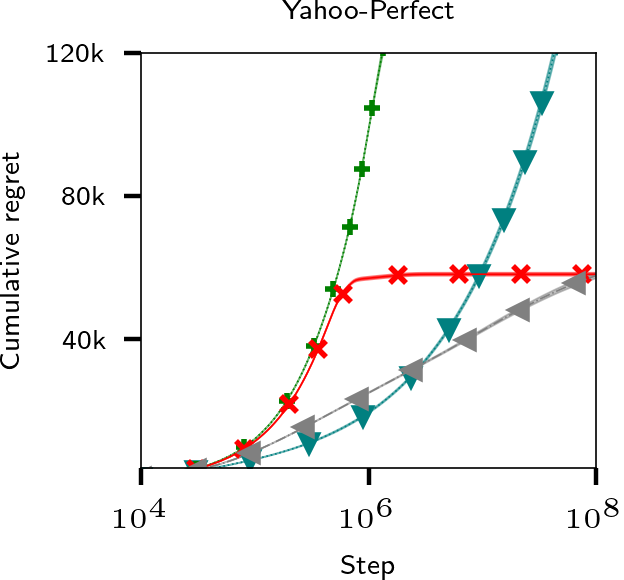}
	\includegraphics{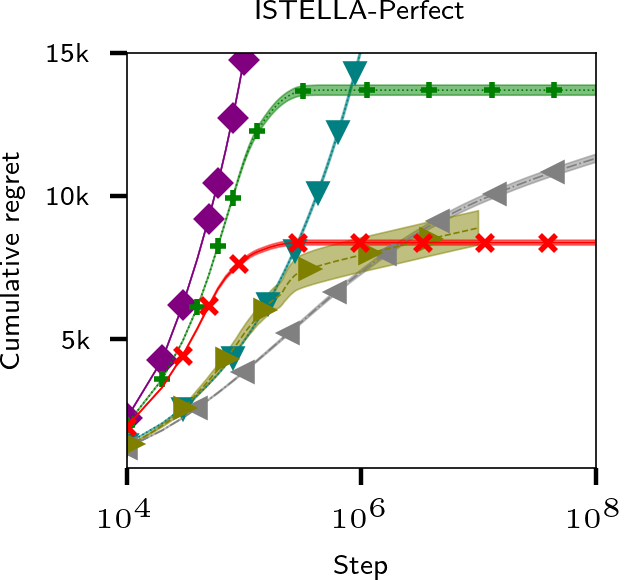}
	\vspace*{1mm}
	\includegraphics{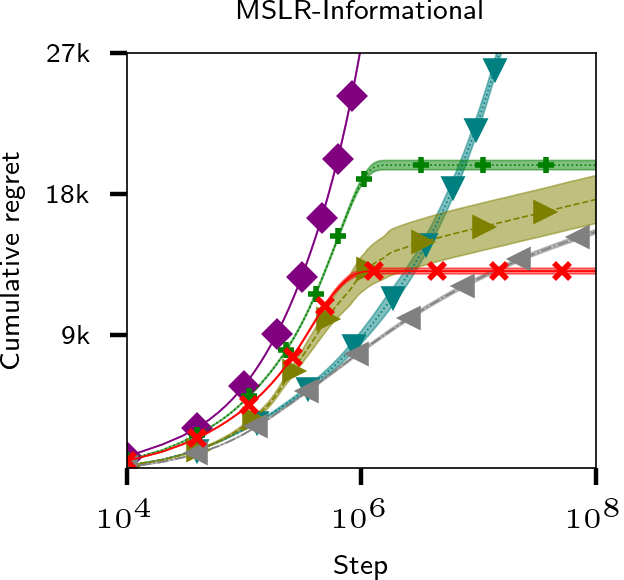}
	\includegraphics{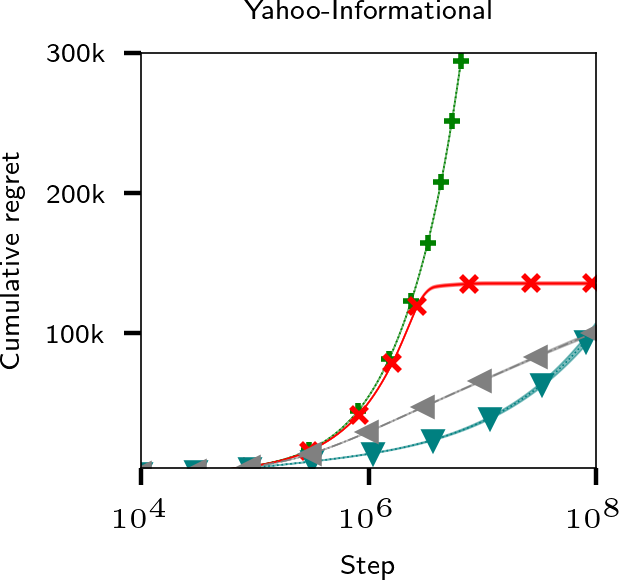}
	\includegraphics{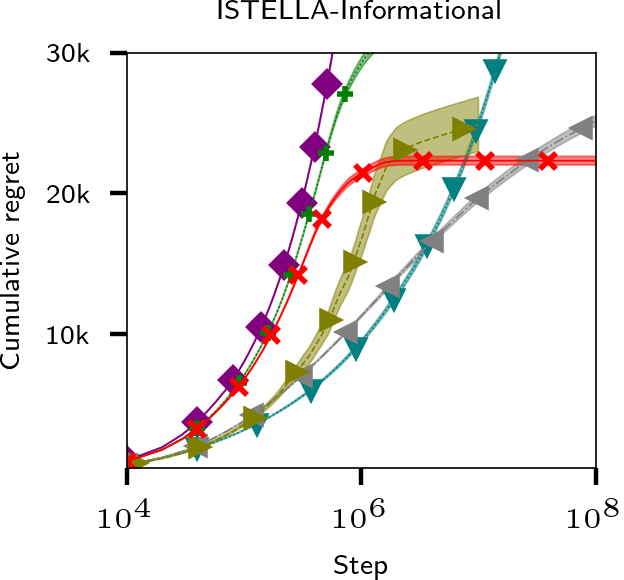}
	\caption{
		Effect  of the level of noise on cumulative regret in click feedback (top row: perfect configuration and bottom row: informational configuration). 
		The results are averaged over $50$ independent runs. (Contrast with figures in \cref{tois:fig:large}.)
	}
	\label{tois:fig:clicks}	
\end{figure}

\subsection{Impact of noise}
\label{tois:sec:noise}

To address 
\textbf{RQ1.3}, 
we run \ac{MergeDTS} in the perfect, navigational and informational configurations (see \cref{tois:sec:clicksimulation}). 
As discussed in \cref{tois:sec:clicksimulation}, the perfect configuration is the easiest one for dueling bandit algorithms to reveal the best ranker, while the informational configuration is the hardest. 
We report the results of the perfect and informational configurations in \cref{tois:fig:clicks}. 
For comparison, we refer readers to plots in \cref{tois:fig:large} for the results of navigational configuration. 

On the MSLR and ISTELLA datasets, in all three configurations, \ac{MergeDTS} with the chosen parameters outperforms the baselines, and the gaps get larger as click feedback gets noisier.
The results also show that Self-Sparring is severely affected by the level of noise. 
This is because Self-Sparring estimates the Borda score~\citep{urvoy13generic} of a ranker and, in our experiments, the noisier click feedback is, the closer the Borda scores are to each other, making it harder for Self-Sparring to identify the winner. 

Results on the Yahoo dataset disagree with results on the MSLR dataset. 
On the Yahoo dataset, \ac{MergeDTS} is affected more severely by the level of noise than Self-Sparring. 
This is because of the existence of uninformative rankers as stated in \cref{tois:assumption:1}. 
In noisier configurations, the gaps between uninformative and informative rankers are smaller, which results in the long time of comparisons for \ac{MergeDTS} to eliminate the uninformative rankers. 
Comparing those uninformative rankers leads to high regret. 

In summary, the performance of MergeDTS is largely affected when the gaps between rankers are small, which is consistent with our theoretical findings.

\begin{figure}
	\centering
	\includegraphics{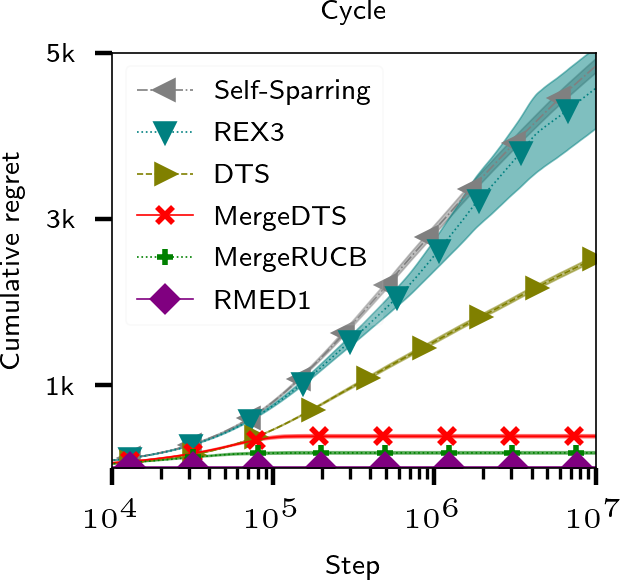}
	\includegraphics{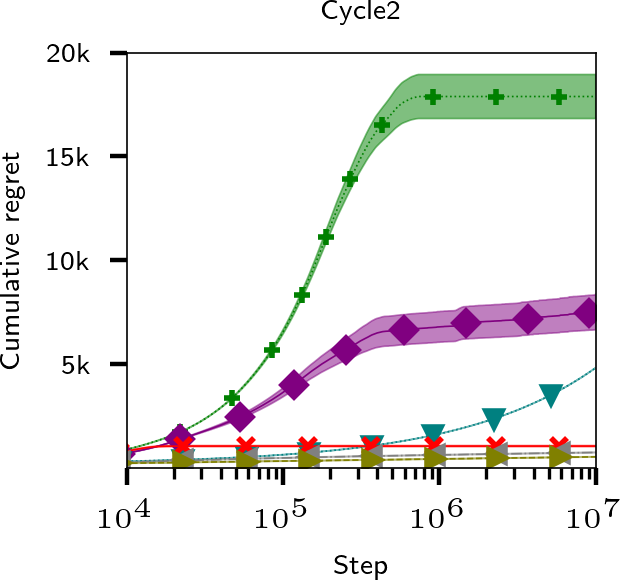}
	\includegraphics{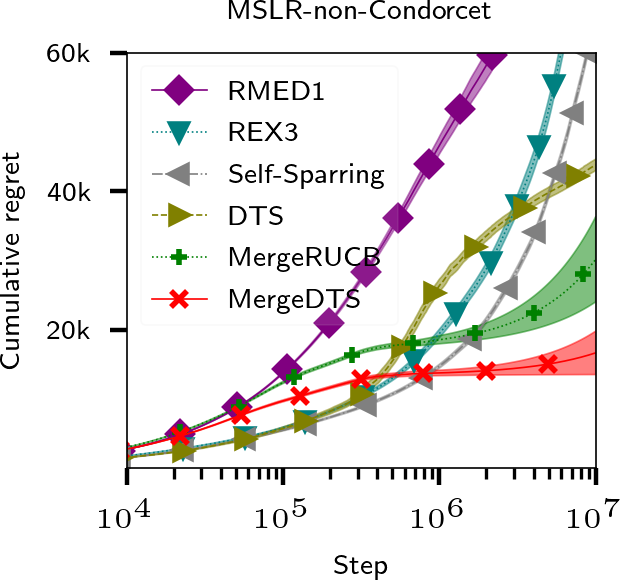}		
 
	\caption{Cumulative regret in cyclic and non-Condorcet setups. The results are averaged over $100$ independent runs.}
	\label{tois:fig:cycle}
\end{figure}	

\subsection{Cycle experiment}
\label{tois:sec:cycle}

We address
\textbf{RQ1.4} 
by running the algorithms that we consider on the Cycle and Cycle2 datasets introduced in \cref{tois:sec:dataset}.
Particularly, we have already observed that Self-Sparring performs well in some cases (see the above experiments and results), but we argue that Self-Sparring may perform poorly when a dueling bandit problem contains cyclic preference relationships. 
This has been identified as a point of grave concern in online evaluation~\citep{zoghi14relativea}.
Therefore, in this section we assess how dueling bandit algorithms behave when a dueling bandit problem contains cycles.

In this section  we conduct experiments for 10 million steps and repeat $100$ times since Merge-style algorithms converge to the Condorcet ranker  within less than 1 million steps and running longer only increases the gaps between \ac{MergeDTS} and baselines. 

For the Cycle dataset (the first plot in \cref{tois:fig:cycle}), 
the cumulative regret of Self-Sparring is an order of magnitude higher than that of \ac{MergeDTS}, 
although it performs well in some cases (see the above experiments). 
As we discussed in \cref{tois:sec:dataset},  Self-Sparring chooses rankers based on their Borda scores  and when 
the Borda scores of different arms become close to each other as in the Cycle dataset, Self-Sparring may perform poorly. 
Also, we  notice that when the gaps in Borda scores of the Condorcet winner and other rankers are large, Self-Sparring performs well, as shown in the middle plot in \cref{tois:fig:cycle}. 

Other than Self-Sparring, we also notice that the other baselines performs quite differently on the two cyclic configurations. 
In the harder configuration of the two, the Cycle dataset, only 
\ac{RMED1} and \ac{MergeRUCB} outperform \ac{MergeDTS}.
\ac{RMED1} excludes rankers from consideration based on relative preferences between two rankers. 
And, in the Cycle dataset, the preferences between suboptimal rankers are large. 
Thus, \ac{RMED1} can easily exclude a ranker based on its relative comparison to another suboptimal ranker. 
For the Cycle2 dataset, where  the relative preferences between two rankers are small, \ac{RMED1} performs worse than \ac{MergeDTS}. 

\ac{MergeRUCB} also slightly outperforms \ac{MergeDTS} on the Cycle dataset. 
This can be explained as follows.
In the Cycle dataset, the preference gap between the Condorcet winner and suboptimal rankers is small (i.e., $0.01$), while the gaps between suboptimal rankers are relatively large (i.e., $1.0$).
Under this setup, \ac{MergeDTS} tends to use the Condorcet winner to eliminate suboptimal rankers in the final stage. 
On the other hand, \ac{MergeRUCB} eliminates a ranker by another ranker who beats it with the largest probability. 
So, \ac{MergeDTS} requires more comparisons to eliminate suboptimal rankers than \ac{MergeRUCB}.
However, the gap between \ac{MergeRUCB} and \ac{MergeDTS} is small.

\subsection{Beyond the Condorcet assumption}
\label{tois:sec:noncondorcet}
To answer 
\textbf{RQ1.5}, 
we evaluate \ac{MergeDTS} on the MSLR-non-Condorcet dataset that does not contain a Condorcet winner. 
Instead, the dataset contains two Copeland winners and this dueling bandit setup is called the Copeland dueling bandit~\citep{zoghi15copeland,wu16double}.
The Copeland winner is selected by the Copeland score 
$\zeta_i = \frac{1}{K-1} \sum_{k\neq i} \mathds{1}(p_{ik} > 1/2)$ that measures the number of rankers beaten by ranker $a_i$. 
The Copeland winner is defined as $\zeta^* = \max_{1\le i \le K}\zeta_i$. 
In the MSLR-non-Condorcet dataset, each Copeland winner beats $30$ other rankers. 
Specifically, one of the Copeland winners beats the other one but is beaten by a suboptimal ranker. 
In the Copeland dueling bandit setup, regret is computed differently from the Condorcet dueling bandit setup. 
Given a pair of rankers $(a_i, a_j)$, regret at step $t$  is computed as: 
\begin{equation}
r_t =  \zeta^* - 0.5(\zeta_i + \zeta_j).
\end{equation}
Among the considered algorithms, only \ac{DTS} can solve the Copeland dueling bandit problem and is the state-of-the-art Copeland dueling bandit algorithm. 
We conduct the experiment for $10$ million steps with which \ac{DTS} converges to the Copeland winners. 
And we run each algorithm $100$ times independently. 
The results are shown in the last plot in \cref{tois:fig:cycle}.

\ac{MergeDTS} has the lowest cumulative regret. 
However, in our experiments, we find that \ac{MergeDTS} eliminates the two Copeland winners one time out of $100$ individual repeats. 
In the other $99$ repeats, we find that \ac{MergeDTS} eliminates one of the two existing winners, which may not be ideal in practice.
Note that we evaluate \ac{MergeDTS} in a relatively easy setup, where only two Copeland winners are considered. 
For more complicated setups, where more than two Copeland winners are considered or the Copeland winners are beaten by several  suboptimal rankers, we speculate that \ac{MergeDTS} can fail more frequently. 
In our experiments, we do not evaluate \ac{MergeDTS} in the more complicated setups, because \ac{MergeDTS} is designed for the Condorcet dueling bandits and is only guaranteed to work under the Condorcet assumption. 
The answer to 
\textbf{RQ1.5}  
is that \ac{MergeDTS} may perform well for some easy setups that go beyond the Condorcet assumption without any guarantees. 

\subsection{Parameter sensitivity }
\label{tois:sec:parameter}

\begin{figure*}
	\centering
	\includegraphics {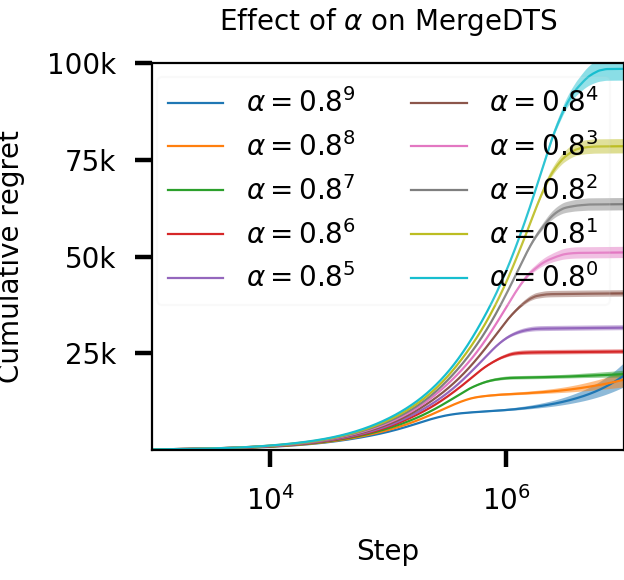}
	\includegraphics {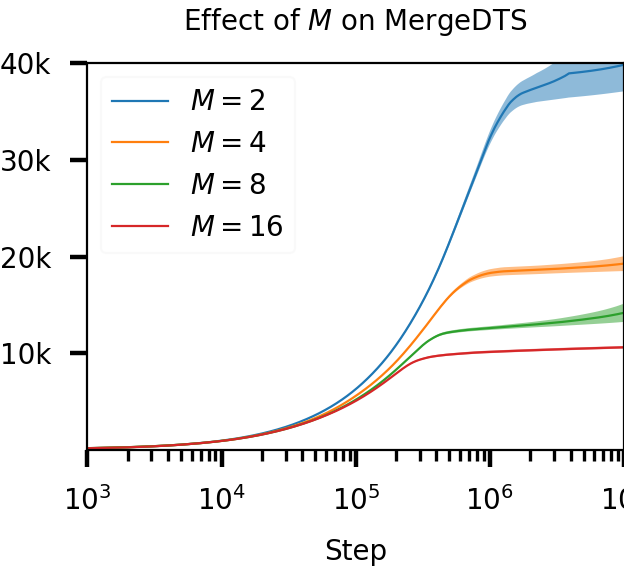}
	\includegraphics {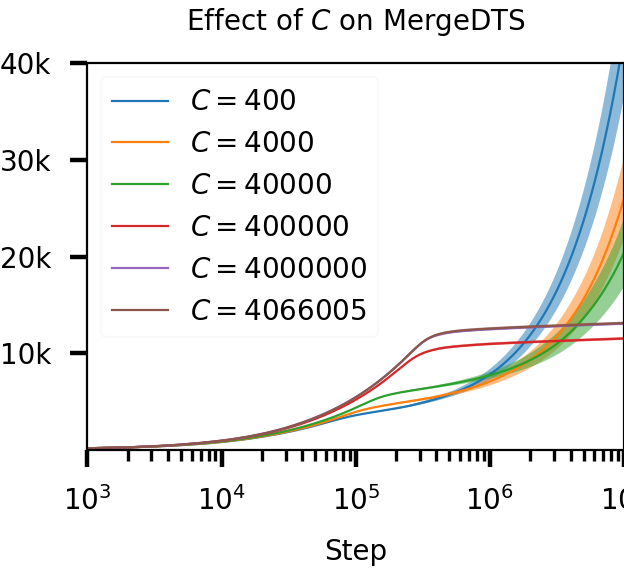}
	\caption{Effect of the parameters $\alpha$, $M$, and $C$ on the performance of \ac{MergeDTS} in the MSLR-Navigational setup. The results are averaged over $100$ independent runs. The shaded areas are $\pm$ standard error.}
	\label{tois:fig:dependence}
\end{figure*}

We answer
\textbf{RQ1.6}
and analyze the parameter sensitivity of \ac{MergeDTS} using the setup described in \cref{tois:sec:parameters setup}. 
Since \ac{MergeDTS} converges to the Condorcet winner within $10$ million steps, we conduct the experiments with $10$ million steps and repeat $100$ times. 
Recall that we conduct the experiments in the MSLR-Navigational setup. 
The  results are reported in \cref{tois:fig:dependence}. 
We also report the standard errors in the plots. 

The left plot in \cref{tois:fig:dependence} shows the effect of the exploration parameter $\alpha$ on the performance of \ac{MergeDTS}.
First, lowering $\alpha$ can significantly increase the performance, e.g., the cumulative regret for $\alpha=0.8^4$ is about one third of the reward for $\alpha=1.0$ (which is close to the theoretically optimal value $\alpha=1.01$).
Second, as we decrease $\alpha$, the number of failures increases, where a failure is an event that \ac{MergeDTS} eliminates the Condorcet winner: 
with $\alpha=\{0.8^9, 0.8^8, 0.8^7\}$ we observe $10$, $4$, $1$ failures, respectively, and, thus, the cumulative regret increases linearly w.r.t.~$T$. 
Since in practice we do not want to eliminate the best ranker, we choose $\alpha = 0.8^6 \approx 0.2621$ in our experiments.

The middle plot in \cref{tois:fig:dependence} shows the effect of the batch size $M$.
The larger the batch size, the lower the regret. 
This can be explained as follows. 
The \ac{DTS}-based strategy uses the full local knowledge in a batch to choose the best ranker.  
A larger batch size $M$ provides more knowledge to \ac{MergeDTS} to make decisions, which leads to a better choice of rankers. 
But the time complexity of \ac{MergeDTS} is $O(TM^2)$, i.e., quadratic in the batch size.
Thus, for realistic scenarios we cannot increase $M$ indefinitely.
We choose $M=16$ as a tradeoff between effectiveness (cumulative regret) and efficiency (running time).

The right plot in \cref{tois:fig:dependence} shows the dependency of \ac{MergeDTS} on $C$.
Similarly to the effect of $\alpha$, lower values of $C$ lead to lower regret, but also to a larger number of failures.
$C=\numprint{4000000}$ is the lowest value that does not lead to any failures, so we choose it in our experiments. 

In summary, the theoretical constraints on the parameters of MergeDTS are rather conservative. 
There is a range of values for the key parameters $\alpha$, $M$ and $C$, where the theoretical guarantees fail to hold, but where MergeDTS performs better than it would if we were to constrain ourselves only to values permitted by theory.

\section{Conclusion}
\label{tois:sec:conclusion}

In this chapter, we have studied the large-scale online ranker evaluation problem under the Condorcet assumption, which can be formalized as a $K$-armed  dueling bandit problem.
We have answered \textbf{RQ1} by introducing a scalable version of the state-of-the-art Double Thompson Sampling algorithm, which we call \ac{MergeDTS}.  
Our experiments have shown that, by choosing the parameter values outside of the theoretical regime, \ac{MergeDTS} is considerably more efficient than \ac{DTS} in terms of computational complexity, and that it significantly outperforms the large-scale state-of-the-art algorithm \ac{MergeRUCB}. 
Furthermore, we have demonstrated the robustness of \ac{MergeDTS} when dealing with difficult dueling bandit problems containing cycles among the arms. 
Lastly, we have shown that the performance of \ac{MergeDTS} is guaranteed if the parameter values fall within the theoretical regime. 

Several interesting directions for future work arise from this chapter:  
\begin{enumerate*}[label=(\arabic*)]
	\item 
		In our experiments, we have shown that there is a large gap between theory and practice. 
		It will be interesting to study this gap and provide a tighter theoretical bound. 
	\item We only study dueling bandits in this chapter. 
	We believe that it is interesting to study a generalization of \ac{MergeDTS}, as well as the theoretical analysis presented here, to the case of online ranker evaluation tasks with a multi-dueling setup.
	\item Since multi-dueling bandits also compare multiple rankers at each step based on relative feedback, it is an interesting direction to compare dueling bandits to multi-dueling bandits in the large-scale setup.
	\item We suspect that the \ac{UCB}-based elimination utilized in \ac{MergeDTS} is too conservative, it might be that more recent minimum empirical divergence based techniques~\citep{komiyama16copeland} may be leveraged to speed up the elimination of the rankers. 
	\item The feature rankers are chosen as arms in our experiments. 
	A more interesting and realistic way of choosing arms is to use well trained learning to rank algorithms. 
\end{enumerate*}


\chapter{Safe Online Learning to Re-Rank}
\label{chapter:uai19}

\footnote[]{This chapter was published as~\citep{li19bubblerank}.}

\acresetall

In this chapter, we answer the following research question: 
\begin{description}
	\item[\textbf{RQ2}] How to achieve safe online learning to re-rank? 
\end{description}
Particularly, we study the problem of safe online learning to re-rank, where user feedback is used to improve the quality of displayed lists. 
We propose $\bubblerank$, a bandit algorithm for safe re-ranking that combines the strengths of both the offline and online settings. 
The algorithm starts with an initial base list and improves it online by gradually exchanging higher-ranked less attractive items for lower-ranked more attractive items. 
We prove an upper bound on the $n$-step regret of $\bubblerank$ that degrades gracefully with the quality of the initial base list. 
Our theoretical findings are supported by extensive experiments on a large-scale real-world click dataset.

\section{Introduction}
\label{uai:sec:introduction}

Learning to rank (LTR) is an important problem in many application domains, such as information retrieval, ad placement, and recommender systems \citep{liu09learning}. More generally, LTR arises in any situation where multiple items, such as web pages, are presented to users. It is particularly relevant when the diversity of users makes it hard to decide which item should be presented to a specific user~\citep{radlinski08learning,yue11linear}.

A traditional approach to LTR is offline learning of rankers from either relevance labels created by judges \citep{qin10letor} or user interactions \citep{joachims02optimizing,mcmahan13ad}. Recent experimental results \citep{zoghi16click} shows that such rankers, even in a highly-optimized search engine, can be improved by online LTR with exploration. Exploration is the key component in multi-armed bandit algorithms \citep{auer02finite}. Many such algorithms have been proposed recently for online LTR in specific user-behavior models \citep{katariya16dcm,kveton15cascading,lagree16multiple}, the so-called \emph{click models} \citep{chuklin15click}. 
Compared to earlier online LTR algorithms \citep{radlinski08learning}, these click model-based algorithms gain in statistical efficiency while giving up on generality. Empirical results indicate that click model-based algorithms are likely to be beneficial in practice.

Yet, existing algorithms for online LTR in click models are impractical for at least three reasons. First, an actual model of user behavior is typically unknown. 
This problem was initially addressed by \citet{zoghi17online}. They showed that the list of items in the descending order of relevance is optimal in several click models and proposed $\batchrank$ for learning it. 
Then \citet{lattimore18toprank} built upon this work and proposed $\toprank$, which is the state-of-the-art online LTR algorithm. 
Second, these algorithms lack safety constraints and explore aggressively by placing potentially irrelevant items at high positions, which may significantly degrade user experience~\citep{wang18position}. A third and related problem is that the algorithms are not well suited for so-called \emph{warm start} scnearios~\citep{vorobev15gathering}, where the offline-trained production ranker already generates a good list, which only needs to be safely improved. 
Warm-starting an online LTR algorithm is challenging since existing posterior sampling algorithms, such as Thompson sampling~\citep{thompson33likelihood}, require item-level priors while only list-level priors are available practically. 

We make the following contributions. First, motivated by the exploration scheme of \citet{radlinski06minimally}, we propose a bandit algorithm for online LTR that addresses all three issues mentioned above. The proposed algorithm gradually improves upon an initial base list by exchanging  higher-ranked less attractive items for lower-ranked more attractive items. The algorithm resembles bubble sort \citep{cormen09introduction}, and therefore we call it $\bubblerank$. Second, we prove an upper bound on the $n$-step regret of $\bubblerank$. The bound reflects the behavior of $\bubblerank$: worse initial base lists lead to a higher regret.
Third, we define our safety constraint, which is based on incorrectly-ordered item pairs in the ranked list, and prove that $\bubblerank$ never violates this constraint with a high probability. Finally, we evaluate $\bubblerank$ extensively on a large-scale real-world click dataset.


\section{Background}
\label{uai:sec:background}

This section introduces our online learning problem. We first review click models \citep{chuklin15click} and then introduce a stochastic click bandit~\citep{zoghi17online}, a learning to rank framework for multiple click models.

The following notation is used in the rest of the chapter. We denote $\set{1, \dots, n}$ by $[n]$. For any sets $A$ and $B$, we denote by $A^B$ the set of all vectors whose entries are indexed by $B$ and take values from $A$. We use boldface letters to denote random variables.

\subsection{Click models}
\label{uai:sec:click models}

A \emph{click model} is a model of how a user clicks on a list of documents. We refer to the documents as \emph{items} and denote the universe of all items by $\cD = [L]$. The user is presented a \emph{ranked list}, an ordered list of $K$ documents out of $L$. We denote this list by $\cR \in \Pi_K(\cD)$, where $\Pi_K(\cD)$ is the set of all $K$-tuples with distinct items from $\cD$. We denote by $\cR(k)$ the item at position $k$ in $\cR$; and by $\cR^{-1}(i)$ the position of item $i$ in $\cR$, if item $i$ is in $\cR$.

Many click models are parameterized by \emph{item-dependent attraction probabilities} $\alpha \in [0, 1]^L$, where $\alpha(i)$ is the \emph{attraction probability} of item $i$. We discuss the two most fundamental click models below.

In the \emph{cascade model} (CM)~\citep{craswell08experimental}, the user scans list $\cR$ from the first item $\cR(1)$ to the last $\cR(K)$. If item $\cR(k)$ is \emph{attractive}, the user \emph{clicks} on it and does not examine the remaining items. If item $\cR(k)$ is not attractive, the user \emph{examines} item $\cR(k + 1)$. The first item $\cR(1)$ is examined with probability one. Therefore, the expected number of clicks is equal to the probability of clicking on any item, and is $r(\cR) = \sum_{k = 1}^K \chi(\cR, k) \alpha(\cR(k))$, where $\chi(\cR, k) = \prod_{i = 1}^{k - 1} (1 - \alpha(\cR(i)))$ is the examination probability of position $k$ in list $\cR$.

In the \emph{position-based model} (PBM)~\citep{richardson07predicting}, the probability of clicking on an item depends on both its identity and position. Therefore, in addition to item-dependent attraction probabilities $\alpha$, the PBM is parameterized by $K$ \emph{position-dependent examination probabilities} $\chi \in [0, 1]^K$, where $\chi(k)$ is the examination probability of position $k$. The user interacts with list $\cR$ as follows. The user \emph{examines} position $k \in [K]$ with probability $\chi(k)$ and then \emph{clicks} on item $\cR(k)$ at that position with probability $\alpha(\cR(k))$. Therefore, the expected number of clicks on list $\cR$ is $r(\cR) = \sum_{k = 1}^K \chi(k) \alpha(\cR(k))$.

CM and PBM are similar models, because the probability of clicking factors into item and position dependent factors. Therefore, both in the CM and PBM, under the assumption that $\chi(1) \geq \dots \geq \chi(K)$, the expected number of clicks is maximized by listing the $K$ most attractive items in descending order of their attraction. More precisely, the most clicked list is
\begin{align}
  \cR^\ast = (1, \dots, K)
  \label{uai:eq:optimal list}
\end{align}
when $\alpha(1) \geq \dots \geq \alpha(L)$. Therefore, perhaps not surprisingly, the problem of learning the optimal list in both models can be viewed as the same problem, a stochastic click bandit \citep{zoghi17online}.

\subsection{Stochastic click bandit}
\label{uai:sec:stochastic click bandit}

An instance of a \emph{stochastic click bandit} \citep{zoghi17online} is a tuple $(K, L, P_\alpha, P_\chi)$, where $K \leq L$ is the number of positions, $L$ is the number of items, $P_\alpha$ is a distribution over binary attraction vectors $\set{0, 1}^L$, and $P_\chi$ is a distribution over binary examination matrices $\set{0, 1}^{\Pi_K(\cD) \times K}$.

The learning agent interacts with the stochastic click bandit as follows. At time $t$, it chooses a list $\rnd{\cR}_t \in \Pi_K(\cD)$, which depends on its history up to time $t$, and then observes \emph{clicks} $\rnd{c}_t \in \set{0, 1}^K$ on all positions in $\rnd{\cR}_t$. A position is clicked if and only if it is examined and the item at that position is attractive. More specifically, for any $k \in [K]$,
\begin{align}
  \rnd{c}_t(k) = \rnd{X}_t(\rnd{\cR}_t, k) \rnd{A}_t(\rnd{\cR}_t(k)),
  \label{uai:eq:click}
\end{align}
where $\rnd{X}_t \in \set{0, 1}^{\Pi_K(\cD) \times K}$ and $\rnd{X}_t(\cR, k)$ is the \emph{examination indicator} of position $k$ in list $\cR \in \Pi_K(\cD)$ at time $t$; and $\rnd{A}_t \in \set{0, 1}^L$ and $\rnd{A}_t(i)$ is the \emph{attraction indicator} of item $i$ at time $t$. Both $\rnd{A}_t$ and $\rnd{X}_t$ are stochastic and drawn i.i.d. from $P_\alpha \otimes P_\chi$.

The key assumption that allows learning in this model is that the attraction of any item is independent of the examination of its position. In particular, for any list $\cR \in \Pi_K(\cD)$ and position $k \in [K]$,
\begin{align}
  \condE{\rnd{c}_t(k)}{\rnd{\cR}_t = \cR} =
  \chi(\cR, k) \alpha(\cR(k)),
\end{align}
where $\alpha = \E{\rnd{A}_t}$ and $\alpha(i)$ is the \emph{attraction probability} of item $i$; and $\chi = \E{\rnd{X}_t}$ and $\chi(\cR, k)$ is the \emph{examination probability} of position $k$ in $\cR$. Note that the above independence assumption is in expectation only. We do not require that the clicks are independent of the position or other displayed items.

The \emph{expected reward} at time $t$ is the expected number of clicks at time $t$. Based on our independence assumption, $\sum_{k = 1}^K \E{\rnd{c}_t(k)} = r(\rnd{\cR}_t, \alpha, \chi)$, where $r(\cR, A, X) = \sum_{k = 1}^K X(\cR, k) A(\cR(k))$ for any $\cR \in \Pi_K(\cD)$, $A \in [0, 1]^L$, and $X \in [0, 1]^{\Pi_K(\cD) \times K}$. The learning agent maximizes the expected number of clicks in $n$ steps. This problem can be equivalently viewed as minimizing the \emph{expected cumulative regret} in $n$ steps, which we define as
\begin{align}
  \mbox{}\hspace*{-2mm}R(n) {=}
  \sum_{t = 1}^n \E{\max_{\cR \in \Pi_K(\cD)} r(\cR, \alpha, \chi) - r(\rnd{\cR}_t, \alpha, \chi)}
  \hspace*{-.75mm}\mbox{}.
  \label{uai:eq:regret}
\end{align}

\section{Online Learning to Re-Rank}
\label{uai:sec:learning to re-rank}

Multi-stage ranking is widely used in production ranking systems \citep{chen17efficient,karmakersantu17application,liu17cascade}, with the re-ranking stage at the very end \citep{chen17efficient}. In the re-ranking stage, a relatively small number of items, typically $10$--$20$, are re-ranked. One reason for re-ranking is that offline rankers are typically trained to minimize the average loss across a large number of queries. Therefore, they perform well on very frequent queries and poorly on infrequent queries. On moderately frequent queries, the so-called \emph{torso queries}, their performance varies. As torso queries are sufficiently frequent, an online algorithm can be used to re-rank so as to optimize their value, such as the number of clicks \citep{zoghi16click}.

We propose an online algorithm that addresses the above problem and adaptively re-ranks a list of items generated by a production ranker with the goal of placing more attractive items at higher positions. We study a non-contextual variant of the problem, where we re-rank a small number of items in a single query. Generalization across queries and items is an interesting direction for future work. We follow the setting in \cref{uai:sec:stochastic click bandit}, except that $\cD = [K]$. Despite these simplifying assumptions, our learning problem remains a challenge. In particular, the attraction of items is only observed through clicks in \cref{uai:eq:click}, which are affected by other items in the list.

\subsection{Algorithm}
\label{uai:sec:algorithm}
\begin{algorithm}[t]
	\begin{algorithmic}[1]
		\REQUIRE Initial list $\cR_0$ over $[K]$ 
		\vspace{0.075in}
		\STATE $\forall i, j \in [K]: \rnd{s}_0(i, j) \gets 0, \ \rnd{n}_0(i, j) \gets 0$
		\STATE $\bar{\rnd{\cR}}_1 \gets \cR_0$ \label{uai:alg:3}
		\FOR{$t = 1, \dots, n$}
		\STATE $h \gets t \ \mathrm{mod} \ 2$
		\vspace{0.075in}
		\STATE $\rnd{\cR}_t \gets \bar{\rnd{\cR}}_t$ \label{uai:alg:6}
		\FOR{$k = 1, \dots, \floors{(K - h) / 2}$}
		\STATE $i \gets \rnd{\cR}_t(2 k - 1 + h), \ j \gets \rnd{\cR}_t(2 k + h)$
		\IF{$\rnd{s}_{t - 1}(i, j) \leq 2 \sqrt{\rnd{n}_{t - 1}(i, j) \log(1 / \delta)}$} \label{uai:alg:10}
		\STATE Randomly exchange items $\rnd{\cR}_t(2 k - 1 + h)$ and $\rnd{\cR}_t(2 k + h)$ in list $\rnd{\cR}_t$
		\ENDIF \label{uai:alg:11}
		\ENDFOR
		\vspace{0.075in}
		\STATE Display list $\rnd{\cR}_t$ and observe clicks $\rnd{c}_t \in \{0, 1\}^K$ \label{uai:alg:13}
		\vspace{0.075in}
		\STATE $\rnd{s}_t \gets \rnd{s}_{t - 1}, \ \rnd{n}_t \gets \rnd{n}_{t - 1}$ \label{uai:alg:15}
		\FOR{$k = 1, \dots, \floors{(K - h) / 2}$}
		\STATE $i \gets \rnd{\cR}_t(2 k - 1 + h), \ j \gets \rnd{\cR}_t(2 k + h)$
		\IF{$\abs{\rnd{c}_t(2 k - 1 + h) - \rnd{c}_t(2 k + h)} = 1$} \label{uai:alg:18}
		\STATE $\rnd{s}_t(i, j) \gets \rnd{s}_t(i, j) + \rnd{c}_t(2 k - 1 + h) - \rnd{c}_t(2 k + h)$
		\STATE $\rnd{n}_t(i, j) \gets \rnd{n}_t(i, j) + 1$
		\STATE $\rnd{s}_t(j, i) \gets \rnd{s}_t(j, i) + \rnd{c}_t(2 k + h) - \rnd{c}_t(2 k - 1 + h)$
		\STATE $\rnd{n}_t(j, i) \gets \rnd{n}_t(j, i) + 1$ \label{uai:alg:22}
		\ENDIF
		\ENDFOR
		\vspace{0.075in}
		\STATE $\bar{\rnd{\cR}}_{t + 1} \gets \bar{\rnd{\cR}}_t$ \label{uai:alg:24}
		\FOR{$k = 1, \dots, K - 1$}
		\STATE $i \gets \bar{\rnd{\cR}}_{t + 1}(k), \ j \gets \bar{\rnd{\cR}}_{t + 1}(k + 1)$
		\IF{$\rnd{s}_t(j, i) > 2 \sqrt{\rnd{n}_t(j, i) \log(1 / \delta)}$} \label{uai:alg:27}
		\STATE Exchange items $\bar{\rnd{\cR}}_{t + 1}(k)$ and $\bar{\rnd{\cR}}_{t + 1}(k + 1)$ in list $\bar{\rnd{\cR}}_{t + 1}$ \label{uai:alg:28}
		\ENDIF
		\ENDFOR
		\ENDFOR
	\end{algorithmic}
	\caption{$\bubblerank$}
	\label{uai:alg:bubble rank}
\end{algorithm}

Our algorithm is presented in \cref{uai:alg:bubble rank}. The algorithm gradually improves upon an \emph{initial base list} $\cR_0$ by ``bubbling up'' more attractive items. Therefore, we refer to it as $\bubblerank$. $\bubblerank$ determines more attractive items by randomly exchanging neighboring items. If the lower-ranked item is found to be more attractive, the items are permanently exchanged and never randomly exchanged again. If the lower-ranked item is found to be less attractive, the items are never randomly exchanged again. We describe $\bubblerank$ in detail below.

$\bubblerank$ maintains a \emph{base list} $\bar{\rnd{\cR}}_t$ at each time $t$. From the viewpoint of $\bubblerank$, this is the best list at time $t$. The list is initialized by the initial base list $\cR_0$ (line~\ref{uai:alg:3}). At time $t$, $\bubblerank$  permutes $\bar{\rnd{\cR}}_t$ into a \emph{displayed list} $\rnd{\cR}_t$ (lines~\ref{uai:alg:6}--\ref{uai:alg:11}). Two kinds of permutations are employed. If $t$ is odd and so $h = 0$, the items at positions $1$ and $2$, $3$ and $4$, and so on, are randomly exchanged. If $t$ is even and so $h = 1$, the items at positions $2$ and $3$, $4$ and $5$, and so on are randomly exchanged. The items are exchanged only if $\bubblerank$ is uncertain regarding which item is more attractive~(line~\ref{uai:alg:10}).

The list $\rnd{\cR}_t$ is displayed and $\bubblerank$ gets feedback (line~\ref{uai:alg:13}). Then it updates its statistics (lines~\ref{uai:alg:15}--\ref{uai:alg:22}). For any exchanged items $i$ and $j$, if item $i$ is clicked and item $j$ is not, the belief that $i$ is more attractive than $j$, $\rnd{s}_t(i, j)$, increases; and the belief that $j$ is more attractive than $i$, $\rnd{s}_t(j, i)$, decreases. The number of observations, $\rnd{n}_t(i, j)$ and $\rnd{n}_t(j, i)$, increases. These statistics are updated only if one of the items is clicked~(line~\ref{uai:alg:18}), not both.

At the end of time $t$, the base list $\bar{\rnd{\cR}}_t$ is improved (lines~\ref{uai:alg:24}--\ref{uai:alg:28}). More specifically, if any lower-ranked item $j$ is found to be more attractive than its higher-ranked neighbor $i$ (line~\ref{uai:alg:27}), the items are permanently exchanged in the next base list $\bar{\rnd{\cR}}_{t + 1}$.

A notable property of $\bubblerank$ is that it explores safely, since any item in the displayed list $\rnd{\cR}_t$ is at most one position away from its position in the base list $\bar{\rnd{\cR}}_t$. Moreover, any base list improves upon the initial base list $\cR_0$, because it is obtained by bubbling up more attractive items with a high confidence. We make this notion of safety more precise in \cref{uai:sec:safety}.

\section{Theoretical Analysis}
\label{uai:sec:analysis}

In this section, we provide theoretical guarantees on the performance of $\bubblerank$, by bounding the $n$-step regret in \cref{uai:eq:regret}.

The content is organized as follows. In \cref{uai:sec:regret bound}, we present our upper bound on the $n$-step regret of $\bubblerank$, together with our assumptions. In \cref{uai:sec:safety}, we prove that $\bubblerank$ is safe. In \cref{uai:sec:discussion}, we discuss our theoretical results. The regret bound is proved in \cref{uai:sec:upper bound proof}. Our technical lemmas are stated and proved in \cref{uai:sec:technical lemmas}.

\subsection{Regret bound}
\label{uai:sec:regret bound}

Before we present our result, we introduce our assumptions\footnote{Our assumptions are slightly weaker than those of  \citet{zoghi17online}. For instance, Assumption~\ref{ass:order-free examination} is on the probability of examination. \citet{zoghi17online} make this assumption on the realization of examination.} and complexity metrics.

\begin{assumption}
\label{uai:sec:assumptions} For any lists $\cR, \cR' \in \Pi_K(\cD)$ and positions $k, \ell \in [K]$ such that $k < \ell$:
\begin{enumerate}[label=A\arabic*.,align=left,leftmargin=*,ref=A\arabic*]
\item \label{ass:optimal list} $r(\cR, \alpha, \chi) \leq r(\cR^\ast, \alpha, \chi)$, where $\cR^\ast$ is defined in \cref{uai:eq:optimal list};
\item \label{ass:order-free examination} $\set{\cR(1), \dots, \cR(k - 1)} = \set{\cR'(1), \dots, \cR'(k - 1)}$ $\implies \chi(\cR, k) = \chi(\cR', k)$;
\item \label{ass:decreasing examination} $\chi(\cR, k) \geq \chi(\cR, \ell)$;
\item \label{ass:examination scaling} If $\cR$ and $\cR'$ differ only in that the items at positions $k$ and $\ell$ are exchanged, then $\alpha(\cR(k)) \leq \alpha(\cR(\ell)) \iff \chi(\cR, \ell) \geq \chi(\cR', \ell)$; and
\item \label{ass:lowest examination} $\chi(\cR, k) \geq \chi(\cR^\ast, k)$.
\end{enumerate}
\end{assumption}

The above assumptions hold in the CM. In the PBM, they hold when the examination probability decreases with the position.

Our assumptions can be interpreted as follows. Assumption~\ref{ass:optimal list} says that the list of items in the descending order of attraction probabilities is optimal. Assumption~\ref{ass:order-free examination} says that the examination probability of any position depends only on the identities of higher-ranked items. Assumption~\ref{ass:decreasing examination} says that a higher position is at least as examined as a lower position. Assumption~\ref{ass:examination scaling} says that a higher-ranked item is less attractive if and only if it increases the examination of a lower position. Assumption~\ref{ass:lowest examination} says that any position is examined the least in the optimal list.

To simplify our exposition, we assume that $\alpha(1) > \dots > \alpha(K) > 0$. Let $\chi_{\max} = \chi(\cR^\ast, 1)$ denote the \emph{maximum examination probability}, $\chi_{\min} = \chi(\cR^\ast, K)$ denote the \emph{minimum examination probability}, and
\begin{align*}
  \Delta_{\min} =
  \min_{k \in [K - 1]} \alpha(k) - \alpha(k + 1)
\end{align*}
be the \emph{minimum gap}. Then the $n$-step regret of $\bubblerank$ can be bounded as follows.

\begin{theorem}
\label{uai:thm:upper bound} In any stochastic click bandit that satisfies \cref{uai:sec:assumptions}, and for any $\delta \in (0, 1)$, the expected $n$-step regret of $\bubblerank$ is bounded as 
\begin{align*}
  R(n) \leq 
  180 K \frac{\chi_{\max}}{\chi_{\min}} \frac{K - 1 + 2 \abs{\cV_0}}{\Delta_{\min}} \log(1 / \delta) +
  \delta^\frac{1}{2} K^3 n^2\,.
\end{align*}
\end{theorem}

\subsection{Safety}
\label{uai:sec:safety}

Let
\begin{align}
  \mbox{}\hspace*{-4mm}
  \cV(\cR) {=}
  \set{(i, j) \in [K]^2: i < j, \cR^{-1}(i) > \cR^{-1}(j)}
  \hspace*{-3mm}\mbox{}
  \label{uai:eq:safety}
\end{align}
be the set of \emph{incorrectly-ordered item pairs} in list $\cR$. Then our algorithm is safe in the following sense.

\begin{lemma}
\label{uai:lem:safety} Let
\begin{align}
  \cV_0 =
  \cV(\cR_0)
  \label{uai:eq:initial safety}
\end{align}
be the incorrectly-ordered item pairs in the initial base list $\cR_0$. Then the number of incorrectly-ordered item pairs in any displayed list $\rnd{\cR}_t$ is at most $\abs{\cV_0} + K / 2$, that is $\abs{\cV(\rnd{\cR}_t)} \leq \abs{\cV_0} + K / 2$ holds uniformly over time with probability of at least $1 - \delta^\frac{1}{2} K^2 n$.
\end{lemma}
\begin{proof}
Our claim follows from two observations. First, by the design of $\bubblerank$, any displayed list $\rnd{\cR}_t$ contains at most $K / 2$ item pairs that are ordered differently from its base list $\bar{\rnd{\cR}}_t$. Second, no base list $\bar{\rnd{\cR}}_t$ contains more incorrectly-ordered item pairs than $\cR_0$ with a high probability. In particular, under event $\cE$ in \cref{uai:lem:concentration}, any change in the base list (line~\ref{uai:alg:28} of $\bubblerank$) reduces the number of incorrectly-order item pairs by one. In \cref{uai:lem:concentration}, we prove that $\mathbb{P}(\cE) \geq 1 - \delta^\frac{1}{2} K^2 n$.
\end{proof}

\subsection{Discussion}
\label{uai:sec:discussion}

Our upper bound on the $n$-step regret of $\bubblerank$ (\cref{uai:thm:upper bound}) is $O(\Delta_{\min}^{-1} \log n)$ for $\delta = n^{-4}$. This dependence is considered to be optimal in gap-dependent bounds. Our gap $\Delta_{\min}$ is the minimum difference in the attraction probabilities of items, and reflects the hardness of sorting the items by their attraction probabilities. This sorting problem is equivalent to the problem of learning $\cR^\ast$. So, a gap like $\Delta_{\min}$ is expected, and is the same as that in \citep{zoghi17online}.

Our regret bound is notable because it reflects two key characteristics of $\bubblerank$. First, the bound is linear in the number of incorrectly-ordered item pairs in the initial base list $\cR_0$. This suggests that $\bubblerank$ should have lower regret when initialized with a better list of items. We validate this dependence empirically in \cref{uai:sec:experiments}. In many domains, such lists exist and are produced by existing ranking policies. They only need to be safely improved.

Second, the bound is $O(\chi_{\max} \chi_{\min}^{-1})$, where $\chi_{\max}$ and $\chi_{\min}$ are the maximum and minimum examination probabilities, respectively. In \cref{uai:sec:sanity check on theorem}, we show that this dependence can be observed in problems where most attractive items are placed at infrequently examined positions. This limitation is intrinsic to $\bubblerank$, because attractive lower-ranked items cannot be placed at higher positions unless they are observed to be attractive at lower, potentially infrequently examined, positions.

The safety constraint of $\bubblerank$ is stated in \cref{uai:lem:safety}. For $\delta = n^{-4}$, as discussed above, $\bubblerank$ becomes a rather safe algorithm, and is unlikely to display any list with more than $K / 2$ incorrectly-ordered item pairs than the initial base list $\cR_0$. More precisely, $\abs{\cV(\rnd{\cR}_t)} \leq \abs{\cV_0} + K / 2$ holds uniformly over time with probability of at least $1 - K^2 / n$. This safety feature of $\bubblerank$ is confirmed by our experiments in \cref{uai:sec:safety results}.

The above discussion assumes that the time horizon $n$ is known. However, in practice, this is not always possible. We can extend $\bubblerank$ to the setting of an unknown time horizon by using the so-called \emph{doubling trick}~\citep[Section 2.3]{cesabianchi06prediction}. Let $n$ be the estimated horizon. Then at time $n + 1$, $\bar{\mathcal{R}}_{n + 1}$ is set to $\mathcal{R}_0$ and $n$ is doubled. The statistics do not need to be reset.

$\bubblerank$  is computationally efficient. The time complexity of $\bubblerank$ is linear in the number of time steps and in each step $O(K)$ operations are required.

In this chapter, we focus on re-ranking. But $\bubblerank$ can be extended to the full ranking problem as follows. Define $s_t(i, j)$ and $n_t(i, j)$ for all item pairs $(i, j)$. For even (odd) $K$ at odd (even) time steps, select a random item below position $K$ that has not been shown to be worse than the item at position $K$, and swap these items with probability $0.5$. The item that is not displayed gets feedback $0$. The rest of $\bubblerank$ remains the same. This algorithm can be analyzed in the same way as $\bubblerank$. 

\subsection{Proof of \cref{uai:thm:upper bound}}
\label{uai:sec:upper bound proof}

In \cref{uai:lem:concentration}, we establish that there exists a favorable event $\cE$ that holds with probability $1 - \delta^\frac{1}{2} K^2 n$, when all beliefs $\rnd{s}_t(i, j)$ are at most $2 \sqrt{\rnd{n}_t(i, j) \log(1 / \delta)}$ from their respective means, uniformly for $i < j$ and $t \in [n]$. Since the maximum $n$-step regret is $K n$, we get that
\begin{align*}
  R(n) \leq
  \E{\hat{R}(n) \I{\cE}} + \delta^\frac{1}{2} K^3 n^2\,,
\end{align*}
where $\hat{R}(n) = \sum_{t = 1}^n r(\cR^\ast, \alpha, \chi) - r(\rnd{\cR}_t, \alpha, \chi)$. We bound $\hat{R}(n)$ next. For this, let
\begin{align*}
  \rnd{\cP}_t = {}
  & \big\{(i, j) \in [K]^2: \,  i < j, \, \abs{\bar{\rnd{\cR}}_t^{-1}(i) - \bar{\rnd{\cR}}_t^{-1}(j)} = 1 \\
  & \rnd{s}_{t - 1}(i, j) \leq 2 \sqrt{\rnd{n}_{t - 1}(i, j) \log(1 / \delta)}\big\}
\end{align*}
be the set of potentially randomized item pairs at time $t$. Then, by \cref{uai:lem:off-base regret} on event $\cE$, which bounds the regret of list $\rnd{\cR}_t$ with the difference in the attraction probabilities of items $(i, j) \in \rnd{\cP}_t$, we have that
\begin{align*}
  &\hat{R}(n) \leq 
   3 K \chi_{\max} \sum_{i = 1}^K \sum_{j = i + 1}^K
  \sum_{t = 1}^n (\alpha(i) - \alpha(j)) \I{(i, j) \in \rnd{\cP}_t}.
\end{align*}
Now note that for any randomized $(i, j) \in \rnd{\cP}_t$ at time $t$,
\begin{align*}
  \chi_{\min} (\alpha(i) - \alpha(j)) &\leq
  \Etp{\rnd{c}_t(\rnd{\cR}_t^{-1}(i)) - \rnd{c}_t(\rnd{\cR}_t^{-1}(j))} \\
  &=
  \Etp{\rnd{s}_t(i, j) - \rnd{s}_{t - 1}(i, j)},
\end{align*}
where $\Etp{\cdot}$ is the expectation conditioned on the history up to time $t$, \break $\rnd{\cR}_1, \rnd{c}_1, \ldots, \rnd{\cR}_{t - 1}, \rnd{c}_{t - 1}$. The inequality is from $\alpha(i) \geq \alpha(j)$, and Assumptions~\cref{ass:order-free examination} and \cref{ass:examination scaling}. The above two inequalities yield
\begin{align*}
  \hat{R}(n)
  & \leq 6 K \frac{\chi_{\max}}{\chi_{\min}} \sum_{i = 1}^K \sum_{j = i + 1}^K
  \sum_{t = 1}^n \Etp{\rnd{s}_t(i, j) - \rnd{s}_{t - 1}(i, j)} \I{(i, j) \in \rnd{\cP}_t} \\
  & \leq 6 K \frac{\chi_{\max}}{\chi_{\min}} \sum_{i = 1}^K \sum_{j = i + 1}^K
  \I{\exists t \in [n]: (i, j) \in \rnd{\cP}_t} \times \\
  & \quad\qquad \sum_{t = 1}^n \Etp{\rnd{s}_t(i, j) - \rnd{s}_{t - 1}(i, j)},
\end{align*} 
where the extra factor of two is because $\bubblerank$ randomizes any pair of items $(i, j) \in \rnd{\cP}_t$ at least once in any two consecutive steps. Moreover, for any $i < j$ on event $\cE$,
\begin{eqnarray*}
  \sum_{t = 1}^n (\rnd{s}_t(i, j) - \rnd{s}_{t - 1}(i, j))  =
  \rnd{s}_n(i, j)
   \leq 15 \frac{\alpha(i) + \alpha(j)}{\alpha(i) - \alpha(j)} \log(1 / \delta) 
 \leq \frac{30}{\Delta_{\min}} \log(1 / \delta).
\end{eqnarray*}
The first inequality is by \cref{uai:lem:cumulative click loss}, which establishes that the maximum difference in clicks of any randomized pair of items is bounded. After that, the better item is found and the pair of items is never randomized again. The last inequality is by $\alpha(i) + \alpha(j) \leq 2$ and $\alpha(i) - \alpha(j) \geq \Delta_{\min}$. Now we chain the above two inequalities and get that
\begin{align*}
  \hat{R}(n) \leq{}
  &180 K \frac{\chi_{\max}}{\chi_{\min}} \frac{1}{\Delta_{\min}} \log(1 / \delta) \times \sum_{i = 1}^K \sum_{j = i + 1}^K \I{\exists t \in [n]: (i, j) \in \rnd{\cP}_t}.
\end{align*}
Finally, let $\rnd{\cP} = \bigcup_{t \in [n]} \rnd{\cP}_t$. Then, on event $\cE$, $\abs{\rnd{\cP}} \leq K - 1 + 2 \abs{\cV_0}$. This follows from the design of $\bubblerank$ (\cref{uai:lem:swap bound}) and completes the proof. \qed

\section{Experimental Results}
\label{uai:sec:experiments}

We conduct four experiments to evaluate $\bubblerank$. 
In \cref{uai:sec:experimental setup}, we describe our experimental setup. In \cref{uai:sec:results with regret}, we report the regret of compared algorithms, which measures the rate of convergence to the optimal list in hindsight. In \cref{uai:sec:safety results}, we validate the safety of $\bubblerank$. In \cref{uai:sec:sanity check on theorem}, we validate the tightness of the regret bound in \cref{uai:thm:upper bound}. 
We report the \emph{Normalized Discounted Cumulative Gain}~(NDCG)
of compared algorithms, which measures the quality of displayed lists, in~\cref{uai:sec:results with ndcg}.

\begin{figure*}[th]
	\includegraphics{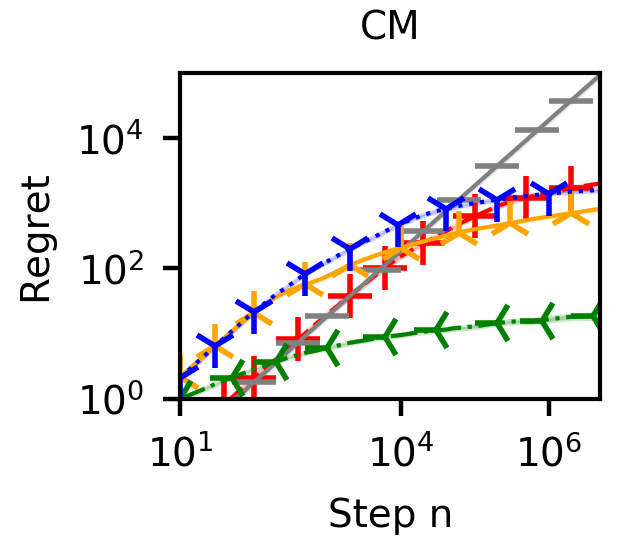}
	\includegraphics{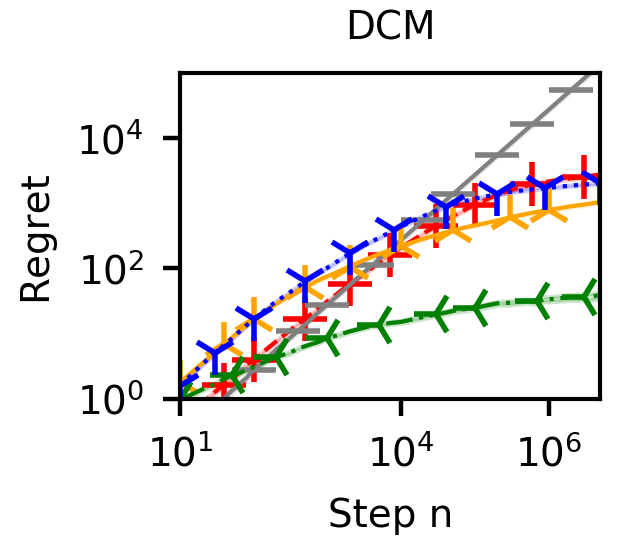}
	\includegraphics{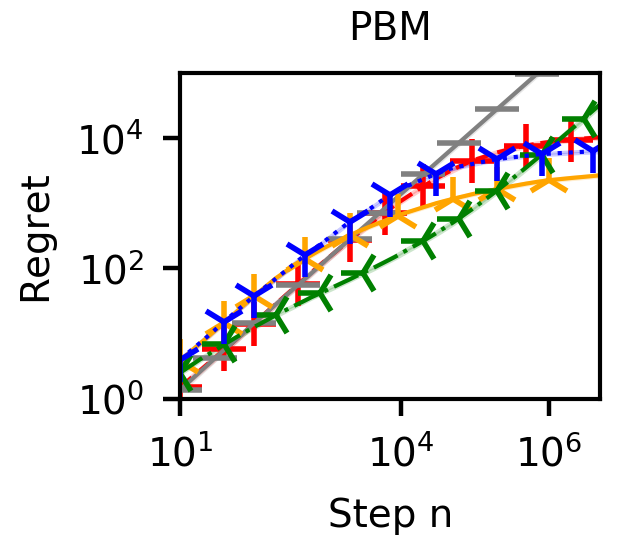}
	\caption{The $n$-step regret of $\bubblerank$ (red), $\cascadeklucb$ (green), $\batchrank$ (blue), $\toprank$ (orange), and $\baseline$ (grey) in the CM, DCM, and PBM in up to $5$ million steps. Lower is better. The results are averaged over all $100$ queries and $10$ runs per query.  The shaded regions represent standard errors of our estimates.}
	\label{uai:fig:all queries}
\end{figure*}

\begin{figure*}[th]
	\includegraphics{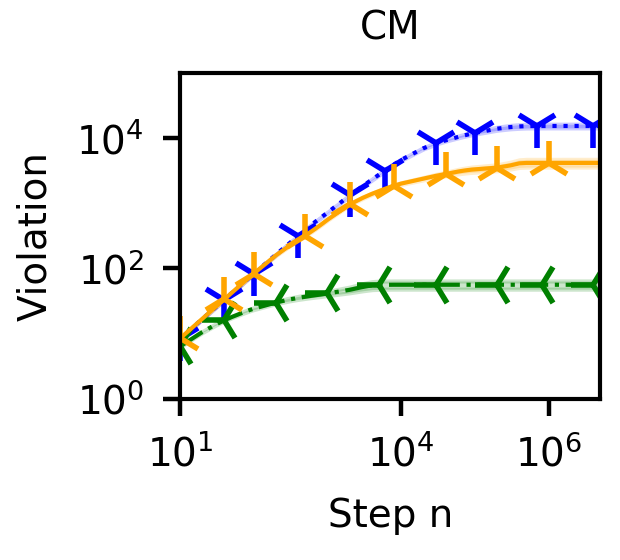}	
	\includegraphics{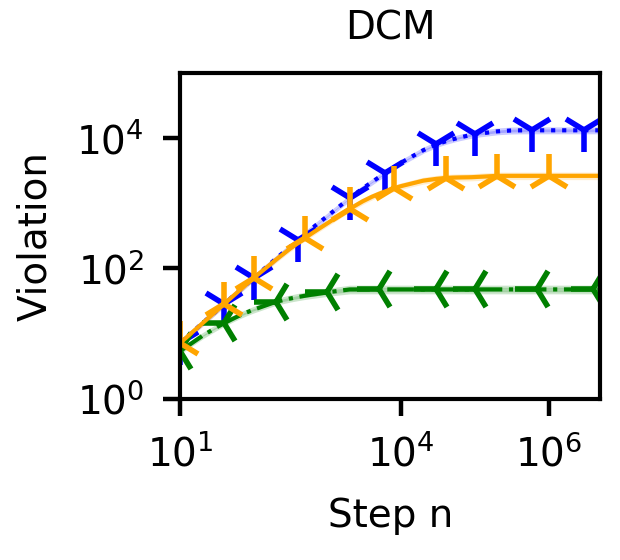}
	\includegraphics{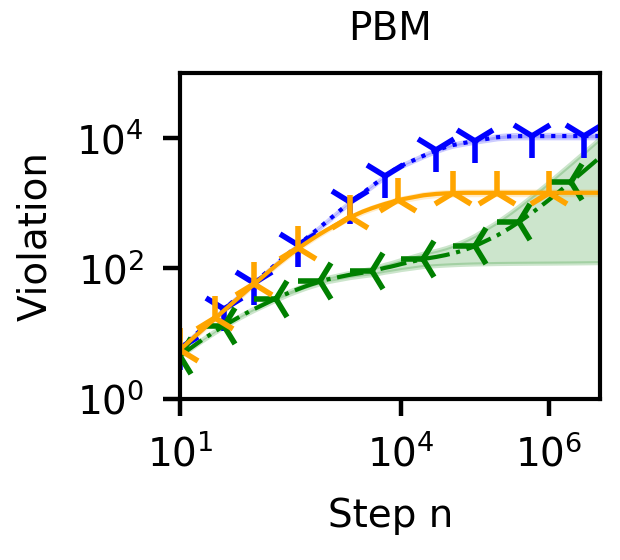}
	\caption{The $n$-step violation of the safety constraint of $\bubblerank$ by $\cascadeklucb$ (green), $\batchrank$ (blue), and $\toprank$ (orange) in the CM, DCM, and PBM in up to $5$ million steps. Lower is better. The shaded regions represent standard errors of our estimates.}
	\label{uai:fig:violations}
\end{figure*}

\subsection{Experimental setup}
\label{uai:sec:experimental setup}

We evaluate $\bubblerank$ on the \emph{Yandex} click  dataset.\footnote{\url{https://academy.yandex.ru/events/data_analysis/relpred2011}} The dataset contains user search sessions from the log of the \emph{Yandex} search engine. It is the largest publicly available dataset containing user clicks, with more than $30$ million search sessions. Each session contains at least one search query together with $10$ ranked items.

We preprocess the dataset as in \citep{zoghi17online}. In particular, we randomly select $100$ frequent search queries, and then learn the parameters of three click models using the PyClick\footnote{\url{https://github.com/markovi/PyClick}} package: CM and PBM, described in \cref{uai:sec:click models}, as well as the \emph{dependent click model (DCM)}~\citep{guo09efficient}. 

The DCM is an extension of the CM~\citep{craswell08experimental} where each position $k$ is associated with an abandonment probability $v(k)$. When the user clicks on an item at position $k$, the user stops scanning the list with probability $v(k)$. Therefore, the DCM can model multiple clicks. Following the work in \citep{katariya16dcm}, we incorporate abandonment into our definition of reward for DCM and define it as the number of abandonment clicks. 
The \emph{abandonment click} is a click after which a user stops browsing the list, and each time step contains at most one abandonment click. 
So, the expected reward for DCM equals the probability of abandonment clicks, which is computed as follows: 
\begin{align*}
r(\cR, \alpha, \chi) & = \sum_{k = 1}^K \chi(\cR, k) v(k) \alpha(\cR(k)), \\
\chi(\cR, k) &= \prod_{i = 1}^{k - 1} (1 - v(i)) \alpha(\cR(i))
\end{align*}
is the examination probability of position $k$ in list $\cR$.  A high reward means a user stops the search session because of clicking on an item with high attraction probability.

The learned CM, DCM, and PBM  are used to simulate user click feedback. We experiment with multiple click models to show the robustness of $\bubblerank$ to multiple models of user feedback.

For each query, we choose $10$ items. The number of positions is equal to the number of items, $K = L = 10$. The objective of our re-ranking problem is to place $5$ most attractive items in descending order of attractiveness at the $5$ highest positions, as in \citep{zoghi17online}. The performance of $\bubblerank$ and our baselines is also measured only at the top $5$ positions.

$\bubblerank$ is compared to three baselines $\tt Cascade\mhyphen$ $\tt KL\mhyphen UCB$~\citep{kveton15cascading}, $\batchrank$~\citep{zoghi17online}, and $\toprank$~\citep{lattimore18toprank}. The former is near optimal in the CM~\citep{kveton15cascading}, but can have linear regret in other click models. Note that linear regret arises when $\cascadeklucb$ erroneously converges to a suboptimal ranked list. $\batchrank$ and $\toprank$ can learn the optimal list $\cR^\ast$ in a wide range of click models, including the CM, DCM, and PBM. However, they can perform poorly in early stages of learning because they randomly shuffles displayed lists to average out the position bias. All experiments are run for $5$ million steps, after which at least two algorithms converge to the optimal ranked list.

In the \emph{Yandex} dataset, each query is associated with many different ranked lists, due to the presence of various personalization features of the production ranker. We take the most frequent ranked list for each query as the initial base list $\cR_0$ in $\bubblerank$, since we assume that the most frequent ranked list is what the production ranker would produce in the absence of any personalization. We also compare $\bubblerank$ to a production baseline, called $\baseline$, where the initial list $\cR_0$ is applied for $n$ steps. 

\subsection{Results with regret}
\label{uai:sec:results with regret}

In the first experiment, we compare $\bubblerank$ to $\cascadeklucb$, $\batchrank$, and $\toprank$ in the CM, DCM, and PBM of all $100$ queries. 
Among them, $\toprank$ is the state-of-the-art online LTR algorithm in multiple click models. 
We evaluate these algorithms by their cumulative regret, which is defined in \cref{uai:eq:regret}, at the top $5$ positions. The regret,  a measure of convergence,  is a widely-used metric in the bandit literature~\citep{auer02finite,katariya16dcm,kveton15cascading,zoghi17online}. In the CM and PBM, the regret is the cumulative loss in clicks when a sequence of learned lists is compared to the optimal list in hindsight. In the DCM, the regret is the cumulative loss in abandonment clicks. We also report the regret of $\baseline$. 

Our results are reported in \cref{uai:fig:all queries}. 
We observe that the regret of  $\baseline$ grows linearly with time $n$, which means that it is not optimal on average. 
$\cascadeklucb$ learns $\cR^\ast$ quickly in both the CM and DCM, but has linear regret in the PBM. 
This is expected since  $\cascadeklucb$ is designed for the CM, and the DCM is an extension of the CM. 
As for the PBM, which is beyond the modeling assumptions of $\cascadeklucb$, there is no guarantee on the performance of $\cascadeklucb$.  
$\bubblerank$, $\batchrank$, and $\toprank$ can learn in all three click models. 
Compared to $\batchrank$ and $\toprank$, $\bubblerank$ has a higher regret in $5$ million steps. 
However, in earlier steps, $\bubblerank$ has a lower regret than $\batchrank$ and $\toprank$, as it takes advantage of the initial base list $\cR_0$. 
In general, these results show that $\bubblerank$ converges to the optimal list slower than $\batchrank$ and $\toprank$. This is expected because $\bubblerank$ is designed to be a safe algorithm, and only learns better lists by exchanging neighboring items in the base list. 

\subsection{Safety results}
\label{uai:sec:safety results}

In the previous experiment, $\bubblerank$ does not learn as fast as $\cascadeklucb$, $\batchrank$, and $\toprank$, because of the additional safety constraint in \cref{uai:lem:safety}. The constraint is that $\bubblerank$ is unlikely to display any list with more than $K / 2$ incorrectly-ordered item pairs than the initial base list $\cR_0$. More precisely, $\abs{\cV(\rnd{\cR}_t)} \leq \abs{\cV(\cR_0)} + K / 2$ holds uniformly over time with probability of at least $1 - K^2 / n$ for $\delta = n^{-4}$ (\cref{uai:sec:discussion}), where $\cV$ is defined in \cref{uai:eq:safety}. In the second experiment, we answer the question how often do $\bubblerank$, $\cascadeklucb$,  $\batchrank$, and $\toprank$ 
violate this constraint empirically.
We define the \emph{safety constraint violation in $n$ steps} as
\begin{align}
  V(n) =
  \sum_{t = 1}^{n} \I{\abs{\cV(\rnd{\cR}_t)} > \abs{\cV(\cR_0)} + K / 2},
  \label{uai:eq:constraint violation}
\end{align}
where $\rnd{\cR}_t$ is the displayed list at time $t$. 

We report the $n$-step safety constraint violation of $\cascadeklucb$, $\batchrank$, and $\toprank$ in \cref{uai:fig:violations}. We do not include results of $\bubblerank$ since $\bubblerank$ never violates the constraint in our experiments. We observe that the safety constraint violations of $\cascadeklucb$ in the first $100$ steps are $24.12 \pm 0.76$, $ 23.33 \pm 0.90$, and $23.63 \pm 0.96$ in the CM, DCM, and PBM, respectively. Translating this to a search scenario, $\cascadeklucb$ may show unsafe results, which are significantly worse than the initial base list $\cR_0$, and may hurt user experience, more than $20\%$ of search sessions in the first $100$ steps. Even worse, the violations of $\cascadeklucb$ grow linearly with time in the PBM. The safety issues of $\batchrank$, and $\toprank$ are more severe than that of $\cascadeklucb$.  More precisely, the violations of $\batchrank$ in the first $100$ steps are $83.01 \pm 0.56$, $71.89 \pm0.92$, and $59.63 \pm 0.98$ in the CM, DCM, and PBM, respectively.
And the violations of $\toprank$ are $83.56 \pm 0.70$, $71.47 \pm1.07$, and $57.00 \pm 1.24$ in the CM, DCM, and PBM, respectively. 
Note that the performance of $\toprank$ is close to that of $\batchrank$ in the first $100$ steps since they both require the ranked lists to be randomly shuffled during the initial stages. 
Thus, $\batchrank$, and $\toprank$  would frequently hurt the user experience  during the early stages of learning.

To conclude, $\bubblerank$ learns without violating its safety constraint, while $\cascadeklucb$, $\batchrank$, and $\toprank$  violate the constraint frequently. 
Together with results in~\cref{uai:sec:results with regret}, $\bubblerank$ is a safe algorithm but, to satisfy the safety constraint, it compromises the performance and learns slower than $\batchrank$ and $\toprank$.  
In~\cref{uai:sec:results with ndcg}, we compare $\bubblerank$ to baselines in NDCG and  show that $\bubblerank$ converges to the optimal lists in hindsight.

\begin{figure*}[t]
	\includegraphics{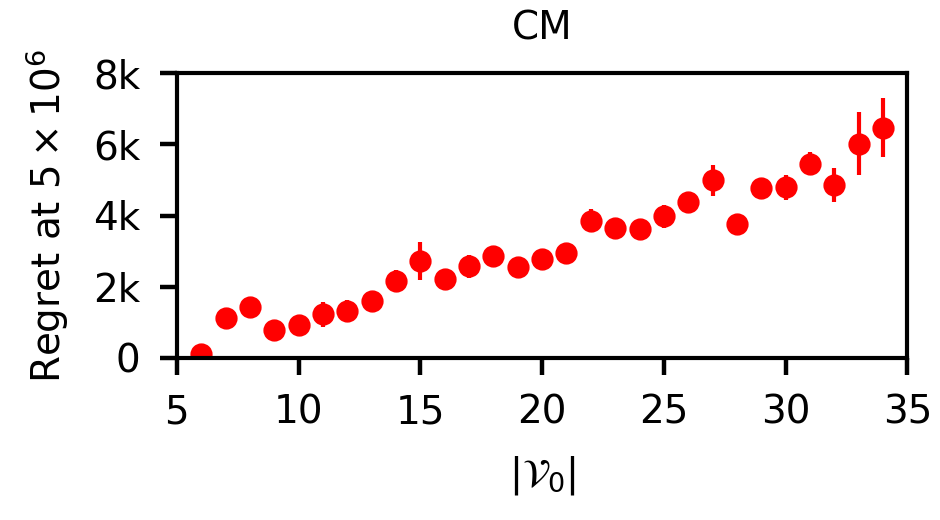}
	\includegraphics{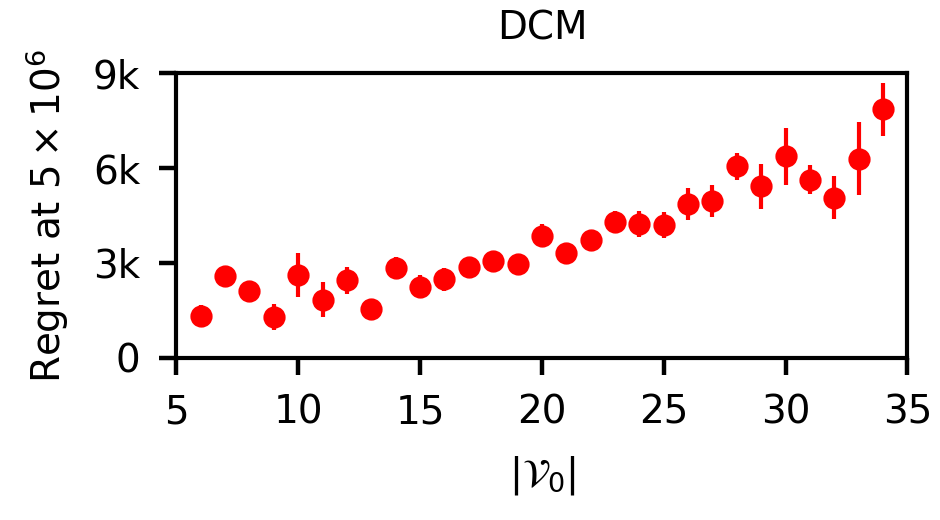}
	\includegraphics{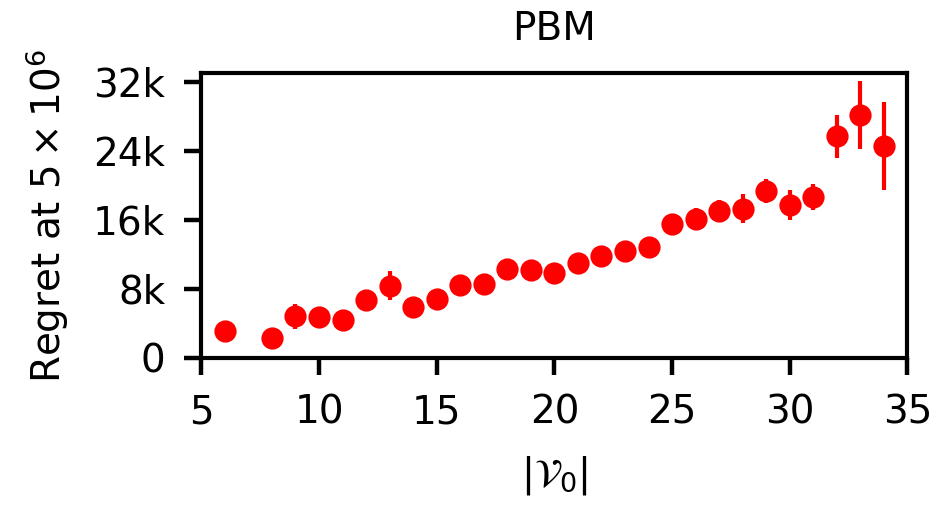}
	\includegraphics{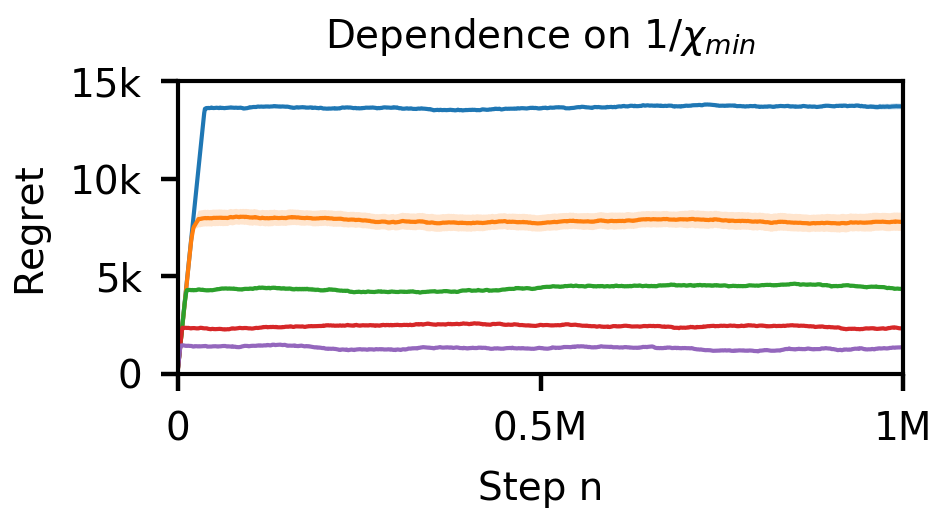}
	\caption{Regret of $\bubblerank$ as a function of the number of  incorrectly-ordered item pairs
		$\abs{\cV_0}$, and the minimal examination probability  $\chi_{\min}$. In the bottom-right plot, the purple, red, green, orange, and blue colors represent $\chi_{\min}$ equals $0.5$, $0.5^2$, $0.5^3$, $0.5^4$, and $0.5^5$, respectively. The shaded regions represent standard errors of our estimates. }
	\label{uai:fig:initialization}
\end{figure*}

\subsection{Sanity check on regret bound}
\label{uai:sec:sanity check on theorem}

We prove an upper bound on the $n$-step regret of $\bubblerank$ in \cref{uai:thm:upper bound}. In comparison to the upper bounds of $\cascadeklucb$ \citep{kveton15cascading} and $\batchrank$ \citep{zoghi17online}, we have two new problem-specific constants: $\abs{\cV_0}$ and $1 / \chi_{\min}$. In this section, we show that these constants are intrinsic to the behavior of $\bubblerank$.

We first study how the number of incorrectly-ordered item pairs in the initial base list $\cR_0$, $\abs{\cV_0}$, impacts the regret of $\bubblerank$. We choose $10$ random initial base lists $\cV_0$ in each of our $100$ queries and plot the regret of $\bubblerank$ as a function of $\abs{\cV_0}$. Our results are shown in \cref{uai:fig:initialization}. We observe that the regret of $\bubblerank$ is linear in $\abs{\cV_0}$ in the CM, DCM, and PBM. This is the same dependence as in our regret bound (\cref{uai:thm:upper bound}). 

We then study the impact of the minimum examination probability $\chi_{\min}$ on the regret of $\bubblerank$. We experiment with a synthetic PBM with $10$ items, which is parameterized by $\alpha = (0.9, 0.5, \dots, 0.5)$ and $\chi = (0.9, \dots, 0.9, 0.5^i, 0.5^i)$ for $i \geq 1$. The most attractive item is placed at the last position in $\cR_0$, $\cR_0 = (2, \dots, K - 1, 1)$. Since this position is examined with probability $0.5^i$, we expect the regret to double when $i$ increases by one. We experiment with $i \in [5]$ in \cref{uai:fig:initialization} and observe this trend in $1$ million steps. This confirms that the dependence on $1 / \chi_{\min}$ in \cref{uai:thm:upper bound} is generally unavoidable.

\subsection{Results with NDCG}
\label{uai:sec:results with ndcg}

In this section, we report the NDCG of compared algorithms, which measures the quality of displayed lists. 
Since $\cascadeklucb$ fails in the PBM and we focus on learning from all types of click feedback, we leave out $\cascadeklucb$ from this section. 

\begin{figure*}[!h]
	\includegraphics{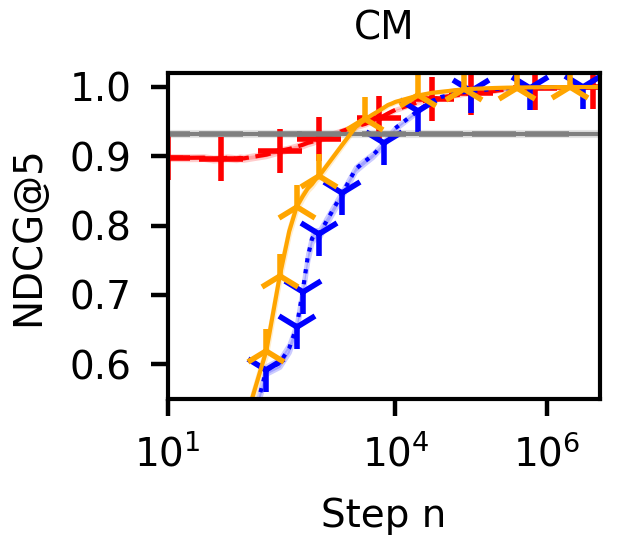}
	\includegraphics{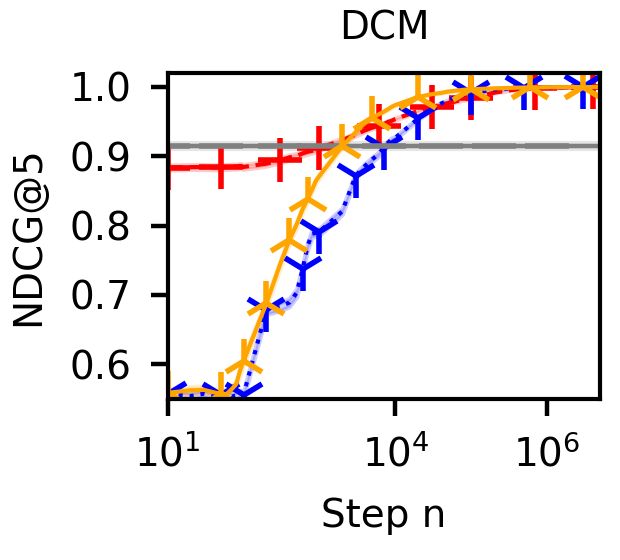}
	\includegraphics{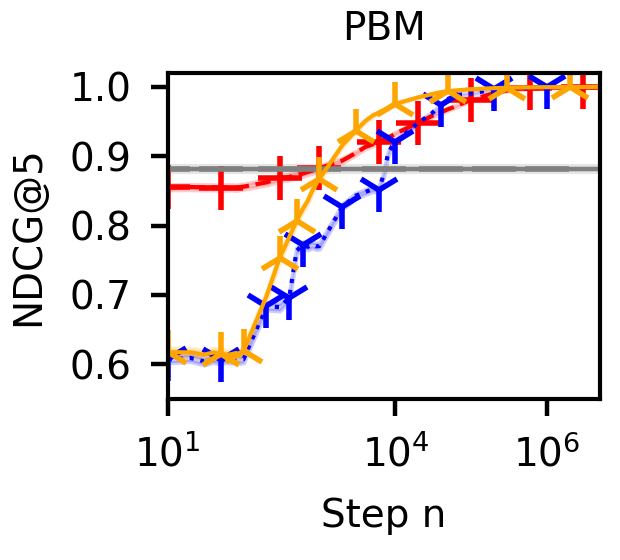}
	\caption{The per-step NDCG@$5$ of $\bubblerank$ (red), $\batchrank$ (blue), $\toprank$ (orange),  and $\baseline$ (grey) in the CM, DCM, and PBM in up to $5$ million steps. Higher is better. The shaded regions represent standard errors of our estimates.}
	\label{fig:ndcg}
\end{figure*}

In the first two experiments, we evaluate algorithms by their regret in \cref{uai:eq:regret} and safety constraint violation in \cref{uai:eq:constraint violation}. Neither of these metrics measure the quality of ranked lists directly. In this experiment, we report the per-step NDCG@$5$ of $\bubblerank$, $\batchrank$, $\toprank$, and $\baseline$ (\cref{fig:ndcg}), which directly measures the quality of ranked lists and is widely used in the LTR literature \citep{jaervelin02cumulated,allan17trec}. Since the \emph{Yandex} dataset does not contain relevance scores for all query-item pairs, we take the attraction probability of the item in its learned click model as a proxy to its relevance score. This substitution is natural since our goal is to rank items in  descending order of their attraction probabilities~\citep{chuklin15click}. We compute the NDCG@$5$ of a ranked list $\cR$ as
\begin{align*}
\mathit{NDCG}@5(\cR) =
\frac{\mathit{DCG}@5(\cR)}{\mathit{DCG}@5(\cR^\ast)}\,, \quad
\mathit{DCG}@5(\cR) =
\sum_{k = 1}^{5} \frac{\alpha(\cR(k))}{\log_{2}(k + 1)}\,, 
\end{align*}
where $\cR^\ast$ is the optimal list and $\alpha(\cR(k)$ is the attraction probability of the $k$-th item in list $\cR$. This is a standard evaluation metric, and is used in TREC evaluation benchmarks~\citep{allan17trec}, for instance. It measures the discounted gain over the attraction probabilities of the $5$ highest ranked items in list $\cR$, which is normalized by the DCG@$5$ of $\cR^\ast$. 

In \cref{fig:ndcg}, we observe that  $\baseline$ has good NDCG@$5$ scores in all click models. 
Yet there is still room for improvement.
$\bubblerank$, $\batchrank$, and $\toprank$ have similar NDCG@$5$ scores after $5$ million steps. But $\bubblerank$ starts with NDCG@$5$ close to that of $\baseline$, while $\batchrank$ and $\toprank$ start with lists with very low NDCG@$5$.

These results validate our earlier findings. As in \cref{uai:sec:results with regret}, we observe that $\bubblerank$ converges to the optimal list in hindsight, since its NDCG@$5$ approaches $1$. As in \cref{uai:sec:safety results}, we observe that $\bubblerank$ is safe, since its NDCG@$5$ is never much worse than that of $\baseline$.

\section{Related Work}
\label{uai:sec:related work}

Online LTR via click feedback has been mainly studied in two approaches: under specific click models~\citep{combes15learning,katariya16dcm,kveton15cascading,kveton15combinatorial,lagree16multiple,zong16cascading}; or without a particular assumption on click models~\citep{lattimore18toprank,radlinski08learning,slivkins13ranked,zoghi17online}. 
Algorithms from the first group efficiently learn optimal rankings in the their considered click models but do not have guarantees beyond their specific click models. 
Algorithms from the second group, on the other hand, learn the optimal rankings in a broader class of click models. 
$\toprank$~\citep{lattimore18toprank} is the state-of-the-art of the second group, which has the regret of $O(K^2\log(n))$ in our re-ranking setup, that is $L=K$. 
$\bubblerank$ also belongs to the second group and the regret of $\bubblerank$ is comparable to that of $\toprank$ given a good initial list, when $| \cV_0 | = O(K)$. 
However, unlike $\bubblerank$, $\toprank$ and all the previous algorithms do not consider safety. 
They explore aggressively in the initial steps and may display irrelevant items at high positions, which may then hurt user experiences~\citep{wang18position}.  

Our safety problem is related to the warm start problem~\citep{strehl10learning}. 
Contextual bandits~\citep{agarwal14taming,li10contextual,moon10online} 
deal with a broader class of models than we do and are used to address the warm start problem. But they are limited to small action sets, and thus unsuitable for the ranking setup that we consider in this chapter.

The warm start LTR has been studied in multiple papers~\citep{hofmann13reusing,vorobev15gathering,yan18practical}, where the goal is to use an online algorithm to fine tune the results generated by an offline-trained ranker. 
In these papers, different methods for learning prior distributions of Thompson sampling based online LTR algorithms from offline datasets have been proposed. 
However, these methods have the following drawbacks. 
First, the offline data may not well align with user preferences~\citep{zoghi16click}, which may result in a biased prior assumption. 
Second, grid search with online A/B tests may alleviate this and find a proper prior assumption~\citep{vorobev15gathering}, but the online A/B test requires additional costs in terms of user experience. 
Third,  there is no safety constraint in these methods. Even with carefully picked priors, they may recommend irrelevant items to users, e.g., new items with little prior knowledge. 
In contrast, $\bubblerank$ starts from the production ranked list and learns under the safety constraint. Thus, $\bubblerank$ gets rid of these drawbacks.  

Another related line of work are conservative bandits~\citep{kazerouni17conservative,wu16conservative}. In conservative bandits, the learned policy is safe in the sense that its expected \emph{cumulative} reward is at least $1 - \alpha$ fraction of that of the baseline policy with high probability. This notion of safety is less stringent than that in our work (\cref{uai:sec:safety}). In particular, our notion of safety is \emph{per-step}, in the sense that any displayed list is only slightly worse than the initial base list with a high probability. We do not compare to conservative bandits in our experiments because existing algorithms for conservative bandits require the action space to be small. The actions in our problem are ranked lists, and their number is exponential in $K$.


\section{Lemmas}
\label{uai:sec:technical lemmas}

\begin{lemma}
\label{uai:lem:list regret} Let $\cR$ be any list over $[K]$. Let
\begin{align}
  \Delta(\cR) =
  \sum_{k = 1}^{K - 1} \I{\alpha(\cR(k + 1)) - \alpha(\cR(k)) > 0} \times
  (\alpha(\cR(k + 1)) - \alpha(\cR(k)))
  \label{uai:eq:attraction gap} 
\end{align}
be the \emph{attraction gap} of list $\cR$. Then the expected regret of $\cR$ is bounded as
\begin{align*}
  \sum_{k = 1}^K (\chi(\cR^\ast, k) \alpha(k) - \chi(\cR, k) \alpha(\cR(k))) \leq
  K \chi_{\max} \Delta(\cR)\,.
\end{align*}
\end{lemma}
\begin{proof}
Fix position $k \in [K]$. Then
\begin{align*}
  \chi(\cR^\ast, k) \alpha(k) - \chi(\cR, k) \alpha(\cR(k))
  & \leq \chi(\cR^\ast, k) (\alpha(k) - \alpha(\cR(k))) \\
  & \leq \chi_{\max} (\alpha(k) - \alpha(\cR(k)))\,,
\end{align*}
where the first inequality follows from the fact that the examination probability of any position is the lowest in the optimal list (Assumption~\ref{ass:lowest examination}) and the second inequality follows from the definition of $\chi_{\max}$. In the rest of the proof, we bound $\alpha(k) - \alpha(\cR(k))$. We consider three cases. First, let $\alpha(\cR(k)) \geq \alpha(k)$. Then $\alpha(k) - \alpha(\cR(k)) \leq 0$ and can be trivially bounded by $\Delta(\cR)$. Second, let $\alpha(\cR(k)) < \alpha(k)$ and $\pi(k) > k$, where $\pi(k)$ is the position of item $k$ in list $\cR$. Then
\begin{align*}
  \alpha(k) - \alpha(\cR(k)) & =
  \alpha(\cR(\pi(k))) - \alpha(\cR(k)) \\
  & \leq  \sum_{i = k}^{\pi(k) - 1} \I{\alpha(\cR(i + 1)) - \alpha(\cR(i)) > 0} \alpha(\cR(i + 1)) - \alpha(\cR(i))).
\end{align*}
From the definition of $\Delta(\cR)$, this quantity is bounded from above by $\Delta(\cR)$. Finally, let $\alpha(\cR(k)) < \alpha(k)$ and $\pi(k) < k$. This implies that there exists an item at a lower position than $k$, $j > k$, such that $\alpha(\cR(j)) \geq \alpha(k)$. Then
\begin{align*}
  \alpha(k) - \alpha(\cR(k)) & \leq
  \alpha(\cR(j)) - \alpha(\cR(k)) \\
  & \leq \sum_{i = k}^{j - 1} \I{\alpha(\cR(i + 1)) - \alpha(\cR(i)) > 0} (\alpha(\cR(i + 1)) - \alpha(\cR(i)))\,.
\end{align*}
From the definition of $\Delta(\cR)$, this quantity is bounded from above by $\Delta(\cR)$. This concludes the proof.
\end{proof}

\begin{lemma}
\label{uai:lem:off-base gap} Let
\begin{align*}
  &\rnd{\cP}_t = \\
  &\big\{(i, j) \in [K]^2: \, & i < j, \, \abs{\bar{\rnd{\cR}}_t^{-1}(i) - \bar{\rnd{\cR}}_t^{-1}(j)} = 1, 
 \rnd{s}_{t - 1}(i, j) \leq 2 \sqrt{\rnd{n}_{t - 1}(i, j) \log(1 / \delta)}\big\}
\end{align*}
be the set of potentially randomized item pairs at time $t$ and $\rnd{\Delta}_t = \max_{\rnd{\cR}_t} \Delta(\rnd{\cR}_t)$ be the \emph{maximum attraction gap} of any list $\rnd{\cR}_t$, where $\Delta(\rnd{\cR}_t)$ is defined in \cref{uai:eq:attraction gap}. Then on event $\cE$ in \cref{uai:lem:concentration},
\begin{align*}
  \rnd{\Delta}_t \leq
  3 \sum_{i = 1}^K \sum_{j = i + 1}^K \I{(i, j) \in \rnd{\cP}_t} (\alpha(i) - \alpha(j))
\end{align*}
holds at any time $t \in [n]$.
\end{lemma}
\begin{proof}
Fix list $\rnd{\cR}_t$ and position $k \in [K - 1]$. Let $i', i, j, j'$ be items at positions $k - 1, k, k + 1, k + 2$ in $\bar{\rnd{\cR}}_t$. If $k = 1$, let $i' = i$; and if $k = K - 1$, let $j' = j$. We consider two cases.

First, suppose that the permutation at time $t$ is such that $i$ and $j$ could be exchanged. Then
\begin{eqnarray*}
 &{}&\alpha(\rnd{\cR}_t^{-1}(k + 1)) - \alpha(\rnd{\cR}_t^{-1}(k)) \\
 &\leq & \I{(\min \set{i, j}, \max \set{i, j}) \in \rnd{\cP}_t} (\alpha(\min \set{i, j}) - \alpha(\max \set{i, j}))
\end{eqnarray*}
holds on event $\cE$ by the design of $\bubblerank$. More specifically, $(\min \set{i, j}, \max \set{i, j}) \notin \rnd{\cP}_t$ implies that $\alpha(\rnd{\cR}_t^{-1}(k + 1)) - \alpha(\rnd{\cR}_t^{-1}(k)) \leq 0$.

Second, suppose that the permutation at time $t$ is such that $i$ and $i'$ could be exchanged, $j$ and $j'$ could be exchanged, or both. Then
\begin{align*}
& \alpha(\rnd{\cR}_t^{-1}(k + 1)) - \alpha(\rnd{\cR}_t^{-1}(k)) \leq {} \\
  & \I{(\min \set{i, i'}, \max \set{i, i'}) \in \rnd{\cP}_t} (\alpha(\min \set{i, i'}) - \alpha(\max \set{i, i'})) + {} \\
  & \alpha(j) - \alpha(i) + {} \\
  & \I{(\min \set{j, j'}, \max \set{j, j'}) \in \rnd{\cP}_t} (\alpha(\min \set{j, j'})  - \alpha(\max \set{j, j'}))
\end{align*}
holds by the same argument as in the first case. Also note that
\begin{align*}
 \alpha(j) - \alpha(i) \leq 
 \I{(\min \set{i, j}, \max \set{i, j}) \in \rnd{\cP}_t} (\alpha(\min \set{i, j}) - \alpha(\max \set{i, j}))
\end{align*}
holds on event $\cE$ by the design of $\bubblerank$. Therefore, for any position $k \in [K - 1]$ in both above cases,
\begin{align*}
  &\alpha(\rnd{\cR}_t^{-1}(k + 1)) - \alpha(\rnd{\cR}_t^{-1}(k)) \leq\!
  \\
  &\sum_{\ell = k - 1}^{k + 1}   \I{\left(\min \set{\bar{\rnd{\cR}}_t^{-1}(\ell), \bar{\rnd{\cR}}_t^{-1}(\ell + 1)},
  \max \set{\bar{\rnd{\cR}}_t^{-1}(\ell), \bar{\rnd{\cR}}_t^{-1}(\ell + 1)}\right) \in \rnd{\cP}_t} \times \\
  & \left(\alpha\left(\min \set{\bar{\rnd{\cR}}_t^{-1}(\ell), \bar{\rnd{\cR}}_t^{-1}(\ell + 1)}\right)- 
  \alpha\left(\max \set{\bar{\rnd{\cR}}_t^{-1}(\ell), \bar{\rnd{\cR}}_t^{-1}(\ell + 1)}\right)\right).
\end{align*}
Now we sum over all positions and note that each pair of $\bar{\rnd{\cR}}_t^{-1}(\ell)$ and $\bar{\rnd{\cR}}_t^{-1}(\ell + 1)$ appears on the right-hand side at most three times, in any list $\rnd{\cR}_t$. This concludes our proof.
\end{proof}

\begin{lemma}
\label{uai:lem:off-base regret} Let $\rnd{\cP}_t$ be defined as in \cref{uai:lem:off-base gap}. Then on event $\cE$ in \cref{uai:lem:concentration},
\begin{eqnarray*}
  &\mbox{}&\sum_{k = 1}^K (\chi(\cR^\ast, k) \alpha(k) - \chi(\rnd{\cR}_t, k) \alpha(\rnd{\cR}_t(k))) \\
  &\leq &
   3 K \chi_{\max} \sum_{i = 1}^K \sum_{j = i + 1}^K \I{(i, j) \in \rnd{\cP}_t} (\alpha(i) - \alpha(j))
\end{eqnarray*}
holds at any time $t \in [n]$.
\end{lemma}
\begin{proof}
A direct consequence of \cref{uai:lem:list regret,uai:lem:off-base gap}.
\end{proof}

\begin{lemma}
\label{uai:lem:swap bound} Let $\rnd{\cP}_t$ be defined as in \cref{uai:lem:off-base gap}, $\rnd{\cP} = \bigcup_{t = 1}^n \rnd{\cP}_t$, and $\cV_0$ be defined as in \cref{uai:eq:initial safety}. Then on event $\cE$ in \cref{uai:lem:concentration},
\begin{align*}
  \abs{\rnd{\cP}} \leq K - 1 + 2 \abs{\cV_0}\,.
\end{align*}
\end{lemma}
\begin{proof}
From the design of $\bubblerank$, $\abs{\rnd{\cP}_1} = K - 1$. The set of randomized item pairs grows only if the base list in $\bubblerank$ changes. When this happens, the number of incorrectly-ordered item pairs decreases by one, on event $\cE$, and the set of randomized item pairs increases by at most two pairs. This event occurs at most $\abs{\cV_0}$ times. This concludes our proof.
\end{proof}

\begin{lemma}
\label{uai:lem:cumulative click loss} For any items $i$ and $j$ such that $i < j$,
\begin{align*}
  \rnd{s}_n(i, j) \leq 15 \frac{\alpha(i) + \alpha(j)}{\alpha(i) - \alpha(j)} \log(1 / \delta)
\end{align*}  
on event $\cE$ in \cref{uai:lem:concentration}.
\end{lemma}
\begin{proof}
To simplify notation, let $\rnd{s}_t = \rnd{s}_t(i, j)$ and $\rnd{n}_t = \rnd{n}_t(i, j)$. The proof has two parts. First, suppose that $\rnd{s}_t \leq 2 \sqrt{\rnd{n}_t \log(1 / \delta)}$ holds at all times $t \in [n]$. Then from this assumption and on event $\cE$ in \cref{uai:lem:concentration},
\begin{align*}
  \frac{\alpha(i) - \alpha(j)}{\alpha(i) + \alpha(j)} \rnd{n}_t - 2 \sqrt{\rnd{n}_t \log(1 / \delta)} \leq
  \rnd{s}_t \leq
  2 \sqrt{\rnd{n}_t \log(1 / \delta)}\,.
\end{align*}
This implies that
\begin{align*}
  \rnd{n}_t \leq
  \left[4 \frac{\alpha(i) + \alpha(j)}{\alpha(i) - \alpha(j)}\right]^2 \log(1 / \delta)
\end{align*}
at any time $t$, and in turn that
\begin{align*}
  \rnd{s}_t \leq
  2 \sqrt{\rnd{n}_t \log(1 / \delta)} \leq
  8 \frac{\alpha(i) + \alpha(j)}{\alpha(i) - \alpha(j)} \log(1 / \delta)
\end{align*}
at any time $t$. Our claim follows from setting $t = n$.

Now suppose that $\rnd{s}_t \leq 2 \sqrt{\rnd{n}_t \log(1 / \delta)}$ does not hold at all times $t \in [n]$. Let $\tau$ be the first time when $\rnd{s}_\tau > 2 \sqrt{\rnd{n}_\tau \log(1 / \delta)}$. Then from the definition of $\tau$ and on event $\cE$ in \cref{uai:lem:concentration},
\begin{align*}
  \frac{\alpha(i) - \alpha(j)}{\alpha(i) + \alpha(j)} \rnd{n}_\tau - 2 \sqrt{\rnd{n}_\tau \log(1 / \delta)}
  & \leq \rnd{s}_\tau \leq
  \rnd{s}_{\tau - 1} + 1 \\
  & \leq 2 \sqrt{\rnd{n}_\tau \log(1 / \delta)} + 1 \\
  & \leq 3 \sqrt{\rnd{n}_\tau \log(1 / \delta)}\,,
\end{align*}
where the last inequality holds for any $\delta \leq 1 / e$. This implies that
\begin{align*}
  \rnd{n}_\tau \leq
  \left[5 \frac{\alpha(i) + \alpha(j)}{\alpha(i) - \alpha(j)}\right]^2 \log(1 / \delta)\,,
\end{align*}
and in turn that
\begin{align*}
  \rnd{s}_\tau \leq
  3 \sqrt{\rnd{n}_\tau \log(1 / \delta)} \leq
  15 \frac{\alpha(i) + \alpha(j)}{\alpha(i) - \alpha(j)} \log(1 / \delta)\,.
\end{align*}
Now note that $\rnd{s}_t = \rnd{s}_\tau$ for any $t > \tau$, from the design of $\bubblerank$. This concludes our proof.
\end{proof}

For some $\cF_t = \sigma(\rnd{\cR}_1, \rnd{c}_1, \dots, \rnd{\cR}_t, \rnd{c}_t)$-measurable event $A$, let $\mathbb{P}_t(A) = \mathbb{P}(A \mid \cF_t)$ be the conditional probability of $A$ given history $\rnd{\cR}_1, \rnd{c}_1, \dots, \rnd{\cR}_t, \rnd{c}_t$. Let the corresponding conditional expectation operator be $\Et{\cdot}$. Note that $\bar{\rnd{\cR}}_t$ is $\cF_{t - 1}$-measurable.

\begin{lemma}
\label{uai:lem:click loss} Let $i, j \in [K]$ be any items at consecutive positions in $\bar{\rnd{\cR}}_t$ and
\begin{align*}
  \rnd{z} = \rnd{c}_t(\rnd{\cR}_t^{-1}(i)) - \rnd{c}_t(\rnd{\cR}_t^{-1}(j))\,.
\end{align*}
Then, on the event that $i$ and $j$ are subject to randomization at time $t$,
\begin{align*}
  \Etp{\rnd{z} \mid \rnd{z} \neq 0} \geq
  \frac{\alpha(i) - \alpha(j)}{\alpha(i) + \alpha(j)}
\end{align*}
when $\alpha(i) > \alpha(j)$, and $\Etp{- \rnd{z} \mid \rnd{z} \neq 0} \leq 0$ when $\alpha(i) < \alpha(j)$.
\end{lemma}
\begin{proof}
The first claim is proved as follows. From the definition of expectation and $\rnd{z} \in \set{-1, 0, 1}$,
\begin{align*}
  \Etp{\rnd{z} \mid \rnd{z} \neq 0}
  & = \frac{\mathbb{P}_{t - 1}(\rnd{z} = 1, \rnd{z} \neq 0) - \mathbb{P}_{t - 1}(\rnd{z} = -1, \rnd{z} \neq 0)}
  {\mathbb{P}_{t - 1}(\rnd{z} \neq 0)} \\
  & = \frac{\mathbb{P}_{t - 1}(\rnd{z} = 1) - \mathbb{P}_{t - 1}(\rnd{z} = -1)}{\mathbb{P}_{t - 1}(\rnd{z} \neq 0)} \\
  & =  \frac{\Etp{\rnd{z}}}{\mathbb{P}_{t - 1}(\rnd{z} \neq 0)}\,,
\end{align*}
where the last equality is a consequence of $\rnd{z} = 1 \implies \rnd{z} \neq 0$ and that $\rnd{z} = -1 \implies \rnd{z} \neq 0$.

Let $\chi_i = \Etp{\chi(\rnd{\cR}_t, \rnd{\cR}_t^{-1}(i))}$ and $\chi_j = \Etp{\chi(\rnd{\cR}_t, \rnd{\cR}_t^{-1}(j))}$ denote the average examination probabilities of the positions with items $i$ and $j$, respectively, in $\rnd{\cR}_t$; and consider the event that $i$ and $j$ are subject to randomization at time $t$. By Assumption~\ref{ass:order-free examination}, the values of $\chi_i$ and $\chi_j$ do not depend on the randomization of other parts of $\bar{\rnd{\cR}}_t$, only on the positions of $i$ and $j$. Then $\chi_i \geq \chi_j$; from $\alpha(i) > \alpha(j)$ and Assumption~\ref{ass:examination scaling}. Based on this fact, $\Etp{\rnd{z}}$ is bounded from below as
\begin{align*}
  \Etp{\rnd{z}} =
  \chi_i \alpha(i) - \chi_j \alpha(j) \geq
  \chi_i (\alpha(i) - \alpha(j))\,,
\end{align*}
where the inequality is from $\chi_i \geq \chi_j$. Moreover, $\mathbb{P}_{t - 1}(\rnd{z} \neq 0)$ is bounded from above as
\begin{align*}
  \mathbb{P}_{t - 1}(\rnd{z} \neq 0)
  & = \mathbb{P}_{t - 1}(\rnd{z} = 1) + \mathbb{P}_{t - 1}(\rnd{z} = -1) \\
  & \leq \chi_i \alpha(i) + \chi_j \alpha(j) \\
  & \leq \chi_i (\alpha(i) + \alpha(j))\,,
\end{align*}
where the first inequality is from inequalities $\mathbb{P}_{t - 1}(\rnd{z} = 1) \leq \chi_i \alpha(i)$ and $\mathbb{P}_{t - 1}(\rnd{z} = -1) \leq \chi_j \alpha(j)$, and the last inequality is from $\chi_i \geq \chi_j$.

Finally, we chain all above inequalities and get our first claim. The second claim follows from the observation that $\Etp{- \rnd{z} \mid \rnd{z} \neq 0} = - \Etp{\rnd{z} \mid \rnd{z} \neq 0}$.
\end{proof}

\begin{lemma}
\label{uai:lem:concentration} 
Let $S_1 = \set{(i, j) \in [K]^2: i < j}$ and $S_2 = \big\{ (i, j) \in [K]^2 : i > j \big\}$. Let
\begin{align*}
  \cE_{t, 1}
  & = \big\{\forall (i, j) \in S_1: 
   \frac{\alpha(i) - \alpha(j)}{\alpha(i) + \alpha(j)} \rnd{n}_t(i, j) - 2 \sqrt{\rnd{n}_t(i, j) \log(1 / \delta)} \leq
  \rnd{s}_t(i, j)\big\}\,, \\
  \cE_{t, 2}
  & = \set{\forall (i, j) \in S_2: \rnd{s}_t(i, j) \leq 2 \sqrt{\rnd{n}_t(i, j) \log(1 / \delta)}}\,.
\end{align*}
Let $\cE = \bigcap_{t \in [n]} (\cE_{t, 1} \cap \cE_{t, 2})$ and $\ccE$ be the complement of $\cE$. Then $\mathbb{P}(\ccE) \leq \delta^\frac{1}{2} K^2 n$.
\end{lemma}
\begin{proof}
First, we bound $\mathbb{P}(\overline{\cE_{t, 1}})$. Fix $(i, j) \in S_1$, $t \in [n]$, and $(\rnd{n}_\ell(i, j))_{\ell = 1}^t$. Let $\tau(m)$ be the time of observing item pair $(i, j)$ for the $m$-th time, $\tau(m) = \min \set{\ell \in [t]: \rnd{n}_\ell(i, j) = m}$ for $m \in [\rnd{n}_t(i, j)]$. Let $\rnd{z}_\ell = \rnd{c}_\ell(\rnd{\cR}_\ell^{-1}(i)) - \rnd{c}_\ell(\rnd{\cR}_\ell^{-1}(j))$. Since $(\rnd{n}_\ell(i, j))_{\ell = 1}^t$ is fixed, note that $\rnd{z}_\ell \neq 0$ if $\ell = \tau(m)$ for some $m \in [\rnd{n}_t(i, j)]$. Let $\rnd{X}_0 = 0$ and
\begin{align*}
  \rnd{X}_\ell =
  \sum_{\ell' = 1}^\ell \condEsub{\rnd{z}_{\tau(\ell')}}{\rnd{z}_{\tau(\ell')} \neq 0}{\tau(\ell') - 1} - \rnd{s}_{\tau(\ell)}(i, j)
\end{align*}
for $\ell \in [\rnd{n}_t(i, j)]$. Then $(\rnd{X}_\ell)_{\ell = 1}^{\rnd{n}_t(i, j)}$ is a martingale, because
\begin{align*}
  \rnd{X}_\ell - \rnd{X}_{\ell - 1} &=
  \condEsub{\rnd{z}_{\tau(\ell)}}{\rnd{z}_{\tau(\ell)} \neq 0}{\tau(\ell) - 1} -
  (\rnd{s}_{\tau(\ell)}(i, j) - \rnd{s}_{\tau(\ell - 1)}(i, j)) \\
  &=
  \condEsub{\rnd{z}_{\tau(\ell)}}{\rnd{z}_{\tau(\ell)} \neq 0}{\tau(\ell) - 1} -
  \rnd{z}_{\tau(\ell)}\,,
\end{align*}
where the last equality follows from the definition of $\rnd{s}_{\tau(\ell)}(i, j) - \rnd{s}_{\tau(\ell - 1)}(i, j)$. Now we apply the Azuma-Hoeffding inequality and get that
\begin{align*}
  P\left(\rnd{X}_{\rnd{n}_t(i, j)} - \rnd{X}_0 \geq 2 \sqrt{\rnd{n}_t(i, j) \log(1 / \delta)}\right) \leq \delta^\frac{1}{2}\,.
\end{align*}
Moreover, from the definitions of $\rnd{X}_0$ and $\rnd{X}_{\rnd{n}_t(i, j)}$, and by \cref{uai:lem:click loss}, we have that 
\begin{align*}
\delta^\frac{1}{2} \geq {}  & P\left(\rnd{X}_{\rnd{n}_t(i, j)} - \rnd{X}_0 \geq 2 \sqrt{\rnd{n}_t(i, j) \log(1 / \delta)}\right) \\
{}={} & P\left(\sum_{\ell' = 1}^{\rnd{n}_t(i, j)} 
  \condEsub{\rnd{z}_{\tau(\ell')}}{\rnd{z}_{\tau(\ell')} \neq 0}{\tau(\ell') - 1} - \rnd{s}_t(i, j) \geq
  2 \sqrt{\rnd{n}_t(i, j) \log(1 / \delta)}\right) \\
  {}\geq{} & P\left(\frac{\alpha(i) - \alpha(j)}{\alpha(i) + \alpha(j)} \rnd{n}_t(i, j) -
  \rnd{s}_t(i, j) \geq 2 \sqrt{\rnd{n}_t(i, j) \log(1 / \delta)}\right) \\
  {}={} & P\left(\frac{\alpha(i) - \alpha(j)}{\alpha(i) + \alpha(j)} \rnd{n}_t(i, j) -
  2 \sqrt{\rnd{n}_t(i, j) \log(1 / \delta)} \geq \rnd{s}_t(i, j)\right).
\end{align*}
The above inequality holds for any $(\rnd{n}_\ell(i, j))_{\ell = 1}^t$, and therefore also in expectation over $(\rnd{n}_\ell(i, j))_{\ell = 1}^t$. From the definition of $\cE_{t, 1}$ and the union bound, we have $\mathbb{P}(\overline{\cE_{t, 1}}) \leq \frac{1}{2} \delta^\frac{1}{2} K (K - 1)$.

The claim that $\mathbb{P}(\overline{\cE_{t, 2}}) \leq \frac{1}{2} \delta^\frac{1}{2} K (K - 1)$ is proved similarly, except that we use $\condEsub{\rnd{z}_{\tau(\ell)}}{\rnd{z}_{\tau(\ell)} \neq 0}{\tau(\ell) - 1} \leq 0$. From the definition of $\ccE$ and the union bound,
\begin{align*}
  \mathbb{P}(\ccE) \leq
  \sum_{t = 1}^n \mathbb{P}(\overline{\cE_{t, 1}}) + \sum_{t = 1}^n \mathbb{P}(\overline{\cE_{t, 2}}) \leq
  \delta^\frac{1}{2} K^2 n\,.
\end{align*}
This completes our proof.
\end{proof}


\section{Conclusions}
\label{uai:sec:conclusions}

In this chapter, we have provide an answer to \textbf{RQ2} by combining the advantages of both offline and online \ac{LTR} algorithms.  
In particular, we fill a gap in the LTR literature by proposing $\bubblerank$, a re-ranking algorithm that gradually improves an initial base list, which we assume to be provided by an offline LTR approach. The improvements are learned from small perturbations of base lists, which are unlikely to degrade the user experience greatly. We prove a gap-dependent upper bound on the regret of $\bubblerank$ and evaluate it on a large-scale click dataset from a commercial search engine.

We leave open several questions of interest. For instance, our chapter studies $\bubblerank$ in the setting of re-ranking. 
Although we explain an approach of extending $\bubblerank$ to the general ranking setup in \cref{uai:sec:discussion}, we expect further experiments to validate this approach. 
Our general topic of interest are exploration schemes that are more conservative than those of existing online LTR methods. Existing methods are not very practical because they can explore highly irrelevant items at frequently examined positions.


\chapter{Cascade Non-Stationary Bandits}
\label{chapter:ijcai19}

\footnote[]{This chapter was published as~\citep{li19cascading}.}

\acresetall

This chapter provides an answer to \textbf{RQ3}:
\begin{description}
	\item[\textbf{RQ3}] How to conduct online learning to rank when users change their preferences constantly? 
\end{description}
We study the online learning to rank problem in a \emph{non-stationary} environment where user preferences change abruptly at an unknown moment in time. 
We consider the problem of identifying the $K$ most attractive items and propose \emph{cascading non-stationary bandits}, an online learning variant of the \emph{cascading model}, where a user browses a ranked list from top to bottom and clicks on the first attractive item. 
We propose two algorithms for solving this non-stationary problem: $\cascadeducb$ and $\cascadeswucb$. 
We analyze their performance and derive gap-dependent upper bounds on the $n$-step regret of these algorithms. 
We also establish a lower bound on the regret for cascading non-stationary bandits and show that both algorithms match the lower bound up to a logarithmic factor.
Finally, we evaluate their  performance on a real-world web search click dataset.

\section{Introduction}
\label{ijcai:sec:introduction}
Learning to rank \acused{LTR}\acs{LTR}~\citep{liu09learning} is a combination of machine learning and information retrieval. 
It is a core problem in many applications, such as web search and recommendation~\citep{liu09learning,zoghi14relative}. 
The goal of \ac{LTR} is to rank items, e.g., documents, and show the top $K$ items to a user. 
Traditional \ac{LTR} algorithms are supervised, offline algorithms; they learn rankers from human annotated data~\citep{qin10letor} and/or users' historical interactions~\citep{joachims02optimizing}. 
Every day billions of users interact with modern search engines and leave a trail of interactions. 
It is feasible and important to design online algorithms that directly learn from such user clicks to help improve users' online experience. 
Indeed, recent studies show that even well-trained production rankers can be optimized by using users' online interactions, such as clicks~\citep{zoghi16click}. 

Generally, interaction data is noisy~\citep{joachims02optimizing}, which gives rise to the well-known exploration vs.\ exploitation dilemma. 
Multi-armed bandit~(\acs{MAB})\acused{MAB}~\citep{auer02finite} algorithms have been designed to balance exploration and exploitation. 
Based on \acp{MAB}, many online \ac{LTR} algorithms have been published~\citep{radlinski08learning,kveton15cascading,katariya16dcm,lagree16multiple,zoghi17online,li19bubblerank}. 
These algorithms address the exploration vs. exploitation dilemma in an elegant way and aim to maximize user satisfaction in a stationary environment where users do not change their preferences over time. 
Moreover, they often come with regret bounds. 

Despite the success of the algorithms mentioned above in the stationary case, they may have linear regret in a non-stationary environment where users may change their preferences abruptly at any unknown moment in time. 
Non-stationarity widely exists in real-world application domains, such as search engines and recommender systems~\citep{yu09piecewise,pereira18analyzing,wu18learning,jagerman19when}.  
Particularly, we consider \emph{abruptly changing environments} where user preferences remain constant in certain time periods, named \emph{epochs}, but change occurs abruptly at unknown moments called \emph{breakpoints}. 
The abrupt changes in user preferences give rise to a new challenge of balancing ``remembering" and ``forgetting"~\citep{besbes14stochastic}: the more past observations an algorithm retains the higher the risk of making a biased estimator, while the fewer observations retained the higher stochastic error it has on the estimates of the user preferences. 

In this chapter, we propose \emph{cascading non-stationary bandits}, an online variant of the \ac{CM}~\citep{craswell08experimental} with the goal of identifying the $K$ most attractive items in a non-stationary environment. 
\ac{CM} is a widely-used model of user click behavior~\citep{chuklin15click,zoghi17online}. 
In \ac{CM}, a user browses the ranked list from top to bottom and clicks the first attractive item.
The items ranked above the first clicked item are browsed but not attractive since they are not clicked. 
The items ranked below the first clicked item are not browsed since the user stops browsing the ranked list after a click. 
Although \ac{CM} is a simple model, it effectively explains user behavior~\citep{kveton15cascading}. 

Our key technical contributions in this chapter are: 
\begin{enumerate*}[label=(\arabic*)]
	\item We formalize a  non-stationary \ac{OLTR} problem as cascading non-stationary bandits. 
	\item  We propose two algorithms, $\cascadeducb$ and $\cascadeswucb$, for solving it. 
	They are  motivated by \emph{discounted UCB} ($\ducb$) and \emph{sliding window UCB} ($\swucb$), respectively~\citep{garivier11upper}. 
	$\cascadeducb$ balances ``remembering" and ``forgetting" by using a discounting factor of past observations, and $\cascadeswucb$ balances the two by using statistics inside a fixed-size sliding window. 
	\item We derive gap-dependent upper bounds on the regret of the proposed algorithms. 
	\item We derive a lower bound on the regret of cascading non-stationary bandits. 
	We show that the upper bounds match this lower bound up to a logarithmic factor. 
	\item We evaluate the performance of $\cascadeswucb$ and  $\cascadeducb$ empirically on a real-world web search click dataset. 
\end{enumerate*}


\section{Background}
\label{ijcai:sec:background}

We define the learning problem at the core of this chapter in terms of cascading non-stationary bandits.
Their definition builds on the \ac{CM} and its online variant \emph{cascading bandits}, which we review in this section.

We write $[n]$ for $\set{1, \dots, n}$. 
For sets $A$ and $B$, we write $A^B$ for the set of all vectors whose entries are indexed by $B$ and take values from $A$. 
We use boldface letters to denote random variables. We denote a set of candidate items by $\cD = [L]$, e.g., a set of preselected documents. 
The presented ranked list is denoted as $\cR \in \Pi_K(\cD)$, where $\Pi_K(\cD)$ denotes the set of all possible  combinations of $K$ distinct items from $\cD$. 
The item at position $k$ in $\cR$ is denoted as $\cR(k)$, and the position of item $a$ in $\cR$ is denoted as $\cR^{-1}(a)$ 

\subsection{Cascade model}
\label{ijcai:sec:cascade model}

We refer readers to \citep{chuklin15click} for an introduction to click models. 
Briefly, a click model models a user's interaction behavior with the search system. 
The user is presented with a $K$-item ranked list $\cR$.  
Then the user browses the list $\cR$ and clicks items that potentially attract him or her. 
Many click models have been proposed and each models a certain aspect of interaction behavior.
We can parameterize a click model by attraction probabilities $\alpha \in [0, 1]^L$ and a click model assumes:
\begin{assumption}
\label{ijcai:ass:1}
The attraction probability $\alpha(a)$ only depends on item $a$ and is independent of other items.
 \end{assumption}
 
\noindent%
\ac{CM} is a widely-used click model~\citep{craswell08experimental,zoghi17online}. 
In the \ac{CM}, a user browses the ranked list $\cR$ from the first item $\cR(1)$ to the last item $\cR(K)$, which is called the \emph{cascading assumption}. 
After the user browses an item $\cR(i)$, he or she clicks on $\cR(i)$ with attraction probability $\alpha(\cR(i))$, and then stops browsing the remaining items. 
Thus, the examination probability of item $\cR(j)$ equals the probability of no click on the higher ranked items: $\prod_{i=1}^{j-1} (1-\alpha(\cR(i)))$. 
The expected number of clicks equals the probability of clicking any item in the list: $1-\prod_{i=1}^{K} (1-\alpha(\cR(i)))$. 
Note that the reward does not depend on the order in $\cR$, and thus, in the \ac{CM}, the goal of ranking is to find the $K$ most attractive items. 

The \ac{CM} accepts at most one click in each search session.  
It cannot explain scenarios where a user may click multiple  items. 
The \ac{CM} has been extended in different ways to capture multi-click cases~\citep{chapelle09dynamic,guo09efficient}. 
Nevertheless, \ac{CM} is still the fundamental click model and fits historical click data reasonably well.  
Thus, in this chapter, we focus on the \ac{CM} and in the next section we introduce  an online variant of \ac{CM}, called \emph{cascading bandits}.

\subsection{Cascading bandits}
\label{ijcai:sec:cascadingbandits}
\emph{Cascading bandits} (\acs{CB})\acused{CB} is defined  by a tuple $B = (\cD, P, K)$, where $\cD = [L]$ is the set of candidate items, $K \leq L$ is the number of positions, $P \in \{0, 1\}^{L}$ is a distribution over binary attractions.

In \ac{CB}, at time $t$, a learning agent builds a ranked list $\vR_t \in \Pi_K(\cD)$ that depends on the historical observations up to $t$ and shows it to the user. 
$\vA_t \in \{0, 1\}^L$ is defined as the \emph{attraction indicator}, which is drawn from $P$ and $\vA_t(\vR_t(i))$ is the attraction indicator of item $\vR_t(i)$. 
The user examines $\vR_t$ from $\vR_t(1)$  to $\vR_t(K)$ and clicks the first attractive item.
Since a \ac{CM} allows at most one click each time, a random variable $\vc_t$ is used to indicate the position of the clicked item, i.e., $\vc_t = \argmin_{i\in [K]}  \mathds{1}\{\vA_t(\vR_t(i))\}$. 
If there is no attractive item, the user will not click, and we set $\vc_t = K+1$ to indicate this case. 
Specifically, if $\vc_t \leq K$, the user clicks an item, otherwise, the user does not click anything. 
After the click or browsing the last item in $\vR_t$, the user leaves the search session. 
The click feedback $\vc_t$ is then observed by the learning agent.
Because of the cascading assumption, the agent knows that items ranked above position $\vc_t$ are observed. 
The reward at time $t$ is defined by the number of clicks:
\begin{equation}
r(\vR_t, \vA_t) = 1 - \prod_{i=1}^{K} (1-\vA_t(\vR_t(i)))\,.
\end{equation}

\noindent%
Under \cref{ijcai:ass:1},  the attraction indicators of each item in $\cD$ are independently distributed.
Moreover, cascading bandits make another assumption. 
\begin{assumption}
	The attraction indicators are distributed as:
	\begin{equation}
	P(\vA) = \prod_{a \in \cD} P_a(\vA(a))\,,
	\end{equation}
	where $P_a$ is a Bernoulli distribution with a mean of $\alpha(a)$.
	\label{ijcai:ass:2}
\end{assumption}

\noindent%
Under \cref{ijcai:ass:1} and~\cref{ijcai:ass:2}, the attraction indicator of item $a$ at time $t$ $\vA_t(a)$ is drawn independently from other items. 
Thus, the expectation of reward of the ranked list at time $t$ can be computed as $\expect{r(\vR_t, \vA_t} = r(\vR_t,  \alpha)$.  
And the goal of the agent is to maximize the expected number of clicks in $n$ steps, which is equivalent to minimizing the $n$-step cumulative regret: 
\begin{equation}
	R(n) = \sum_{t=1}^{n}\expect{\max_{R \in \Pi_K(\cD)} r(R, \alpha) - r(\vR_t, \alpha)}.
\end{equation}
Cascading bandits are designed for a stationary environment, where the attraction probability $P$ remains constant. 
However, in real-world applications, users change their preferences constantly~\citep{jagerman19when}, which is called a \emph{non-stationary environment}, and learning algorithms proposed for cascading bandits, e.g., $\cascadeklucb$ and $\cascadeucb$~\citep{kveton15cascading}, may have linear regret in this setting. 
In the next section, we propose cascading non-stationary bandits, the first non-stationary variant of cascading bandits, and then propose two algorithms for solving this problem.

\section{Cascading Non-Stationary Bandits}
\label{ijcai:sec:cascading non-stationary bandits}
We first define our non-stationary online learning setup, and then we propose two algorithms learning in this setup. 

\subsection{Problem setup}
\label{ijcai:sec:setup}

The learning problem we study is called \emph{cascading non-stationary bandits}, a variant of \ac{CB}. 
We define it by a tuple $B = (\cD, P, K, \Upsilon_n)$, where $\cD = [L]$ and $K \leq L$ are the same as in \ac{CB} bandits, $P \in \{0, 1\}^{n\times L}$ is a distribution over binary attractions and $\Upsilon_n$ is the number of abrupt changes in $P$ up to step $n$.
We use $P_t(\vR_t(i))$ to indicate the attraction probability distribution of item $ \vR_t(i)$ at time $t$. 
If $\Upsilon_n = 0$, this setup is  same as \ac{CB}.
The difference is that  we consider a non-stationary learning setup in which $\Upsilon_n > 0$ and the non-stationarity in attraction probabilities characterizes our learning problem.

In this chapter, we consider an abruptly changing environment, where  the attraction probability $P$ remains constant within an epoch but can change at any unknown moment in time and the number of abrupt changes up to $n$ steps is $\Upsilon_n$.
The learning agent interacts with cascading non-stationary bandits in the same way as with \ac{CB}. 
Since the agent is in a non-stationary environment, we write $\alpha_t$ for the mean of the attraction probabilities at time $t$ and we evaluate the agent by the expected cumulated regret expressed as:  
\begin{equation}
	R(n) = \sum_{t=1}^{n}\expect{\max_{R \in \Pi_K(\cD)} r(R, \alpha_t) - r(\vR_t, \vA_t)}.
\end{equation}
The goal of the agent it to minimizing the $n$-step regret. 

\subsection{Algorithms}
\label{ijcai:sec:algorithms}

\begin{algorithm}[t]
	\caption{UCB-type algorithm for Cascading non-stationary bandits.}
	\label{alg:cascadingbandits}
	\begin{algorithmic}[1]
		\STATE \textbf{Input}: discounted factor $\gamma$ or sliding window size $\tau$
		\vspace{0.05in}
		
		\STATE // Initialization
		\STATE \label{alg:initial1} $\forall a \in \cD: \vN_0(a) = 0$
		\STATE \label{alg:initial2} $\forall a \in \cD: \vX_0(a) = 0$
		
		\vspace{0.05in}
		\FOR{$t = 1, 2, \ldots, n$}
		\FOR{$a \in \cD$}
		\STATE // Compute UCBs
		\STATE \label{alg:ucbs}
		$\vU_t(a) \leftarrow  
		\begin{cases}
		\text{ \cref{ijcai:eq:ducb}} & (\cascadeducb)\\
		\text{\cref{ijcai:eq:swucb}} & (\cascadeswucb)
		\end{cases}
		$ 
		\ENDFOR
		\vspace{0.05in}
		
		\STATE // Recommend top $K$ items and receive clicks
		\STATE $\vR_t \leftarrow \argmax_{\vR\in\Pi_K(\cD)} r(\vR, \vU_t)$
		\STATE Show $\vR_t$ and receive clicks $\vc_t \in \{1, \ldots, K+1 \}$
		\vspace{0.05in}
		
		\STATE // Update statistics
		\IF{$\cascadeducb$}  \label{alg:statistics1}
		\STATE // for $\cascadeducb$
		\STATE \label{alg:ducb1} $\forall a \in \cD: \vN_t(a) =  \gamma \vN_{t-1}(a)$
		\STATE \label{alg:ducb2} $\forall a \in \cD: \vX_t(a) =  \gamma \vX_{t-1}(a)$
		\ELSE
		\STATE // for $\cascadeswucb$
		\STATE\label{alg:swucb1} $\forall a \in \cD: \vN_t(a) =  \sum_{s=t-\tau +1}^{t-1}
		\mathds{1}\{a \in \vR_s\}$
		\STATE \label{alg:swucb2}$\forall a \in \cD: \vX_t(a) =  \sum_{s=t-\tau +1}^{t-1}
		\mathds{1}\{\vR_s^{-1}(a) = \vc_s\}$
		
		\ENDIF
		\FOR{$i = 1, \ldots, \min\{\vc_t, K\}$}
		\STATE $a \leftarrow \vR_t(i)$
		\STATE  $\vN_t(a) =  \vN_t(a)  + 1 $
		\STATE  $\vX_t(a) = \vX_t(a) + \mathds{1}\{i = \vc_t \}$
		\label{alg:statistics2}
		\ENDFOR
		\ENDFOR
	\end{algorithmic}
\end{algorithm}

We propose two algorithms, $\cascadeducb$ and $\cascadeswucb$, for solving cascading non-stationary bandits. 
The former one is inspired by $\ducb$ and the later one is inspired by $\swucb$~\citep{garivier11upper}.
We summarize the pseudocode of both algorithms in~\cref{alg:cascadingbandits}.

$\cascadeducb$ and  $\cascadeswucb$ learn in a similar pattern. 
They differ in the way they estimate the \acf{UCB} $\vU_t(\vR_t(i))$ of the attraction probability of item $\vR_t(i)$ as time $t$, as discussed later in this section. 
After estimating the UCBs (line~\ref{alg:ucbs}), both algorithms construct $\vR_t$ by including the top $K$ most relevant items by UCB. 
Since the order of top $K$ items only affects the observation but does not affect the payoff of  $\vR_t$, we construct $\vR_t$ as follows:
\begin{equation}
 \vR_t = \argmax_{\cR \in \Pi_K(\cD)} r(\cR, \vU_t). 
\end{equation}
After receiving the user's click feedback $\vc_t$, both algorithms update their statistics (lines~\ref{alg:statistics1}--\ref{alg:statistics2}). 
We use $\vN_t(i)$ and $\vX_t(i)$  to indicate the number of items $i$ that have been observed and clicked up to $t$ step, respectively.

To tackle the challenge of non-stationarity, $\cascadeducb$  penalizes old observations with a discount factor $\gamma \in (0, 1)$. 
Specifically, each of the previous statistics is discounted by $\gamma$ (lines~\ref{alg:ducb1}--\ref{alg:ducb2}).
The UCB of item $a$ is  estimated as: 
\begin{equation}
\label{ijcai:eq:ducb}
	\vU_t(a) = \bar{\valpha}_t(\gamma, a)+ c_t(\gamma, a),
\end{equation}
where $\bar{\valpha}_t(\gamma, a) = \frac{\vX_t(a)}{\vN_t(a)}$ is the  average of discounted attraction indicators  of item $i$ and
\begin{equation}
c_t(\gamma, a) = 2\sqrt{\frac{\epsilon\ln{N_t(\gamma)}}{\vN_t(a)}}
\end{equation}
is the confidence interval around $\bar{\valpha}_t(i)$ at time $t$. 
Here, we compute $N_t(\gamma) = \frac{1 - \gamma^t}{1 - \gamma}$ as the discounted time horizon. 
As shown in \citep{garivier11upper}, $\alpha_t(a) \in [ \bar{\valpha}_t(\gamma, a) - c_t(\gamma, a) ,  \bar{\valpha}_t(\gamma, a)+ c_t(\gamma, a)] $ holds with high probability.

As to $\cascadeswucb$, it estimates UCBs by observations inside a sliding window with size $\tau$. 
Specifically, it only considers the observations in the previous $\tau$ steps~(lines~\ref{alg:swucb1}--\ref{alg:swucb2}). 
The UCB of item $i$ is estimated as
\begin{equation}
\label{ijcai:eq:swucb}
\vU_t(a) = \bar{\valpha}_t(\tau, a)+ c_t(\tau, a),
\end{equation}
where $\bar{\valpha}_t(\tau, a)=\frac{\vX_t(a)}{\vN_t(a)} $ is the  average of observed attraction indicators of item $a$ inside the sliding window and
\begin{equation}
c_t(\tau, a) = \sqrt{\frac{\epsilon\ln{( t \land \tau)}}{\vN_t(a)}} 
\end{equation}
is the confidence interval, and $t \land \tau = \min(t, \tau)$.

\paragraph{Initialization.}
In the initialization phase, we set all the statistics to $0$ and define  $\frac{x}{0}:=1$ for any $x$ (lines~\ref{alg:initial1}--\ref{alg:initial2}).
Mapping back this to UCB, at the beginning, each item has the optimal assumption on the attraction probability with an optimal bonus on uncertainty. 
This is a common initialization strategy for UCB-type bandit algorithms~\citep{li20mergedts}.

\section{Analysis}
\label{ijcai:sec:analysis}

In this section, we analyze the $n$-step regret of $\cascadeducb$ and $\cascadeswucb$. 
We  first derive regret upper bounds on $\cascadeducb$ and $\cascadeswucb$, respectively.
Then we  derive a regret lower bound on cascading non-stationary bandits.
Finally, we discuss our theoretical  results.

\subsection{Regret upper bound}
\label{ijcai:sec:upper bound on cascadeducb}
We refer to $\cDst \subseteq [L]$ as the set of the $K$ most attractive  items in set $\cD$ at time $t$ and $\bar{\cD}_t$ as the complement of $\cDst$, i.e.,  $\forall a \in \cD_t^*\,, \forall a^* \in \bar{\cD}_t: \alpha_t(a) \geq \alpha_t(a^*)$ and $\cDst \cup \cDct = \cD\,, \cDst \cap \cDct = \emptyset$. 
At time $t$, we say an item $a^*$ is optimal if $a^* \in \cDst$ and an item $a$ is suboptimal if $a \in \cDct$. 
The regret at time $t$ is caused by the case that $\vR_t$ includes at least one suboptimal and examined items. 
Let $\Delta_{a, a^*}^t$ be the gap of attraction probability between a suboptimal item $a$ and an optimal $a^*$ at time $t$: $\Delta_{a, a^*}^t = \alpha_t(a^*) - \alpha_t(a)$.
Then we refer to $\Delta_{a, K}$ as the smallest gap of between item $a$ and the $K$-th most attractive item in all $n$ steps when $a$ is not the optimal items: 
$\Delta_{a, K} = \min_{t \in [n], a^* \in \cDst}  \alpha_t(a^*) - \alpha_t(a) $.

\begin{theorem}
	\label{ijcai:th:upperboundcascadeducb}
	Let $\epsilon \in (1/2, 1)$ and $\gamma \in (1/2, 1)$, the expected $n$-step regret of $\cascadeducb$ is bounded as: 
	\begin{equation}
	\label{ijcai:eq:upperboundcascadeducb}
	\begin{split}
		R(n) \leq 
		 L \Upsilon_n\frac{\ln[(1-\gamma)\epsilon]}{\ln\gamma} 
		+ 
		\sum_{a \in \cD}  C(\gamma, a) \lceil n (1-\gamma) \rceil \ln{\frac{1}{1-\gamma}} \,,
	\end{split}
	\end{equation}
	where 
	\begin{equation}
	\begin{split}
	C(\gamma, a) = 
	\frac{4}{1-1/e}\ln{(1+4\sqrt{1-1/2\epsilon})} + \frac{32\epsilon}{\Delta_{a, K}\gamma^{1/(1-\gamma)}} \,.
	\end{split}
	\end{equation}
\end{theorem}

\noindent%
We outline the proof in $4$ steps below; the full version is in \cref{ijcai:sec:proofoftheorem2}.\footnote{\url{https://arxiv.org/abs/1905.12370}}

\begin{proof}
Our proof is adapted from the analysis in \citep{kveton15cascading}. 
The novelty of the proof comes from the fact that, in a non-stationary environment, the discounted estimator $\barvalphat(\gamma, a)$ is now a biased estimator of $\alpha_t(a)$ (\emph{Step 1, 2} and \emph{4}).

\emph{Step 1.} We bound the regret of the event that estimators of the attraction probabilities are biased by $L\Upsilon\frac{\ln[(1-\gamma)\epsilon]}{\ln\gamma}$. 
This event happens during the steps following a breakpoint. 

\emph{Step 2.} We bound the regret of the event that $\alpha_t(a)$ falls outside of the confidence interval around $\barvalphat(\gamma, a)$ by $\frac{4}{1-1/e}\ln{(1+4\sqrt{1-1/2\epsilon})} n(1-\gamma) \ln\frac{1}{1-\gamma}$. 

\emph{Step 3.} We decompose the regret at time $t$ based on \cite[Theorem 1]{kveton15cascading}.

\emph{Step 4.} For each item $a$, we bound the number of times that item $a$ is chosen when $a \in \cDct$ in $n$ steps and get the term $\frac{32\epsilon \lceil n (1-\gamma) \rceil  \ln\frac{1}{1-\gamma}}
{\Delta_{a, K}\gamma^{1/(1-\gamma)}}  $. 
Finally, we sum up all the regret. 
\end{proof}

\noindent%
The bound depends on step $n$ and the number of breakpoints $\Upsilon_n$. 
If they are known beforehand, we can choose $\gamma$ by minimizing the right hand side of ~\cref{ijcai:eq:upperboundcascadeducb}. 
Choosing $\gamma = 1 - 1/4 \sqrt{(\Upsilon_n/n)}$ leads to $R(n) = O(\sqrt{n\Upsilon_n}\ln n)$. 
When $\Upsilon_n$ is independent of $n$, we have $R(n) = O(\sqrt{n\Upsilon}\ln n)$. 

\begin{theorem}
	\label{ijcai:th:upperboundcascadeswucb}
	Let $\epsilon \in (1/2, 1)$. For any integer $\tau$, the expected $n$-step regret of $\cascadeswucb$ is bounded as: 
	\begin{equation}
	\label{ijcai:eq:upperboundcascadeswucb}
	\begin{split}
	R(n) \leq 
	L  \Upsilon_n\tau \!+\! \frac{L\ln^2\tau}{\ln(1+4\sqrt{(1-1/2\epsilon)})} \!+\! \sum_{a \in \cD} C(\tau, a) \frac{n\ln \tau}{\tau} ,\mbox{}\hspace*{-2mm}
	\end{split}
	\end{equation}
	where 
	\begin{equation}
	\begin{split}
	C(\tau, a) = 
	\frac{2}{\ln\tau} \left\lceil \frac{\ln \tau}{\ln(1+4\sqrt{(1-1/2\epsilon)})}\right\rceil + \frac{8\epsilon}{\Delta_{a, K}} \frac{\lceil n/\tau \rceil}{n/\tau}.
	\end{split}
	\end{equation}
	When $\tau$ goes to infinity and $n/\tau$ goes to $0$, 
	\begin{equation}
	 C(\tau, a) = \frac{2}{\ln(1+4\sqrt{(1-1/2\epsilon)})}+ \frac{8\epsilon}{\Delta_{a, K}} .
	\end{equation}
\end{theorem}

\noindent%
We outline the proof in $4$ steps below and the full version is in \cref{ijcai:sec:proofoftheorem3}. 

\begin{proof}
The proof follows the same lines as the proof of \cref{ijcai:th:upperboundcascadeducb}. 

\emph{Step 1.} We bound the regret of the event that estimators of the attraction probabilities are biased by $L  \Upsilon_n\tau$. 

\emph{Step 2.} We bound the regret of the event that $\alpha_t(a)$ falls outside of the confidence interval around $\barvalphat(\tau, a)$ by 
\begin{equation}
\ln^2\tau + 2n \left\lceil \frac{\ln \tau}{\ln(1+4\sqrt{(1-1/2\epsilon)})}\right\rceil. 
\end{equation}

\emph{Step 3.} We decompose the regret at time $t$ based on \cite[Theorem 1]{kveton15cascading}.

\emph{Step 4.} For each item $a$, we bound the number of times that item $a$ is chosen when $a \in \cDct$ in $n$ steps and get the term $ \frac{8\epsilon}{\Delta_{a, K}} \lceil \frac{ n}{\tau}\rceil $. 
Finally, we sum up all the regret. 
\end{proof}

\noindent%
If we know $\Upsilon_n$ and $n$ beforehand, we can choose the window size $\tau$ by minimizing the right hand side of ~\cref{ijcai:eq:upperboundcascadeswucb}. 
Choosing $\tau = 2\sqrt{n\ln(n) / \Upsilon_n}$ leads to $R(n) = O(\sqrt{n\Upsilon_n\ln n})$. 
When $\Upsilon_n$ is independent of $n$, we have $R(n) = O(\sqrt{n\Upsilon\ln n})$. 

\subsection{Regret lower bound}
\label{ijcai:sec:lower bound}

We consider a particular cascading non-stationary bandit and refer to it as  $B_{L}=(L, K, \Delta, p, \Upsilon)$. 
We have a set of $L$ items  $\cD = [L]$ and $K = \frac{1}{2} L$ positions. 
At any time $t$, the distribution of attraction probability of each item $a\in\cD$ is parameterized by: 
\begin{equation}
\alpha_t(a)= \begin{cases}
	p \qqquad   \text{if }a \in \cDst\\
	p-\Delta  \qquad \text{if }a \in \cDct, 
\end{cases}
\end{equation}
where $ \cDst$ is the set of optimal items at time $t$, $\cDct$ is the set suboptimal items at time $t$,  and $\Delta \in (0, p]$ is the gap between optimal items and suboptimal items.
Thus, the attraction probabilities only take two values: $p$ for optimal items and $p-\Delta$ for suboptimal items up to $n$-step. 
$\Upsilon$ is the number of breakpoints when the attraction probability of an item changes from $p$ to $p-\Delta$ or other way around. 
Particularly, we consider a simple variant that  the distribution of attraction probability of each item is piecewise constant and has two breakpoints. 
And we assume another constraint on the number of optimal items that $|\cDst| = K$ for all time steps $t\in[n]$.
Then, the regret that any learning policy can achieve when interacting with $B_L$ is lower bounded by \cref{ijcai:th:lowerbound}.

\begin{theorem}
	\label{ijcai:th:lowerbound}
	The $n$-step regret of any learning algorithm inter\-acting with 
	cascading non-stationary bandit $B_{L}$ is lower bounded as follows:
	\begin{equation}
		\label{ijcai:eq:lowerbound}
		\liminf\limits_{n\rightarrow \infty} R(n) \geq L \Delta (1-p)^{K-1} \sqrt{\frac{2 n}{3D_{KL}(p-\Delta || p)}}, 
	\end{equation}
	where $D_{KL}(p-\Delta || p)$ is the Kullback-Leibler (KL) divergence between two Bernoulli distributions with means $p-\Delta$ and $p$. 
\end{theorem}

\begin{proof}
The proof is based on the analysis in \citep{kveton15cascading}. 
We first refer to  $\cR_t^*$ as the optimal list at time $t$ that includes $K$ items. 
For any time step $t$, any item $a \in \cDct$ and any item $a^* \in \cDst$, we define the event that item $a$ is included in $\vR_t$ instead of item $a^*$ and  item $a$ is examined but not clicked at time step $t$ by: 
\begin{equation}
\begin{split}
G_{t, a, a^*} =
 \quad\{ \exists 1\leq k <\vc_t ~s.t.~ \vR_t(k) = a \,  , \cR_t(k)  = a^*\}.
\end{split}
\end{equation}
By \cite[Theorem 1]{kveton15cascading}, the regret at time $t$ is decomposed as:
\begin{equation}
	\mathbb{E}[r(\vR_t, \valpha_t)] \geq  \Delta (1-p)^{K-1}\sum_{a \in \cDct} \sum_{a* \in \cDst} \mathds{1}\{ G_{a, a^*, t} \}.
\end{equation}
Then, we bound the $n$-step regret as follows:
\begin{equation}
\begin{split}
	R(n) &\geq  \Delta (1-p)^{K-1}  \sum_{t=1}^{n} \sum_{a \in \cDct} \sum_{a^* \in \cDst} \mathds{1}\{G_{t, a, a^*}\} \\ 
	&\geq  \Delta (1-p)^{K-1} \sum_{a \in \cD}\sum_{t=1}^{n} \mathds{1}\{a\in \cDct, a \in\vR_t\}\\
	 &= \Delta (1-p)^{K-1}  \sum_{a \in \cD} \vT_n(a),
\end{split}	 
\end{equation}
where $\vT_n(a) =  \sum_{t=1}^{n}\mathds{1}\{a\in \cDct, a\in\vR_t, \vR_t^{-1}(a) \leq \vc_t \}$. 
The second inequality is based on the fact that, at time $t$,  the event $G_{t, a, a^*}$ happens if and only if item $a$ is suboptimal and examined. 
By the results of \cite[Theorem 3]{garivier11upper}, if a suboptimal item $a$ has not been examined enough times, the learning policy may play this item for a long period after a breakpoint. 
And we get: 
\begin{equation}
		\liminf\limits_{n\rightarrow \infty} \vT(n) \geq  \sqrt{\frac{2 n}{3D_{KL}(p-\Delta || p)}}.
\end{equation}
We sum up all the inequalities and obtain: 
\begin{equation*}
\liminf\limits_{n\rightarrow \infty} R(n) \geq L \Delta (1-p)^{K-1} \sqrt{\frac{2 n}{3D_{KL}(p-\Delta || p)}} . \qedhere
\end{equation*}
\end{proof}

\subsection{Discussion}
\label{ijcai:sec:discussion}
We have shown that the $n$-step regret upper bounds of $\cascadeducb$  and  $\cascadeswucb$ have the order of $O(\sqrt{n} \ln n)$and $O(\sqrt{n\ln n})$, respectively. 
They match the lower bound proposed in \cref{ijcai:th:lowerbound} up to a logarithmic factor. 
Specifically, the upper bound of $\cascadeducb$ matches the lower bound up to $\ln n$. 
The upper bound of $\cascadeswucb$ matches the lower bound up to $\sqrt{\ln n}$, an improvement over $\cascadeducb$, as confirmed by experiments in~\cref{ijcai:sec:experiments}. 

We have assumed that step $n$ is know beforehand. 
This may not always be the case. 
We can extend $\cascadeducb$ and $\cascadeswucb$ to the case where $n$ is unknown by using the \emph{doubling trick}~\citep{garivier11upper}. 
Namely, for $t > n$ and any $k$, such that $2^k \leq t < 2^{k+1}$, we reset $\gamma = 1 - \frac{1}{4\sqrt{2^k}}$ and $\tau = 2\sqrt{2^k\ln(2^k)}$. 

$\cascadeducb$ and $\cascadeswucb$ can be computed efficiently. 
Their complexity is linear in the number of time steps. 
However, $\cascadeswucb$ requires extra memory to remember past ranked lists and rewards to update $\vX_t$ and $\vN_t$. 

\section{Experimental Analysis}
\label{ijcai:sec:experiments}

We evaluate $\cascadeducb$ and $\cascadeswucb$ on the \emph{Yandex} click  dataset,\footnote{\url{https://academy.yandex.ru/events/data_analysis/relpred2011}} which is the largest public click collection. 
It contains  more than $30$ million search sessions, each of which  contains at least one search query. 
We process the queries in the same manner as  in~\citep{zoghi17online,li19bubblerank}. 
Namely, we randomly select $100$ frequent search queries with the $10$ most attractive items in each query, and then learn a CM for each query using PyClick.\footnote{\url{https://github.com/markovi/PyClick}}
These CMs are used to generate click feedback. 
In this setup, the size of  candidate items is $L=10$ and we choose $K=3$ as the number of positions. 
The objective of the learning task is to identify $3$ most attractive items and then maximize the expected number of clicks at the $3$  highest positions. 

We consider a simulated non-stationary environment setup, where we take the learned attraction probabilities as the default and change the attraction probabilities periodically. 
Our simulation can be described in $4$ steps: 
\begin{inparaenum}[(1)]
		\item For each query, the attraction probabilities of the top $3$ items remain constant over time. 
		\item We  randomly choose three suboptimal items and set their attraction probabilities to $0.9$ for the next $m_1$ steps. 
		\item Then we reset the attraction probabilities and keep them constant for the next $m_2$ steps.  
		\item We repeat step (2) and step (3) iteratively.
\end{inparaenum}
This simulation mimics abrupt changes in user preferences and is widely used in previous work on non-stationarity~\citep{garivier11upper,jagerman19when}. 
In our experiment, we set $m_1 = m_2 = 10$k and choose $10$ breakpoints. 
In total, we run experiments for $100$k steps.
Although the non-stationary aspects in our setup are simulated, the other parameters of a CM are learned from the Yandex click dataset.

We compare $\cascadeducb$ and $\cascadeswucb$,  to $\rankedexpthree$~\citep{radlinski08learning},\break $\cascadeklucb$~\citep{kveton15cascading} and $\batchrank$~\citep{zoghi17online}. 
We describe the baseline algorithms in slightly more details in \cref{ijcai:sec: related work}. 
Briefly, $\rankedexpthree$, a variant of ranked bandits, is based on an adversarial bandit algorithm $\expthree$~\citep{auer95gambling}; it is the earliest bandit-based ranking algorithm and is popular in practice. 
$\cascadeklucb$~\citep{kveton15cascading} is a near optimal algorithm in CM. 
$\batchrank$~\citep{zoghi17online} can learn in a wide range of click models. 
However, these algorithms only learn in a stationary environment. 
We choose them as baselines to show the superiority of our algorithms in a non-stationary environment.
In experiments,  we set $\epsilon=0.5$, $\gamma = 1 - 1/(4 \sqrt{n})$ and $\tau= 2\sqrt{n\ln(n)}$, the values that roughly minimize the upper bounds.

\begin{figure}[t]
\centering
	\includegraphics[width=.9\textwidth]{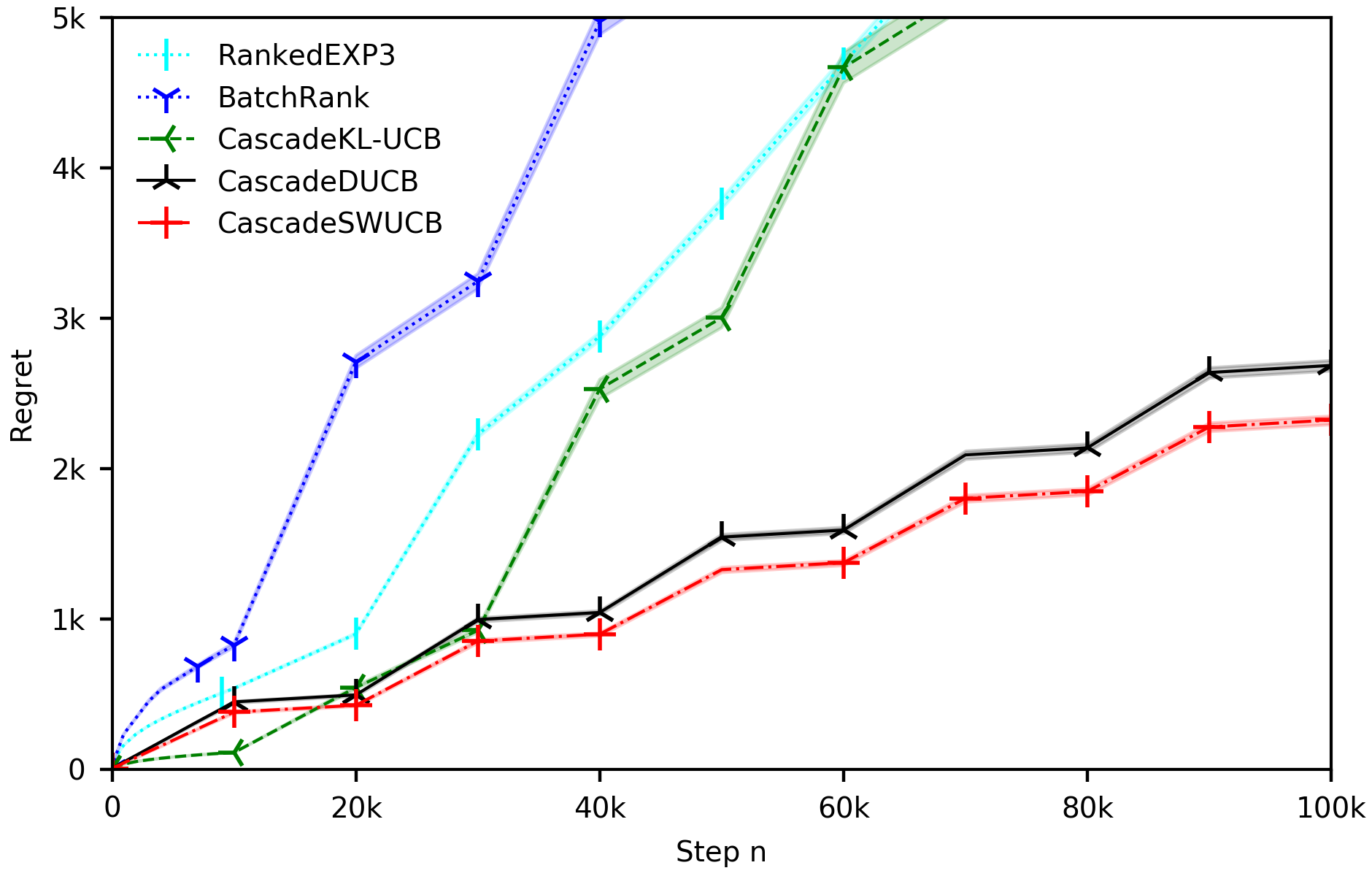}
	\caption{The n-step regret in up to $100$k steps. Lower is better. The results are averaged over all $100$ queries and $10$ runs per query. The shaded regions represent standard errors of our estimates.}\label{ijcai:fig:regret}
\end{figure}

We report the $n$-step regret averaged over $100$ queries and $10$ runs per query in~\cref{ijcai:fig:regret}. 
All baselines have linear regret in time step. 
They fail in capturing the breakpoints. 
Non-stationarity makes the baselines perform even worse during epochs where the attraction probability are set as  the default.
E.g., $\cascadeklucb$ has $111.50 \pm 1.12$ regret in the first $10$k steps but has $447.82\pm137.16$ regret between step $80$k and $90$k. 
Importantly, the attraction probabilities equal the default and remain constant inside these two epochs. 
This is caused by the fact that the baseline algorithms rank items based on all historical observations, i.e., they do not balance ``remembering" and ``forgetting." 
Because of the use of a discounting factor and/or a sliding window, $\cascadeducb$ and $\cascadeswucb$ can detect breakpoints and show convergence. 
$\cascadeswucb$ outperforms $\cascadeducb$ with a small gap; this is consistent with our theoretical finding that $\cascadeswucb$ outperforms $\cascadeducb$ by a $\sqrt{\ln n}$ factor. 


\section{Related Work}
\label{ijcai:sec: related work}

The idea of directly learning to rank from user feedback has been widely studied in a stationary environment.
\emph{Ranked bandits}~\citep{radlinski08learning} are among the earliest \ac{OLTR} approaches. 
In ranked bandits, each position in the list is modeled as an individual underlying \acp{MAB}. 
The ranking task is then solved by asking each individual \ac{MAB} to recommend an item to the attached position. 
Since the reward, e.g., click, of a lower position is affected by higher positions, the underlying \ac{MAB} is typically adversarial, e.g., $\expthree$~\citep{auer95gambling}.
$\batchrank$ is a  recently proposed \ac{OLTR} method~\citep{zoghi17online}; it is an elimination-based algorithm: once an item is found to be inferior to $K$ items, it will be removed from future consideration. 
$\batchrank$ outperforms ranked bandits in the stationary environment.
In our experiments, we include  $\batchrank$ and $\rankedexpthree$, the $\expthree$-based ranked bandit algorithm, as baselines.

Several \ac{OLTR} algorithms have been proposed in specific click models~\citep{kveton15cascading,lagree16multiple,katariya16dcm}. 
They can efficiently learn an optimal ranking given the click model they consider. 
Our work is related to cascading bandits and we compare our algorithms to $\cascadeklucb$, an algorithm proposed for soling cascading bandits~\citep{kveton15cascading}, in \cref{ijcai:sec:experiments}. 
Our work differs from cascading bandits in that we consider learning in a non-stationary environment.

Non-stationary bandit problems have been widely studied~\citep{slivkins08adapting,yu09piecewise,garivier11upper,besbes14stochastic,liu18change}.  
However, previous work requires a small action space. 
In our setup, actions are (exponentially many) ranked lists.
Thus, we do not consider them as baselines in our experiments. 

In \emph{adversarial bandits} the reward realizations, in our case attraction indicators, are selected by an \emph{adversary}. 
Adversarial bandits originate from game theory~\citep{blackwell56analog} and have been widely studied, cf.~\citep{auer95gambling,cesabianchi06prediction} for an overview. 
Within  adversarial bandits, the performance of a policy is often measured by comparing to a static oracle which always chooses a single best arm that is obtained after seeing all the reward realizations up to step $n$. 
This static oracle can perform poorly in a non-stationary case when the single best arm is suboptimal for a long time between two breakpoints.
Thus, even if a policy performs closely to the static oracle, it can still perform sub-optimally in a non-stationary environment. 
Our work differs from adversarial bandits in that we compare to a dynamic oracle that can balance the dilemma of ``remembering" and ``forgetting" and chooses the per-step best action.

\section{Proofs}

In this proofs, we refer to  $\cR_t^*$ as the optimal list at time $t$ that includes $K$ items sorted by the decreasing order of their attraction probabilities. 
We refer to $\cDst \subseteq [L]$ as the set of the $K$ most attractive  items in set $\cD$ at time $t$ and $\bar{\cD}_t$ as the complement of $\cDst$, i.e.  $\forall a^* \in \cD_t^*, \forall a \in \bar{\cD}_t: \alpha_t(a^*) \geq \alpha_t(a)$ and $\cDst \cup \cDct = \cD, \cDst \cap \cDct = \emptyset$. 
At time $t$, we say an item $a^*$ is optimal if $a^* \in \cDst$ and an item $a$ is suboptimal if $a \in \cDct$. 
We denote $\vr_t =\max_{R \in \Pi_K(\cD)} r(R, \alpha_t) - r(\vR_t, \vA_t)$ to be the regret at time $t$ of the learning algorithm. 
Let $\Delta_{a, a^*}^t$ be the gap of attraction probability between a suboptimal item $a$ and an optimal $a^*$ at time $t$: $\Delta_{a, a^*}^t = \alpha_t(a^*) - \alpha_t(a)$.
Then we refer to $\Delta_{a, K}$ as the smallest gap between item $a$ and the $K$-th most attractive item in all $n$ time steps when $a$ is not the optimal item: 
$\Delta_{a, K} = \min_{t \in [n], a \notin \cDst}  \alpha_t(K) - \alpha_t(a) $.

\subsection{Proof of \cref{ijcai:th:upperboundcascadeducb}}
\label{ijcai:sec:proofoftheorem2}

Let $M_{t} = \{\exists a\in\cD~s.t.~|\alpha_t(a) - \barvalphat(\gamma, a)| > c_t(\gamma, t)\}$ be the event that $\alpha_t(a)$ falls out of the confidence interval around $\barvalphat(\gamma, a)$ at time t, and  $\bar{M}_{t}$ be the complement of ${M}_{t}$. 
We re-write the $n$-step regret of $\cascadeducb$ as follows: 
\begin{equation}
\label{ijcai:eq:R}
R(n)  = \expect{\sum_{t=1}^{n} \mathds{1}\{M_{t}\} \vr_t} + \expect{\sum_{t=1}^{n} \mathds{1}\{\bar{M}_{t}\} \vr_t}. 
\end{equation}
We then bound both terms above separately. 

We refer to $\cT$ as the set of all time steps such that for $t \in \cT$, $s \in [t - B(\gamma), t]$ and any item $a\in\cD$ we have $\alpha_s(a) = \alpha_t(a)$, where $B(\gamma) = \lceil \log_{\gamma}(\epsilon (1-\gamma)) \rceil$. 
In other words, $\cT$ is the set of time steps that do not follow too close after breakpoints. 
Since for any time step $t \notin \cT$ the estimators of attraction probabilities are biased, $\cascadeducb$ may recommend suboptimal items constantly. Thus, we get the following bound:
\begin{equation}
\label{ijcai:eq:ducbdecompose}
	 \expect{\sum_{t=1}^{n} \mathds{1}\{M_{t}\} \vr_t}  \leq 
	 L\Upsilon_n B(\gamma) + \expect{\sum_{t\in\cT} \mathds{1}\{M_{t}\} \vr_t}.
\end{equation}
Then, we show that for steps $t \in \cT$, the attraction probabilities are well estimated for all items with high probability. 
For an item $a$, we consider the bias and variance of $\barvalphat(\gamma, a)$ separately. 
We denote:  
\[	\vX_t(\gamma, a) = \sum_{s=1}^{t} \gamma^{(t-s)} 		  	\mathds{1}\{a\in\vR_s, \vR_s(\vc_s)=a\}, \quad
	\vN_t(\gamma, a) = \sum_{s=1}^{t} \gamma^{(t-s)} \mathds{1}\{a\in\vR_s\},
\]
as the discounted number of being clicked, and the discounted number of being examined.

First, we  consider the bias. 
The ``bias" at time $t$ can be written as $x_t(\gamma, a)/\vN_t(\gamma, a) - \alpha_t$, where $x_t(\gamma, a) = \sum_{s=1}^{t} \gamma^{(t-s)} \mathds{1}\{a\in\vR_s\} \alpha_s(a)$. 
For $t\in\cT$: 
\begin{eqnarray*}
	|x_t(\gamma, a) - \alpha_t(a) \vN_t(\gamma, a)| &=& \abs{\sum_{s=1}^{t-B(\gamma)} \gamma^{(t-s)}(\alpha_s(a)-\alpha_t(a)\mathds{1}\{a \in \vR_t\}} \\ 
	&\leq& \sum_{s=1}^{t-B(\gamma)}\gamma^{(t-s)} \abs{(\alpha_s(a)-\alpha_t(a)}\mathds{1}\{a \in \vR_t\}\\
	&\leq& \gamma^{B(\gamma)} \vN_{t-B(\gamma)}(\gamma, a)
	\leq  \gamma^{B(\gamma)} \frac{1}{1-\gamma},
\end{eqnarray*}
where the last inequality is due to the fact that $\vN_{t-B(\gamma)}(\gamma, a) \leq 1/(1-\gamma)$. 
Thus, we get:
\begin{eqnarray*}
	\abs{\frac{x_t(\gamma, a)}{\vN_t(\gamma, a)} - \alpha_t(a)} 
	&\leq& \frac{\gamma^{B(\gamma)}}{(1-\gamma)\vN_t(\gamma, a)}
	\leq \left(1 \land  \frac{\gamma^{B(\gamma)}}{(1-\gamma)\vN_t(\gamma, a)}\right) \\
	&\leq& 
	\sqrt{\frac{\gamma^{B(\gamma)}}{(1-\gamma)\vN_t(\gamma, a)}}  
	\leq \sqrt{\frac{\epsilon \ln N_t(\gamma)}{\vN_t(\gamma, a)}} \leq \frac{1}{2} c_t(\gamma, a) , 
\end{eqnarray*} 
where the third inequality is due to the fact that $1\land x \leq \sqrt{x}$ and the last inequality is due to the definition of $B(\gamma)$. 
So, $B(\gamma)$ time steps after a breakpoint, the ``bias" is at most half as large as the confidence bonus;
and the second half of the confidence interval is used for the variance. 

Second, we consider the variance. 
For $a\in \cD$ and $t\in \cT$, let $M_{t,a} = \{|\alpha_t(a) - \barvalphat(\gamma, a)| > c_t(\gamma, t)\}$ be the event that $\alpha_t(a)$ falls out of the confidence interval around $\barvalphat(\gamma, a)$ at time $t$. 
By using a Hoeffding-type inequality \cite[Theorem 4]{garivier11upper}, for an item $a\in\cD$,  $t\in \cT$, and any $\eta > 0$, we get:
\begin{eqnarray*}
P(M_{t,a}) 
&\leq&
P\left( \frac{\vX_t(\gamma, a) - x_t(\gamma, a)}{\sqrt{\vN_t(\gamma^2, a)}} > \sqrt{\frac{\epsilon \ln N_t(\gamma)}{\vN_t(\gamma^2, a)}}\right) \\
&\leq&
P\left( \frac{\vX_t(\gamma, a) - x_t(\gamma, a)}{\sqrt{\vN_t(\gamma^2, a)}} > \sqrt{\epsilon \ln N_t(\gamma)}\right) \\
&\leq&
\left\lceil \frac{\ln{N_t(\gamma)}}{\ln{(1+\eta)}} \right\rceil 
N_t(\gamma)^{-2\epsilon(1-\frac{\eta^2}{16})} .
\end{eqnarray*}
Thus, we get the following bound: 
\[
\expect{\sum_{t\in\cT} \mathds{1}\{M_{t}\} \vr_t} 
\leq 
2L\sum_{t\in\cT} 
\left\lceil \frac{\ln{N_t(\gamma)}}{\ln{(1+\eta)}} \right\rceil 
N_t(\gamma)^{-2\epsilon(1-\frac{\eta^2}{16})}.
\]
By taking $\eta = 4\sqrt{1-1/2\epsilon}$ such that  $1-\frac{\eta^2}{16} = 1$, and with $t_0 = (1-\gamma)^{(-1)}$
we get: 
\begin{eqnarray*}
&\mbox{}&
\sum_{t\in\cT} 
\left\lceil \frac{\ln{N_t(\gamma)}}{\ln{(1+\eta)}} \right\rceil 
N_t(\gamma)^{-2\epsilon(1-\frac{\eta^2}{16})} 
\\
&\leq& 
t_0 + \sum_{t\in\cT, t \geq t_0} 
\left\lceil \frac{\ln{N_{t_0}(\gamma)}}{\ln{(1+\eta)}} \right\rceil 
N_{t_0}(\gamma)^{-1}  
\\
&\leq& 
t_0 +
\left\lceil \frac{\ln{N_{t_0}(\gamma)}}{\ln{(1+\eta)}} \right\rceil 
\frac{n}{N_{t_0}(\gamma)} 
\\
&\leq& 
\frac{1}{1-\gamma} +
\left\lceil \frac{\ln{N_{t_0}(\gamma)}}{\ln{(1+\eta)}} \right\rceil 
\frac{n(1-\gamma)}{1-\gamma^{1/(1-\gamma)}} .
\end{eqnarray*}
We sum up and get the upper bound: 
\begin{equation}
	\label{ijcai:eq:bound for outside}
	\expect{\sum_{t=1}^{n} \mathds{1}\{\bar{M}_{t}\} \vr_t} 
	\leq 
	L \Upsilon_n B(\gamma) 
	+
	2L\frac{1}{1-\gamma} +
	2L\left\lceil \frac{\ln{N_{t_0}(\gamma)}}{\ln{(1+\eta)}} \right\rceil 
	\frac{n(1-\gamma)}{1-\gamma^{1/(1-\gamma)}} .
\end{equation}


Third, we upper bound the second term in \cref{ijcai:eq:R}. 
The regret is caused by recommending a suboptimal item to the user and the user examines but does not click the item.
Since there are $\Upsilon_n$ breakpoints, we refer to $[t_1, \ldots, t_{\Upsilon_n}]$ as the time step of a breakpoint that occurs. 
We consider the time step in the individual epoch that does not contain a breakpoint. 
For any epoch and any time $t \in \{t_e, t_e+1, \ldots, t_{e+1}-1\}$, any item $a \in \cDci$ and any item $a^* \in \cDsi$, we define the event that item $a$ is included in $\vR_t$ instead of item $a^*$, and  item $a$ is examined but not clicked at time $t$ by: 
\begin{equation*}
\begin{split}
G_{t, a, a^*} =\{ \exists 1\leq k <\vc_t ~s.t.~ \vR_t(k) = a, \cR_t(k)  = a^*\}.
\end{split}
\end{equation*} 
Since the attraction probability remains constant in the epoch, we refer to $\cDsi$ as the optimal items and $\cDci$ as the suboptimal items in epoch $e$. 
By \cite[Theorem 1]{kveton15cascading}, the regret at time $t$ is decomposed as:
\begin{equation}
\label{ijcai:eq:ducbrdecompose}
\mathbb{E}[\vr_t] \leq  \Delta^t_{a, a^*}\sum_{a \in \cDci} \sum_{a* \in \cDsi} \mathds{1}\{ G_{a, a^*, t} \} .
\end{equation}
Then we have:
\begin{equation}
\label{ijcai:eq:inside}
\expect{\sum_{t=t_i}^{t_{i+1}-1} \mathds{1}\{\bar{M}_{t}\} \vr_t} 
= 
\sum_{t=t_i}^{t_{i+1}-1} \mathds{1}\{\bar{M}_{t}\} \expect{\vr_t}
\leq 
\sum_{a \in \cDci} \expect{\sum_{a^*\in\cDsi} \sum_{t=t_i}^{t_{i+1}-1}  
\Delta^t_{a, a^*} \mathds{1}\{ G_{a, a^*, t} \}} ,
\end{equation}
where the first equality is due to the tower rule, and the inequality is due to \cref{ijcai:eq:ducbrdecompose}. 

Now, for any suboptimal item $a$ in epoch $e$, we upper bound \break$ \expect{\sum_{a^*\in\cDsi} \sum_{t=t_i}^{t_{i+1}-1}  
	\Delta^t_{a, a^*} \mathds{1}\{ G_{a, a^*, t} \}} $. 
At time $t$,  event $\mathds{1}\{\bar{M}_{t}\}$ and event $ \mathds{1}\{a \in \vR_t, a \in \cDct\}$ happen when there exists an optimal item $a^*\in \cDsi$ such that:
 \[
 \alpha_t(a) + 2 c_t(\gamma, a) \geq \vU_t(a) \geq \vU(a^*) \geq \alpha_t(a^*) , 
\]
which implies that $2 c_t(\gamma, a) \geq \alpha_t(a^*) - \alpha_t(a)$. 
Taking the definition of the confidence interval, we get: \[\vN_t(\gamma, a) \leq \frac{16\epsilon \ln N_{t}(\gamma)}{\Delta^{2}_{t, a, a^*}},\] 
where we set $\Delta_{t, a, a^*} = \Delta^{t}_{ a, a^*}$.

Together with \cref{ijcai:eq:inside}, we get: 
\begin{eqnarray}
&\mbox{}&
\expect{
	\sum_{t=t_i}^{t_{i+1}-1} \mathds{1}\{\bar{M}_{t}\} \vr_t
}\notag 
\\
	&\leq& 
	\sum_{a \in \cDci} \expect{\sum_{a^*\in\cDsi} 
	 \frac{16\epsilon \ln N_{t}(\gamma)}{\Delta_{t, a, a^*}}}\notag
	 \\
	&\leq&
	16\epsilon \ln N_{t}(\gamma)
	 \left[
	 	\Delta_{t, a, 1}\frac{1}{\Delta^2_{t, a, 1}} + \sum_{a^*=2}^{K} \Delta_{t, a, a^*} \left( \frac{1}{\Delta^2_{t, a, a^*}} - \frac{1}{\Delta^2_{t, a, a^*-1}} \right)
	 \right] \notag
	 \\
	 &\leq&
	 \frac{32\epsilon \ln N_{t}(\gamma)}{\Delta_{t, a, K}} ,
\end{eqnarray}
where the last inequality is due to \cite[Lemma 3]{kveton14matroid}. 
Let $\Delta_{a, K} = \min_{t \in [n]} \Delta_{t, a, K}$ be the smallest gap between the suboptimal item $a$ and an optimal item in all time steps.  
When $\vN_t(\gamma, a) > \frac{32\epsilon \ln N_t(\gamma)}{\Delta^2_{a, K}}$, $\cascadeducb$ will not select item $a$ at time $t$.
Thus we get: 
\begin{equation}
\label{ijcai:eq:ducbinside2}
\begin{split}
	&\sum_{a \in \cD} \expect{ \sum_{t=1}^{n} \mathds{1}\{\bar{M}_{t}\} \mathds{1}\{a \in \vR_t, a \in \cDct\}}
	\\ \leq&
	\sum_{e\in [\Upsilon_n]} \sum_{a \in \cDci} \frac{32\epsilon \ln N_{t}(\gamma)}{\Delta_{t, a, K}}   
	\\
	\leq&
	\sum_{a \in \cD} \lceil n (1-\gamma) \rceil 
	 \frac{32\epsilon \ln N_n(\gamma)}{\Delta_{a, K}}
	\gamma^{1/(1-\gamma)} ,
\end{split}
\end{equation}
where the last inequality is based on \cite[Lemma 1]{garivier11upper}.

Finally, together with \cref{ijcai:eq:R,ijcai:eq:ducbdecompose,ijcai:eq:bound for outside,ijcai:eq:ducbrdecompose,ijcai:eq:inside,ijcai:eq:ducbinside2}, we get \cref{ijcai:th:upperboundcascadeducb}.

\subsection{Proof of \cref{ijcai:th:upperboundcascadeswucb}} 
\label{ijcai:sec:proofoftheorem3}

Let $M_{t} = \{\exists a\in\cD : |\alpha_t(a) - \barvalphat(\tau, a)| > c_t(\tau, t)\}$ be the event that $\alpha_t(a)$ falls out of the confidence interval around $\barvalphat(\tau, a)$ at time t, and  let $\bar{M}_{t}$ be the complement of ${M}_{t}$.  
We re-write the $n$-step regret of $\cascadeswucb$ as follows: 
\begin{equation}
\label{ijcai:eq:swucbR}
R(n)  = \expect{\sum_{t=1}^{n} \mathds{1}\{M_{t}\} \vr_t} + \expect{\sum_{t=1}^{n} \mathds{1}\{\bar{M}_{t}\} \vr_t} . 
\end{equation}
We then bound both terms in \cref{ijcai:eq:swucbR} separately. 

First, we refer to $\cT$ as the set of all time steps such that for $t \in \cT$, $s \in [t - \tau, t]$ and any item $a\in\cD$ we have $\alpha_s(a) = \alpha_t(a)$.  
In other words, $\cT$ is the set of time steps that do not follow too close after breakpoints. 
Obviously, for any time step $t \notin \cT$ the estimators of attraction probabilities are biased and $\cascadeswucb$ may recommend suboptimal items constantly. Thus, we get the following bound:
\begin{equation}
\label{ijcai:eq:swucbdecompose}
\expect{\sum_{t=1}^{n} \mathds{1}\{M_{t}\} \vr_t}  \leq 
L\Upsilon_n \tau + \expect{\sum_{t\in\cT} \mathds{1}\{M_{t}\} \vr_t}  .
\end{equation}
$\tau$ time steps after a breakpoint, the estimators of the attraction probabilities are not biased.

Then, we consider the variance. 
By using a Hoeffding-type inequality \cite[Corollary 21]{garivier08upper}, for an item $a\in\cD$, $t\in\cT$, and any $\eta > 0$, we get:
\begin{eqnarray*}
P\left( \abs{\barvalphat(\tau, a) -\alpha_t(a)} > c_t(\tau, t)\right) 
	&\leq&
	P\left( \barvalphat(\tau, a) > \alpha_t(a) +\sqrt{\frac{\epsilon \ln{(t\land \tau)}}{\vN_t(\tau, a)}}\right) \\
	&& {}+ 	P\left( \barvalphat(\tau, a) < \alpha_t(a) -\sqrt{\frac{\epsilon \ln{(t\land \tau)}}{\vN_t(\tau, a)}}\right)
	\\
	&\leq&
	2\left\lceil \frac{\ln{(t\land \tau)}}{\ln(1+\eta)} \right\rceil \exp\left(-2\epsilon \ln(t\land \tau)(1-\frac{\eta}{16}) \right)
	\\
	&=&
	2\left\lceil \frac{\ln{(t\land \tau)}}{\ln(1+\eta)} \right\rceil
	(t \land \tau)	^{-2\epsilon(1-\eta^2/16)} .
\end{eqnarray*}
Taking $\eta = 4\sqrt{1-\frac{1}{2\epsilon}}$, we have: 
$P\left( \abs{\barvalphat(\tau, a) -\alpha_t(a)}\right) \leq 2 \frac{\left\lceil \frac{\ln{(t\land \tau)}}{\ln(1+\eta)} \right\rceil}{t\land\tau} .
$
Thus, we get the following bound: 
\begin{equation*}
\begin{aligned}
	\expect{\sum_{t\in\cT} \mathds{1}\{M_{t}\} \vr_t} 
	&\leq 
	2L\sum_{t\in\cT} 
	\frac{\left\lceil \frac{\ln{(t\land \tau)}}{\ln(1+\eta)} \right\rceil}{t\land\tau}
	\\
	&\leq
	\frac{L\ln^2(\tau)}{\ln(1+4\sqrt{1-1/2\epsilon)}} + \frac{2Ln\ln \tau}{\tau \ln(1+4\sqrt{1-1/2\epsilon})}.
\end{aligned}
\end{equation*}
We sum up and get the upper bound: 
\begin{equation}
\begin{split}
\expect{\sum_{t=1}^{n} \mathds{1}\{\bar{M}_{t}\} \vR_t} 
\leq 
L\Upsilon_n \tau 
& +
\frac{L\ln^2(\tau)}{\ln(1+4\sqrt{1-1/2\epsilon)}} \\
& + 
\frac{2Ln\ln \tau}{\tau \ln (1+4\sqrt{1-1/2\epsilon})} .
\end{split}
\label{ijcai:eq:swucb bound for outside}
\end{equation}

Third, we upper bound the second term in \cref{ijcai:eq:swucbR}. 
The regret is caused by recommending a suboptimal item to the user and the user examines but does not click the item.
Since there are $\Upsilon_n$ breakpoints, we refer to $[t_1, \ldots, t_{\Upsilon_n}]$ as the time steps of a breakpoint. 
We consider the time step in the individual epoch that does not contain a breakpoint. 
For any epoch and any time $t \in \{t_e, t_e+1, \ldots, t_{e+1}-1\}$, any item $a \in \cDci$ and any item $a^* \in \cDsi$, we define the event that item $a$ is included in $\vR_t$ instead of item $a^*$ and  item $a$ is examined but not clicked at time $t$ by: 
\begin{equation*}
\begin{split}
G_{t, a, a^*} =  \{ \exists 1\leq k <\vc_t ~s.t.~ \vR_t(k) = a   , \cR_t(k)  = a^*\}.
\end{split}
\end{equation*} 
Since the attraction probabilities remain constant in the epoch, we refer to $\cDsi$ as the optimal items and $\cDci$ as the suboptimal items in epoch $e$. 
By \cite[Theorem 1]{kveton15cascading}, the regret at time $t$ is decomposed as:
\begin{equation}
\label{ijcai:eq:swucbrdecompose}
\mathbb{E}[\vr_t] \leq  \Delta^t_{a, a^*}\sum_{a \in \cDci} \sum_{a* \in \cDsi} \mathds{1}\{ G_{a, a^*, t} \} .
\end{equation}
Then we have:
\begin{equation}
\begin{split}
\expect{\sum_{t=t_i}^{t_{i+1}-1} \mathds{1}\{\bar{M}_{t}\} \vr_t} 
& = 
\sum_{t=t_i}^{t_{i+1}-1} \mathds{1}\{\bar{M}_{t}\} \expect{\vr_t} \\
& \leq 
\sum_{a \in \cDci} \expect{\sum_{a^*\in\cDsi} \sum_{t=t_i}^{t_{i+1}-1}  
	\Delta^t_{a, a^*} \mathds{1}\{ G_{a, a^*, t} \}} ,
\end{split}
\label{ijcai:eq:swucbinside1}
\end{equation}
where the first equality is due to the tower rule, and the inequality if due to \cref{ijcai:eq:swucbrdecompose}. 

Now, for any suboptimal item $a$ in epoch $e$, we upper bound 
\[ \expect{\sum_{a^*\in\cDsi} \sum_{t=t_i}^{t_{i+1}-1}  
	\Delta^t_{a, a^*} \mathds{1}\{ G_{a, a^*, t} \}}.
\] 
At time $t$,  event $\mathds{1}\{\bar{M}_{t}\}$ and event $ \mathds{1}\{a \in \vR_t, a \in \cDct\}$ happen when there exists an optimal item $a^*\in \cDsi$ such that:
\[
\alpha_t(a) + 2 c_t(\tau, a) \geq \vU_t(a) \geq \vU(a^*) \geq \alpha_t(a^*) , 
\]
which implies that $2 c_t(\tau, a) \geq \alpha_t(a^*) - \alpha_t(a)$. 
Taking the definition of the confidence interval, we get: \[\vN_t(\tau, a) \leq \frac{4\epsilon \ln N_{t}(\tau)}{\Delta^{2}_{t, a, a^*}},\] 
where we set $\Delta_{t, a, a^*} = \Delta^{t}_{ a, a^*}$.

Together with \cref{ijcai:eq:swucbinside1}, we get: 
\begin{eqnarray}
&\mbox{}&\expect{
	\sum_{t=t_i}^{t_{i+1}-1} \mathds{1}\{\bar{M}_{t}\} \vr_t
} \notag
\\
&\leq&
\sum_{a \in \cDci} \expect{\sum_{a^*\in\cDsi} 
	\frac{4\epsilon \ln N_{t}(\gamma)}{\Delta_{t, a, a^*}}}\notag
\\
&\leq&
4\epsilon \ln N_{t}(\gamma)
\left[
\Delta_{t, a, 1}\frac{1}{\Delta^2_{t, a, 1}} + \sum_{a^*=2}^{K} \Delta_{t, a, a^*} \left( \frac{1}{\Delta^2_{t, a, a^*}} - \frac{1}{\Delta^2_{t, a, a^*-1}} \right)
\right] \notag
\\
&\leq& 
\frac{8\epsilon \ln N_{t}(\gamma)}{\Delta_{t, a, K}} ,
\end{eqnarray}
where the last inequality is due to \cite[Lemma 3]{kveton14matroid}. 
Let $\Delta_{a, K} = \min_{t \in [n]} \Delta_{t, a, K}$ be the smallest gap between the suboptimal item $a$ and an optimal item in all time steps.  
When $\vN_t(\tau, a) > \frac{8\epsilon \ln N_t(\tau)}{\Delta^2_{a, K}}$, $\cascadeducb$ will not select item $a$ at time $t$.
Thus we get: 
\begin{equation}
\label{ijcai:eq:swucbinside2}
\begin{split}
\mbox{}&\sum_{a \in \cD} \expect{ \sum_{t=1}^{n} \mathds{1}\{\bar{M}_{t}\} \mathds{1}\{a \in \vR_t, a \in \cDct\}} 
\notag\\
\leq&
\sum_{e\in [\Upsilon_n]} \sum_{a \in \cDci} \frac{8\epsilon \ln N_{t}(\tau)}{\Delta_{t, a, K}} 
\notag\\
\leq&
\sum_{a \in \cD} \left\lceil \frac{n}{\tau} \right\rceil 
\frac{8\epsilon \ln (n\land\tau)}{\Delta_{a, K}}  ,
\end{split}
\end{equation}
where the last inequality is based on \cite[Lemma 25]{garivier08upper}.

Finally, together with \cref{ijcai:eq:swucbR,ijcai:eq:swucbdecompose,ijcai:eq:swucb bound for outside,ijcai:eq:swucbinside1,ijcai:eq:swucbinside2}, we get \cref{ijcai:th:upperboundcascadeswucb}. 

\section{Additional Experiments}
\begin{figure}[t]
	\includegraphics{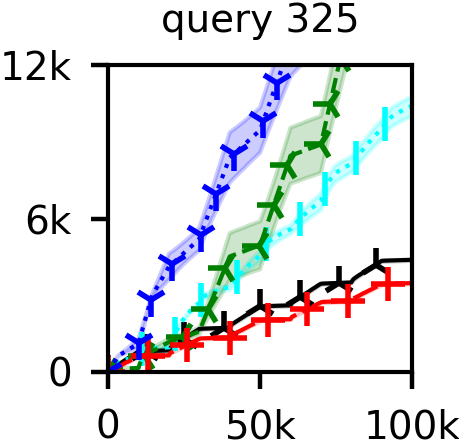}
	\includegraphics{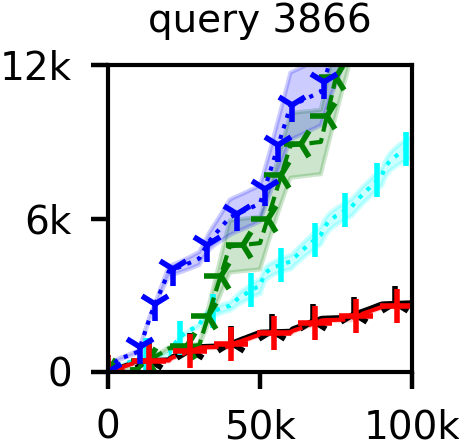}
	\includegraphics{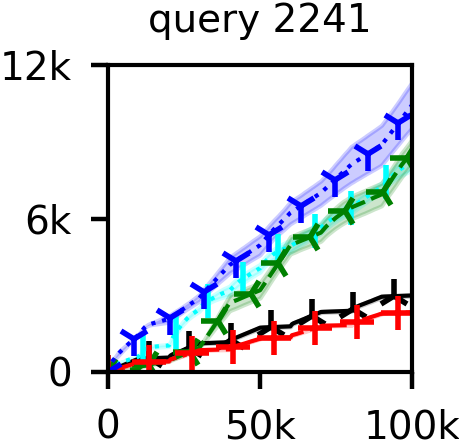}
	\includegraphics{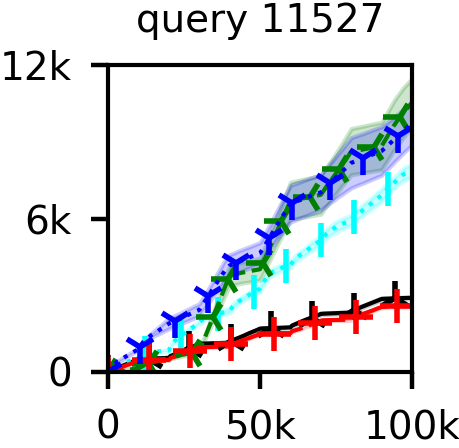}
	\vspace{2mm}
	\includegraphics{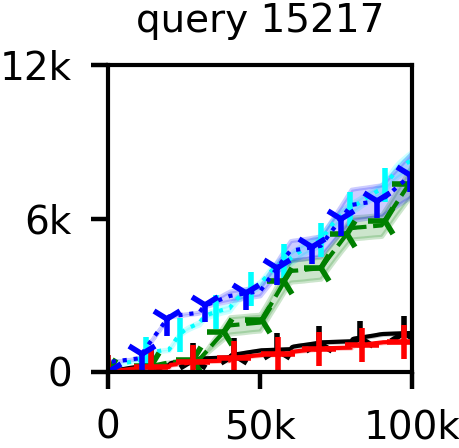}
	\includegraphics{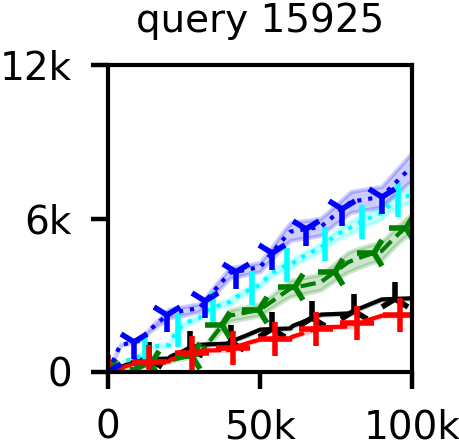}
	\includegraphics{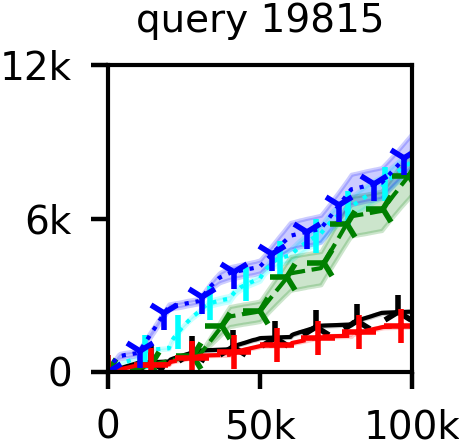}
	\includegraphics{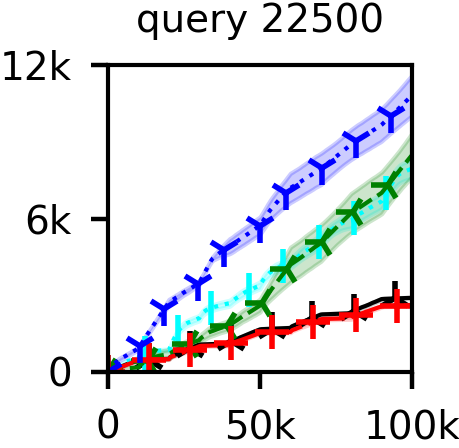}		\vspace{2mm}
	\includegraphics{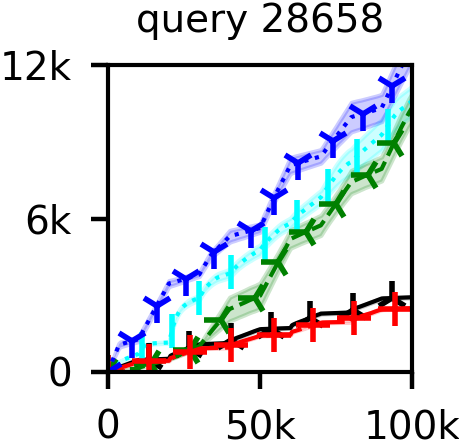}
	\includegraphics{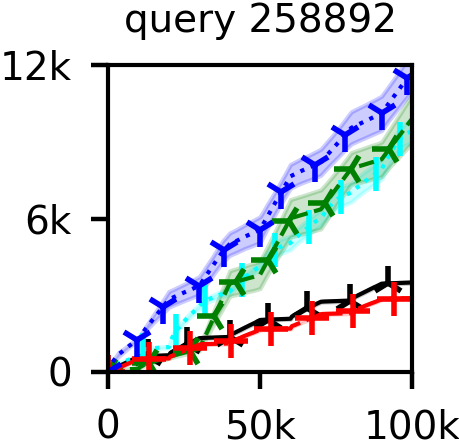}
	\includegraphics{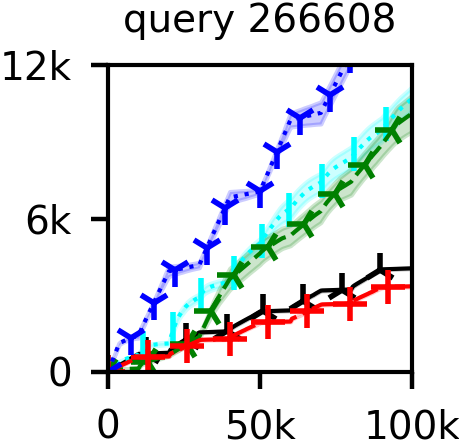}
	\includegraphics{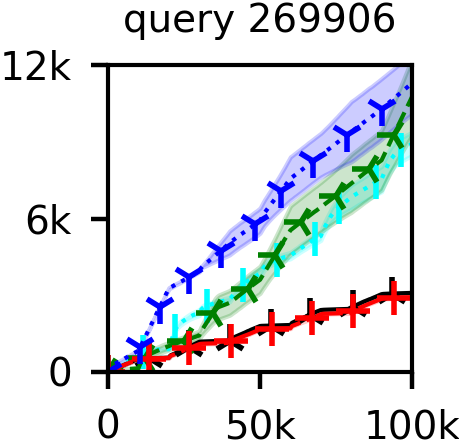}
	\vspace{2mm}
	\includegraphics{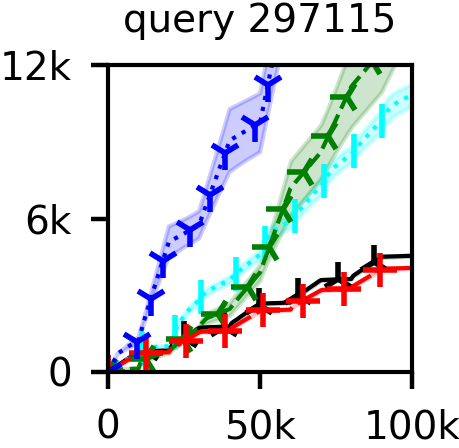}
	\includegraphics{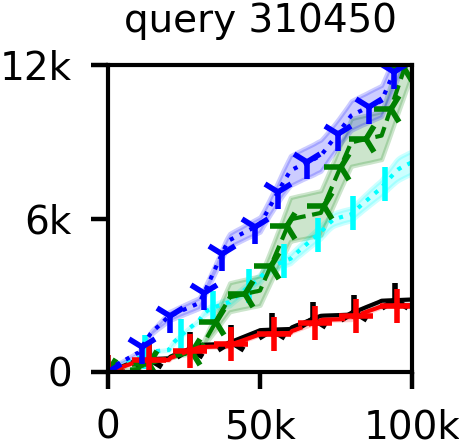}
	\includegraphics{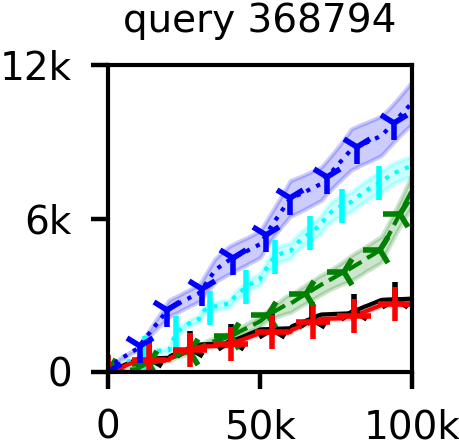}
	\includegraphics{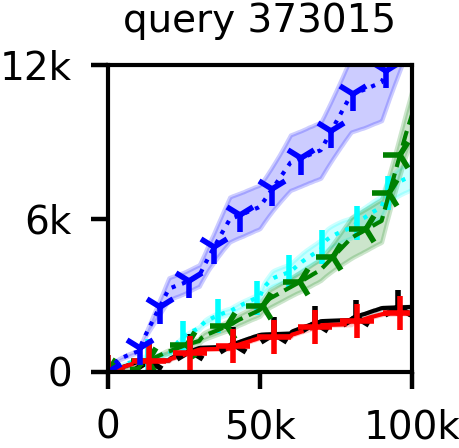}
	\vspace{2mm}
	\includegraphics{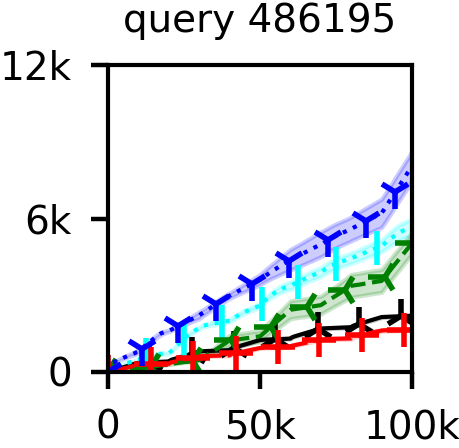}
	\includegraphics{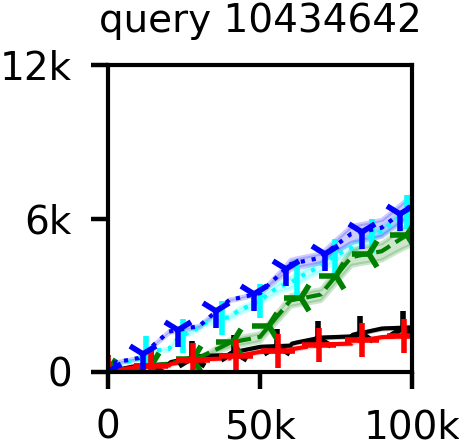}
	\includegraphics{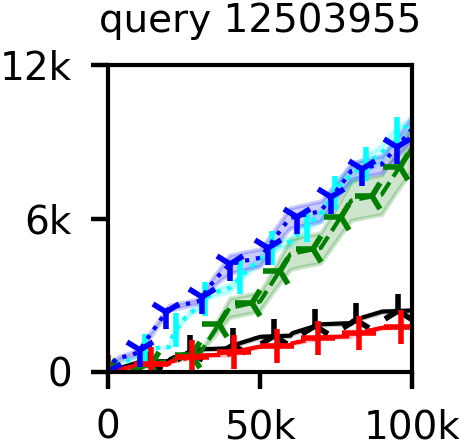}
	\includegraphics{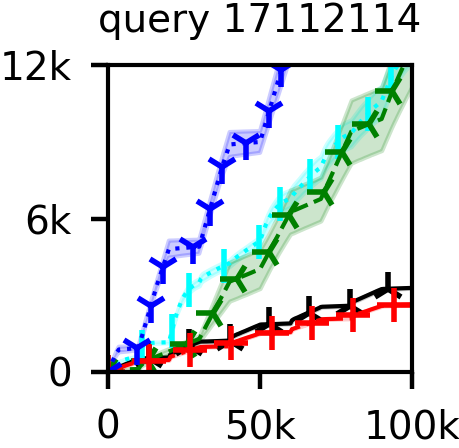}
	\caption{The n-step regret of $\cascadeducb$ (black), $\cascadeswucb$ (red), $\rankedexpthree$ (cyan), $\cascadeklucb$ (green) and $\bubblerank$ (blue)  on single queries in up to $100$k steps. Lower is better. The x-axis is step $n$. 
	The results are averaged over $10$ runs per query. The shaded regions represent standard errors of our estimates.}\label{ijcai:fig:singleregret}
\end{figure}

In this section, we compare  $\cascadeducb$, $\cascadeswucb$ and baselines on single queries. 
We pick 20 queries and report the results in~\cref{ijcai:fig:singleregret}. 
The results exemplify that $\cascadeducb$ and $\cascadeswucb$ have sub-linear regret while other baselines have linear regret. 


\section{Conclusion}
\label{ijcai:sec:conclusion}

In this chapter, we have answered \textbf{RQ3} by studying the \acf{OLTR} problem in a non-stationary environment where user preferences change abruptly. 
We focus on a widely-used user click behavior model \acf{CM} and have proposed an online learning variant of it called \emph{cascading non-stationary bandtis}.
Two algorithms, $\cascadeducb$ and $\cascadeswucb$, have been proposed for solving it.  
Our theoretical  have shown that they have sub-linear regret.  
These theoretical findings have been confirmed by our experiments on the Yandex click dataset. 
We open several future directions for non-stationary \ac{OLTR}. 

First, we have only considered the \ac{CM} setup. 
Although a \ac{CM} is powerful in explanaining user behavior, it can only learn up to the first click, which may ignore part of user feedback. 
Other click models that can handle multiple clicks such as DCM~\citep{guo09efficient} and DBN~\citep{chapelle09dynamic} may be considered as part of future work. 
We believe that our analysis can be adapted to those cases easily. 
Second, we focus on $\ucb$-based policy in building rankings. 
Another possibility is to use the family of softmax policies, which has original been designed for adversarial bandits~\citep{auer03nonstochastic,besbes14stochastic}.
Along this line, one may obtain upper bounds independent from the number of breakpoints.


\newcommand{\cA}{\mathcal{A}}
\newcommand{\cO}{\mathcal{O}}
\newcommand{\cM}{\mathcal{M}}
\newcommand{\cS}{\mathcal{S}}
\newcommand{\cB}{\mathcal{B}}
\newcommand{\cU}{\mathcal{U}}
\newcommand{\cH}{\mathcal{H}}

\newcommand{\vXt}{\myvec{X}_t}
\newcommand{\vOmegat}{\myvec{\Omega}_t}
\newcommand{\vZt}{\myvec{Z}_t}
\newcommand{\vQt}{\myvec{Q}_t}
\newcommand{\vOt}{\myvec{O}_t}
\newcommand{\vEt}{\myvec{E}_t}

\chapter{Online Learning to Rank for Relevance and Diversity}
\label{chapter:recsys20}

\footnote[]{This chapter was published as~\citep{li20cascading}.}

\acresetall

This chapter is devoted to answering the following question:
\begin{description}
	\item[\textbf{RQ4}] How to learn a ranker online considering both relevance and diversity? 
\end{description}
We study an online learning setting that aims to recommend a ranked list with $K$ items that maximizes the ranking utility, i.e., a list whose items are relevant and whose topics are diverse. 
We formulate it as the \emph{cascade hybrid bandits}~(\acs{CHB})\acused{CHB} problem.  
\ac{CHB} assumes a cascading user behavior, where a user browses the displayed list  from top to bottom, clicks the first attractive item, and stops browsing the rest. 
We propose a hybrid contextual bandit approach, called $\cascadehybrid$, for solving this problem.
$\cascadehybrid$ models item relevance and topical diversity using two independent functions and simultaneously learns those functions from user click feedback.
We conduct experiments to evaluate $\cascadehybrid$ on two real-world recommendation datasets: MovieLens and Yahoo music datasets.
Our experimental results show that $\cascadehybrid$ outperforms the baselines.
In addition, we prove theoretical guarantees on the $n$-step performance demonstrating the soundness of $\cascadehybrid$.

\section{Introduction}
\label{recsys:sec:introduction}

Ranking is at the heart of modern interactive systems, such as recommender and search systems. 
Learning to rank~(\acs{LTR})\acused{LTR} addresses the ranking problem in such systems by using machine learning approaches~\citep{liu09learning}. 
Traditionally, \ac{LTR} has been studied in an offline fashion, in which human labeled data is required~\citep{liu09learning}. 
Human labeled data is expensive to obtain, cannot capture future changes in user preferences, and may not well align with user needs~\citep{hofmann11balancing}.
To circumvent these limitations, recent work has shifted to learning directly from users' interaction feedback, e.g., clicks~\citep{hofmann13reusing,zoghi16click,jagerman19model}. 

User feedback is abundantly available in interactive systems and is a valuable source for training online \ac{LTR} algorithms~\citep{grotov16online}. 
When designing an algorithm to learn from this source, three challenges need to be addressed: 
\begin{enumerate*}[label=(\arabic*)]
	\item The learning algorithm should address position bias (the phenomenon that higher ranked items are more likely be observed than lower ranked items); 
	\item The learning algorithm should infer item relevance from user feedback and recommend lists containing relevant items (relevance ranking);
	\item The recommended list should contain no redundant items and cover a broad range of topics (result diversification).
\end{enumerate*}

To address the position bias, a common approach is to make assumptions on the user's click behavior and model the behavior using a click model~\citep{chuklin15click}.
The cascade model~(\acs{CM})\acused{CM}~\citep{craswell08experimental} is a simple but effective click model to explain user behavior.
It makes the so-called \emph{cascade assumption}, which assumes that a user browses the list from the first ranked item to the last one and clicks on the first attractive item and then stops browsing. 
The clicked item is considered to be positive, items before the click are treated as negative and items after the click will be ignored. 
Previous work has shown that the cascade assumption can explain the position bias effectively and several algorithms have been proposed under this assumption~\citep{kveton15cascading,zong16cascading,hiranandani19cascading,li19cascading}. 

In online \ac{LTR}, the implicit signal that is inferred from user interactions is noisy~\citep{hofmann11balancing}. 
If the learning algorithm only learns from these signals, it may reach a suboptimal solution where the optimal ranking is ignored simply because it is never exposed to users.
This problem can be tackled by exploring new solutions, where the learning algorithm displays some potentially ``good'' rankings to users and obtains more signals. 
This behavior is called \emph{exploration}. 
However, exploration may hurt the user experience. 
Thus, learning algorithms face an exploration vs.\ exploitation dilemma. 
Multi-armed bandit~(\acs{MAB}) \acused{MAB}~\citep{auer02finite,lattimore20bandit} algorithms are commonly used to address this dilemma. 
Along this line, multiple algorithms have been proposed~\citep{li10contextual,hofmann13reusing,li19bubblerank}. 
They all address the dilemma in elegant ways and aim at recommending the top-$K$ most relevant items to users. 
However, only recommending the most relevant items may result in a list with redundant items, which diminishes the utility of the list and decreases user satisfaction~\citep{agrawal09diversifying,yue11linear}. 

The submodular coverage model~\citep{nemhauser78analysis} can capture the pattern of diminishing utility and has been used in online \ac{LTR} for diversified ranking.
One assumption in this line of work is that items can be represented by a set of topics.\footnote{In general, each topic may only capture a tiny aspect of the information of an item, e.g., a single phrase of a news title or a singer of a song~\citep{yue11linear,agrawal09diversifying}.} 
The task, then, is to recommend a list of items that ensures a maximal coverage of topics. 
\citet{yue11linear} develop an online feature-based diverse \ac{LTR} algorithm by optimizing submodular utility models~\citep{yue11linear}. 
\citet{hiranandani19cascading} improve online diverse \ac{LTR} by bringing the cascading assumption into the objective function. 
However, we argue that not all features that are used in a \ac{LTR} setting can be represented by topics~\citep{liu09learning}.
Previous online diverse \ac{LTR} algorithms tend to ignore the relevance of individual items and may recommend a diversified list with less relevant items. 

In this chapter, we address the aforementioned challenges and make four contributions: 
\begin{enumerate}[label=(\arabic*)]
	\item We focus on a novel online \ac{LTR} setting that targets both relevance ranking and result diversification.
	We formulate it as a \acf{CHB} problem, where the goal is to select $K$ items from a large candidate set that maximize the utility of the ranked list (\cref{recsys:sec:problem formulation}).
	\item We propose $\cascadehybrid$, which utilizes a hybrid model, to solve this problem~(\cref{recsys:sec:cascadehybrid}).  
	\item We evaluate $\cascadehybrid$ on two real-world recommendation datasets: MovieLens and Yahoo and show that  $\cascadehybrid$ outperforms state-of-the-art baselines (\cref{recsys:sec:experiments}). 
	\item We theoretically analyze the performance of $\cascadehybrid$ and provide guarantees on its proper behavior; moreover, we are the first to show that the regret bounds on feature-based ranking algorithms with the cascade assumption are linear in $\sqrt{K}$. 
\end{enumerate}

\noindent%
The rest of the chapter is organized as follows. 
We recapitulate the background knowledge in~\cref{recsys:sec:background}. 
In~\cref{recsys:sec:algorithm}, we formulate the learning problem and propose our $\cascadehybrid$ algorithm that optimizes both item relevance and list diversity. \cref{recsys:sec:experiments} contains our empirical evaluations of $\cascadehybrid$, comparing it with several state-of-the-art baselines. 
An analysis of the upper bound on the $n$-step performance of $\cascadehybrid$ is presented in~\cref{recsys:sec:analysis}. 
In~\cref{recsys:sec:related work}, we review related work. 
Conclusions are formulated in~\cref{recsys:sec:conclusion}.

\section{Background}
\label{recsys:sec:background}

In this section, we recapitulate the \acf{CM}, \ac{CB}, and the submodular coverage model.
Throughout the chapter, we consider the ranking problem of $L$ candidate items and $K$ positions with $K \leq L$. 
We denote $\{1, \ldots, n\}$ by $[n]$ and for the collection of items we write $\cD = [L]$.  
A ranked list contains $K \leq L$ items and is denoted by $\cR \in \Pi_K(\cD)$, where $\Pi_K{(\cD)}$ is the set of all permutations of $K$ distinct items from the collection $\cD$. 
The item at the $k$-th position of the list is denoted by $\cR(k)$ and, if $\cR$ contains an item $i$, the position of this item in $\cR$ is denoted by $\cR^{-1}(i)$. 
All vectors are column vectors.
We use bold font to indicate a vector and bold font with a capital letter to indicate a matrix.
We write $\vI_{d}$ to denote the $d\times d$ identity matrix and $\mathbf{0}_{d\times m}$ the $d\times m$ zero matrix.

\subsection{Cascade model}
\label{recsys:sec:cascade models}
Click models have been widely used to interpret user's interactive click behavior;  cf.~\citep{chuklin15click}.  
Briefly, a user is shown a ranked list $\cR$, and then browses the list and leaves click feedback.  
Every click model makes unique assumptions and models a type of user interaction behavior. 
In this chapter, we consider a simple but widely used click model, the \emph{cascade model}~\citep{craswell08experimental,kveton15cascading,li19cascading}, which makes the cascade assumption about user behavior. 
Under the cascade assumption, a user browses a ranked list $\cR$ from the first item to the last one by one and clicks the first attractive item. 
After the click, the user stops browsing the remaining items. 
A click on an examined item $\cR(i)$ can be modeled as a Bernoulli random variable with a probability of $\alpha(\cR(i))$, which is also called the \emph{attraction probability}.
Here, the \acf{CM} assumes that each item attracts the user independent of other items in $\cR$. 
Thus, the \ac{CM} is parametrized by a set of attraction probabilities $\valpha \in [0, 1]^L$. 
The examination probability of item $\cR(i)$ is $1$ if $i=1$, otherwise $1-\prod_{j=1}^{i-1}(1-\alpha(\cR(j)))$. 

With the \ac{CM}, we translate the implicit feedback to training labels as follows: 
Given a ranked list, items ranked below the clicked item are ignored since none of them are browsed. 
Items ranked above the clicked item are negative samples and the clicked item is the positive sample. 
If no item is clicked, we know that all items are browsed but not clicked. 
Thus, all of them are negative samples. 

The vanilla \ac{CM} is only able to capture the first click in a session, and there are various extensions of \ac{CM} to model multi-click scenarios; cf.~\citep{chuklin15click}. 
However,  we still focus on the \ac{CM}, because it has been shown in multiple publications that the \ac{CM} achieves good performance in both online and offline setups~\citep{chuklin15click,kveton15cascading,li19bubblerank}.

\subsection{Cascading bandits}
\label{recsys:sec:cascading click bandits}

Cascading bandits (\acs{CB})\acused{CB} are a type of online variant of the \ac{CM}~\citep{kveton15cascading}. 
A \ac{CB} is represented by a tuple $(\cD, K, P)$, where $P$ is a binary distribution over  $\{0, 1\}^L$.  
The learning agent interacts with the \ac{CB} and learns from the feedback. 
At each step $t$, the agent generates a ranked list $\cR_t \in \Pi_K(\cD)$ depending on observations in the previous $t-1$ steps and shows it to the user. 
The user browses the list with cascading behavior and leaves click feedback. 
Since the \ac{CM} accepts at most one click, we write $c_t \in [K+1]$ as the click indicator,  where $c_t$ indicates the position of the click and $c_t = K+1$ indicates no click. 
Let $A_t \in \{0, 1\}^L$ be the attraction indicator, where $A_t$ is drawn from $P$ and $A_t(\cR_t(i))=1$ indicates that item $\cR_t(i)$ attracts the user at step $t$. 
The number of clicks at step $t$ is considered as the reward and computed as follows: 
\begin{equation}
\label{recsys:eq:reward}
r(\cR_t, A_t) = 1 - \prod_{i=1}^{K} (1-A_t(\cR_t(i))).
\end{equation}
Then, we assume that the attraction indicators of items are distributed independently as Bernoulli variables: 
\begin{equation}
\label{recsys:eq:bernoulli}
	P(A) = \prod_{i \in \cD} P_{\alpha(i)}(A(i)),
\end{equation}
where $P_{\alpha(i)}(\cdot)$ is the Bernoulli distribution with mean $\alpha(i)$.
The expected number of clicks at step $t$ is computed as $\expect{r(\cR_t, A_t)} = r(\cR_t, \valpha)$. 
The goal of the agent is to maximize the expected number of clicks in $n$ steps or minimize the expected $n$-step regret: 
\begin{equation}
\label{recsys:eq:regret}
	R(n) = \sum_{t=1}^{n}\expect{  \max_{\cR \in \Pi_K(\cD)} r(\cR, \valpha) - r(\cR_t, A_t) }.
\end{equation}
\ac{CB} has several variants depending on assumptions on the attraction probability $\valpha$.
Briefly, cascade linear bandits~\citep{zong16cascading} assume that an item $a$ is represented by a feature vector $\vz_a \in \real^{m}$ and that the attraction probability of an item $a$ to a user is a linear combination of features: $ \alpha(a)\approx \vz_a^T \beta^*$, where $\beta^* \in \real^m$ is an unknown parameter. 
With this assumption, the attraction probability of an item is independent of other items in the list, and  this assumption is used in relevance ranking problems. 
$\cascadelinucb$ has been proposed to solve this problem. 
For other problems, \citet{hiranandani19cascading} assume the attraction probability to be submodular, and propose $\cascadelsb$ to solve result diversification. 

\subsection{Submodular coverage model}
\label{recsys:sec:submodular coverage model}

Before we recapitulate the submodular function, we introduce two properties of a diversified ranking. 
Different from the relevance ranking, in a diversified ranking, the utility of an item depends on other items in the list. 
Suppose we focus on news recommendation.
Items that we want to rank are news itms, and each news item covers a set of topics, e.g., weather, sports, politics, a celebrity, etc. 
We want to recommend a list that covers a broad range of topics. 
Intuitively, adding a news item to a list does not decrease the number of topics that are covered by the list, but adding a news item to a list that covers highly overlapping topics might not bring much extra benefit to the list. 
The first property can be thought of as a monotonicity property, and the second one is the notion of diminishing gain in the utility. 
They can be captured by the submodular function~\citep{yue11linear}. 

We introduce two properties of submodular functions. 
Let $g(\cdot)$ be a set function, which maps a set to a real value. 
We say that $g(\cdot)$ is monotone and submodular if given two item sets $\cA$ and $\cB$, where  $\cB \subseteq \cA$, and an item $a$, $g(\cdot)$ has the following two properties:  
\begin{equation*}
\begin{aligned}
&\emph{monotonicity}: g(\cA \cup \{a\}) \geq g(\cA);  \\
&\emph{submodularity}: g(\cB \cup \{a\}) - g(\cB) \geq	g(\cA \cup \{a\}) - g(\cA). 
\end{aligned}
\end{equation*}

\noindent
In other words, the gain in utility of adding an item $a$ to a subset of $\cA$ is larger than or equal to that of adding an item to $\cA$, and adding an item $a$ to $\cA$ does not decreases the utility. 
Monotonicity and submodularity together provide a natural framework to capture the properties of a diversified ranking.
The shrewd reader may notice that a linear function is a special case of submodular functions, where only the inequalities in \emph{monotonicity} and \emph{submodularity} hold. 
However, as discussed above, the linear model assumes that the attraction probability of an item is independent of other items: it cannot capture the diminishing gain in the result diversification. 
In the rest of this section, we introduce the \emph{probabilistic coverage model}, which 
is a widely used submodular function for result diversification~\citep{agrawal09diversifying,yue11linear,qin13promoting,ashkan15optimal,hiranandani19cascading}.

Suppose that an item $a\in\cD$ is represented by a $d$-dimensional vector $\vx_a \in [0, 1]^d$. 
Each entry of the vector $\vx_a(j)$ describes the probability of item $a$ covering topic $j$.
Given a list $\cA$, the probability of $\cA$ covering topic $j$ is 
\begin{equation}
	\label{recsys:eq:topic coverage}
g_j(\cA) = 1-\prod_{a \in \cA} (1-\vx_a(j)). 
\end{equation}
The gain in topic coverage of adding an item $a$ to $\cA$ is:
\begin{equation}
\label{recsys:eq:topic gain}
\Delta(a\mid \cA) = (\Delta_1(a\mid \cA), \ldots, \Delta_d(a\mid \cA)), 
\end{equation}
where $\Delta_j(a\mid \cA) = g_j(\cA\cup\{a\}) - g_j(\cA)$. 
With this model, the attraction probability of the $i$-th item in a ranked list $\cR$ is defined as: 
\begin{equation}
	\label{recsys:eq:topic coverage attration probability}
	\alpha(\cR(i)) = \vomega_{\cR(i)}^T \vtheta^*, 
\end{equation}
where $\vomega_{\cR(i)} = \Delta(\cR(i) \mid (\cR(1), \ldots, \cR(i-1)))$ and $\vtheta^*$ is the unknown user preference to different topics~\citep{hiranandani19cascading}. 
In \cref{recsys:eq:topic coverage attration probability},  the attraction probability of an item depends on the items ranked above it; $\alpha(\cR(i))$ is small if $\cR(i)$ covers similar topics as higher ranked items.  
$\cascadelsb$~\citep{hiranandani19cascading} has been proposed to solve cascading bandits with this type of attraction probability and aims at building diverse ranked lists.

\section{Algorithm}
\label{recsys:sec:algorithm}
In this section, we first formulate our online learning to rank problem, and then propose $\cascadehybrid$ to solve it. 

\subsection{Problem formulation}
\label{recsys:sec:problem formulation}
We study a variant of cascading bandits, where the attraction probability of 
an item in a ranked list depends on two aspects: item relevance and item novelty. 
Item relevance is independent of other items in the list.
Novelty of an item depends on the topics covered by higher ranked items;
a novel item brings a large gain in the topic coverage of the list, i.e., a large value in \cref{recsys:eq:topic coverage attration probability}.
Thus, given a ranked list $\cR$, the attraction probability of item $\cR(i)$ is defined as follows: 
\begin{equation}
	\label{recsys:eq:hybridy attraction probability}
	\alpha(\cR(i)) =  \vz_{\cR(i)}^T\vbeta^* +  \vomega_{\cR(i)}^T\vtheta^*, 
\end{equation} 
where $\vomega_{\cR(i)}= \Delta(\cR(i) \mid (\cR(1), \ldots, \cR(i-1))$ is the topic coverage gain discussed in \cref{recsys:sec:submodular coverage model}, and $\vz_{\cR(i)} \in \real^m$ is the relevance feature, $\vtheta^* \in \real^d$ and $\vbeta^* \in \real^m$ are two unknown parameters that characterize the user preference. 
In other words, the attraction probability is a hybrid of a modular (linear) function parameterized by $\vbeta^*$ and a submodular function parameterized by $\vtheta^*$. 

Now, we define our learning problem, \acf{CHB}, as a tuple $(\cD, \vtheta^*, \vbeta^*,  K)$. 
Here, $\cD = [L]$ is the item candidate set and each item $a$ can be represented by a feature vector $[\vx_a^T, \vz_a^T]^T$, where $\vx_a \in [0, 1]^d$ is the topic coverage of item $a$ discussed in \cref{recsys:sec:submodular coverage model}.
$K$ is the number of positions.
The action space for the problem are all permutations of $K$ individual items from $\cD$, $\Pi_K(\cD)$. 
The reward of an action at step $t$ is the number of clicks, defined in \cref{recsys:eq:reward}. 
Together with \cref{recsys:eq:reward,recsys:eq:hybridy attraction probability,recsys:eq:bernoulli},  the expectation of reward at step $t$ is computed as follows:
\begin{equation}
\expect{r(\cR_t, A_t)} = 1-\prod_{a \in \cR_t} (1- \vz_a^T\vbeta^* -  \vomega_a^T\vtheta^*). 
\end{equation}
In the rest of the chapter, we write $r(\cR_t) = \expect{r(\cR_t, A_t)}$ for short. 
And the goal of the learning agent is to maximize the  reward or, equivalently, to minimize the  $n$-step regret defined as follow: 
\begin{equation}
\label{recsys:eq:our regret}
R(n) = \sum_{t=1}^{n} \left[\max_{\cR\in \Pi_K(\cD)}r(\cR) - r(\cR_t)\right].
\end{equation}

\noindent%
The previously proposed  $\cascadelinucb$~\citep{zong16cascading} cannot solve \ac{CHB} since it only handles the linear part of the attraction probability. 
$\cascadelsb$~\citep{hiranandani19cascading} cannot solve  \ac{CHB}, either, because it only uses one submodular function and handles the submodular part of the attraction probability. 
Thus, we need to extend the previous models or, in other words, propose a new hybrid model that can handle both linear and submodular properties in the attraction probability.  

\subsection{Competing with a greedy benchmark}
\label{recsys:eq: optimal list}

Finding the optimal set that maximizes the utility of a submodular function is an NP-hard problem~\citep{nemhauser78analysis}. 
In our setup, the attraction probability of each item also depends on the order in the list. 
To the best of our knowledge, we cannot find the optimal ranking 
\begin{equation}
\cR^* = \argmax_{\cR\in \Pi_K(\cD)}r(\cR)
\end{equation} 
efficiently. 
Thus, we compete with a greedy benchmark that approximates the optimal ranking $\cR^*$. 
The greedy benchmark chooses the items that have the highest attraction probability given the higher ranked items:  for any positions $k \in [K]$, 
\begin{equation}
\tilde{\cR}(k) = \argmax\limits_{a \in \cD\backslash \{\tilde{\cR}(1),\ldots, \tilde{\cR}(k-1)\}} \vz_a^T\vbeta^* +  \vomega_a^T\vtheta^*,
\end{equation}
where $\tilde{\cR}(k)$ is the ranked list generated by the benchmark. 

This greedy benchmark has been used in previous literature~\citep{yue11linear,hiranandani19cascading}. 
As shown by~\citet{hiranandani19cascading}, in the \ac{CM}, the greedy benchmark is at least a $\eta$-approximation of $\cR^*$.
That is, 
\begin{equation}
	\label{recsys:eq:greedy guarantees}
	r(\tilde{\cR}) \geq \eta r(\cR^*), 
\end{equation}
where $\eta = (1-\frac{1}{e}) \max\{\frac{1}{K}, 1-\frac{K-1}{2}\alpha_{max}\}$ with $\alpha_{max} = \max_{a \in \cD} \vz_a^T\vbeta^* +  \vx_a^T\vtheta^*$.
In the rest of the chapter, we focus on competing with this greedy benchmark.

\begin{algorithm}[t]
	\begin{algorithmic}[1]
		\setcounter{ALC@unique}{0}
		\REQUIRE $\gamma$
		\STATE 	\COMMENT{Initialization} 
		\STATE \label{recsys:alg:initialization}
		$\vH_1 \leftarrow \vI_{d}, 
		\vu_1 \leftarrow \mathbf{0}_{d}, 
		\vM_1 \leftarrow \vI_{m}, 
		\vy_1 \leftarrow \mathbf{0}_{m}, 
		\vB_1 = \mathbf{0}_{d\times m} $
		
		\FOR{$t = 1, 2, \ldots, n$}
		\STATE \Comment{Estimate parameters}
		\STATE \label{recsys:alg:estimate parameters}
		$\hat{\vtheta}_{t} \leftarrow \vH_{t}^{-1}\vu_{t}$, 
		$\hat{\vbeta}_{t}\leftarrow \vM_{t}^{-1}(\vy_{t} - \vB_{t}^T  \vH_{t}^{-1}\vu_{t})$
		
		\STATE \Comment{Build ranked list}
		\STATE \label{recsys:alg:start building}
		$\cS_0 \leftarrow \emptyset$
		\FOR{$k = 1, 2, \ldots K$}
		\FOR{$a \in \cD \setminus \cS_{k-1}$} \label{recsys:alg:begin estimate}
		\STATE $\vomega_a \leftarrow \Delta(\vx_a | \cS_{k-1})$
		\Comment{Recalculate the topic coverage gain. }
		\STATE $\mu_{a} \leftarrow$  \cref{recsys:eq:ucb}  \Comment{Compute UCBs.}
		\ENDFOR \label{recsys:alg:end estimate}
		
		\STATE \label{recsys:alg:argmax}
		$a_k^t \leftarrow \argmax\limits_{a\in \cD \setminus \cS_{k-1}} \mu_{a} $
		\STATE $\cS_k \leftarrow \cS_{k-1} + {a_k^t}$
		\ENDFOR
		\STATE \label{recsys:alg:end building}
		$\cR_t = (a_1^t, \ldots, a_K^t)$ \Comment {Ranked list}
		
		\STATE \label{recsys:alg:display ranking}
		Display $\cR_t$ and observe click feedback $c_t \in [K+1]$
		\STATE $k_t \leftarrow \min(K, c_t)$
		\STATE \Comment{Update statistics}
		\STATE $\vH_{t} \leftarrow \vH_{t} + \vB_{t} \vM_{t}^{-1}\vB_{t}^T$, $\vu_{t}\leftarrow \vu_{t} + \vB_{t}\vM_{t}^{-1}\vy_{t}$
		\FOR{$a \in \cR_t(1:k_t)$}
		\STATE $\vM_{t+1} \leftarrow \vM_{t} + \vz_a \vz_a^T $, 
		$\vB_{t+1} \leftarrow \vB_{t} + \vomega_a\vz_a^T$, 
		$\vH_{t} \leftarrow \vH_{t} +  \vomega_a \vomega_a^T$			
		\ENDFOR
		\IF{$c_t \leq K$}
		\STATE $\vy_{t+1} \leftarrow \vy_{t} + \vz_{\cR_t(c_t)}$, 
		$\vu_{t} \leftarrow \vu_{t} + \vomega_{\cR_t(c_t)}$
		\ENDIF
		\STATE \label{recsys:alg:end click collection} \hspace{-0.11in}$\vH_{t+1} \leftarrow \vH_{t} - \vB_{t+1} \vM_{t+1}^{-1}\vB_{t+1}^T$, $\vu_{t+1} \leftarrow \vu_{t} - \vB_{t+1}\vM_{t+1}^{-1}\vy_{t+1}$
		\ENDFOR
	\end{algorithmic}
	\caption{$\cascadehybrid$}
	\label{recsys:alg:cascadehybrid}
\end{algorithm}
 
\subsection{\textsf{CascadeHybrid}}
 \label{recsys:sec:cascadehybrid}
We propose $\cascadehybrid$ to solve the \ac{CHB}.
As the name suggests, the algorithm is a hybrid of a linear function and a submodular function. 
The linear function is used to capture item relevance and the submodular function to capture diversity in topics. 
$\cascadehybrid$ has access to item features, $[\vx_a^T, \vz_a^T]^T$, and uses the probabilistic coverage model to compute the gains in topic coverage. 
The user preferences $\vtheta^*$ and $\vbeta^*$ are unknown to $\cascadehybrid$. 
They are estimated from interactions with users. 
The only tunable hyperparameter for $\cascadehybrid$ is $\gamma \in \real_+$, which controls exploration: a larger value of $\gamma$ means more exploration. 

The details of $\cascadehybrid$ are provided in \cref{recsys:alg:cascadehybrid}. 
At the beginning of each step $t$ (line~\ref{recsys:alg:estimate parameters}), $\cascadehybrid$ estimates the user preference as $\hat{\vtheta}_t$ and $\hat{\vbeta}_{t}$ based on the previous $t-1$ step observations. 
$\hat{\vtheta}_t$ and $\hat{\vbeta}_{t}$ can be viewed as maximum likelihood estimators on the rewards,\footnote{The derivation is based on matrix block-wise inversion. We omit the derivation since it is not a major contribution of this chapter. 
} 
where $\vM_{t}, \vH_{t}, \vB_{t}$ and $\vy_t, \vu_t$ summarize the features and click feedback of all observed items in the previous $t-1$ steps.
Then, $\cascadehybrid$ builds the ranked list $\cR_t$, sequentially~(lines~\ref{recsys:alg:start building}--\ref{recsys:alg:end building}). 
In particular, for each position $k$, we recalculate the topic coverage gain of each item (line~\ref{recsys:alg:start building}). 
The new gains are used to estimate the attraction probability of items. 
$\cascadehybrid$ makes an optimistic estimate of the attraction probability of each item (lines~\ref{recsys:alg:begin estimate}--\ref{recsys:alg:end estimate}) and chooses the one with the highest  estimated attraction probability (line~\ref{recsys:alg:argmax}). 
This is known as the principle of  optimism in the face of uncertainty~\citep{auer02finite}, and the estimator for an item $a$ is called the \ac{UCB}: 
\begin{equation}
\label{recsys:eq:ucb}
	\mu_{a} =  \vomega_a^T \hat{\vtheta}_t + \vz_a^T \hat{\vbeta}_t  + \gamma\sqrt{s_{a}},  
\end{equation}
with 
\begin{equation}
\label{recsys:eq:cb}
\begin{aligned}
s_{a}  =&  \vomega_a^T\vH_{t}^{-1}\vomega_a - 2\vomega_a^T\vH_{t}^{-1}\vB_{t}\vM_{t}^{-1}\vz_a +  \vz_a \vM_{t}^{-1} \vz_a + \\
&\quad \vz_a \vM_{t}^{-1}\vB_{t}^T\vH_{t}^{-1}\vB_{t} \vM_{t}^{-1} \vz_a.
\end{aligned}
\end{equation}
Finally, $\cascadehybrid$ displays the ranked list $\cR_t$ to the user and collects click feedback (lines~\ref{recsys:alg:display ranking}--\ref{recsys:alg:end click collection}). 
Since $\cascadehybrid$ only accepts one click, we use $c_t \in [K+1]$ to indicate the position of the click;\footnote{
	For multiple-click cases, we only consider the first click and keep the rest of $\cascadehybrid$ the same. }
$c_t = K+1$ indicates that no item in $\cR_t$ is clicked. 

\subsection{Computational complexity}
\label{recsys:eq:computational complexity}

The main computational cost of \cref{recsys:alg:cascadehybrid} is incurred by computing matrix inverses, which is cubic in the dimensions of the matrix. 
However, in practice, we can use the Woodbury matrix identity~\citep{golub96matrix} to update $\vH_{t}^{-1}$ and $\vM_{t}^{-1}$ instead of $\vH_{t}$ and $\vM_{t}$, which is square in the dimensions of the matrix. 
Thus, computing the \ac{UCB} of each item is $O(m^2+d^2)$. 
As $\cascadehybrid$ greedily  chooses $K$ items out of $L$, the per-step computational complexity of $\cascadehybrid$ is $O(LK(m^2+d^2))$.

\section{Experiments}
\label{recsys:sec:experiments}

\begin{figure*}[t]
	\centering
	\includegraphics{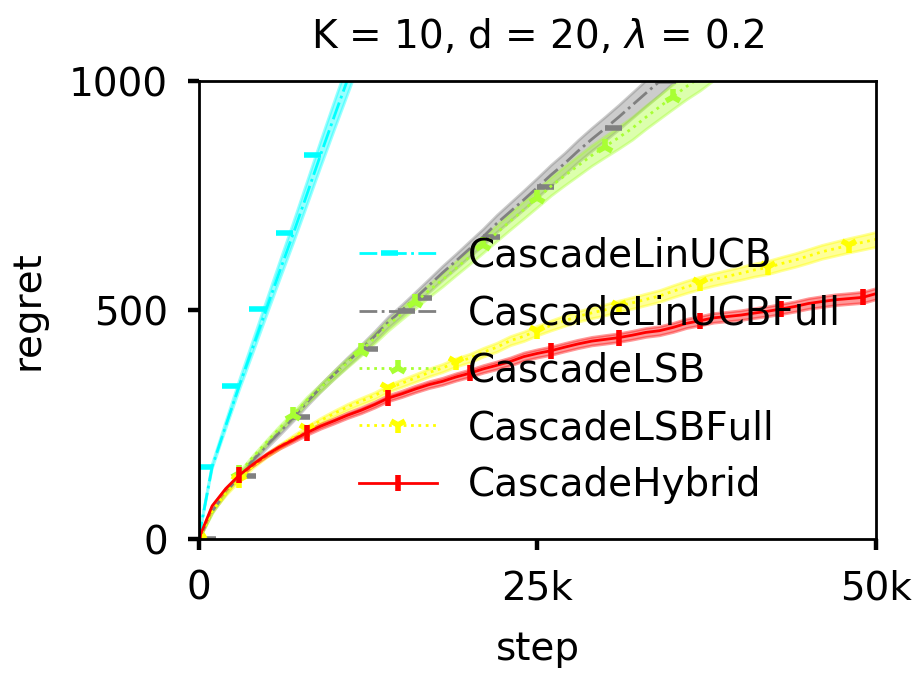}
	\includegraphics{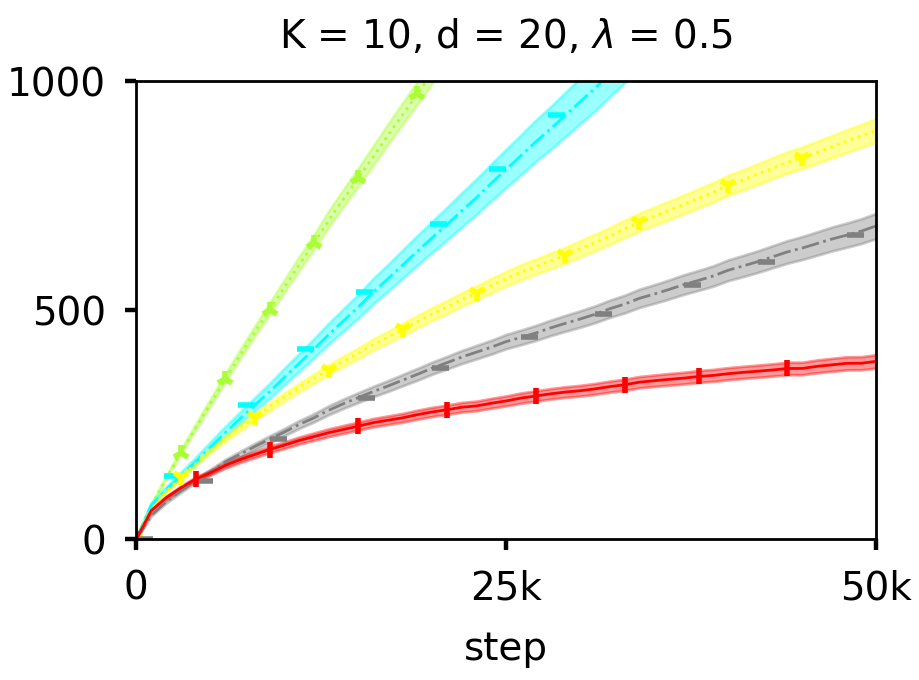}
	\includegraphics{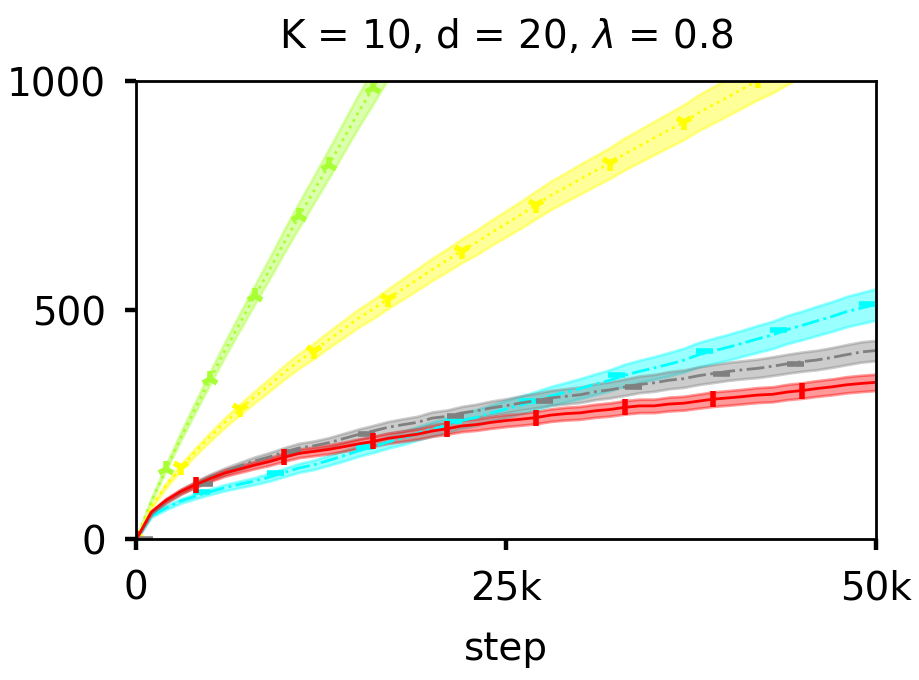}
	\includegraphics{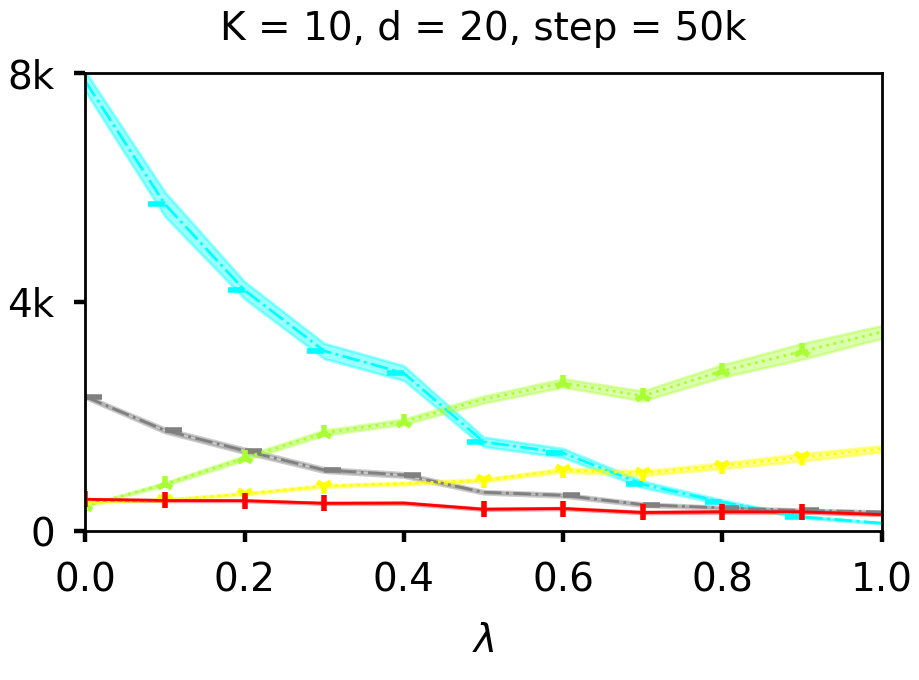}
	\caption{$n$-step regret on the MovieLens dataset with different $\lambda$. Results are averaged over $500$ users with $2$ repeats per user. Lower regret means more clicks received by the algorithm during the online learning.
		Shaded areas are the standard errors of  estimates. }  
	\label{recsys:fig:movie lambda}
\end{figure*}

This section starts with the experimental setup, where we first introduce the datasets, click simulator and baselines.
After that we report our experimental results.  

\subsection{Experimental setup}
\label{recsys:sec:experimental setup}

Off-policy evaluation~\citep{li10contextual} is an approach to evaluate interaction algorithms without live experiments. 
However, in our problem, the action space is exponential in $K$, which is too large for  commonly used off-policy evaluation methods. 
As an alternative, we evaluate the $\cascadehybrid$ in a simulated interaction environment, where the simulator is built based on offline datasets. 
This is a commonly used evaluation setup in the literature~\citep{kveton15cascading,zong16cascading,hiranandani19cascading}.

\myparagraph{Datasets}
We evaluate $\cascadehybrid$ on two datasets: MovieLens 20M~\citep{harper15movielens} and  Yahoo.\footnote{R2 - Yahoo! Music User Ratings of Songs with Artist, Album, and Genre Meta Information, v. 1.0~\url{https://webscope.sandbox.yahoo.com/catalog.php?datatype=r}} 
The MovieLens dataset contains $20$M ratings on $27$k movies by $138$k users, with $20$ genres.\footnote{In both datasets, one of the $20$ genres is called \emph{unknown}. } 
Each movie belongs to at least one genre. 
The Yahoo dataset contains  over $700$M ratings of $136$k songs given by $1.8$M users and genre attributes of each song; we consider the top level attribute, which has $20$ different genres; each song belongs to a single genre. 
All the ratings in the two datasets are on a $5$-point scale. 
All movies and songs are considered as items and genres are considered as topics.  

\myparagraph{Data preprocessing} 
We follow the data preprocessing approach in~\citep{zong16cascading,hiranandani19cascading,li19online}. 
First, we extract the $1$k most  active users and the $1$k most rated items. 
Let $\cU = [1000]$ be the user set, and $\cD = [1000]$ be the item set.  
Then, the ratings are mapped onto a binary scale: 
rating $5$ is converted to $1$ and others to $0$. 
After this mapping, in the MovieLens dataset, about $7\%$ of the user-item pairs get rating $1$, and, in the Yahoo dataset, about $11\%$ of user-item pairs get rating $1$.
Then, we use the matrix  $\vF \in \{0, 1\}^{|\cU|\times |\cD|}$ to capture the converted ratings and $\vG \in \{0, 1\}^{|\cD|\times d}$ to record the items and topics, where $d$ is the number of topics and each entry $\vG_{jk}=1$ indicates that item $j$ belongs to topic $k$.

\myparagraph{Click simulator}
In our experiments, the click simulator follows the cascade assumption, and considers both  item relevance and diversity of the list.
To design such a simulator, we combine the simulators used in \citep{li19online} and \citep{hiranandani19cascading}.
Because of the cascading assumption, we only need to define the way of computing attraction probabilities of items in a list. 

We first divide the users into training and test groups evenly, i.e., $\vF_{train}$ and $\vF_{test}$. 
The training group  is used to estimate features of items used by online algorithms, while the test group are used to define the click simulator. 
This is to mimic the real-world scenarios that online algorithms estimate user preferences without knowing the perfect topic coverage of items.
Then, we follow \citep{li19online} to obtain the relevance part of the attraction probability, i.e., $\vz$ and $\vbeta^*$, and the process in \citep{hiranandani19cascading} to get the topic coverage of items, $\vx$, and the user preferences on topics, $\vtheta^*$.

In particular,  the relevance features $\vz$ are obtained by conducting singular-value decomposition on $\vF_{train}$. 
We pick the  $10$ largest singular values and thus the dimension of relevance features is $m=10$.  
Then, we normalize each relevance feature by the transformation: $\vz_a \leftarrow \frac{\vz_a}{\norm{\vz_a}}$, where $\norm{\vz_a}$ is the L2 norm of $\vz_a$. 
The user preference $\vbeta^*$ is computed by solving the least square on $\vF_{test}$ and then $\vbeta^*$ is normalized by the same transformation. 
Note that $\forall a \in \cD: \vz_a^T \vbeta^*\in [0, 1]$,  since $ \norm{\vz_a}=1$ and $\norm{\vbeta^*} = 1$.

Then, we follow the process in \citep{hiranandani19cascading}. 
If item $a$ belongs to topic $j$, we compute the topic coverage of item $a$ to topic $j$ as the quotient of the number of users rating item $a$ to be attractive to the number of users who rate at least one item in topic $j$ to be attractive: 
\begin{equation}
\label{recsys:eq:topic_features}
\vx_{a, j} = \frac{\sum_{u \in \cU} F_{u, a} G_{a, j}}	
{\sum_{u \in \cU} \mathds{1}\{\exists a' \in \cD: F_{u, a'}G_{a', j} >0 \}}. 
\end{equation}
Given user $u$, the preference for topic $j$ is computed as the number of items rated to be attractive in topic $j$ over the number of items in all topics rated by $u$ to be attractive: 
\begin{equation}
\label{recsys:eq:topic_preferences}
\vtheta_j^* = \frac{\sum_{a \in \cD} F_{u, a} G_{a, j}}{\sum_{j'\in[d]} \sum_{a' \in \cD} F_{u, a'} G_{a', j'}} .
\end{equation}
For some cases, we may have $\exists a:~ \sum_{j \in [d]} \vx_{a, j} >  1$ and thus $\vx_{a}^T \vtheta^* > 1$. 
However, given the high sparsity in our datasets,  we have $\vx_{a}^T \vtheta^* \in[0, 1]$ for all items during our experiments. 

Finally, we combine the two parts and obtain the attraction probability used in our click simulator. 
To simulate different types of user preferences, we introduce a trade-off parameter $\lambda \in [0, 1]$, which is unknown to online algorithms, and compute the attraction probability of the $i$th item in $\cR$ as follows:
\begin{equation}
\alpha(\cR(i)) = \lambda \vz_{\cR(i)}^T\vbeta^* +   (1-\lambda)  \vomega_{\cR(i)}^T\vtheta^*. 
\end{equation}
By changing the value of $\lambda$, we simulate different types of user preference: a larger value of $\lambda$ means that the user prefers items to be relevant; a smaller value of $\lambda$ means that the user prefers the topics in the ranked list to be diverse. 

\myparagraph{Baselines}
We compare $\cascadehybrid$ to two online algorithms, each of which has two configurations. 
In total, we have four baselines, namely 
$\cascadelinucb$ and $\cascadelinucbfull$~\citep{zong16cascading}, and $\cascadelsb$ and  $\cascadelsbfull$~\citep{hiranandani19cascading}. 
The first two only consider relevance ranking. 
The differences are that 
$\cascadelinucb$ takes $\vz$ as the  features,  while $\cascadelinucbfull$ takes $\{\vx, \vz\}$ as the features. 
$\cascadelsb$ and  $\cascadelsbfull$ only consider the result diversification, where $\cascadelsb$ takes $\vx$ as features, while $\cascadelsbfull$ takes $\{\vx, \vz\}$ as features. 
$\cascadelinucbfull$ and $\cascadelinucb$  are expected to  perform well when $\lambda \to  1$, and that $\cascadelsb$ and $\cascadelsbfull$ perform well when $\lambda \to 0$. 
For all baselines, we set the exploration parameter $\gamma = 1$ and the learning rate to $1$. 
This parameter setup is used in \citep{yue11linear}, which leads to better empirical performance. 
We also set $\gamma=1$ for $\cascadehybrid$. 

We report the cumulative regret, \cref{recsys:eq:our regret}, within $50$k steps, called \emph{$n$-step regret}.  
The $n$-step regret is commonly used to evaluate bandit algorithms~\citep{yue11linear,zong16cascading,hiranandani19cascading,li19cascading}. 
In our setup, it measures the difference in number of received clicks between the oracle that knows the ideal $\beta^*$ and $\theta^*$ and the online algorithm, e.g., $\cascadehybrid$, in $n$ steps. 
The lower regret means the more clicks received by the algorithm. 
We conduct our experiments with $500$ users from the test group and $2$ repeats per user.
In total, the results are averaged over $1$k repeats. 
We also include the standard errors of our estimates. 
To show the impact of different factors on the performance of online \ac{LTR} algorithms, we choose $\lambda \in \{0.0, 0.1, \ldots, 1.0\}$, the number of positions $K\in\{5, 10, 15, 20\}$, and the number of topics $d \in \{5, 10, 15, 20\}$. 
For the number of topics $d$, we choose the topics with the maximum number of items.

\begin{figure*}[t]
	\centering
	\includegraphics{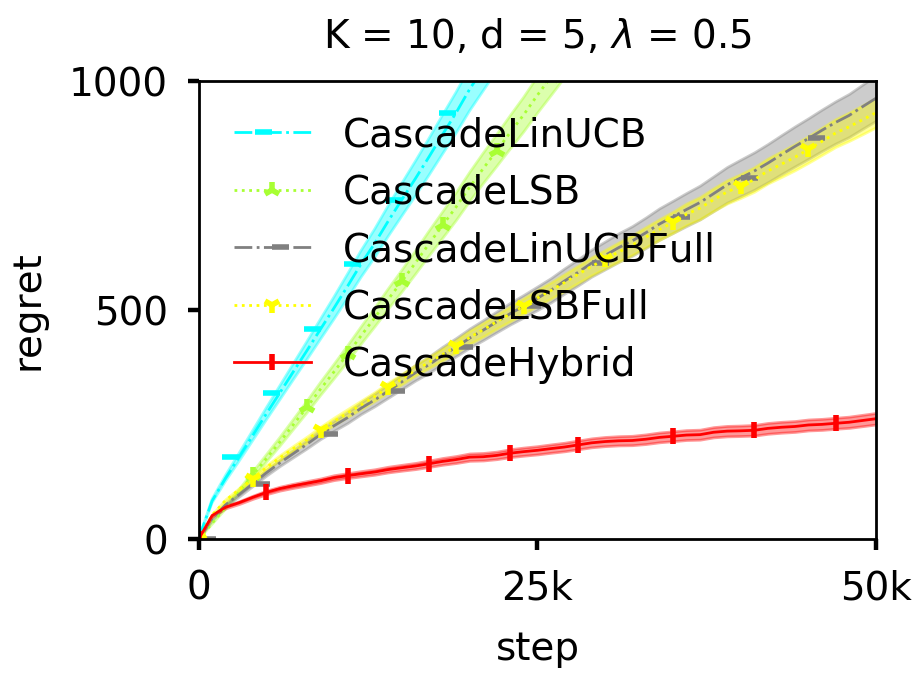}
	\includegraphics{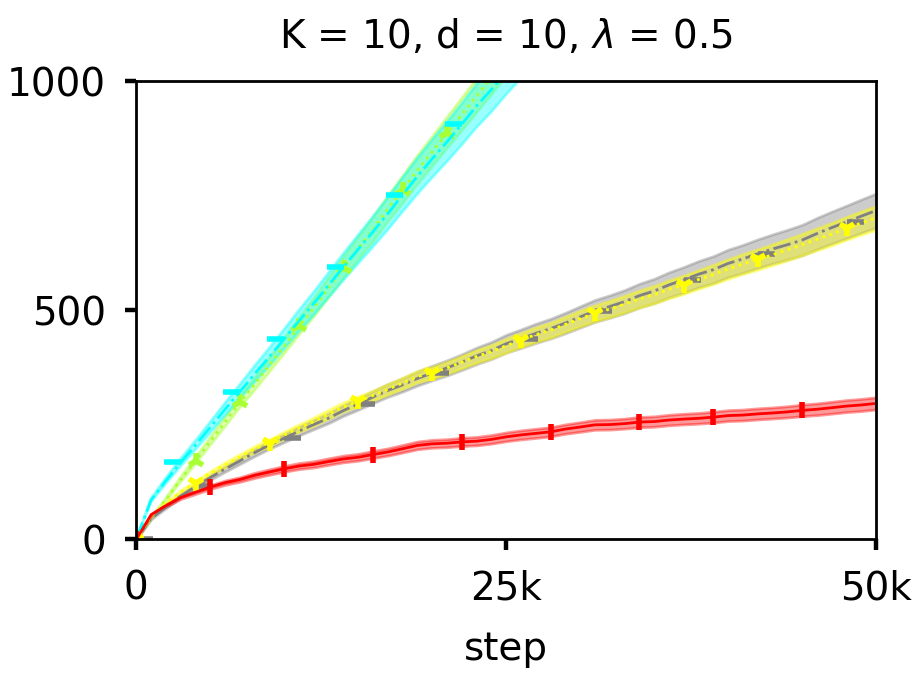}
	\includegraphics{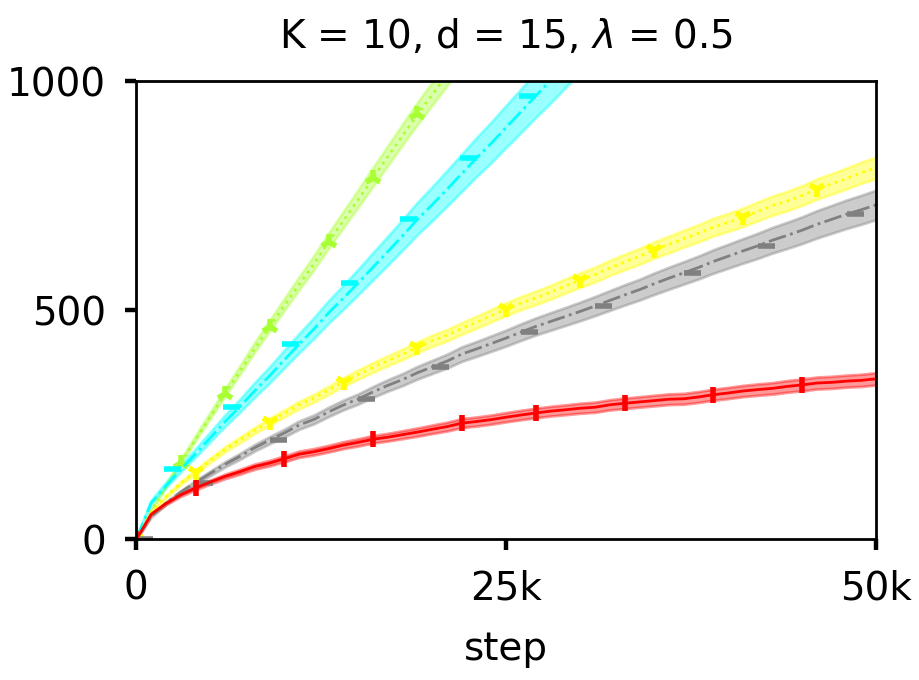}
	\includegraphics{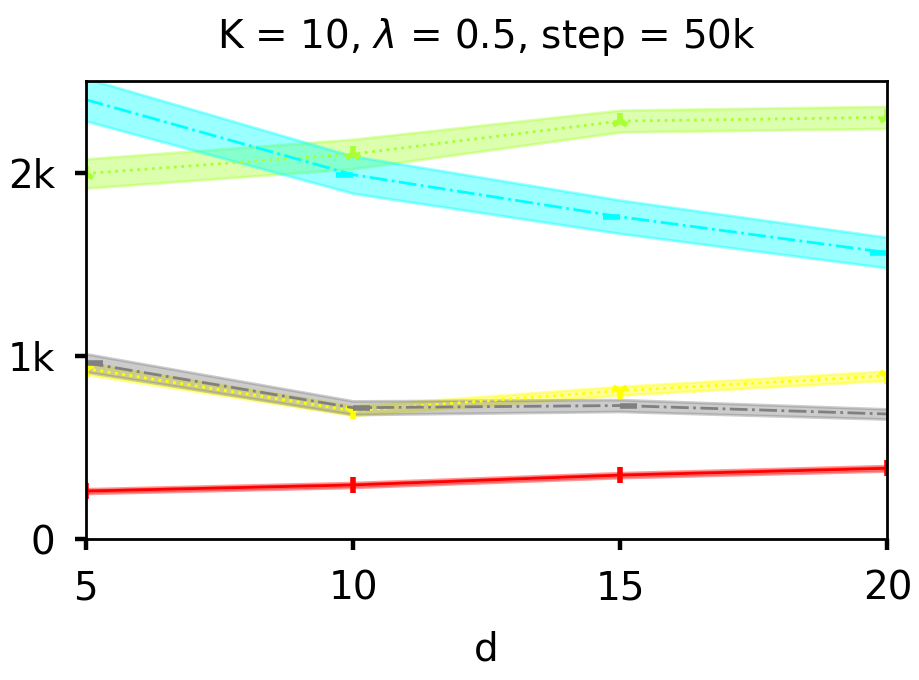}
	\caption{$n$-step regret on the MovieLens dataset with different topics $d$.}  
	\label{recsys:fig:movie topic}
\end{figure*}

\begin{figure*}[t]
	\centering
	\includegraphics{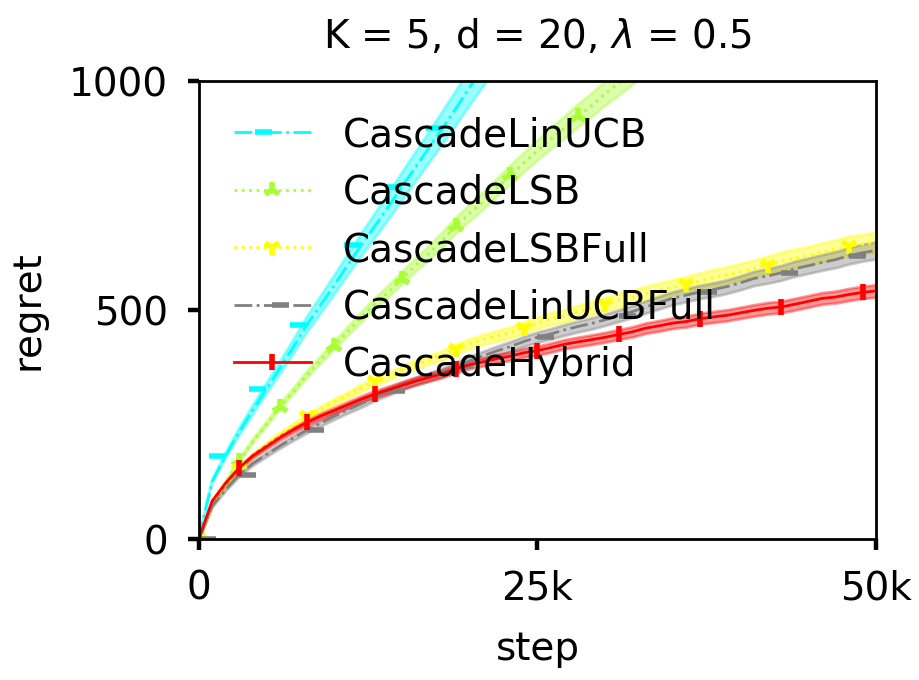}
	\includegraphics{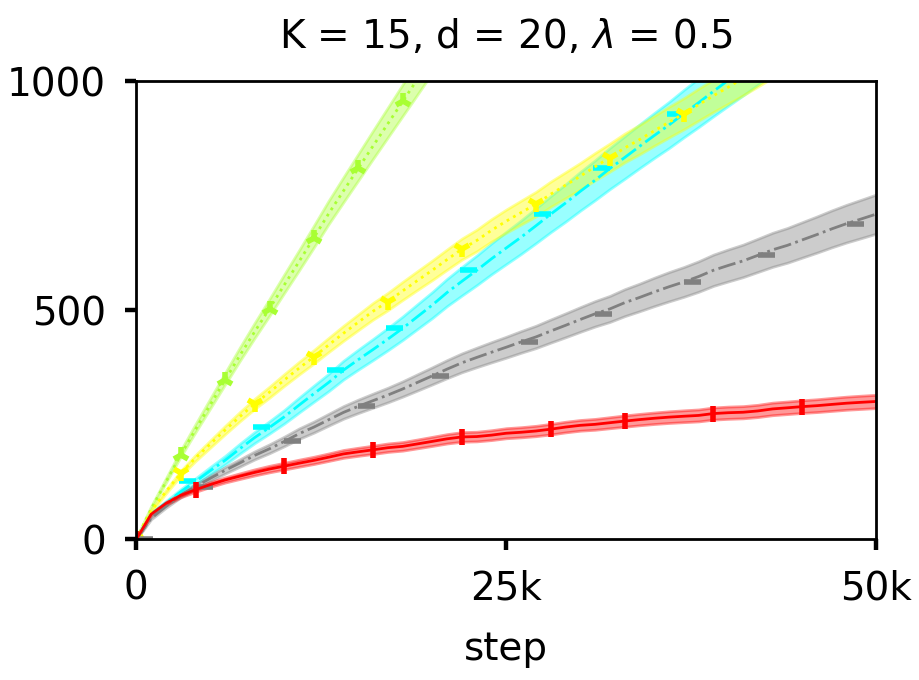}
	\includegraphics{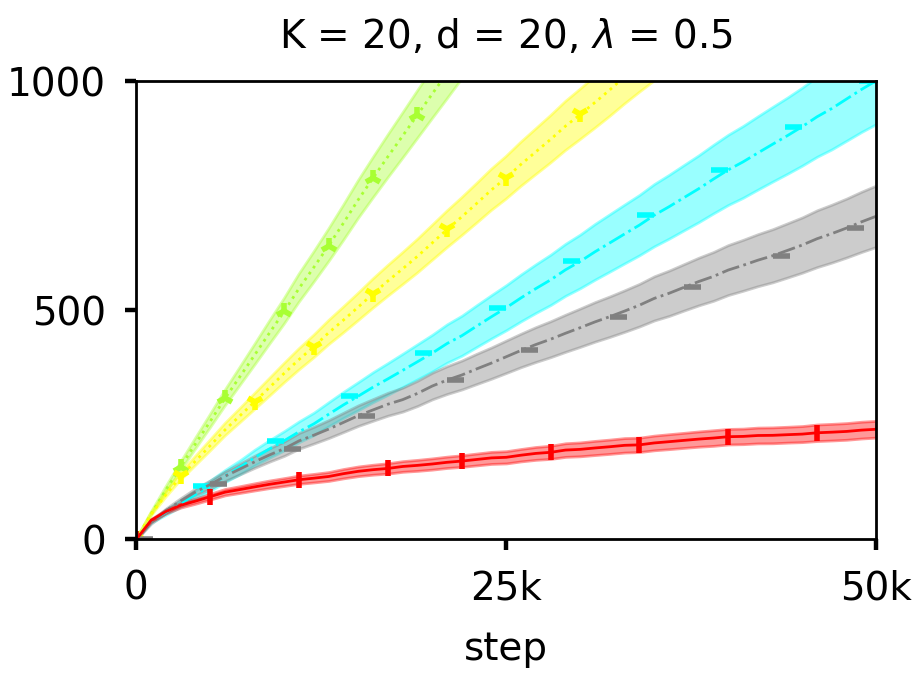}
	\includegraphics{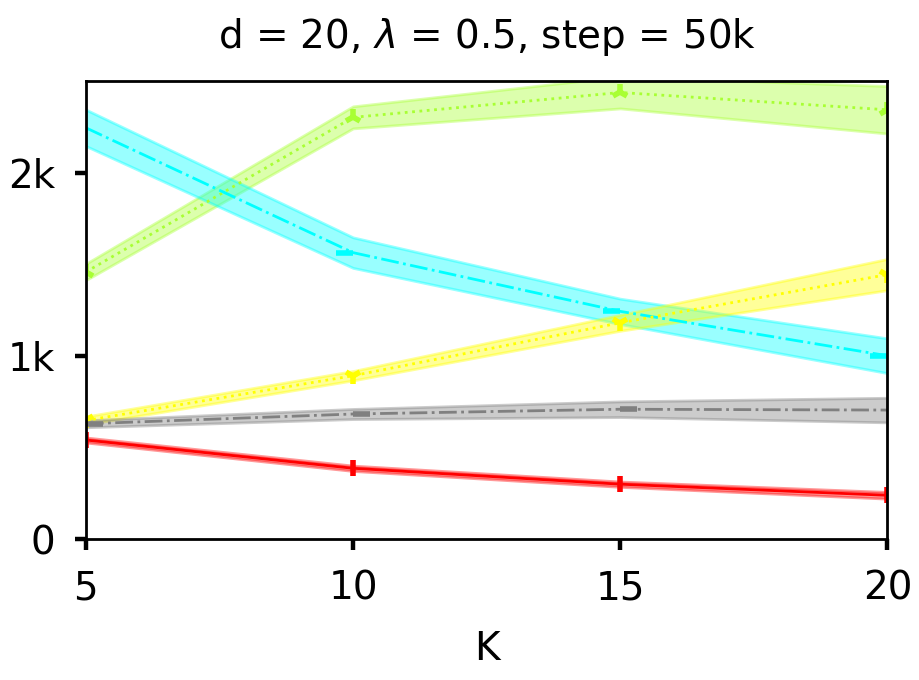}
	\caption{$n$-step regret on the MovieLens dataset with positions $K$. }  
	\label{recsys:fig:movie position}
\end{figure*}

\subsection{Experimental results}
\label{recsys:sec:experimental results}

\begin{figure*}[t]
	\centering
	\includegraphics{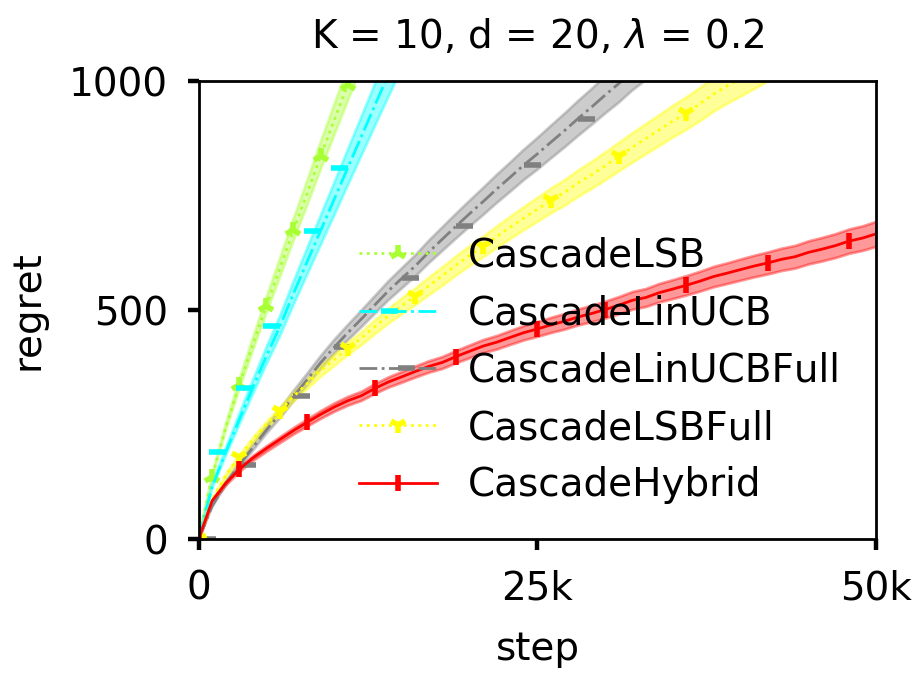}
	\includegraphics{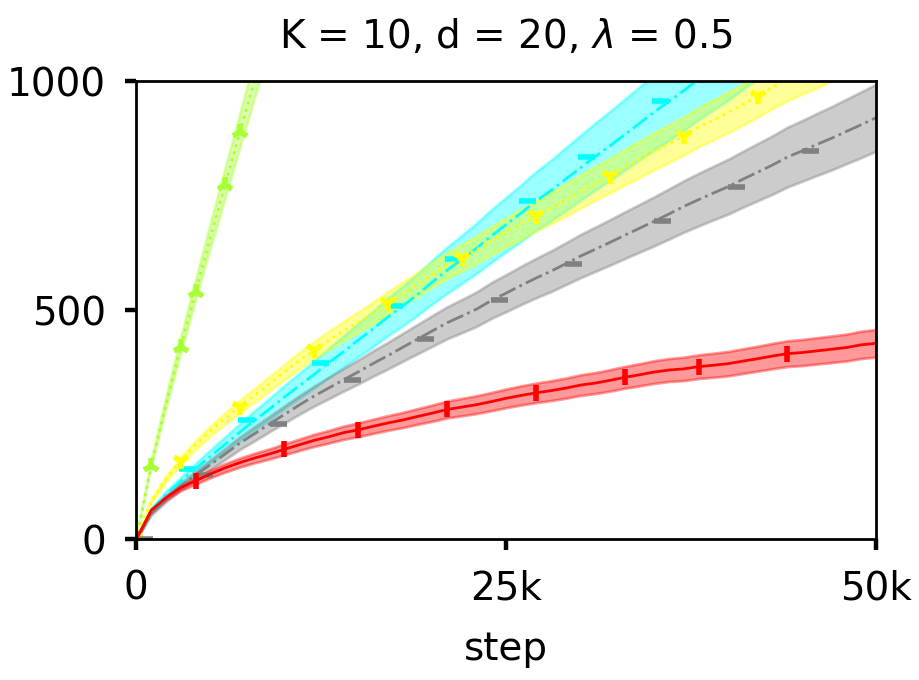}
	\includegraphics{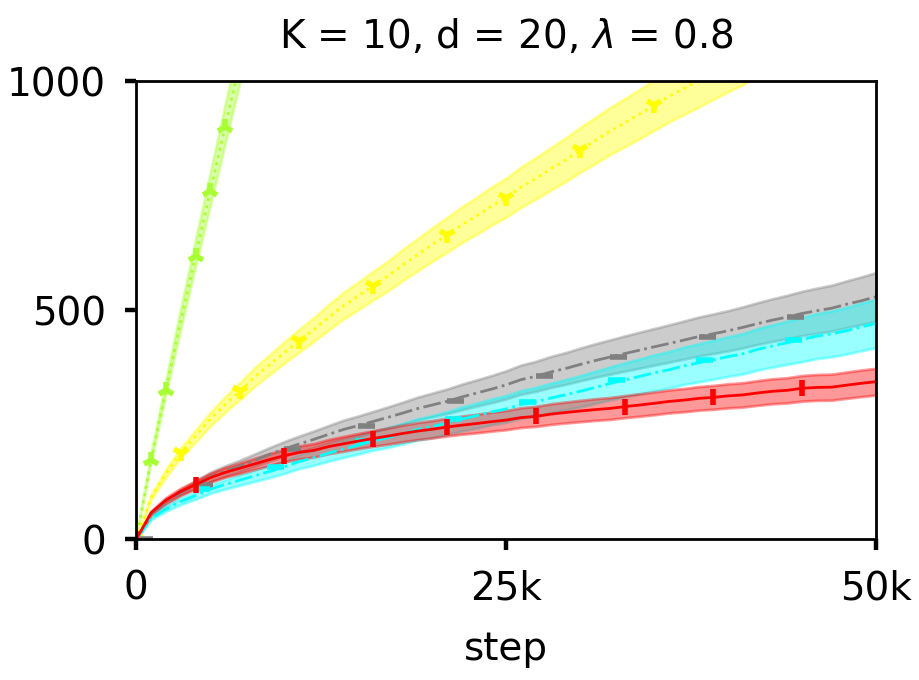}
	\includegraphics{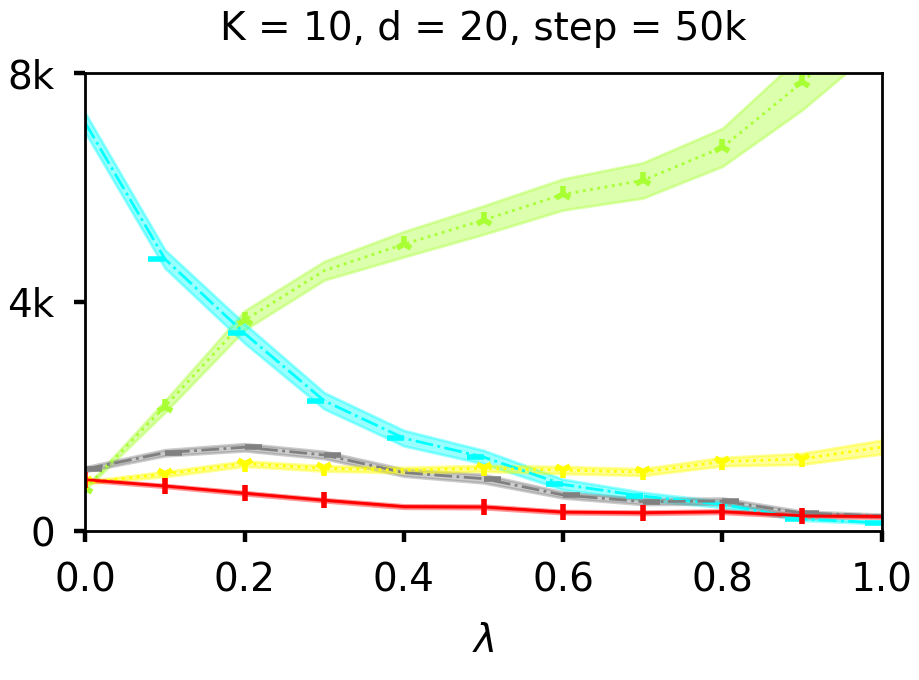}
	\caption{$n$-step regret on the Yahoo  dataset with different $\lambda$. Results are averaged over $500$ users with $2$ repeats per user.}  
	\label{recsys:fig:yahoo lambda}	
\end{figure*}

\begin{figure*}[t]
	\centering
	\includegraphics{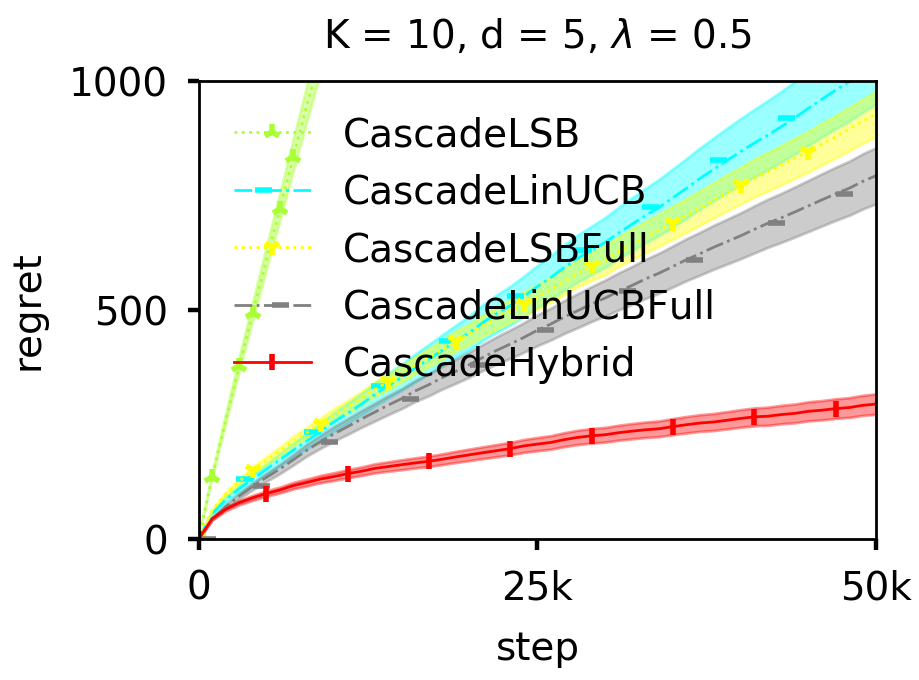}
	\includegraphics{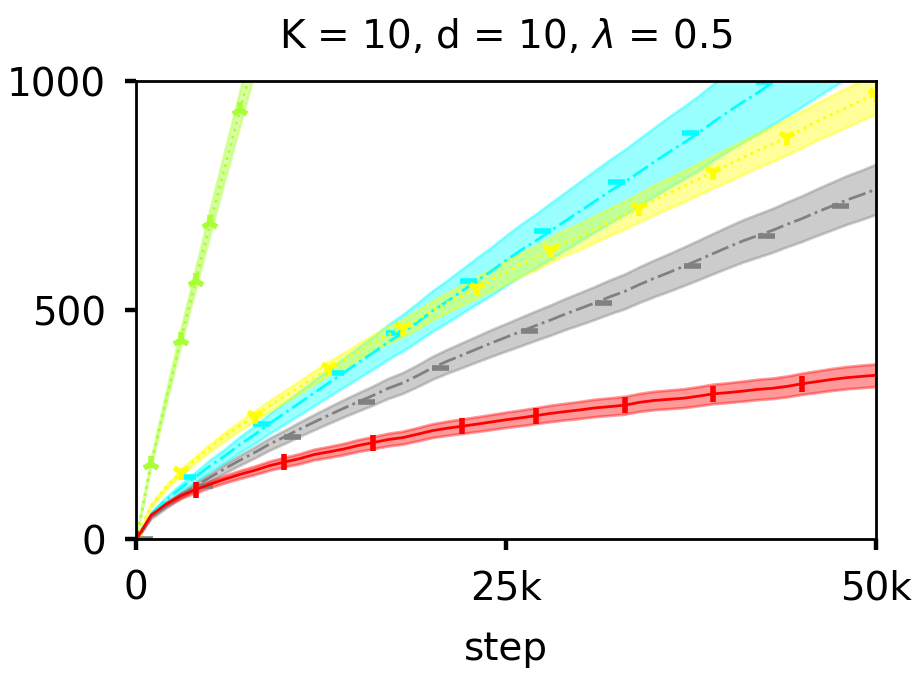}
	\includegraphics{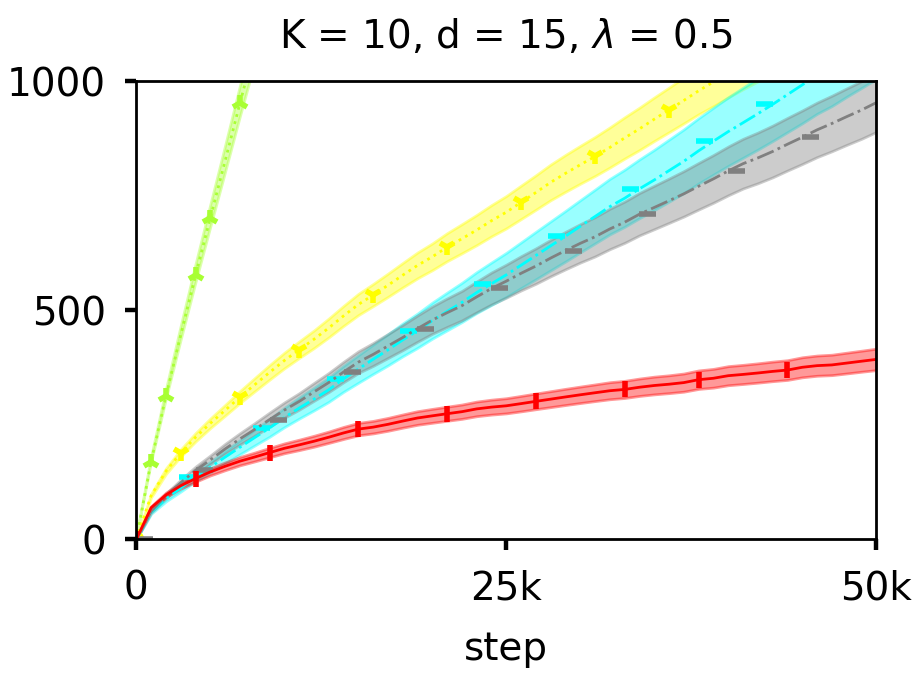}
	\includegraphics{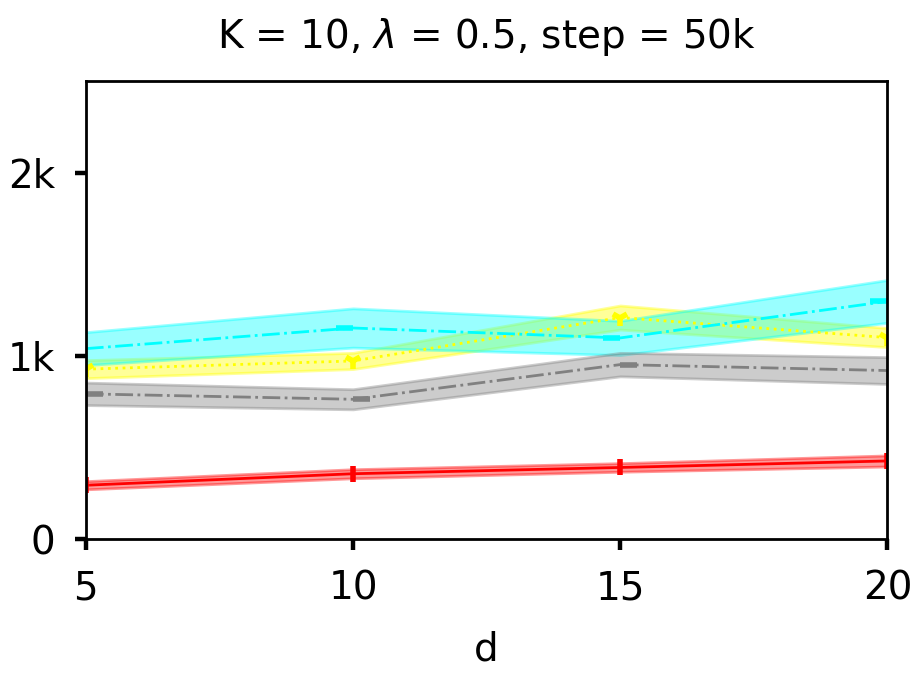}
	\caption{$n$-step regret on the Yahoo  dataset with different topics $d$.}  
	\label{recsys:fig:yahoo topic}	
\end{figure*}

\begin{figure*}[t]
	\centering
	\includegraphics{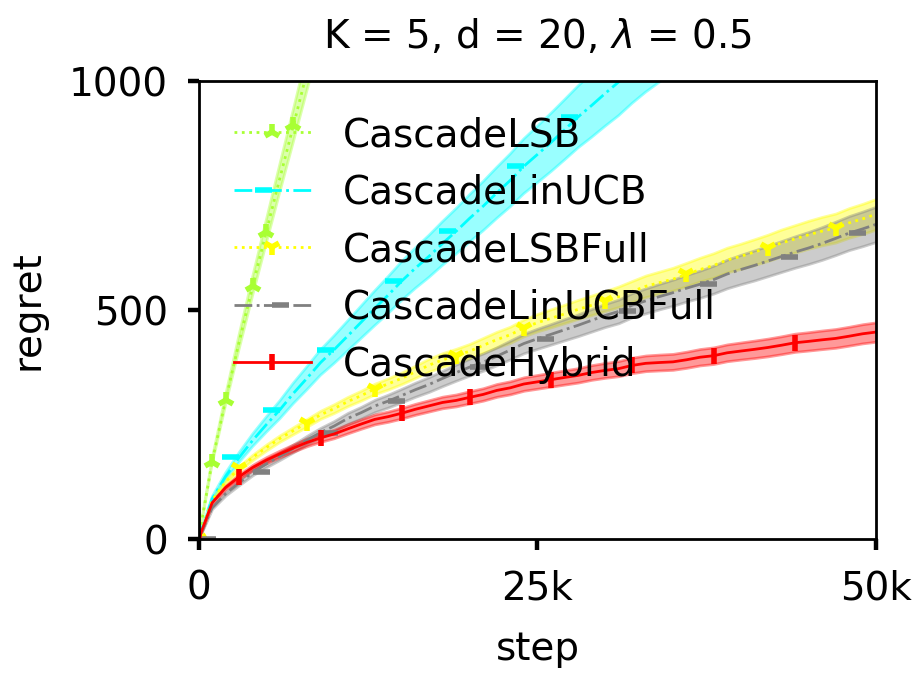}
	\includegraphics{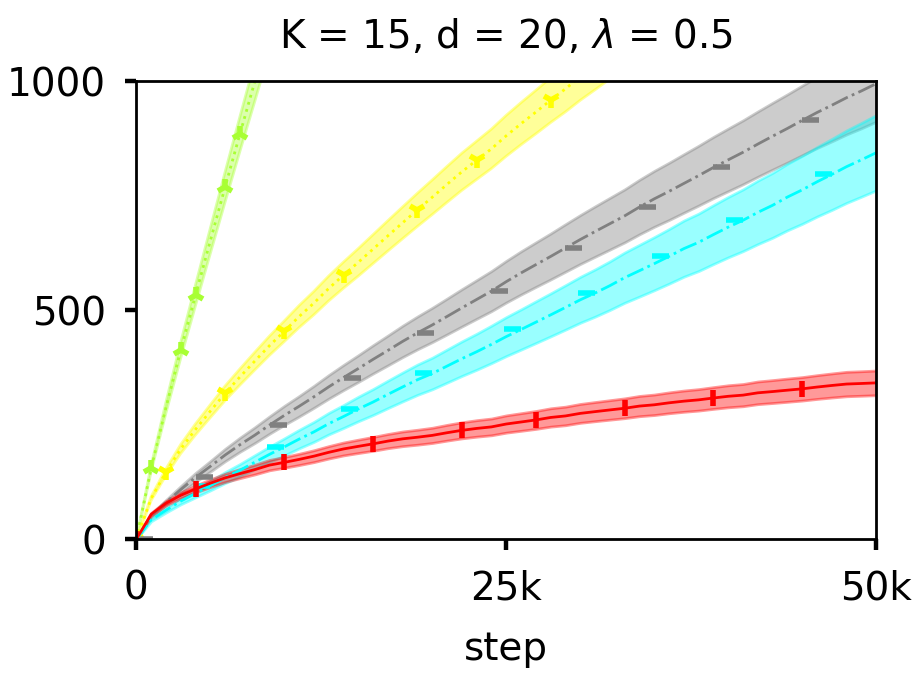}
	\includegraphics{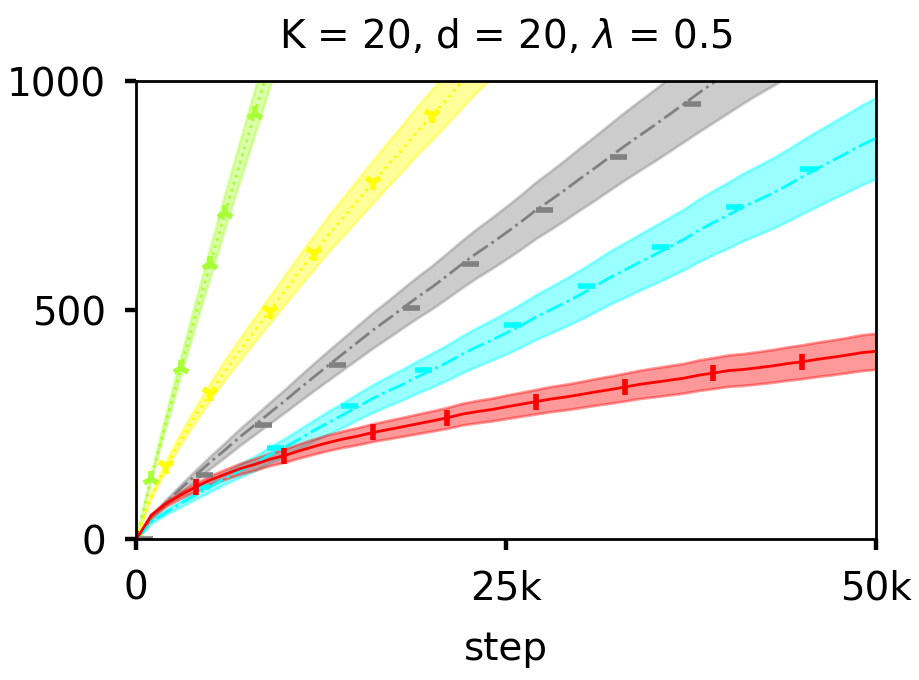}
	\includegraphics{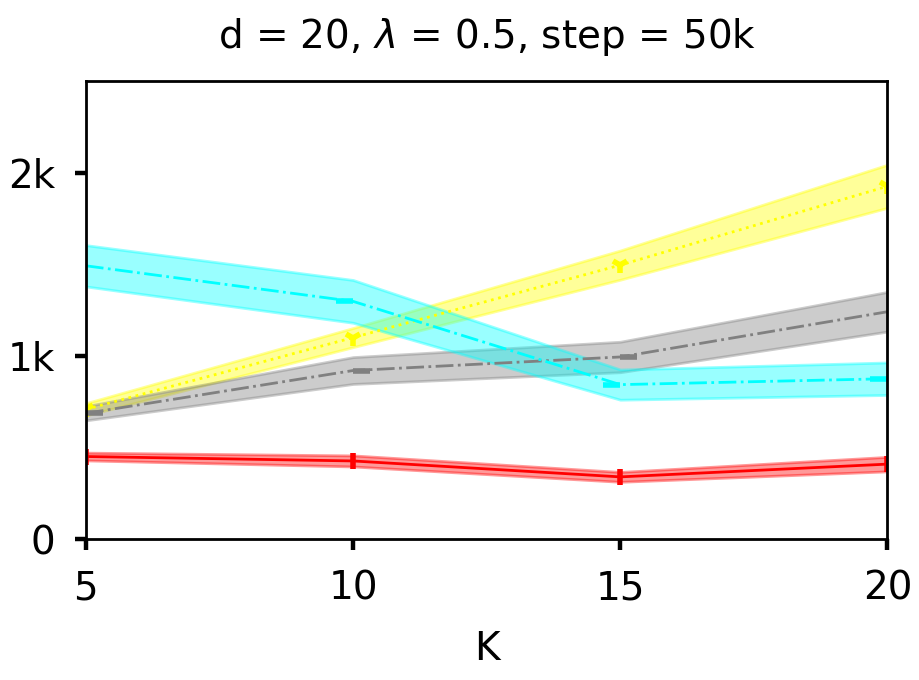}
	\caption{$n$-step regret on the Yahoo  dataset with different positions $K$.}  
	\label{recsys:fig:yahoo position}	
\end{figure*}

We first study the movie recommendation task on the MovieLens dataset and show the results in \cref{recsys:fig:movie lambda,recsys:fig:movie topic,recsys:fig:movie position}.
\cref{recsys:fig:movie lambda} shows the impact of $\lambda$, where we fix $K=10$ and $d=20$. 
$\lambda$ is a trade-off parameter  in our simulation. 
It balances the relevance and diversity, and is unknown to online algorithms. 
Choosing a small $\lambda$, the simulated user prefers recommended movies in the list to be relevant, while choosing a large $\lambda$, the simulated user prefers the recommended movies to be diverse. 
As shown in the top row, $\cascadehybrid$ outperforms all baselines when $\lambda \in \{0.1, 0.2, \ldots 0.8\}$, and only loses to the particularly designed baselines ($\cascadelsb$ and $\cascadelinucb$) with small gaps in some extreme cases where they benefit most. 
This is reasonable since they have fewer parameters to be estimated than $\cascadehybrid$. 
In all cases, $\cascadehybrid$ has lower regret than $\cascadelinucbfull$ and $\cascadelsbfull$ that work with the same features as $\cascadehybrid$. 
This result indicates that including more features in $\cascadelsb$ and $\cascadelinucb$ is not sufficient to capture both item relevance and result diversification.  

\cref{recsys:fig:movie topic} shows the impact of different numbers of topics, where we fix $\lambda=0.5$ and $K=10$. 
We see that $\cascadehybrid$ outperforms all baselines with large gaps. 
In the last plot of the middle row, we see that the gap of regret between $\cascadehybrid$ and $\cascadelinucbfull$ decreases with larger values of $d$. 
This is because, on the MovieLens dataset, when $d$ is small, a user tends to prefer a diverse ranked list: when $d$ is small, an item is more likely to  belong to only one topic, and each entry of $\vtheta_j^*$ becomes relatively larger since $\sum_{j\in[d]}\vtheta_j^* = 1$. 
And given an item $a$ and a set $\cS$, the difference between $\Delta(a | \cS)^T \vtheta^*$ and $\Delta(a | \emptyset)^T \vtheta^*$ is large. 
This behavior is also confirmed by the fact that $\cascadelinucbfull$ outperforms $\cascadelsbfull$ for large $d$ while they perform similarly for small $d$. 
Finally, we study the impact of the number of positions on the regret. 
The results are displayed in \cref{recsys:fig:movie position}, where we choose $\lambda=0.5$ and $d=20$. 
Again, we see that $\cascadehybrid$ outperforms baselines with large gaps. 

Next, we report the results on the Yahoo dataset in \cref{recsys:fig:yahoo lambda,recsys:fig:yahoo position,recsys:fig:yahoo topic}. 
We follow the same setup as for the MovieLens dataset and observe a similar behavior. 
$\cascadehybrid$ has slightly higher regret than the best performing baselines in three cases: $\cascadelsb$ when $\lambda=0$ and $\cascadelinucb$ when $\lambda\in\{0.9, 1\}$. 
Note that these are relatively extreme cases, where the particularly designed baselines can benefit most. 
Meanwhile, $\cascadelsb$ and $\cascadelinucb$ do not generalize well with different $\lambda$s. 
In all setups, $\cascadehybrid$ has lower regret than $\cascadelsbfull$ and $\cascadelinucbfull$, which confirms our hypothesis that the hybrid model has benefit in capturing both relevance and diversity.

\section{Analysis}
\label{recsys:sec:analysis}

\subsection{Performance guarantee}
Since $\cascadehybrid$ competes with the greedy benchmark, we focus on the $\eta$-scaled expected $n$-step regret which is defined as: 
\begin{equation}
\label{recsys:eq:scaled regret}
R_\eta(n) = \sum_{t=1}^{n}\expect{  \eta r(\cR^*, \valpha) - r(\cR_t, A_t) }, 
\end{equation} 
where $\eta = (1-\frac{1}{e}) \max\{\frac{1}{K}, 1-\frac{K-1}{2}\alpha_{max}\}$.
This is a reasonable metric, since computing the optimal $\cR^*$ is computationally inefficient. 
A similar scaled regret has previously been used in diversity problems~\citep{yue11linear,qin14contextual,hiranandani19cascading}. 
For simplicity, we write $\vw^* = [\vtheta^{*T}, \vbeta^{*T}]^T$.
Then, we bound the $\eta$-scaled regret of $\cascadehybrid$ as follows: 

\begin{thm}
\label{th: regret}
For  $\norm{\vw^*} \leq 1$ and any 
\begin{equation}
\label{recsys:eq:gamma}
\gamma \geq \sqrt{(m+d)\log\left(1+\frac{nK}{m+d}\right) + 2\log(n)} + \norm{\vw^*}, 
\end{equation}
we have 
\begin{equation}
	\label{recsys:eq:regret bound}
	R_{\eta}(n) \leq 
2\gamma \sqrt{{2nK(m+d)\log{\left(1+\frac{nK}{m+d}\right)}}}
	+ 1.	
\end{equation}
\end{thm}

\noindent%
Combining \cref{recsys:eq:gamma,recsys:eq:regret bound}, we have $R_\eta(n) = \tilde{O}((m+d)\sqrt{Kn})$, where the $\tilde{O}$ notation ignores logarithmic factors.   
Our bound has three characteristics: 
\begin{enumerate*}[label=(\roman*)]
	\item \cref{th: regret} states a gap-free bound, where the factor $\sqrt{n}$ is considered near optimal;
	\item  This bound is linear in the number of features, which is a common dependence in learning bandit algorithms~\citep{abbasiyadkori11improved}; and 
	\item Our bound is $\tilde{O}(\sqrt{K})$ lower than other bounds for linear bandit algorithms in \ac{CB}~\citep{zong16cascading,hiranandani19cascading}. 
\end{enumerate*}
We include a proof of \cref{th: regret} in \cref{recsys:sec:proof of theorem}. 
We use a similar strategy to decompose the regret as in~\citep{zong16cascading,hiranandani19cascading}, but we have a better analysis on how to sum up the regret of individual items. 
Thus, our bound depends on  $\tilde{O}(\sqrt{K})$ rather than $\tilde{O}(K)$. 
We believe that our analysis can be applied to both $\cascadelinucb$ and $\cascadelsb$, and then show that their regret is actually bounded by $\tilde{O}(\sqrt{K})$ rather than $\tilde{O}(K)$.

\subsection{Proof of~\cref{th: regret}} 
\label{recsys:sec:proof of theorem}
We first define some additional notation. 
We write $\vw^* = [\vtheta^{*T}, \vbeta^{*T}]^T$. 
Given a ranked list $\cR$ and $a = \cR(i)$, we write $\vphi_a  = [\vomega_{a}^T, \vz_{a}^T]^T$.
With the $\vphi_a$ and $\vw^*$ notation, $\cascadehybrid$ can be viewed as an extension of $\cascadelsb$, where two submodular functions instead of one are used in a single model. 
We write $\vOt = \vI_{m+d} + \sum_{i=1}^{t-1}\sum_{a\in\cO_i} \vphi_a \vphi_a^T$ as the collected features in $t$ steps,  $\cH_t$ as the collected features and clicks up to step $t$, and $\cR^i = (\cR(1), \ldots, \cR(i))$.  
Then, the confidence bound in \cref{recsys:eq:cb} on the $i$-th item in $\cR$ can be re-written as:  
\begin{equation}
\label{recsys:eq:new cb}
s(\cR^i) = \vphi_{\cR(i)}^T \vOt^{-1} \vphi_{\cR(i)}. 
\end{equation}
Let $\Pi(\cD) = \bigcup_{i=1}^L\Pi_i(\cD)$ be the set of all ranked lists of $\cD$ with length $[L]$, and $\kappa: \Pi(\cD) \rightarrow [0, 1]$ be an arbitrary list function. 
For any $\cR \in \Pi(\cD)$ and any $\kappa$, we  define 
\begin{equation}
\label{recsys:eq:new f}
f(\cR, \kappa) = 1 - \prod_{i=1}^{|\cR|} (1-\kappa(\cR^i)).
\end{equation}
We define upper and lower confidence bounds, and $\kappa$ as: 
\begin{equation}
\label{recsys:eq:ucb lcb}
\begin{split}
u_t(\cR) ={}& \text{F}_{[0, 1]}[\vphi_{\cR(l)}^T\hat{\vw}_t + s(\cR^l)]  \\
l_t(\cR) = {}&\text{F}_{[0, 1]}[\vphi_{\cR(l)}^T\hat{\vw}_t - s(\cR^l)] \\
\kappa(R) = {}&\vphi_{\cR(l)}^T\vw^*, 
\end{split}
\end{equation}
where $l = |\cR|$ and  $\text{F}_{[0,1]}[\cdot] =  \max(0, \min(1, \cdot))$. 
With the definitions in \cref{recsys:eq:ucb lcb}, $f(\cR, \kappa) = r(\cR, \valpha)$ is the reward of list $\cR$.

\begin{proof}
	Let $g_t = \{ l_t(\cR) \leq \kappa(\cR) \leq u_t(\cR), \forall \cR \in \Pi(\cD)\}$ be the event that the attraction probabilities are bounded  by the lower and upper confidence bound, and $\bar{g}_t$ be the complement of $g_t$. 
	We have 
	\begin{eqnarray}
	\label{recsys:eq:regret to f}
	\begin{split}
	\expect{\eta r(\cR^*, \valpha) - r(\cR_t, \vA_t)} & =	\expect{\eta f(\cR^*, \kappa) - f(\cR_t, \kappa)} \\
	&
	\stackrel{(a)}{\leq} P(g_t) \expect{\eta f(\cR^*, \kappa) - f(\cR_t, \kappa)} + P(\bar{g}_t) \\
	&
	\stackrel{(b)}{\leq} P(g_t) \expect{\eta f(\cR^*, u_t) - f(\cR_t, \kappa)} + P(\bar{g}_t)  
	\\
	&
	\stackrel{(c)}{\leq} P(g_t) \expect{ f(\cR_t, u_t) - f(\cR_t, \kappa)} + P(\bar{g}_t),  
	\end{split}
	\end{eqnarray}
	where $(a)$ holds because $\expect{\eta f(\cR^*, \kappa) - f(\cR_t, \kappa)} \leq 1$,  $(b)$ holds because under event $g_t$ we have $f(\cR, l_t) \leq f(\cR, \kappa) \leq f(\cR, u_t)$, $\forall \cR \in \Pi(\cD)$, and $(c)$ holds by the definition of the $\eta$-approximation, where we have  
	\begin{equation}
	\eta f(\cR^*, u_t) \leq  \max_{\cR \in \Pi_K(\cD)} \eta f(\cR, u_t) \leq f(\cR_t, u_t). 
	\end{equation}
	By the definition of the list function $f(\cdot, \cdot)$ in \cref{recsys:eq:new f}, we have 
		\begin{equation}
		\begin{split}
		&f(\cR_t, u_t) - f(\cR_t, \kappa)  \\
		&= \prod_{k=1}^{K} (1- \kappa(\cR_t^k)) -  \prod_{k=1}^{K} (1- u_t(\cR_t^k)) 
		\\
		&\stackrel{(a)}{=} \sum_{k=1}^{K} \!\!\left[ \prod_{i=1}^{k-1} (1- \kappa(\cR_t^i)) \right] \!\!(u_t(\cR_t^k) - \kappa(\cR_t^k))\!\!
		\left[ \prod_{j=k+1}^{K} (1- u(\cR_t^j)) \right]   
		\\&
		\stackrel{(b)}{\leq}  \sum_{k=1}^{K}\!\! \left[ \prod_{i=1}^{k-1} (1- \kappa(\cR_t^i)) \right] \!\!(u_t(\cR_t^k) - \kappa(\cR_t^k)),
		\end{split}
		\end{equation}%
	where $(a)$ follows from Lemma 1 in~\citep{zong16cascading} and $(b)$ is because of the fact that $0\leq \kappa(\cR_t) \leq u_t(\cR_t) \leq 1$. 
	We then define the event $h_{ti} = \{\text{item } \cR_t(i) \text{ is observed} \}$, where we have $\expect{\ind{h_{ti}}} = \prod_{k=1}^{i-1} (1- \kappa(\cR_t^k))$.
	For any $\cH_t$ such that $g_t$ holds, we have 
	\begin{equation}
	\label{recsys:eq:f to sqrt}
	\begin{split}
	&  \expect{f(\cR_t, u_t) - f(\cR_t, \kappa) \mid \cH_t} \\  
	& \leq \sum_{i=1}^K \expect{\ind{h_{ti}} \mid \cH_t }(u_t(\cR_t^i) - l_t(\cR_t^i))  \\
	&\stackrel{(a)}{\leq} 2\gamma \expect{\ind{h_{ti}} \sum_{i=1}^{K}\sqrt{s(\cR_t^i)}   \mid \cH_t}  \\
	&\stackrel{(b)}{\leq} 2\gamma \expect{\sum_{i=1}^{\min(K, c_t)}\sqrt{s(\cR_t^i)}   \mid \cH_t},  
	\end{split}	
	\end{equation}
	where inequality $(a)$ follows from the definition of $u_t$ and $l_t$ in \cref{recsys:eq:ucb lcb}, and inequality $(b)$ follows from the definition of $h_{ti}$.
	Now, together with \cref{recsys:eq:scaled regret,recsys:eq:regret to f,recsys:eq:f to sqrt}, we have 
	\begin{equation}
	\label{recsys:eq:proof cumulative regret}
	\begin{split}
	R_{\eta}(n)& = \sum_{t=1}^{n}\expect{\eta r(\cR^*, \valpha) - r(\cR_t, \vA_t)  }  \\
	&\leq  \sum_{t=1}^{n} \left[ 2\gamma  \expect{\sum_{i=1}^{\min(K, c_t)}\sqrt{s(\cR_t^i)}   \mid g_t} P(g_t) 
	+ P(\bar{g}_t) 
	\right] 
	\\ &
	\leq     2\gamma  \expect{\sum_{t=1}^{n}\sum_{i=1}^{K}\sqrt{s(\cR_t^i)} }
	+ \sum_{t=1}^{n}p(\bar{g}_t).  
	\end{split}
	\end{equation}
	For the first term in \cref{recsys:eq:proof cumulative regret}, we have
	\begin{equation}
	\begin{split}
	\label{recsys:eq:to det}
	&\sum_{t=1}^{n}\sum_{i=1}^{K}\sqrt{s(\cR_t^i)}
	\stackrel{(a)}{\leq} \sqrt{nK \sum_{t=1}^{n}\sum_{i=1}^{K}s(\cR_t^i)} 
	\stackrel{(b)}{\leq} 
	\sqrt{nK 2\log det(\vO_t) },  
	\end{split}	
	\end{equation}
	where inequality $(a)$ follows from the Cauchy-Schwarz inequality and $(b)$ follows from Lemma 5 in~\citep{yue11linear}. 
	Note that $\log det(\vO_t) \leq (m+d) \log(K(1+n/(m+d)))$, which can be obtained by the determinant and trace inequality, and together with \cref{recsys:eq:to det}:
	\begin{equation}
	\label{recsys:eq:to log}
	\mbox{}\hspace*{-2mm}
	\sum_{t=1}^{n}\sum_{i=1}^{K}\sqrt{s(\cR_t^i)} 
	\leq 
	\sqrt{2nK(m+d)\log(K(1+\frac{n}{m+d}))  }. 
	\hspace*{-2mm}\mbox{}
	\end{equation}
	For the second term in \cref{recsys:eq:proof cumulative regret}, by Lemma 3 in~\citep{hiranandani19cascading}, we have $P(\bar{g}_t) \leq 1/n$ for any $\gamma$ that satisfies \cref{recsys:eq:gamma}. 
	Thus, together with \cref{recsys:eq:proof cumulative regret,recsys:eq:to det,recsys:eq:to log}, we have 
	\begin{equation*}
	R_{\eta}(n) \leq 2\gamma\sqrt{2nK(m+d)\log(1+\frac{nK}{m+d}) }  +1. 
	\end{equation*}
	This concludes the proof of~\cref{th: regret}.
\end{proof}

\section{Related Work}
\label{recsys:sec:related work}

The literature on offline \acf{LTR} methods that account for position bias and diversity is too broad to review in detail. 
We refer readers to \citep{aggarwal16recommender} for an overview. 
In this section, we mainly review online \ac{LTR} papers that are closely related to our work, i.e., stochastic click bandit models.  

Online \ac{LTR} in a stochastic click models has been well-studied~\citep{radlinski08learning,slivkins10learning,yue11linear, kveton15cascading,katariya16dcm,zong16cascading,zoghi17online,li19online,li19cascading,li19bubblerank}. 
Previous work can be categorized into two groups: feature-free models and feature-rich models. 
Algorithms from the former group use a tabular representation on items and maintain an estimator for each item. 
They learn inefficiently and are limited to the problem with a small number of item candidates. 
In this chapter, we focus on the ranking problem with a large number of items. 
Thus, we do not consider feature-free model in the experiments. 

Feature-rich models learn efficiently in terms of the number of items.
They are suitable for large-scale ranking problems. 
Among them, ranked bandits~\citep{radlinski08learning,slivkins10learning} are early approaches to online \ac{LTR}. 
In ranked bandits, each position is model as a \ac{MAB} and diversity of results is addressed in the sense that items ranked at lower positions are less likely to be clicked than those at higher positions, which is different from the topical diversity as we study. 
Also, ranked bandits do not consider the position bias and are suboptimal in the problem where a user browse different possition unevenly, e.g., \ac{CM}~\citep{kveton15cascading}.
$\lsbgreedy$~\citep{yue11linear} and $\ccucb$~\citep{qin14contextual} use submodular functions to solve the online diverse \ac{LTR} problem. 
They assume that the user browses all displayed items and, thus, do not consider the position bias either.

Our work is closely related to $\cascadelinucb$~\citep{zong16cascading} and  $\cascadelsb$~\citep{hiranandani19cascading},  the baselines, and can be viewed as a combination of both. 
$\cascadelinucb$ solves the relevance ranking in the \ac{CM} and assumes the attraction probability is a linear combination of features. 
$\cascadelsb$ is designed for result diversification and assumes that the attraction probability is computed as a submodular function; see \cref{recsys:eq:topic coverage attration probability}. 
In our $\cascadehybrid$, the attraction probability is a hybrid of both; see \cref{recsys:eq:hybridy attraction probability}. 
Thus, $\cascadehybrid$ handles both relevance ranking and result diversification. 
$\recurrank$~\citep{li19online} is a recently proposed algorithm that aims at learning the optimal list in term of item relevance in most click models. 
However, to achieve this task, $\recurrank$ requires a lot of randomly shuffled lists and is outperformed by $\cascadelinucb$ in the \ac{CM}~\citep{li19online}. 

The hybrid of a linear function and a submodular function has been used in solving combinatorial semi-bandits.
\citet{perrault19exploiting} use a linear set function to model the expected reward of arm, and use the submodular function to compute the \emph{exploration bonus}. 
This is different from our hybrid model, where both the linear and submodular functions are used to model the attraction probability and the confident bound is used as the exploration bonus.

\section{Conclusion}
\label{recsys:sec:conclusion}

In real world interactive systems, both relevance of individual items and topical diversity of result lists are critical factors in user satisfaction. 
This chapter has provide one solution to this problem, i.e., \textbf{RQ4}. 
In order to better meet users' information needs, we propose a novel online \ac{LTR} algorithm that optimizes both factors in a hybrid fashion.  
We formulate the problem as \acf{CHB}, where the attraction probability is a hybrid function that combines a function of relevance features and a submodular function of topic features. 
$\cascadehybrid$ utilizes a hybrid model as a scoring function and the \ac{UCB} policy for exploration. 
We provide a gap-free bound on the $\eta$-scaled $n$-step regret of $\cascadehybrid$, and conduct experiments on two real-world datasets. 
Our empirical study shows that $\cascadehybrid$ outperforms two existing online \ac{LTR} algorithms that exclusively consider either relevance ranking or result diversification.

In future work, we intend to conduct experiments on live systems, where feedback is obtained from multiple users so as to test whether $\cascadehybrid$ can learn across users. 
Another direction is to adapt \ac{TS}~\citep{thompson33likelihood} to our hybrid model, since \ac{TS} generally outperforms \ac{UCB}-based algorithms~\citep{zong16cascading,li20mergedts}.

\bookmarksetup{startatroot} 
\addtocontents{toc}{\bigskip} 


\chapter{Conclusions}
\label{chapter:conclusions}

We set up this thesis to study the online optimization of ranking systems as bandit problems.
We consider two typical  ranking tasks:  online ranker evaluation and online learning to rank.
A typical challenge in these tasks is that the feedback is implicit and partially-observed. 
Specifically, in the online evaluation task, the feedback is the relative comparison of two rankers; and in \ac{OLTR}, the feedback is the click but due to the position bias, the examination of each item is unobserved. 
In this thesis, we have considered four research directions in the online ranker optimization and formulated each of them as a bandit problem and proposed the corresponding algorithm. 
In the rest of this chapter, we first summarize the main results of this thesis and then outline some potential directions for future research. 

\section{Results}
We first revisit the research questions that are asked in \cref{introduction:rqs}. 
\begin{description}
	\item[\textbf{RQ1}]  How to conduct effective large-scale online ranker evaluation? 
\end{description}
The question has been answered in \cref{chapter:tois}, where we introduce the \acf{TS} based algorithm, called MergeDTS. 
Although in theory, the regret of MergeDTS has the same order as some baselines, i.e., MergeRUCB and Self-Sparring, MergeDTS outperforms those baselines with large gaps in our large-scale online ranker evaluation experiments, as shown in \cref{tois:sec:experimental results}. 
Specifically, we have conducted experiments on three large-scale datasets, each of which has been combined with $3$ different types of click behavior. 
MergeDTS has lower regret than the baselines in $7$ out of $9$ configurations, and only has slightly higher regret than Self-Sparring in $2$ configurations. 
Meanwhile, the time efficiency of MergeDTS is better than that of Self-Sparring.

To answer \textbf{RQ1}, we conduct an effective large-scale online ranker evaluation by combining Thompson sampling and the divide-and-conquer idea in MergeRUCB. 
\begin{description}
	\item[\textbf{RQ2}] How to achieve safe online learning to re-rank? 
\end{description}
\cref{chapter:uai19} has provided one answer to this question, where the $\bubblerank$ algorithm is introduced. 
$\bubblerank$ is inspired by the bubble sort algorithm and the pairwise comparison in dueling bandits. 
$\bubblerank$ starts from the ranked list produced by a production ranker, and conducts pairwise comparisons between an item with its neighbors.
To get rid of the position bias, the position of an item is randomly exchanged with its neighbors, during the comparisons. 
Thus, during the exploration, an item will not displayed too far from its original position. 
On the other hand, if one item is considered better than its upper neighbor with a high confidence, the positions of two items are exchanged permanently. 
The safety feature of $\bubblerank$ has been theoretically proved in \cref{uai:lem:safety}. 
The $n$-step regret of $\bubblerank$ is $O(K^3\log(n))$, where $K$ is the number of ranking positions and $n$ is the time horizon.
This regret bound is only $O(K)$ higher than the regret of $\toprank$, the state-of-the-art but unsafe baseline. 
We have conducted an empirical evaluation on a large-scale click dataset, the \emph{Yandex} click dataset, and the empirical results have confirmed the theoretical findings. 
\begin{description}
	\item[\textbf{RQ3}] How to conduct online learning to rank when users change their preference constantly? 
\end{description}
Our answer to this question has been provided in \cref{chapter:ijcai19}. 
Firstly, the considered non-stationary \ac{OLTR} problem has been formulated as a cascading non-stationary bandit problem, where we assume that users follow the cascading browsing behavior. 
Then, two algorithms, $\cascadeducb$ and $\cascadeswucb$, have been proposed to solve this problem.
To deal with the non-stationarity, the former algorithm uses the discount factor, and the latter uses a sliding window. 
The $n$-step regret of $\cascadeducb$ and $\cascadeswucb$  is $O(\sqrt{n}\ln(n))$ and $O(\sqrt{n\ln(n)})$, respectively, where $n$ is the time horizon.  
They both match the lower bound up to logarithmic factors. 
Finally, we have evaluated the proposed algorithms in a semi-synthetic experiment on the \emph{Yandex} click dataset. 
The experimental results are consistent with our theoretical findings. 
\begin{description}
	\item[\textbf{RQ4}] How to learn a ranker online considering both relevance and diversity? 
\end{description}
This question is answered in \cref{chapter:recsys20}. 
As a follow up to \cref{chapter:ijcai19}, the cascade click model has also been used to interpret the click feedback. 
We have first formulated the \ac{OLTR} problem for relevance and diversity as a cascade hybrid bandit problem, and then introduced the $\cascadehybrid$ algorithm to solve the proposed problem. 
The term \emph{hybrid} comes from the fact that we use a linear function for relevance and a submodular function for diversity. 
The $n$-step regret of the proposed $\cascadehybrid$ is with the form of $\tilde{O}(\sqrt{Kn})$, where $\tilde{O}(\cdot)$ ignores logarithmic terms, $K$ is the number of positions, and $n$ is the time horizon. 
This bound is comparable to other linear bandit approaches for \ac{OLTR}. 
In our experiments, we have evaluated $\cascadehybrid$ on two tasks: movie recommendation on the MovieLens dataset and music recommendation on the \emph{Yahoo} music dataset. 
The experimental results indicate that with the underlying hybrid model, $\cascadehybrid$ learns a ranker that considers both  item relevance and result diversification.

\section{Future Work}

As with any research, this thesis has some limitations, and we believe that more questions have been generated than answered by our results.
in addition to the suggestions for future work provided at the end of each of the research chapters, we believe there are two important future directions.

\paragraph{Beyond the linear model} 
The proposed algorithms in this thesis are based on linear or tabular models. 
Although they learn efficiently, the relatively low learning capacity limits the application scenarios of these algorithms. 
To model more complicated problems, it is appealing to consider non-linear models, e.g., tree-based models and deep models, which have a more powerful learning capacity. 
However, training non-linear models online with implicit feedback is not without difficulty. 
One of the challenges is exploration with non-linear models. 

Let us step back and take a look at the exploration strategies in this thesis, i.e., \ac{UCB} and \ac{TS}. 
They both require closed-form solutions of the underlying models to estimate the reward of each arm.
This requirement can easily be satisfied when linear or tabular models are used. 
However, there are generally no closed-form solutions for  non-linear models. 
To use  \ac{UCB} and \ac{TS} for non-linear models,  additional approximation methods, e.g., the Laplacian approximation, are required~\citep{riquelme18deep,zhou19neural}, which can be computationally expensive. 
On the other hand, $\epsilon$-greedy is one of the basic exploration strategies, and can be used for non-linear models. 
Although $\epsilon$-greedy is far from optimal in \acl{MAB} (the tabular case) and linear bandits (the linear case), for some non-linear bandit setups, $\epsilon$-greedy is shown to be rather competitive~\citep{riquelme18deep,kveton20randomized}. 

A more sophisticated policy is bootstrapped sampling or \emph{bootstrapping}~\citep{riquelme18deep,kveton19garbage,kveton20randomized,kveton20differentiable}. 
Bootstrapping is a statistical technique that estimates the distribution of random variables by conducting sampling with replacement. 
Thus, it can be used to estimate the variance of the predictions of non-linear models. 
However, there is no conclusive study to indicate which policy is the optimal one for the ranking task.  
We believe that it is an interesting direction to study how to conduct exploration for non-linear \ac{OLTR}.

\paragraph{Privacy}

This direction comes from the recently regulations on data protection, e.g., the \ac{GDPR}\footnote{\url{https://gdpr-info.eu/}} by European Union and consumer data protection in the White House report~\citep{house12consumer}. 
The algorithms we have proposed in this thesis depend on the collection of user interaction data, which inevitably brings privacy risks. 
One way to provide protection w.r.t.\ user privacy is \acfi{FL}~\citep{konevcny16federated,bonawitz19towards,li21federated}, where the learning happens on the local devices under the coordination of a central server, called \emph{federator}. 

Importantly, \acl{FL} is easily combined with randomization techniques from differential privacy~\citep{dwork14algorithmic}, and provides theoretically sound approaches to protect user privacy. 
In \acl{FL}, the private user data is not collected by any central server but the learned model gradients are.
Briefly, the federator maintains a global model that is trained by the participating clients, e.g., local mobile devices. 
The training goes in rounds. 
At the start of each round, the federator broadcasts the global model to local clients.
Each client trains the model based on the local dataset, and then sends back the local gradient or the updated model to the federator. 
The federator updates the global model according to all local updates. 
This process continues until a certain termination criterion has been met, e.g., the loss stops dropping.

Federated learning also works in an online fashion. 
That is, after each round, the new global model is updated based on user requests. 
However, in \ac{FL}, the update at each round consists of several local updates,  while in online learning, there is only one update per round. 
To combine the two learning paradigms, we have to deal with batched feedback or updates, which again brings new challenges to exploration. 
As federated learning to rank is a rather new but important topic, we expect to see more algorithms published in the future.

\backmatter



\renewcommand{\bibsection}{\chapter{Bibliography}}
\renewcommand{\bibname}{Bibliography}
\markboth{Bibliography}{Bibliography}
\renewcommand{\bibfont}{\footnotesize}
\setlength{\bibsep}{0pt}

\bibliographystyle{abbrvnat}
\bibliography{thesis}



\chapter{Summary}

People use interactive systems, such as search engines, as the main tool to obtain information. 
To satisfy the information needs, such systems usually provide a list of items that are selected out of a large candidate set and then sorted in the decreasing order of their usefulness. 
The result lists are generated by a ranking algorithm, called ranker, which takes the request of user and candidate items as the input and decides the order of candidate items. 
The quality of these systems depends on the underlying rankers. 

There are two main approaches to optimize  the ranker in an interactive system: using data annotated by humans or using the interactive user feedback. 
The first approach has been widely studied in history, also called offline learning to rank, and is the industry standard. 
However, the annotated data  may not well represent  information needs of users and are not timely. 
Thus, the first approaches may lead to suboptimal rankers. 
The second approach optimizes rankers by using interactive feedback. 
This thesis considers the second approach, learning from the interactive feedback. 
The reasons are two-fold: 
\begin{enumerate*}[label=(\arabic*)]
	\item Everyday, millions of users interact with the interactive systems and generate a huge number of interactions, from which we can extract the information needs of users. 
	\item Learning from the interactive data have more potentials to assist in designing the online algorithms. 
\end{enumerate*}

Specifically, this thesis considers the task of learning from the user click feedback. 
The main contribution of this thesis is proposing a safe online learning to re-rank algorithm, named $\bubblerank$, which addresses one main disadvantage of online learning, i.e., the safety issue, by combining the advantages of both offline and online learning to rank algorithms. 
The thesis also proposes three other online algorithms, each of which solves unique online ranker optimization problems. 
All the proposed algorithms are theoretically sound and empirically effective.

\chapter{Samenvatting}

Mensen gebruiken interactiesystemen, zoals een zoekmachine, als hun belangrijkste hulpmiddel om informatie te verkrijgen. Om aan de informatiebehoefte te voldoen, bieden dergelijke zoekmachines meestal een lijst met  items aan. Deze items zijn geselecteerd uit een grote set van kandidaten en zijn vervolgens gesorteerd in afnemende volgorde van relevantie. Deze resultatenlijst wordt gegenereerd door een ranking-algoritme, \emph{ranker} genaamd, dat de informatiebehoefte van de gebruiker en de kandidaat-items als input neemt en vervolgens de volgorde van de kandidaat-items bepaalt. De kwaliteit van deze systemen is afhankelijk van de onderliggende rankers.

Er zijn twee belangrijke benaderingen om de rankers in een interactiesysteem te optimaliseren: De eerste manier is door gebruik te maken van de gegevens die door mensen zijn geannoteerd. De tweede is door gebruik te maken van de interactie-feedback van gebruikers. De eerste benadering is uitgebreid bestudeerd, ook wel bekend als offline-learning-to-rank, en is de standaard binnen de industrie. Echter, geannoteerde gegevens geven mogelijk niet de juiste informatiebehoeften van de gebruikers weer en zijn niet actueel. De eerste benadering kan dus leiden tot sub-optimale rankers.
De tweede benadering optimaliseert rankers door gebruik te maken van interactie-feedback. Dit proefschrift behandelt de tweede benadering. Namelijk, het leren van feedback van gebruikers-interactie.
De redenen zijn tweeledig:
\begin{enumerate*}[label=(\arabic*)]
	\item Elke dag hebben miljoenen gebruikers interactie met interactie-systemen waarmee zij een enorm aantal interacties genereren. Uit deze interactie-data van de gebruikers kan de informatiebehoeften van de gebruikers worden afgeleid.
	
	\item Het leren van de interactie-gegevens biedt meer mogelijkheden bij het ontwerpen van de online-algoritmen.
\end{enumerate*}

Dit proefschrift gaat specifiek in op de taak om te leren van feedback van gebruikers op basis van clicks. De belangrijkste bijdrage van dit proefschrift is het introduceren van een algoritme voor veilig online learning to re-rank, genaamd $\bubblerank$, dat een belangrijk nadeel van online leren aanpakt, namelijk het veiligheidsprobleem. Het doet dit door de voordelen van zowel offline als online learning to rank te combineren. Het proefschrift stelt ook drie andere online algoritmen voor, die elk een uniek optimalisatieprobleem voor online rankers oplossen.
Alle voorgestelde algoritmen zijn theoretisch correct en empirisch effectief.

\end{document}